\PassOptionsToPackage{dvipsnames}{xcolor}
\documentclass[acmtog,nonacm]{acmart}

\usepackage[dvipsnames]{xcolor}
\usepackage{overpic}

\usepackage{algpseudocode}
\usepackage{float}
\usepackage{xspace}
\usepackage{booktabs} %
\usepackage{soul}
\usepackage{amsmath}
\usepackage{tcolorbox}

\tcbuselibrary{breakable, listings}

\authorsaddresses{}
\renewcommand\keywordsname{Keywords}

\usepackage{booktabs} %
\usepackage{units} %

\usepackage[linesnumbered,ruled,vlined]{algorithm2e}

\SetAlFnt{\normalsize}
\SetAlCapFnt{\normalsize}
\SetAlCapNameFnt{\normalsize}
\SetAlCapHSkip{0pt}
\usepackage{hhline}
\usepackage{bbold} %
\usepackage{amsfonts} 
\usepackage{amsmath} 
\usepackage{pgfplots}
\usepackage{pgf,tikz}
\usepackage{tkz-euclide}
\usepackage{bm}
\usepackage{wrapfig}
\usepackage{multirow}
\usepackage{mathtools}
\usepackage{enumerate}
\usepackage{enumitem}
\usepackage{verbatim}
\usepackage{xcolor} 

\let\tempProof\proof
\let\tempEndProof\endproof
\let\proof\relax
\let\endproof\relax
\usepackage{amsthm}  
\let\proof\tempProof
\let\endproof\tempEndProof

\usepackage{dsfont}

\newcommand{\z}{\matr{z}}

\usepackage{subcaption}

\newcommand{\matr}[1]{\mathit{\mathbf{#1}}}%

\usepackage{soul}

\usepackage{listings}
\usepackage{color, colortbl} %
\definecolor{mygreen}{RGB}{28,172,0} %
\definecolor{mylilas}{RGB}{170,55,241}
\definecolor{ffzzcc}{rgb}{1,0.6,0.8}
\definecolor{mytbcol}{RGB}{175,227,246}
\definecolor{mypink}{RGB}{254, 197, 187}
\definecolor{tabyellow}{HTML}{7AAAF2} %

\usepackage{xurl}

\theoremstyle{definition}

\usepackage{lipsum}

\DeclareFontFamily{U}{mathx}{\hyphenchar\font45}
\DeclareFontShape{U}{mathx}{m}{n}{<-> mathx10}{}
\DeclareSymbolFont{mathx}{U}{mathx}{m}{n}

\usepackage{tipa}
\UndeclareTextCommand{\!}{T3}
\DeclareTextCommand{\tipaEXCLAM}{T3}{}
\DeclareRobustCommand{\!}{%
  \ifmmode\mskip-\thinmuskip\else\expandafter\tipaEXCLAM\fi
}

\usepackage{booktabs}       %
\usepackage{mathrsfs}
\usetikzlibrary{arrows}
\usetikzlibrary{matrix}
\usetikzlibrary{positioning,calc,fadings}
\usetikzlibrary{backgrounds, fit}
\pgfplotsset{compat=1.14}
\usepackage{pgfplotstable}

\definecolor{myblue}{RGB}{23, 195, 178}
\definecolor{myyellow}{RGB}{255, 186, 8}
\definecolor{mygray}{rgb}{0.5,0.5,0.5}
\definecolor{myred}{RGB}{254, 109, 115}

\definecolor{colblue}{HTML}{7fc8f8}
\definecolor{colyellow}{HTML}{ffe45e}

\definecolor{colvu}{HTML}{99d98c}
\definecolor{colvp}{HTML}{7fc8f8}
\definecolor{coleu}{HTML}{ff8fab}
\definecolor{colep}{HTML}{ffd000} %
\definecolor{coled}{HTML}{99d98c}

\newif\ifdraft

\drafttrue

\ifdraft

\newcommand{\todo}[1]{}

\fi

\newcommand{\secref}[1]{Sec.~\ref{#1}}

\newcommand{\figref}[1]{Fig.~\ref{#1}}

\newcommand{\tabref}[1]{Table~\ref{#1}}

\usepackage{pifont}%
\newcommand{\cmark}{\ding{52}}%
\newcommand{\xmark}{{\ding{56}}}%
\usepackage{accsupp}

\usepackage{graphicx}
\usepackage{arydshln}

\newcommand{\imgwithlabeltiny}[3]{%
\begin{tikzpicture}
  \node[inner sep=0pt, anchor=south west] (im) at (0,0)
    {\includegraphics[#1]{#2}};
  \node[anchor=south,
        font=\small\bfseries,
        text=black,
        text opacity=1,
        rounded corners=1pt,
        inner xsep=1.5pt, inner ysep=0.8pt,
        yshift=1pt] at (im.north) {#3};
\end{tikzpicture}%
}

\newcommand{\AddAutoInsetExact}[6]{%
  \begin{tikzpicture}[baseline]
    \node[anchor=south west, inner sep=0] (main) at (0,0)
      {\includegraphics[width=0.9\linewidth]{#1}};

    \begin{scope}[x={(main.south east)}, y={(main.north west)}]
      \pgfmathsetmacro{\half}{#4/2}
      \coordinate (roi_center) at (#2,#3);
      \coordinate (roi_sw) at ($(roi_center)+(-\half,-\half)$);
      \coordinate (roi_ne) at ($(roi_center)+(\half,\half)$);
      \draw[red, thick] (roi_sw) rectangle (roi_ne);
    \end{scope}

    \coordinate (inset_anchor) at ($(main.south) - (0,#5)$);

    \node[anchor=north, inner sep=2pt, draw=black, line width=0.8pt] (inset)
      at (inset_anchor)
    {%
      \begin{tikzpicture}[x=#6, y=#6]
        \clip (-0.5,-0.5) rectangle (0.5,0.5);
    
        \pgfmathsetmacro{\z}{1/#4}
        
        \pgfmathsetlengthmacro{\ScaledW}{\z*#6}
    
        \pgfmathsetmacro{\ax}{-\z*#2}
        \pgfmathsetmacro{\ay}{-\z*#3}
    
        \node[anchor=south west, inner sep=0] at (\ax,\ay)
          {\includegraphics[width=\ScaledW]{#1}};
      \end{tikzpicture}%
    };

    \draw[black, thick] (roi_sw) -- (inset.north west);
    \draw[black, thick] (roi_ne) -- (inset.north east);
  \end{tikzpicture}%
}

\newcommand{\CornerAutoInsetExact}[6]{%
  \begin{tikzpicture}[baseline]
    \node[anchor=north east, inner sep=0] (main) at (0,0)
      {\includegraphics[width=\linewidth]{#1}};
    \begin{scope}[
      shift={(main.south west)},
      x={($(main.south east)-(main.south west)$)},
      y={($(main.north west)-(main.south west)$)}
    ]
      \pgfmathsetmacro{\half}{#4/2}
      \coordinate (roi_center) at (#2,#3);
      \coordinate (roi_sw) at ($(roi_center)+(-\half,-\half)$);
      \coordinate (roi_ne) at ($(roi_center)+(\half,\half)$);
      \coordinate (roi_nw) at ($(roi_center)+(-\half,\half)$);
      \coordinate (roi_se) at ($(roi_center)+(\half,-\half)$);
      \draw[red, thick] (roi_sw) rectangle (roi_ne);
    \end{scope}
    \coordinate (inset_anchor) at ($(main.south west) + (-#6+#5, -#6+#5)$);
    \node[anchor=south west, inner sep=0pt, draw=black, line width=0.5pt] (inset)
      at (inset_anchor)
    {%
      \begin{tikzpicture}[x=#6, y=#6]
        \clip (-0.5,-0.5) rectangle (0.5,0.5);
        \pgfmathsetmacro{\z}{1/#4}
        \pgfmathsetlengthmacro{\ScaledW}{\z*#6}
        \pgfmathsetmacro{\ax}{-\z*#2}
        \pgfmathsetmacro{\ay}{-\z*#3}
        \node[anchor=south west, inner sep=0] at (\ax,\ay)
          {\includegraphics[width=\ScaledW]{#1}};
      \end{tikzpicture}%
    };
    \draw[black, thin] (roi_nw) -- (inset.north west);
    \draw[black, thin] (roi_se) -- (inset.south east);
  \end{tikzpicture}%
}

\newcommand{\TeaserCellW}{0.29\linewidth}
\newcommand{\TeaserCellGap}{\hspace{0.05\linewidth}}

\newcommand{\TeaserCell}[2]{%
  \begin{minipage}[t]{\TeaserCellW}
    \centering
    {\footnotesize #1}\par\nointerlineskip
    \CornerAutoInsetExact{#2}{0.7}{0.3}{0.35}{0.4\linewidth}{0.50\linewidth}
  \end{minipage}%
}

\newcolumntype{C}[1]{>{\centering\arraybackslash}m{#1}}

\DeclareRobustCommand{\ours}{\textsc{QDF}\xspace}

\newcommand{\methodlink}[1]{\hyperref[tab:methods]{#1}\xspace}
\newcommand{\FEG}{\methodlink{FEG}}
\newcommand{\VTP}{\methodlink{VTP}}
\newcommand{\DGGVTP}{\methodlink{DGG-VTP}}
\newcommand{\DGG}{\methodlink{DGG}}
\newcommand{\GSP}{\methodlink{GSP}}
\newcommand{\FMM}{\methodlink{FMM}}
\newcommand{\FIM}{\methodlink{FIM}}
\newcommand{\HM}{\methodlink{HM}}
\newcommand{\DFA}{\methodlink{DFA}}
\newcommand{\QDF}{\methodlink{\textsc{QDF}}}

\providecommand{\tabref}{}\renewcommand{\tabref}[1]{\hyperref[#1]{Table~\ref*{#1}}}
\providecommand{\figref}{}\renewcommand{\figref}[1]{\hyperref[#1]{Fig.~\ref*{#1}}}
\providecommand{\secref}{}\renewcommand{\secref}[1]{\hyperref[#1]{Sec.~\ref*{#1}}}

\newcommand{\imglabel}[2]{%
\begin{tikzpicture}
    \node[inner sep=0pt, outer sep=0pt] (img) {\includegraphics[width=0.9\linewidth]{#1}};
    \node[anchor=south east, fill opacity=0.7, text opacity=1, inner sep=1pt, xshift=-1pt, yshift=-8pt] 
        at (img.south east) {\small #2};
\end{tikzpicture}%
}

\usepackage[export]{adjustbox}

\makeatletter
\def\num#1{\expandafter\parseE#1e\relax}

\def\parseE#1e#2\relax{%
  \if\relax\detokenize{#2}\relax
    $#1$%
  \else
    \parseScientific#1e#2\relax
  \fi
}

\def\parseScientific#1e#2e\relax{%
  $#1 \times 10^{\the\numexpr#2\relax}$%
}
\makeatother

\begin{document}
\title{Convex Quadratic Distance Field Computation}

\author{Yue Ruan}
\affiliation{%
	\institution{University of Edinburgh}
	\country{United Kingdom}
}

\author{Albert Chern}
\affiliation{%
	\institution{University of California San Diego}
	\country{United States}
}

\author{Tzu-Mao Li}
\affiliation{%
	\institution{University of California San Diego}
	\country{United States}
}

\author{Kartic Subr}
\affiliation{%
	\institution{University of Edinburgh}
	\country{United Kingdom}
}

\author{Amir Vaxman}
\affiliation{%
	\institution{University of Edinburgh}
	\country{United Kingdom}
}
\begin{abstract}

Methods for computing distances from sources on discrete meshes commonly either compute geodesics directly on polyhedral surfaces or approximate the distance in a finite-element framework. 
Exact window-based polyhedral methods are highly accurate on clean manifold surfaces, but their unfolding construction does not extend to tetrahedral volumes and relies on manifold connectivity. Because their results are tied to the input polyhedron, geometric noise directly affects the computed distance field. 
Finite-element methods extend naturally to triangle surfaces and tetrahedral volumes, but state-of-the-art methods represent the distance field as a piecewise-linear (PL) function, limiting accuracy on coarse or poorly shaped meshes.

We argue that the PL representation itself, rather than the algorithm built on top of it, limits the result. Geodesic distance exhibits cone-like behaviour at its source and is not piecewise linear even on flat domains. In contrast, the squared distance is exactly quadratic in such domains.

Therefore, our Quadratic Distance Field method represents the \emph{squared} distance using piecewise-quadratic (PQ) elements, reproducing flat squared distances exactly. We show that simply increasing the element order does not improve existing algorithms, and develop a convex formulation for the squared distance over PQ elements. The same formulation applies to both triangle surfaces and tetrahedral volumes, supports anisotropic metrics and nonmanifold connectivity, and remains robust under noise.

Finally, we present an efficient solver based on the alternating direction method of multipliers and demonstrate its robustness and accuracy on a benchmark of thousands of real-world models.

\end{abstract}


\begin{teaserfigure}
    \centering
    \begin{tabular}{@{}c@{\hspace{6pt}}c@{\hspace{6pt}}c@{}}
        \begin{minipage}[b]{0.455\textwidth}
            \centering
            \TeaserCell{Mesh}{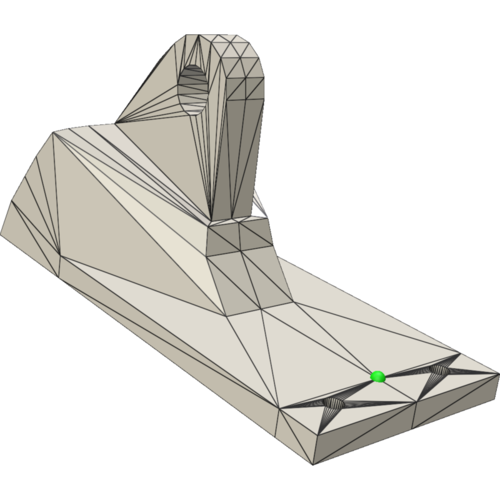}\TeaserCellGap
            \TeaserCell{Ground truth}{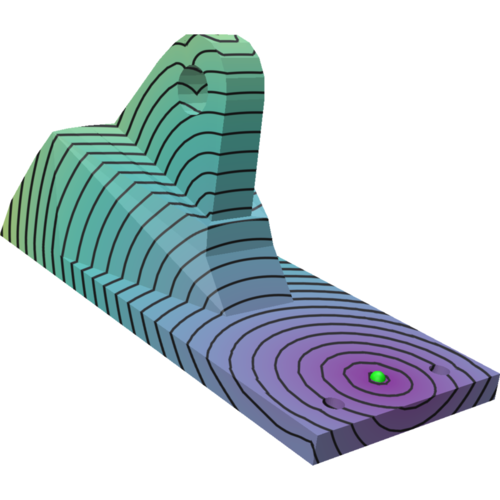}\TeaserCellGap
            \TeaserCell{\HM}{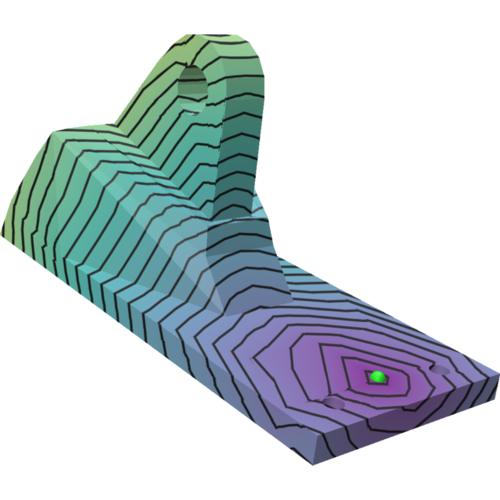}\TeaserCellGap
           \\[0.8ex]
            \TeaserCell{\FMM}{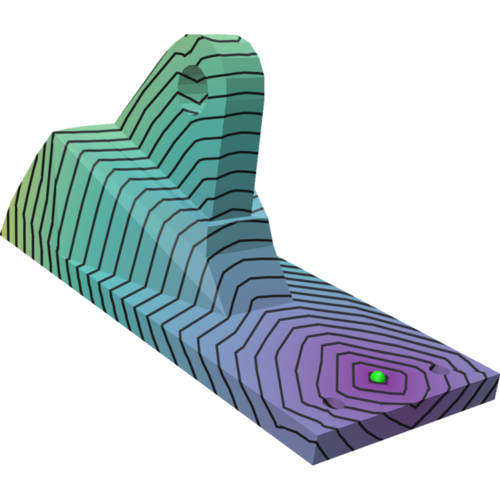}\TeaserCellGap
            \TeaserCell{\DFA}{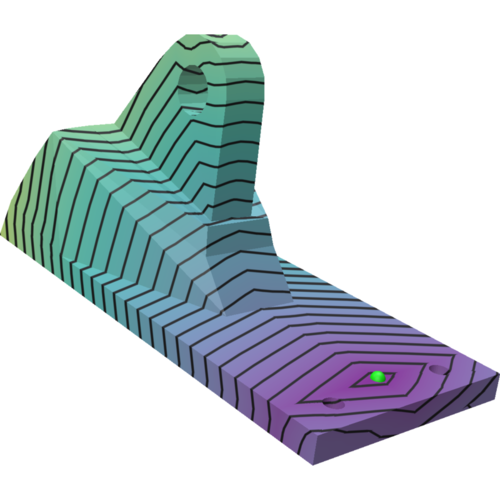}\TeaserCellGap
            \TeaserCell{Ours (\ours)}{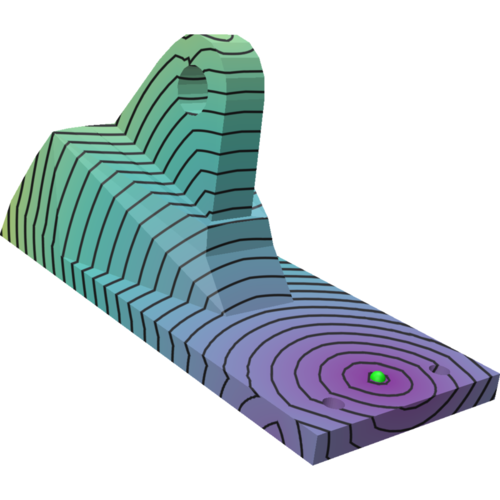}
        \end{minipage} &
        \begin{tabular}{@{}c@{\hspace{2pt}}c@{}}
            \includegraphics[width=0.171\textwidth]{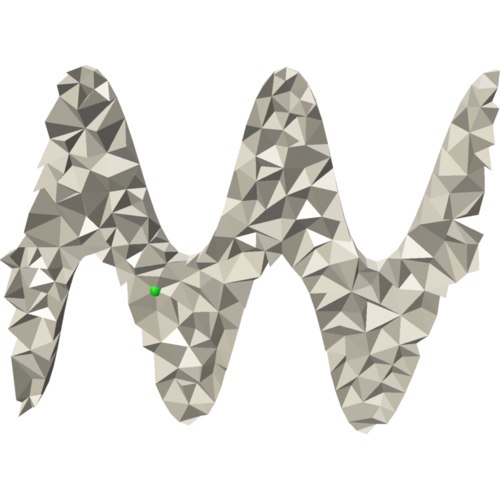} &
            \includegraphics[width=0.171\textwidth]{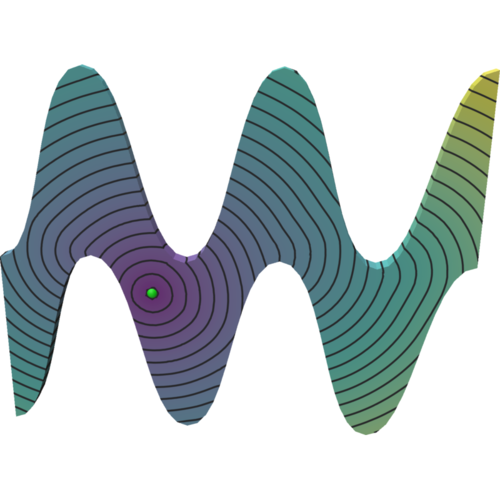} \\
            \includegraphics[width=0.171\textwidth]{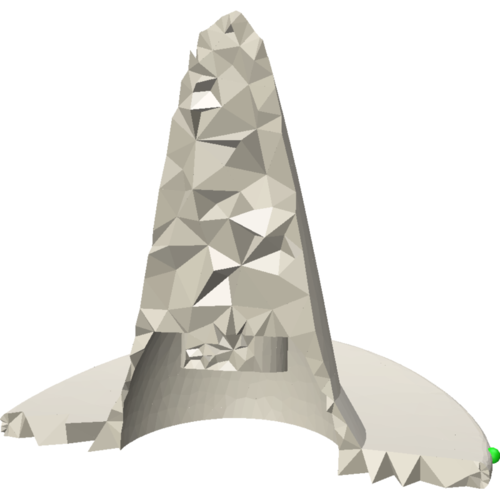} &
            \includegraphics[width=0.171\textwidth]{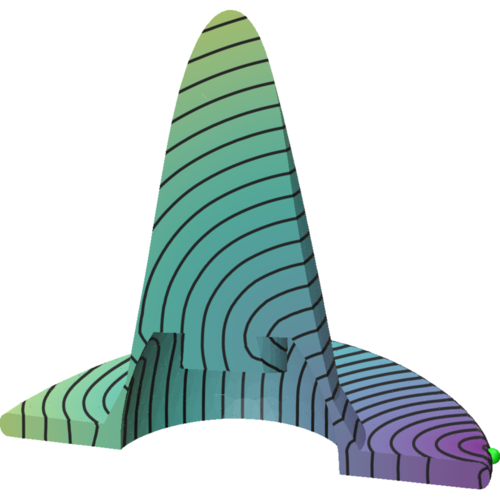} \\
        \end{tabular} &
        \begin{tabular}{@{}c@{}}
            \includegraphics[width=0.171\textwidth]{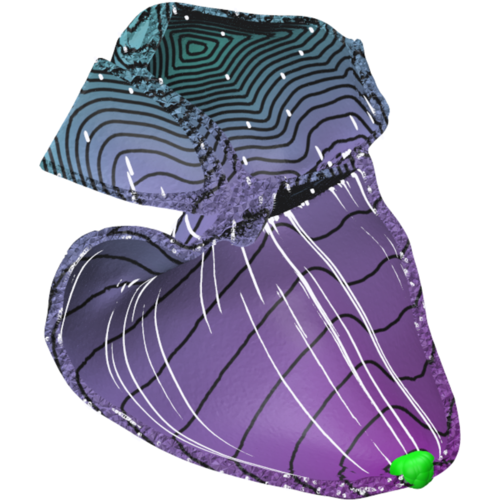} \\
            \includegraphics[width=0.171\textwidth]{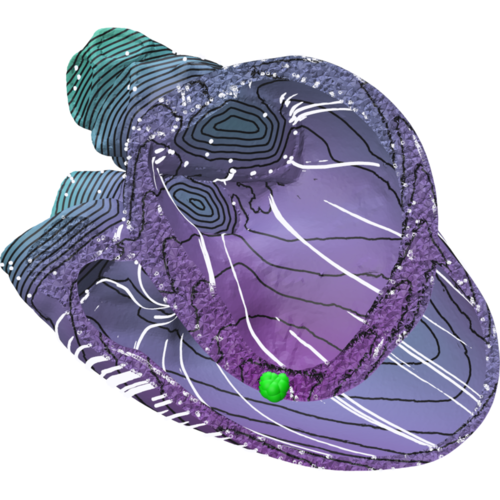} \\
        \end{tabular} \\[2pt]
        {\small (a) surfaces} & {\small (b) volumes} & {\small (c) anisotropic metrics} \\
    \end{tabular}
    \caption{We compute piecewise-quadratic squared geodesics on geometric domains. Our method is a finite-element method that work both on surfaces (a) and in volumes (b), unlike polyhedral-exact methods (e.g., \VTP~\cite{Qin2016}). Our method is considerably more accurate than low-order finite-element methods (e.g., \HM~\cite{geodesics_in_heat}, \FMM~\cite{kimmel1998computing}, \DFA~\cite{Belyaev-Fayolle}), especially on coarse and uneven meshes. Our method is flexible and allows for simple extensions such as using anisotropic metrics (c), exemplified by stretching along the white muscle-fiber streamlines of a four-chamber heart~\cite{strocchi2020publicly}. }
    \label{fig:teaser}
\end{teaserfigure}

\maketitle

\section{Introduction}

Computing geodesic distances on discrete surfaces and inside volumes is a fundamental task in geometry processing, and decades of research have produced a rich literature of methods (Surveyed in, e.g.,~\cite{crane2020survey}).

Given a collection of sources on a geometric domain, our task is to compute a scalar field which value at each point is its shortest distance from the closest source. There are two main approaches in the literature. The first is the polyhedral approach, which computes exact or approximate discrete geodesics directly on triangle meshes~\cite{Qin2016, Surazhsky_2005, trettner2021geodesic}. 
For well-connected $2$-manifolds, these methods are typically ideal, giving the true discrete distance. However, as such, they are susceptible to geometric and topological artifacts. Furthermore, Exact methods do not directly extend to tetrahedral volumes. In fact, 
exact geodesics are conjectured to be NP-hard~\cite{canny1987new}, which means it's even difficult to verify the exactness of exact solutions. 
Finally, support for anisotropic metrics varies across polyhedral methods, even though such metrics are vital in several applications (e.g.,~\secref{sec:cardiac}).

The second approach computes geodesics by posing them as partial differential equations, often using a finite-element discretization (e.g.,~\cite{geodesics_in_heat, Belyaev-Fayolle, kimmel1998computing}). 
Although approximate, these methods are simple, work seamlessly on both triangle and tetrahedral meshes (and potentially other geometric representations, including non-manifold), generalizable to any metric, and degrade more gracefully with noise or artifacts. However, such methods have been defined with low-order piecewise-linear elements. This representation does not match the local structure of a distance field, and cannot exactly reproduce its behaviour at or near the source, as shown in \figref{fig:teaser}.

Our method follows this PDE-based approach. We however compute the \emph{squared} geodesic distance using piecewise-quadratic (PQ) elements. On any star-shaped flat Euclidean domain with the source in its kernel, the squared distance is exactly quadratic and can therefore be represented exactly by PQ elements. On curved surfaces and within volumes, the PQ elements also provides substantially higher accuracy.
Nevertheless, simply replacing linear elements with quadratic ones does not improve every existing algorithm, as demonstrated in \secref{subsec:limitation-PL}. We therefore develop a convex formulation specifically for the squared distance.
The same formulation also accommodates anisotropic metrics, demonstrated on a fibre-aligned metric used in our cardiac activation example (\figref{fig:heart_app}). We denote our method and representation as a Quadratic Distance Field (\textbf{\ours}).

Our contributions are:
\begin{itemize}
    \item We represent geodesic distances as square roots of piecewise-quadratic (PQ) functions, reproducing closed-form solutions in star-shaped flat Euclidean volumes, and achieving superior accuracy across a benchmark of triangle and tetrahedral meshes with widely varying quality (\secref{sec:method}).
    \item We formulate the computation as a convex second-order cone program over PQ elements (\secref{subsec:quadratic-ours}) that is robust on defect-riddled, unevenly triangulated, and non-manifold meshes, and easily adapts to anisotropic metrics.
    \item We develop an efficient solver based on the Alternating Direction Method of Multipliers (ADMM~\cite{boyd2011distributed}) with runtimes comparable to linear methods  (\secref{subsec:admm-algorithm}).
\end{itemize}

\section{Related Work}
\label{sec:related-work}

\subsection{Geodesic distances on discrete manifolds} A complete survey of this decades-old problem is beyond our scope, and we refer readers to recent surveys~\cite{crane2020survey, peyre2010geodesic}. We discuss only the work most relevant to our setting. \tabref{tab:capability} summarizes the capabilities of these methods in relation to ours.

\subsection{Polyhedral methods}
\label{subsec:discrete-polyhedral-methods}
Polyhedral methods compute geodesic distances directly on the discrete polyhedral domain, without introducing a PDE discretization. There are broadly two classes of such methods:

\paragraph*{Wavefront propagation methods.} The classic Mitchell \emph{et al.}~\shortcite{mitchell1987discrete} (\textbf{MMP}) computes geodesic distances as a ``wave'', usually using a windowed approach. They essentially work by unfolding the mesh from the prescribed sources, which allows computation with high accuracy, limited only by the numerical accuracy and robustness of the unfolding process itself. Among these, we note \textbf{\FEG}~\cite{Surazhsky_2005}, which uses a windowing strategy to advance the front, and Vertex-oriented Triangle Propagation (\textbf{\VTP}~\cite{Qin2016}). The latter considerably optimizes the propagation by trimming many redundant windows while maintaining the same accuracy, substantially reducing runtime and memory consumption.  

Geodesic Source Propagation (\textbf{\GSP}~\cite{trettner2021geodesic}) propagates virtual sources rather than scalar distance values and provides fast approximate solutions. The paper also describes extensions to anisotropic metrics and nonstandard geometric inputs, including triangle soups and point clouds (we note only triangle-mesh implementations have been released). %

Wavefront methods are theoretically exact and achieve machine-precision accuracy on well-connected 2-manifold triangle meshes. They can be considerably optimized in terms of memory and time cost, and allow computing distances from any point to any other point in the mesh. As such, we mostly use them as ground truth reference solutions for triangle meshes. Nevertheless, they suffer from several structural limitations. First, exact window-based wavefront methods developed for polyhedral surfaces do not directly extend to volumetric domains. This might be even theoreticall impossible or extremely inefficient, since exact shortest-path computation in three-dimensional polyhedral domains is conjectured to be NP-hard~\cite{canny1987new}. Second, they are sensitive to geometric and topological defects and to numerical inaccuracies because they rely on exact geometric predicates, as demonstrated by \citet[Sec.~5.3]{sharp2020flipout} and in (our) \secref{subsec:robustness}. In practice, we also observe failures traced to implementation details, highlighting the importance of careful numerical handling when implementing these methods (\secref{subsec:robustness}). Finally, the unfolding operation is not well-defined on non-manifold configurations.

\paragraph*{Graph-based methods.} One family of exact methods computes shortest paths on a graph, with different methods distinguished mainly by whether they add nodes to the original mesh. \citet{lanthier1997shortest} inserted \emph{Steiner points} along mesh edges and searched the resulting augmented graph, so the graph grows with the sampling density. \citet{meng2021geodesic} make this more efficient by spacing the Steiner points evenly and independently of the source, treating each as a simplified window, and pruning most broadcast events with filtering rules. The surviving propagation is organised into \emph{geodesic tracks}, which are intersection-free subregions of each triangle within which the distance field is approximately linear. Another line of work keeps the mesh vertices as its nodes and lets one graph edge stand for an entire short geodesic. The Discrete Geodesic Graph (\textbf{DGG}~\cite{Wang:2017,Adikusuma:2020}) computes those edges by window propagation and answers queries with a label-correcting shortest-path sweep, at the cost of a precomputation that pays off only when many sources are queried. \textbf{DGG-VTP}~\cite{Adikusuma:2022} avoids this precomputation by building only the DGG edges reachable from the given source, on the fly, inside \VTP's window organisation. Short-Term Vector Dijkstra (\textbf{STVD}~\cite{campen2013}) does not need an auxiliary graph, running Dijkstra on the mesh edge graph itself and enriching the state rather than the graph: each update replaces the summed lengths of the last $k$ edges by the length of their vector sum. Because the metric enters only through that sum, anisotropy is native, and in volumes the unfolding step is simply dropped and the 3D edge vectors summed directly.

Graph-based methods are potentially more generalizable than wavefront-propagation methods. However,  \citet{Wang:2017} excludes a wide range of discrete domains and anisotropic metrics, and  \citet{Adikusuma:2020} remove non-manifold meshes
from their test set. This restriction is inherited, since their graph edges are computed by MMP- or CH-style window propagation. Steiner-point methods~\cite{lanthier1997shortest,meng2021geodesic} do not have these restrictions, where  \citet{meng2021geodesic} specifically reports non-manifold results by keeping one instance of each Steiner point per incident face with two instances in the manifold case. Apart from STVD, none of these methods addresses tetrahedral volumes, and
we could not find an implementation supporting them.

\subsection{PDE-based methods}

Given the shortcomings of exact methods, other methods sought to treat the problem as a PDE and provide approximate smooth solutions in a robust framework with good accuracy nonetheless.
\paragraph*{Viscosity solutions} The defining equation for geodesic distance is the eikonal equation $|\nabla d| = 1$. Nevertheless, it is only well defined within the exponential map (away from the medial axis and the sources themselves). A class of methods computes the \emph{viscosity} solution by propagating it locally with upwind updates. The Fast Marching Method (\textbf{\FMM}~\cite{kimmel1998computing}) traced fronts with upwinding, resulting in a viscosity solution to the geodesic distance problem. Obtuse elements break its upwind condition, which \FMM\ handles by splitting the obtuse angle through a neighbouring triangle. The Fast Iterative Method (\textbf{\FIM}~\cite{fim_fu_2013}) solved the same equation on tetrahedra, lifting that construction to solid angles and replacing the heap that orders \FMM's updates with an unordered active list, which buys fine-grained parallelism. Its local solver also covers anisotropic speed functions. Fast marching has also been extended to unstructured tetrahedral grids for first-arrival traveltime computation~\cite{lelievre2011computing}. We use \FIM as our volumetric propagation baseline because it is designed for tetrahedral meshes and anisotropic speed functions and provides a released reference implementation.

Recent methods posed the geodesic problem variationally, solved by global optimization. Cao \emph{et al.}~\shortcite{cao2020qgdf} compute a quasi-geodesic field by quadratic programming, trading accuracy for smoothness. Chen \emph{et al.}~\shortcite{Chen:2024} later converted the eikonal problem into a dense Convex Quadratic Programming system.
Générau \textit{et al.}~\shortcite{generau2022numerical} computed the cut locus by minimizing a convex functional. 
Belyaev and Fayolle~\shortcite{Belyaev-Fayolle} also discuss the Poisson problem $\Delta d=-1$ as a simple linearized surrogate, but its solution generally provides only a loose approximation of the distance. Instead, their Distance Function Approximation (\textbf{\DFA}) uses Ishii's extension of Perron's method for Hamilton-Jacobi equations~\cite{perron} to compute the maximal viscosity subsolution, resulting in a convex system. Our method follows similar reasoning by also solving a maximal subsolution problem.

The Geodesics in Heat method (\textbf{\HM}~\cite{geodesics_in_heat}), also extended to signed distance~\cite{Feng:2024:SHM}, computes the geodesics as the limit of the short-time heat kernel (per Varadhan's formula~\cite {varadhan1967behavior}), which linearizes the problem and results in high-quality solutions fast.  Later variants scale it to large meshes with parallel iterative solvers~\cite{tao2019parallel} and generalize the diffusion to non-uniform metrics~\cite{sun2022nonuniform}.
Methods like ViscoGrid~\cite{pumarola2022visco} use \emph{vanishing viscosity} terms (e.g., adding a small $\epsilon \cdot \Delta f$) to their loss functions. 
This is generalized by Edelstein et al.~\cite{edelstein2023convex} for a family of regularizers that compute smooth geodesics. 
The vanishing viscosity term is related to Varadhan's formula through the Cole-Hopf transformation and is used to design losses for training neural distance fields~\cite{lipman2021phase,wang2025hotspot, HeatSDF}. These neural methods can potentially be used for computing geodesic distances by training a neural network on a mesh~\cite{costabal2024delta}, but have mostly been demonstrated in the Euclidean domain. Finally, we note that there are recent methods for data- and learning-driven geodesic computation (e.g.~\cite{Zhang:2023,Adikusuma:2026}). They are usually extrinsic (to allow for a learnable representation), and fast at inference time, but have no guarantee of adherence to the eikonal equation.

\subsection{Higher-order geometry processing} A considerable limiting factor of most methods above, as also demonstrated in all our comparisons, is that they work on piecewise-linear elements with degrees of freedom on vertices. This assumes good and reasonably fine-grained mesh quality, which is rare ``in the wild''~\cite{Thingi10K}, and also limits the sources to lie on these vertices. Some methods (e.g.,~\cite{xin2012constant, sharp2020flipout, liu2015delaunay}) changed the triangulation (intrinsically or otherwise) to improve the situation. Our method avoids remeshing operations and works with the original discretization. Drawing on classical FEM practice, recent methods in geometry processing~\cite{high-orderdf, 10.1145/3272127.3275067, Wang2025Polar} work with piecewise-quadratic and higher-order representations, which considerably alleviate such limitations in the context of geometry processing applications. Our results empirically show that PQ elements are a natural representation for (squared) geodesic distances, but also that this is not a trivial generalization (\figref{fig:heat_method_p2}). We demonstrate that our algorithm uses the representation well to deliver a robust and accurate solution on challenging meshes.

\section{Background}
\label{sec:background}
\begin{table}[t]
  \centering
  \small
  \setlength{\tabcolsep}{4pt}
  \renewcommand{\arraystretch}{1.15}
  \begin{tabular}{@{}>{\centering\arraybackslash}p{0.1\linewidth}
                    >{\raggedright\arraybackslash}p{0.38\linewidth}
                    >{\raggedright\arraybackslash}p{0.5\linewidth}@{}}
    \toprule
    \multicolumn{3}{c}{\textbf{Polyhedral Methods}} \\[2pt]
     \textbf{FEG} & \cite{Surazhsky_2005}
      & Fast exact geodesics on meshes \\
    \textbf{VTP} & \cite{Qin2016}
      & Vertex-oriented triangle propagation \\
    \textbf{DGG} & \cite{Wang:2017}
      & Discrete geodesic graph (DGG) for computing geodesic distances on polyhedral surfaces \\
    \textbf{DGG-VTP} & \cite{Adikusuma:2022}
      & Accuracy-controllable geodesic distances on triangle meshes \\
    \textbf{GSP} & \cite{trettner2021geodesic}
      & Geodesic distance computation via virtual source propagation \\
    \midrule
    \multicolumn{3}{c}{\textbf{PDE Methods}} \\[2pt]
    \textbf{FMM} & \cite{kimmel1998computing}
      & Fast marching method \\
    \textbf{FIM} & \cite{fim_fu_2013}
      & Fast iterative method \\
    \textbf{HM} & \cite{geodesics_in_heat}
      & Heat method \\
    \textbf{DFA} & \cite{Belyaev-Fayolle}
      & Distance function approximation \\
    \textbf{\ours} & (ours)
      & Quadratic distance field  \\
    \bottomrule
  \end{tabular}
  \caption{Referenced methods and acronyms. Clicking an acronym within the paper leads to this table.}
  \label{tab:methods}
\end{table}

We formally discuss representations and algorithms for solving the discrete geodesic problem in the PDE setting, and highlight the challenge that we address in this work. For convenience, we collect the acronyms for all the methods we refer to in \tabref{tab:methods}. All acronyms are hereby referenced (by clickable links) to the table.

\subsection{Geodesic distances on discrete manifolds}
\paragraph{The continuous setting.} We consider a Riemannian $k$-manifold $(\mathcal{M},g)$ for $k=2,3$ with a metric $g$, and a set $\mathcal{B} \subset \mathcal{M}$ of source points. Consider the pairwise geodesic distance function $\overline{d}(x,y):\mathcal{M} \times \mathcal{M}\rightarrow [0, \infty)$. Our task is to compute the minimal geodesic (shortest) distance from $\mathcal{B}$. That is, reduce $\overline{d}$ to a scalar function $d:\mathcal{M} \rightarrow [0, \infty)$ so that $d(x) = \min_{b \in \mathcal{B}}(\overline{d}(b,x))$. Geodesic distances are always continuous and well-defined for single-component domains. However, they are only $C^1$-smooth away from conjugate points (or: the medial-axis), and the source set $\mathcal{B}$.
\paragraph{The discrete setting.} We tessellate a domain into a simplicial complex, either a triangle mesh $\mathcal{M}=(\mathcal{V}, \mathcal{E}, \mathcal{F})$ in 2D, or a tetrahedral mesh $\mathcal{M}=(\mathcal{V}, \mathcal{E}, \mathcal{F}, \mathcal{T})$ in 3D, where $\mathcal{V}$ is the set of vertices, $\mathcal{E}$ the set of edges, $\mathcal{F}$ the set of faces, and $\mathcal{T}$ the set of tetrahedra. We do not assume any restriction on the genus or boundaries. We will generally assume the mesh is a $k$-manifold (that is, an edge-manifold for triangle meshes, and a face-manifold for tetrahedral meshes), unless explicitly stated.

\subsection{Linear finite-element methods}

The goal of these methods is to solve the eikonal equation discretely with finite-element operators, which can be interpreted as solving the problem in a weak sense over the PL space. %
The eikonal equation $|\nabla d| = 1$ is only defined away from the cut locus and the sources. A classical way to approach this PDE is by looking for a \emph{viscosity solution}. Intuitively, a viscosity solution satisfies the equation almost everywhere and ``auto-completes'' the solution at points where the PDE isn't well-defined continuously (The classic example is $d = |x|$ with the natural completion $d(0)=0$). \DFA achieves this by computing:
\begin{equation}
d = \text{argmax}{\int_\mathcal{M}d},\ \ s.t.\ \ |\nabla d| \leq 1,\  d(b \in \mathcal{B}) = 0
\label{eq:linear-d-viscosity}
\end{equation}
Formally, this is a special case of using Perron's method for computing the maximal subsolution for such a Hamilton-Jacobi type equation~\cite{perron}, leading to a viscosity solution. This formulation is appealing since the problem becomes convex, as the inequality constraint is a convex cone. Furthermore, the nonlinear constraint is pointwise-local, whereas the global objective is linear, which enables the definition of an efficient ADMM solver. In the discrete setting, $d$ is a piecewise-linear conforming function with nodal values on $\mathcal{V}$, and $\nabla d$ is a piecewise-constant face-based vector field. 

\begin{figure*}[t]
	\centering
	\setlength{\tabcolsep}{5pt}
	\begin{tabular}{ccccc}
		\includegraphics[width=0.166\linewidth]{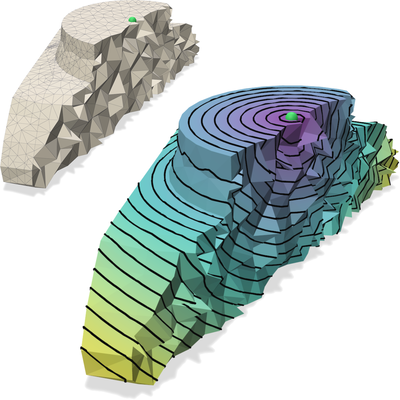} &
		\includegraphics[width=0.166\linewidth]{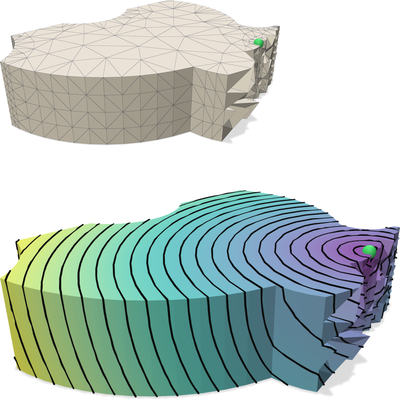} &
		\includegraphics[width=0.166\linewidth]{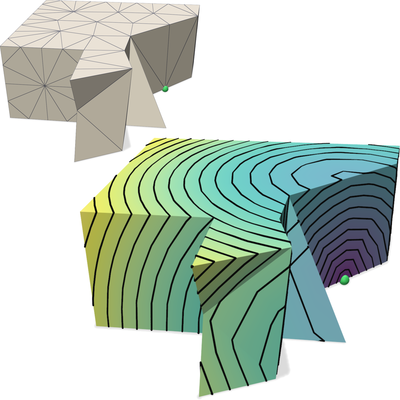} &
		\includegraphics[width=0.166\linewidth]{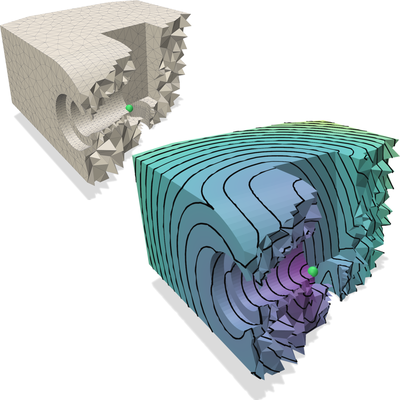} &
		\includegraphics[width=0.166\linewidth]{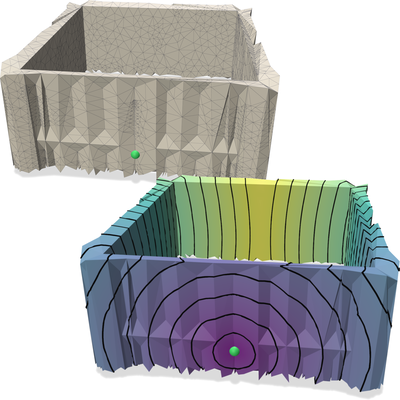} \\
		
		\includegraphics[width=0.166\linewidth]{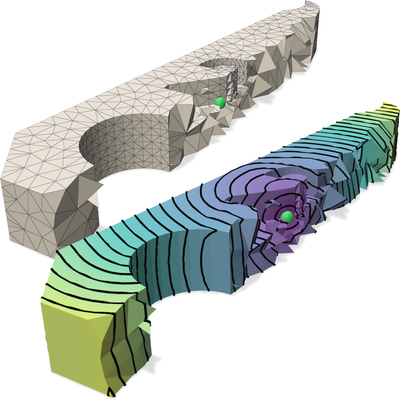} &
		\includegraphics[width=0.166\linewidth]{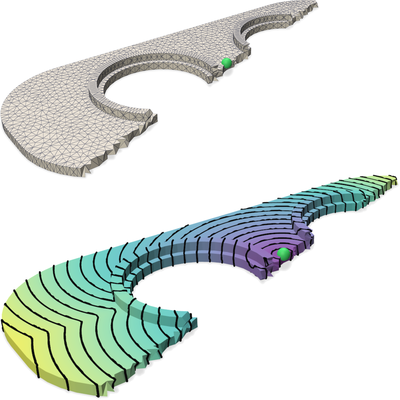} &
		\includegraphics[width=0.166\linewidth]{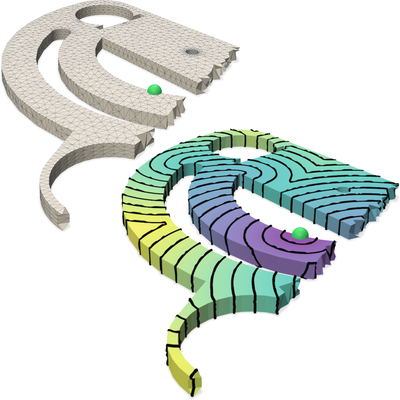} &
		\includegraphics[width=0.166\linewidth]{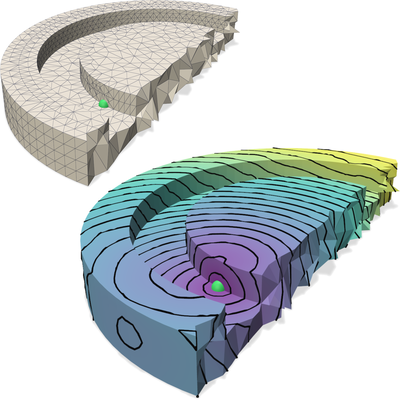} &
		\includegraphics[width=0.166\linewidth]{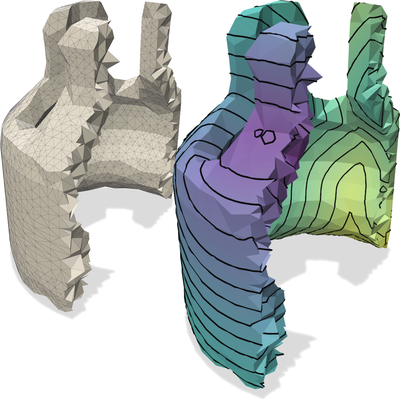} \\
		
		\includegraphics[width=0.166\linewidth]{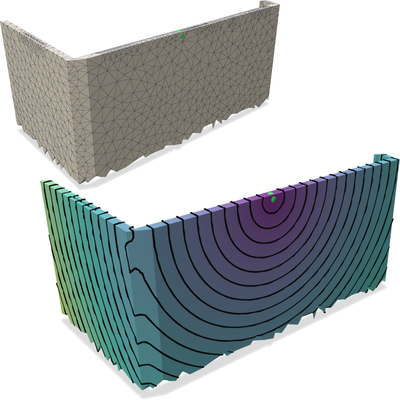} &
		\includegraphics[width=0.166\linewidth]{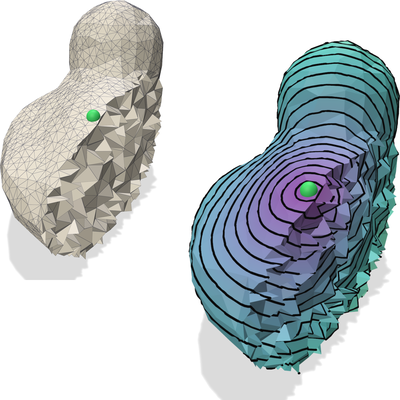} &
		\includegraphics[width=0.166\linewidth]{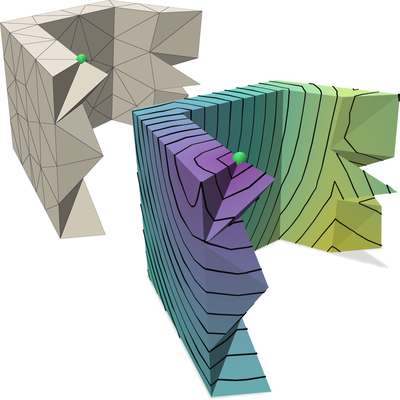} &
		\includegraphics[width=0.166\linewidth]{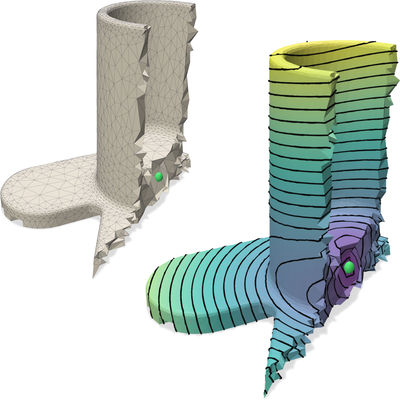} & 
		\includegraphics[width=0.166\linewidth]{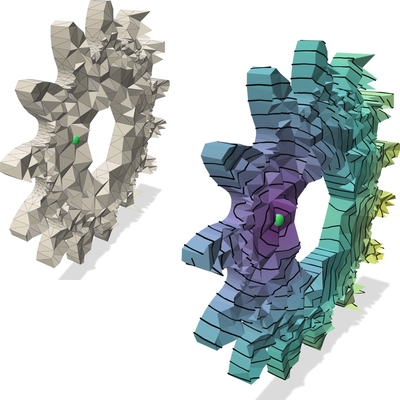} \\
	\end{tabular}
	\caption{Geodesic distance gallery for tetrahedral (volumetric) meshes with large variety in resolution and tet shapes.}
	\label{fig:gallery_3d}
\end{figure*}

\HM computes the geodesic distance as the limit of the short-time heat kernel $k_t(x,y)$ through Varadhan's formula.
\begin{align}
\label{eq:VaradhanFormula}
d(x,y) = \lim_{t\to 0} \sqrt{-4t \log k_t(x,y)}
\end{align}
The method implicitly diffuses heat, normalizes the resulting gradient, and reintegrates the direction field in the least-squares sense. This requires two linear solves and yields a fast method that is simple to implement.
While only approximating the discrete distance, and thus resulting in less accurate results than vertex-exact methods, FE distance methods enjoy several advantages over exact methods. First, they extend trivially to 3D by considering the appropriate discrete operators (\figref{fig:gallery_3d}). 
Second, global finite element formulations are more robust to mesh defects and noise (\figref{fig:gaussian_noise} and \figref{fig:robust_holes}), and can be extended to nonmanifold meshes once the discrete operators are suitable generalized (\figref{fig:nonmanifold}).
FE methods can also incorporate arbitrarily prescribed metrics (that alter the gradient and Laplacian operators). Finally, they are usually simpler to implement with only linear solves, and without the need for exact predicates. As such, FE methods provide value where exact methods cannot. Nevertheless,  as evident in our comparisons (\tabref{tab:comparison}),linear FE methods are still sensitive to the mesh, as they rely on PL discretizations of the gradient and Laplace-Beltrami operators. We note that this can be mitigated by using intrinsic triangulations~\cite{Sharp:2021:GPI, gillespie2021integer} or flip operations~\cite{sharp2020flipout}, which compute the solution on an alternative connectivity for the same geometry. 
However, \citet[Sec.~4.2]{campen2013} identify several limitations of intrinsic triangulations for geodesic computation. We do not explore these issues further and instead work with the original mesh.

\section{Squared Geodesic Distance on PQ Elements}
\label{sec:method}

\subsection{Choice of discrete representation}
\label{subsec:limitation-PL}

\begin{figure}[t] 
  \centering
  \setlength{\tabcolsep}{1pt} 
  \renewcommand{\arraystretch}{0} 

  \begin{tabular}{@{}cccc@{}} 
  	\begin{minipage}{0.24\linewidth}\centering
  		 \includegraphics[width=\linewidth]{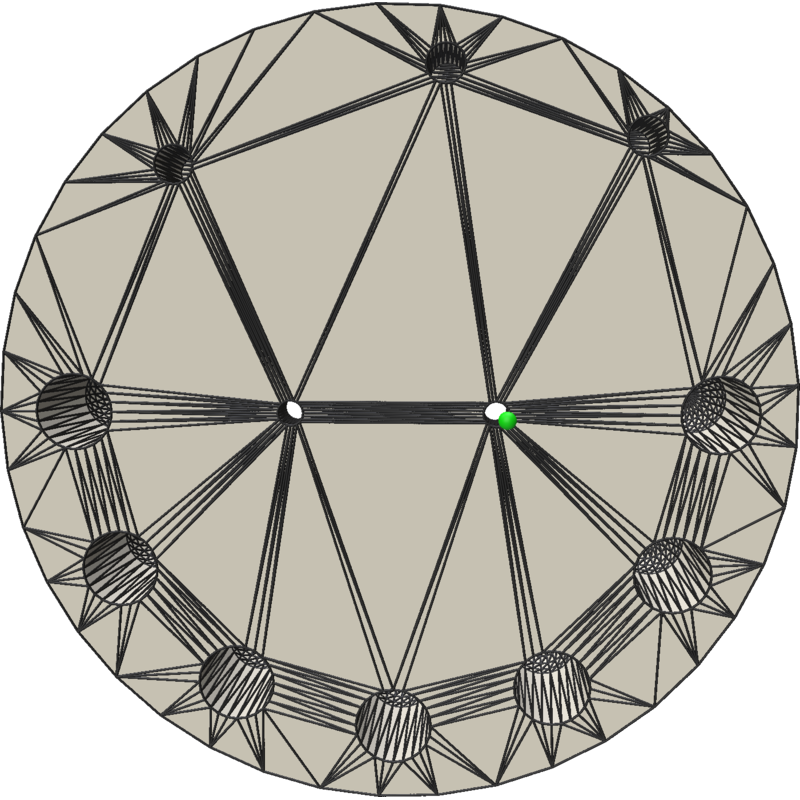} \\
  		 \small (a) Coarse
  	\end{minipage}
  	
  	\begin{minipage}{0.24\linewidth}\centering
  	\includegraphics[width=\linewidth]{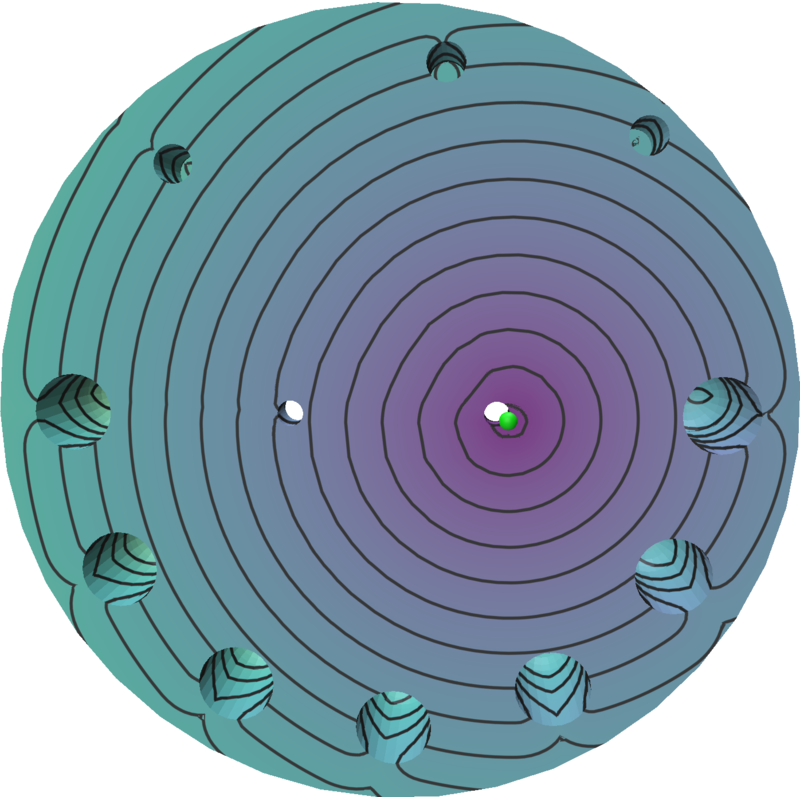} \\
  	\small (b) reference
  	\end{minipage}
  	
  	\begin{minipage}{0.24\linewidth}\centering
  	\includegraphics[width=\linewidth]{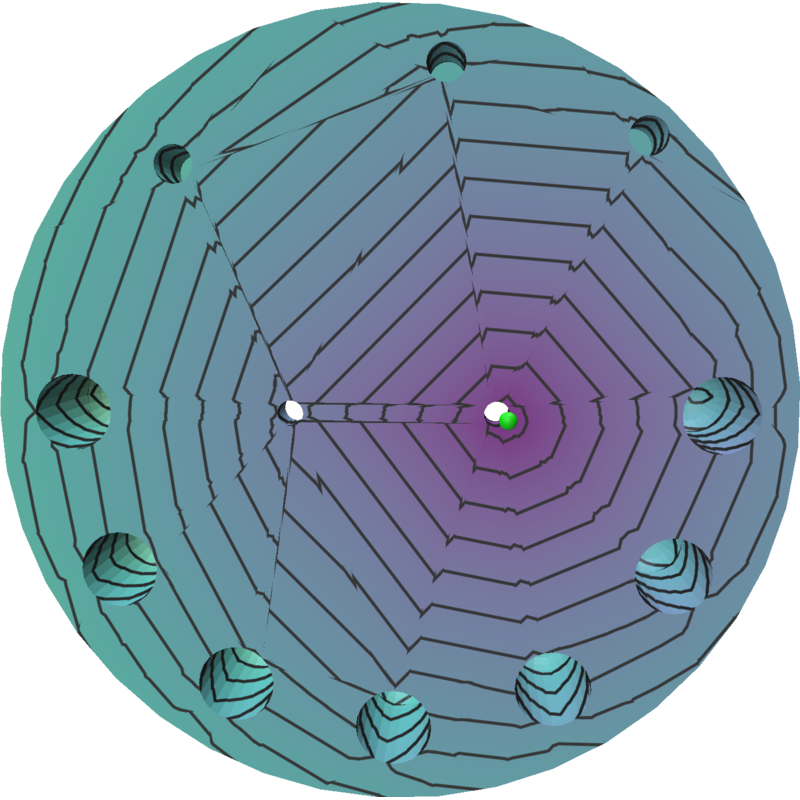} \\
  	\small (c) PL solution
  	\end{minipage}

   	\begin{minipage}{0.24\linewidth}\centering
   	\includegraphics[width=\linewidth]{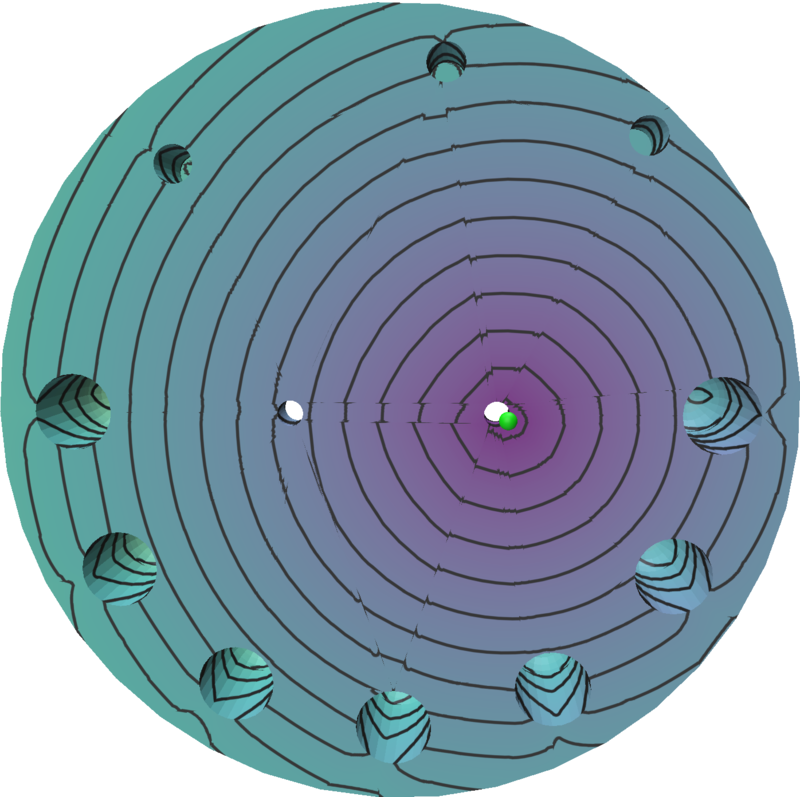} \\
   	\small (d) PQ solution
   \end{minipage}
  \end{tabular}

  \caption{Even the $L_2$-optimal PL solution is substantially worse than its PQ counterpart, which bounds the accuracy of the linear \emph{representation} itself.}
  \label{fig:best_solution}
\end{figure}

All aforementioned PDE methods use piecewise-linear (P1) elements. Our main insight is that this choice, \emph{even before considering} the algorithm built on top of it, limits the achievable accuracy, because:
\begin{enumerate}
    \item It does not reproduce continuous geodesic distances. As a primary example (e.g., Fig~\ref{fig:closed_form}), it does not even reproduce the closed-form flat distance $d(x,y) = \sqrt{\lVert x-y \rVert^2}$ on star-shaped flat meshes with sources at their kernels. In curved meshes, one can expect even lower accuracy.
    \item Even if the method allows sources to be prescribed anywhere inside triangles, the linear interpolant does not reproduce their local environment correctly ( \figref{fig:disc}).
    \item As low-order interpolants, PL elements are sensitive to mesh quality, and introduce an error even when the vertex-based distances are exact or optimal (\figref{fig:best_solution} and \secref{subsec:robustness}).
\end{enumerate}
Furthermore, the function $d$ itself is problematic. Near the source, it forms a convex cone that is continuous but not differentiable at $d=0$. Although a viscosity solution exists, PL elements approximate this behaviour poorly and produce ``tent-like'' artifacts around the source (Fig.~\ref{fig:meshing}).

To demonstrate the limitation of the \emph{representation} independently of the algorithm, we compute the optimal PL field in the least-squares sense (\figref{fig:best_solution}). We subdivide the mesh three times and compute the exact fine solution using \VTP. Let $A_\text{fine}$ be the piecewise-linear upsampling matrix and $d_\text{fine}$ the fine scalar field. We compute
$$d_\text{coarse} = \text{argmin}\lVert Ad_\text{coarse} - d_\text{fine} \rVert^2$$
to project the fine solution onto the coarse mesh in least squares. That results in coarse solutions that are optimal in the $L_2$ sense. The resulting PL field has considerable artifacts. Doing the same for the optimal PQ field (in P2 elements) results in a fields that is almost indistinguishable from the fine solution. This demonstrates the accuracy limit imposed by the PL representation.

\paragraph{Squared distances} Consider the squared distance $u=d^2$ represented using PQ elements. On a flat Euclidean manifold, this is a quadratic polynomial, which resolves two major limitations of PL elements. First, squared distances can be represented exactly using PQ elements, \emph{regardless} of mesh quality or resolution. Second, sources where $u=0$ can be represented explicitly with zeros in the quadratic function anywhere on the mesh, where the function is trivially smooth (e.g., Fig.~\ref{fig:moving_source}). Thus, it is by definition a better choice of representation.

\begin{figure}[!htbp]
	\centering
	\setlength{\tabcolsep}{1pt} 
	
	\begin{tabular}{ccc}
		\begin{minipage}{0.30\linewidth}\centering
			\includegraphics[width=\linewidth]{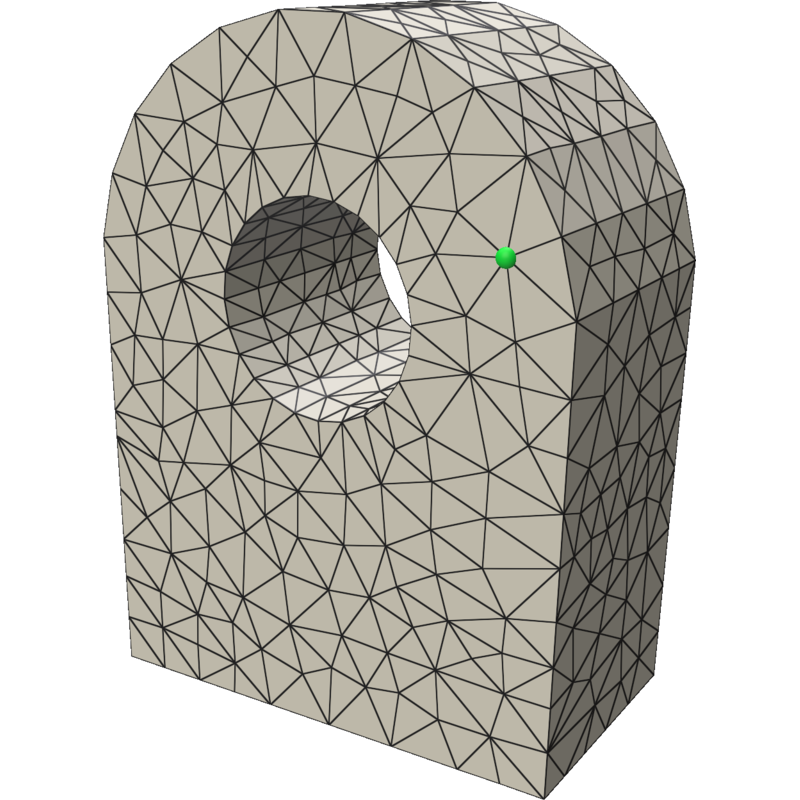} \\
		\end{minipage}
		&
		\begin{minipage}{0.30\linewidth}\centering
			\begin{tabular}{@{}c@{}}
				\includegraphics[width=0.75\linewidth]{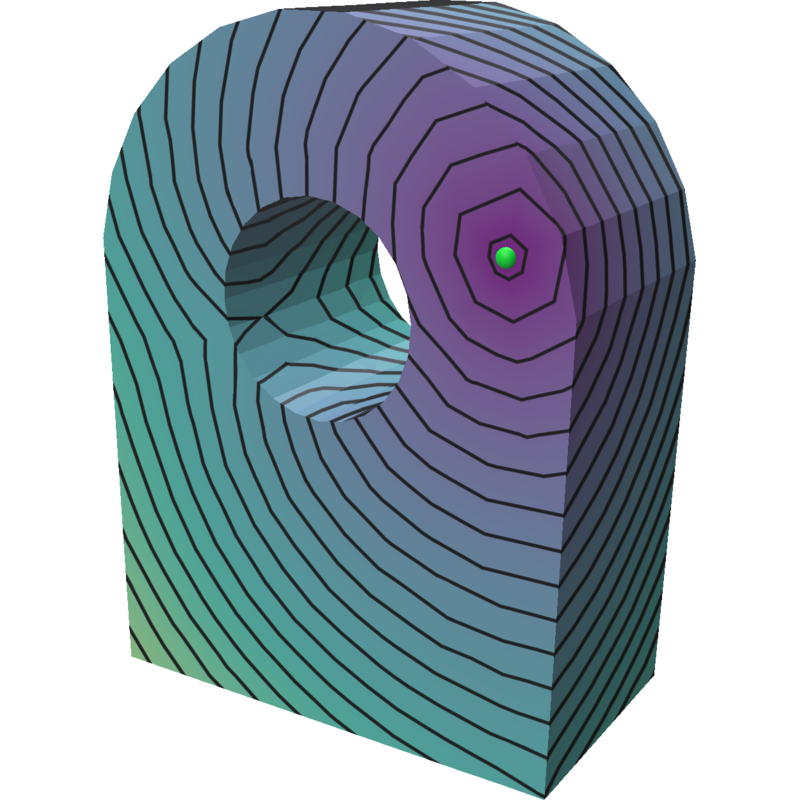} 
				\includegraphics[width=0.2\linewidth]{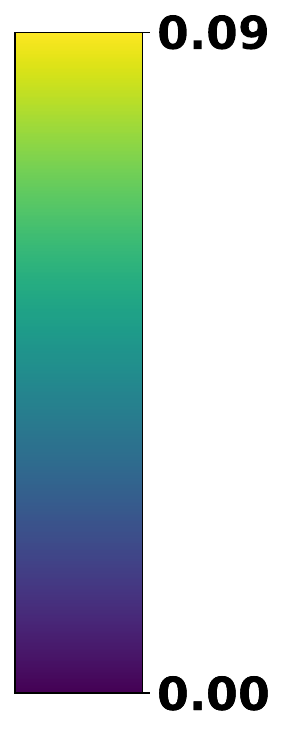} \\
			\end{tabular} \\
			\small (a) HM with PL operators
		\end{minipage}
		&
		\begin{minipage}{0.30\linewidth}\centering
			\begin{tabular}{@{}c@{}}
				\includegraphics[width=0.75\linewidth]{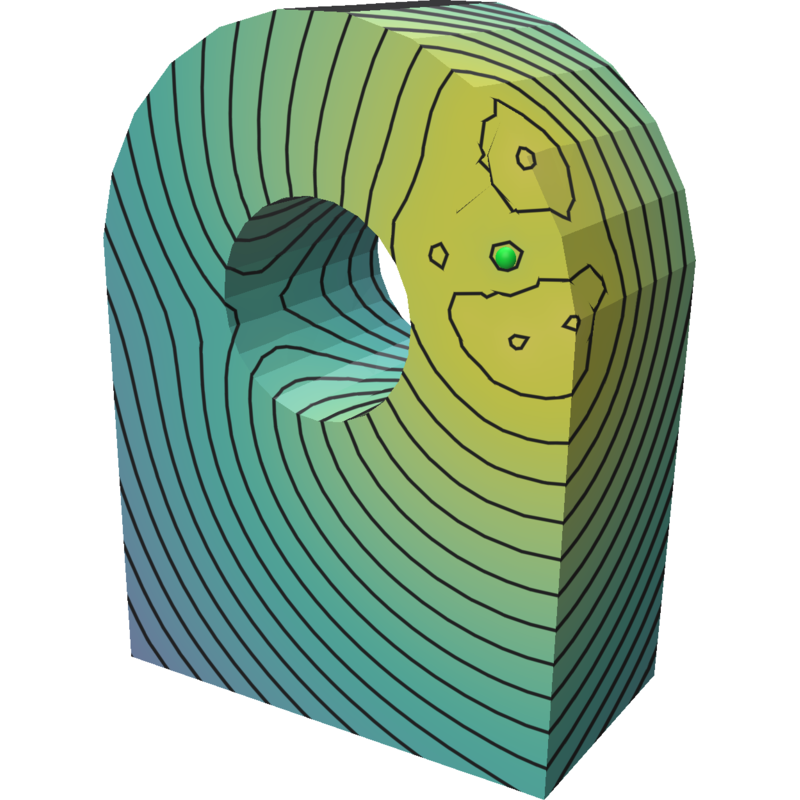} 
				\includegraphics[width=0.2\linewidth]{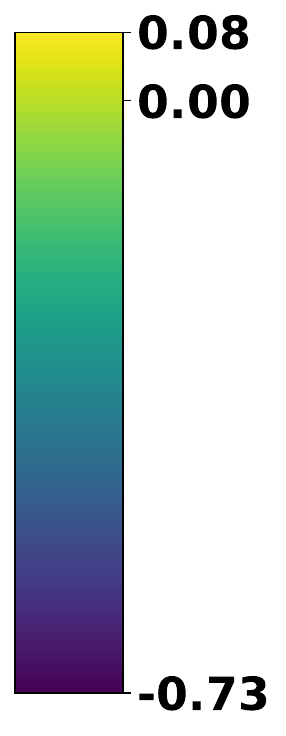} \\
			\end{tabular} \\
			\small (b) HM with PQ operators
		\end{minipage}
	\end{tabular}

    \caption{Using PQ operators for the heat method degrades the result, even causing negative values.}
    \label{fig:heat_method_p2}
\end{figure}

Unfortunately, simply raising the degree of the polynomial and using a FE method to compute distance fields does not guarantee better results. As an example (Fig.~\ref{fig:heat_method_p2}), computing \HM with PQ operators \emph{degrades} the PL solution and even produces negative distances, because PQ basis functions are partly negative and the short-time diffusion of \HM is poorly approximated by them. Choosing a method that provides a robust result in PQ elements is then a challenge in itself.

\subsection{Quadratic Distance Field (\protect\ours) geodesics}
\label{subsec:continuous-pde}
Our solution, denoted \QDF (for Quadratic Distance Field) redefines the approach of \DFA to the PQ setting. First, we formulate the equivalent of the eikonal equation directly for the squared distance. Setting $u = d^2$, we get:
\begin{align}
1 = \nonumber \lVert \nabla d \rVert &= \lVert \nabla (\sqrt{u}) \rVert = \frac{1}{2\sqrt{u}} \lVert \nabla u \rVert\\
\implies\quad u &= \frac{1}{4}\lVert \nabla u\rVert^2 .
\end{align}
This PDE can be interestingly derived directly from Varadhan's formula, and we include a proof in Appendix~\ref{sec:squared_geodesic_proof}, although this is a known result. Similarly to \DFA, we can formulate this as a convex maximal viscosity subsolution problem:
\begin{equation}
u = \text{argmax} \int_\mathcal{M}{u},\ \ s.t.\ \ \ u \geq \frac{1}{4}\Vert \nabla u\rVert ^2,\ \ u(b\in \mathcal{B}) = 0.
\label{eq:quadratic-convex-formulation}
\end{equation}
This system retains the convexity of \DFA, with the cone inequalities. In fact, this formulation is \emph{equivalent} to that of Eq.~\ref{eq:linear-d-viscosity} (we include a proof in Appendix~\ref{sec:dfa-equivalence}) for the continuous setting. Note that this guarantees that $u \geq 0$ by design.

\subsection{PQ operators}
\label{subsec:pq-operators}
To define a discretization of Eq.~\ref{eq:quadratic-convex-formulation}, we provide a brief overview of the PQ setup, where in-depth details of the matrices for reproduction are in Appendix~\ref{sec:pq_matrices}. We represent $u$ using standard Galerkin conforming PQ elements $\left\{\psi_\mathcal{V},\psi_\mathcal{E}\right\}$ with degrees of freedom (nodal values) on vertices and mid-edges, including boundaries. That means that a quadratic function is defined by $6$ degrees of freedom in a triangle, and $10$ degrees of freedom in a tetrahedron.

\paragraph*{Operators and mass matrices}
We consider the following (sparse) matrices:

\begin{enumerate}
    \item The conforming PQ gradient matrix $G$, taking values on the PQ nodes and producing piecewise linear (tangent continuous, or $H(curl)$-valued) vector fields inside simplices. For triangle meshes, it is a $G: 6|\mathcal{F}| \times \left(|\mathcal{V}|+|\mathcal{E}|\right)$ matrix, and for tet meshes, it is $G: 10|\mathcal{T}| \times \left(|\mathcal{V}|+|\mathcal{E}|\right)$.
    \item  Mass matrix for scalar per-simplex PQ inner products. For both triangle and tet meshes they are $M_2:(|\mathcal{V}|+|\mathcal{E}|) \times (|\mathcal{V}|+|\mathcal{E}|)$.
    \item A sampling matrix $Q_\mathcal{S}$ that interpolates nodal values to obtain values corresponding to any prescribed point set $\mathcal{A}$ within the mesh, with dimensions $|\mathcal{A}|\times(|\mathcal{V}|+|\mathcal{E}|)$ for a PQ function . When we sample a vector-valued quantity (e.g., a gradient field) at the same points, we denote it in bold, $\mathbf{Q}_\mathcal{A}$, with the column count scaled accordingly ($\times 2$ on triangle meshes, $\times 3$ on tet meshes). %
\end{enumerate}

\subsection{Discrete formulation}
\label{subsec:quadratic-ours}

We represent the source set $\left\{b \in\mathcal{B}\right\}$ as a set of points on the mesh, parameterized by barycentric coordinates $\lambda_i(b), \lambda_j(b), \lambda_k(b)(,\lambda_l(b))$ in the single triangle $ijk$ (resp. tetrahedron $ijkl$) containing $b$. In case $b$ is on a boundary between two simplices, we arbitrarily choose one of the adjacent simplices as the representative (it does not affect the result).

We next discretize Eq.~\ref{eq:quadratic-convex-formulation}. The \emph{unknown} is the PQ function nodal values $u:|\mathcal{V}|+|\mathcal{E}| \rightarrow [0, \infty)$. Our discrete formulation then becomes:
\begin{align}
    \nonumber u &= \text{argmin} (- \mathbf{1}^TM_2u)\\
    \nonumber 
    s.t. \ \ Q_\mathcal{S}u &\geq \frac{1}{4}|\lVert \mathbf{Q}_\mathcal{S} Gu \rVert^2\\
    Q_\mathcal{B} u &= 0
    \label{eq:discrete-convex-formulation}
\end{align}
$\mathbf{1}$ is the vector of all $1$'s of the appropriate length. Our constraints are enforced \emph{only} at a finite set of points define by $\mathcal{S}$: we define it as the PQ nodes of each simplex together with its barycenter, resulting in $7$ points per triangle ($3$ vertices, $3$ edge midpoints, and the barycenter) and $11$ per tetrahedron ($4$ vertices, $6$ edge midpoints, and the barycenter). This means that, rather than satisfying the inequalities pointwise everywhere, we satisfy them in an appropriate weak sense~\cite{citeulike:8936200}. While not nominally guaranteeing pointwise positivity everywhere (being positive at the nodal points and the barycenter is necessary but not sufficient), we demonstrate empirically in all our examples, and specifically in ~\secref{subsec:benchmark-and-comparison}, that this is a reasonable relaxation that does not harm accuracy. We conjecture that the adversarial maximization of $u$ and the nature of the constraint encourage the constraints to also be satisfiable pointwise, but leave a concrete proof to future work. To obtain the final distance $d$, we simply take the square root in-place after interpolating $u$ with PQ elements to any chosen point.

\subsection{ADMM solver}
\label{subsec:admm-algorithm}
Eq.~\ref{eq:discrete-convex-formulation} is a second-order cone program: the objective $c^\top u$ (with $c:=-M_2\mathbf{1}$) and the source condition are both linear, and every sample $s\in\mathcal{S}$ contributes one \emph{local} rotated cone constraint $\tfrac14\Vert(G_\mathcal{S}u)_s\Vert^2\le(Q_\mathcal{S}u)_s$, where $G_\mathcal{S}:=\mathbf{Q}_\mathcal{S}G$ is the gradient sampled at $\mathcal{S}$. An off-the-shelf conic solver such as MOSEK~\cite{mosek} could solve this problem directly, but every one of its iterations couples the entire mesh at once (Appendix~\ref{sec:scalability-curves}), so it does not scale to the resolutions we target. We instead exploit the structure of the problem to build an efficient iterative solver.

\paragraph*{Eliminating sources by anchoring}
The source condition $Q_\mathcal{B}u=0$ is a Dirichlet boundary condition. Rather than keep it as a constraint in the optimization, we eliminate it: let $\overline{\mathcal{B}}$ be the nodal values of $u$ belonging to every simplex incident to $\mathcal{B}$, and prescribe them directly to their exact squared Euclidean distance from the source,
$$
u(\overline{b}) = \min_{b\in\mathcal{B}}\,\bigl|\overline{b} - b\bigr|^{2}, \ \  \forall \overline{b}\in\overline{\mathcal{B}},
$$
for simplicity, we restrict this to sources $b$ sharing a simplex with $\overline{b}$, so it reduces to a closed-form Euclidean computation. A scattering of nearby point sources could in principle hurt this approximation, but such a configuration would already produce a medial axis too complex for PQ elements to represent accurately near the sources. We partition $u=(u_\mathcal{I},u_{\overline{\mathcal{B}}})$ into free and fixed blocks, with $u_{\overline{\mathcal{B}}}$ known. Denote $G_\mathcal{S}u=G_{\mathcal{S},\mathcal{I}}u_\mathcal{I}+g_0$, $c^\top u = c^\top_\mathcal{I}u_\mathcal{I}+c^\top_{\overline{\mathcal{B}}}u_{\overline{\mathcal{B}}}$, and $Q_\mathcal{S}u=Q_{\mathcal{S},\mathcal{I}}u_\mathcal{I}+q_0$, where the offsets $g_0$ and $q_0$ depend only on the fixed block and are computed once, before the iteration starts, and we only solve for $u_\mathcal{I}$.

\paragraph*{Splitting and iterating}
To effectively split the problem into local and global components, we introduce the auxiliary vector field $g$, and introduce the linear coupling constraint $g=G_\mathcal{S}u\in\mathbb{R}^{3|\mathcal{S}|}$. This turns one globally coupled problem into a single global linear solve plus $|\mathcal{S}|$ independent, pointwise projections. The associated Lagrangian multiplier is $y\in\mathbb{R}^{3|\mathcal{S}|}$. For simplicity of notation, we denote $r:=G_\mathcal{S,I}u_\mathcal{I}+g_0-g$ for the primal residual. With a penalty coefficient $\rho\in\mathbb{R}^+$, the scaled dual variable $w=y/\rho$, and the diagonal mass matrix $M_\mathcal{S}$ of triangle areas associated with the samples in $\mathcal{S}$, our augmented Lagrangian is:
\begin{equation}
	\mathcal{L}_\rho(u_\mathcal{I},g,w) = c_\mathcal{I}^\top u_\mathcal{I}
	+\frac{\rho}{2}(r+w)^\top M_\mathcal{S}(r+w)
	-\frac{\rho}{2}\, w^\top M_\mathcal{S}\, w,
\end{equation}
subject to the inequality cone constraints of Eq.~\eqref{eq:discrete-convex-formulation}. 
To accelerate convergence, we over-relax the $u$-update by extrapolating it with the previous iterate before the projection and dual steps, a standard ADMM technique~\cite{boyd2011distributed}.
The resulting ADMM iterations are: 
\begin{equation}
	\begin{aligned}
		&\mathrlap{u\text{-update:}}\\
		&&\bigl(\rho\,G_\mathcal{S,I}^\top M_\mathcal{S} G_\mathcal{S,I}\bigr)\,u^{k+1}_\mathcal{I}
		&= \rho\,G_\mathcal{S,I}^\top M_\mathcal{S}\bigl(g^{k}-w^{k}-g_0\bigr)-c_\mathcal{I},\\[3pt]
		&\mathrlap{\text{over-relaxation:}} &\hat g^{k+1}
		&= \alpha\,\bigl(G_\mathcal{S,I}u^{k+1}_\mathcal{I}+g_0\bigr)+(1-\alpha)\,g^{k},\\[3pt]
		&\mathrlap{g\text{-update:}} &g^{k+1}_j
		&= \min\!\Bigl(1,\ 2\sqrt{\max(q^{k+1}_j,0)}\,\big/\,\Vert v_j\Vert\Bigr)\,v_j,\\[3pt]
		&\mathrlap{w\text{-update:}} &w^{k+1}
		&= w^{k}+\hat g^{k+1}-g^{k+1},
	\end{aligned}
	\label{eq:admm-updates}
\end{equation}
Where $v_j=\hat g^{k+1}_j+w^{k}_j$ and
 $q^{k+1}=Q_\mathcal{S,I}u^{k+1}_\mathcal{I}+q_0$ is the field value at the samples. $\alpha$ is the over-relaxation parameter. At $\alpha=1$, the $w$-update reduces to the usual scaled dual ascent $w^{k+1}=w^{k}+(G_\mathcal{S,I}u^{k+1}_\mathcal{I}+g_0)-g^{k+1}$, accumulating the residual of $g=G_\mathcal{S}u$ over the iterations. Table~\ref{tab:admm-settings} lists our parameter values for both mesh types across the entire benchmark. %

Splitting only on the gradient pays off because the $u$-update matrix $\rho\,G_\mathcal{S,I}^\top M_\mathcal{S} G_\mathcal{S,I}$ never changes across iterations, we factor it once as a preprocessing step, and every subsequent $u$-update is a single back-substitution. The $g$-update is just as cheap: it decouples completely across samples, clamping each $g_j$ radially onto a ball of radius $2\sqrt{q_j^{+}}$ set by the field value at that sample. An entire ADMM iteration therefore costs one back-substitution plus $O(|\mathcal{S}|)$ elementwise projections.

\paragraph*{Initialization and termination}
On triangle meshes, we warm-start $u^0$ with \HM\ below a vertex-count threshold and with \FMM\ above it. We found that the choice doesn't matter for accuracy and convergence; the choice is guided by computational cost: \HM is cheaper on coarser meshes, where its prefactorization step dominates in finer ones, where \FMM is cheaper. On tetrahedral meshes, we warm-start with \HM\ throughout the range of sizes we consider. %
We also initialize $g^0 = G_\mathcal{S}u^0$ and the scaled dual-variable 
$w^0 = G_\mathcal{F}\bigl(\rho\,G_\mathcal{F}^\top M_\chi G_\mathcal{F}\bigr)^{-1}(-c)$. We terminate on a fixed iteration cap or once the relative field change drops below a tolerance; both, along with every other setting, are given in \tabref{tab:admm-settings}.

\paragraph*{Adaptive penalty scaling}
On tetrahedral meshes, we scale the ADMM penalty per element by the inverse of the local warm-start magnitude. This keeps the cone constraint active near the source; see Appendix~\ref{sec:adaptive_penalty_scaling} for the exact scheme.
On surfaces it makes little difference, so we keep the uniform penalty.

\paragraph*{Hardware}
On surfaces we run entirely on the CPU, as do all the surface baselines. On volumes we use a hybrid implementation that keeps the sequential factorization and triangular solves on the CPU and moves the independent per-sample projections to the GPU. All our volumetric timings come from this implementation. Among the volumetric baselines, \FIM\ is likewise GPU-accelerated, while \HM\ and \DFA\ run on the CPU. Details are in Appendix~\ref{sec:gpu-implementation}.

\begin{figure}[htbp] 
  \centering
  \setlength{\tabcolsep}{1pt} 
  \renewcommand{\arraystretch}{0} 

  \begin{tabular}{@{}cccc@{}} 
    \includegraphics[width=0.24\linewidth]{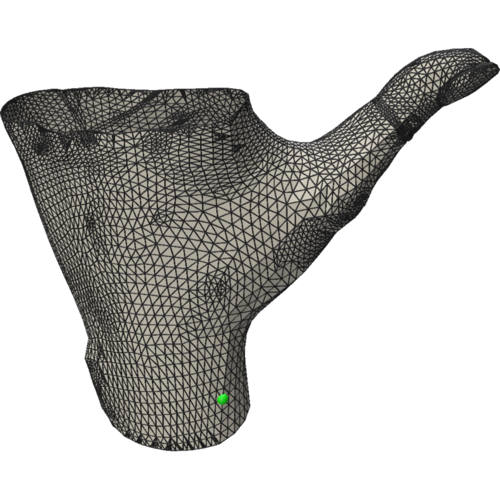} &
    \includegraphics[width=0.24\linewidth]{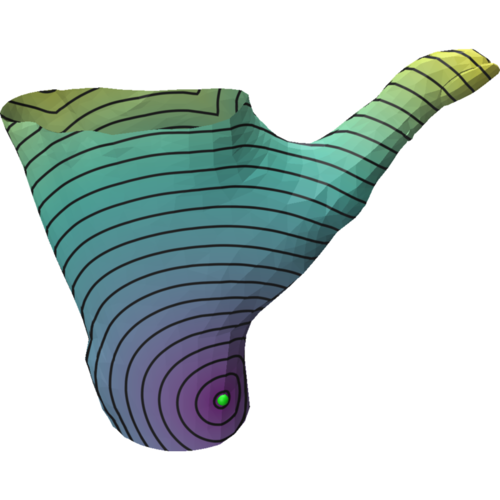} &
    \includegraphics[width=0.24\linewidth]{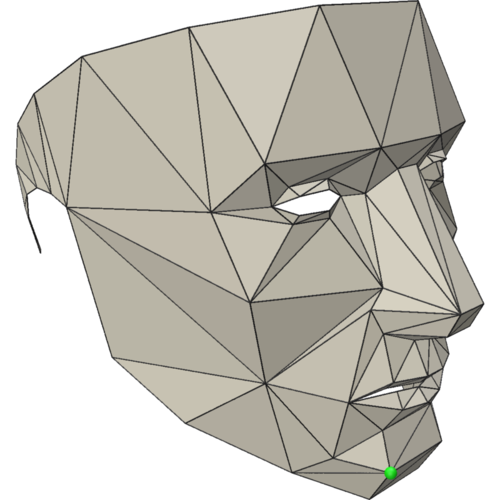} &
    \includegraphics[width=0.24\linewidth]{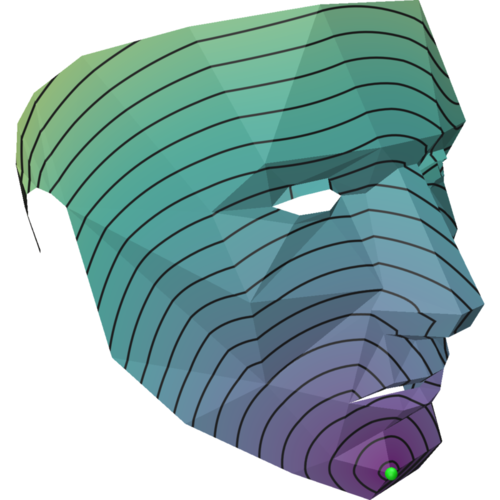} \\[2pt]
   \includegraphics[width=0.24\linewidth]{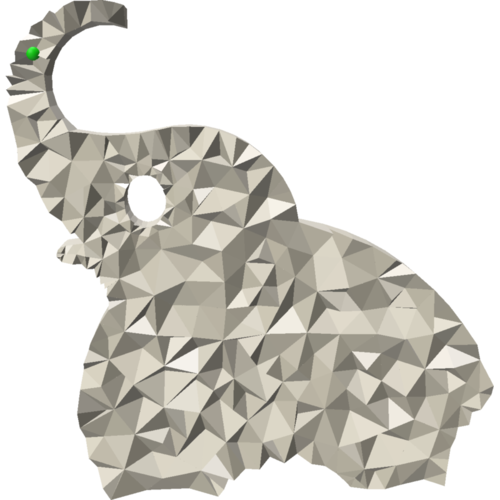} &
   \includegraphics[width=0.24\linewidth]{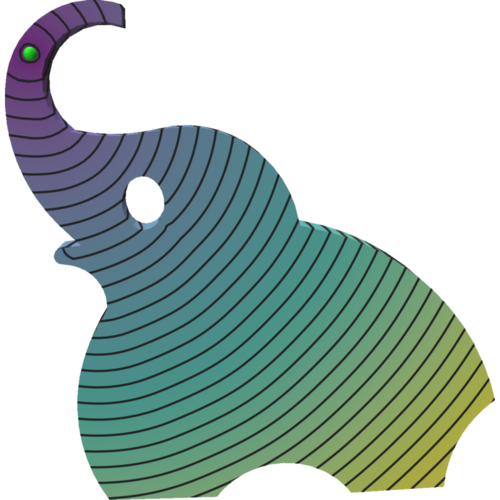} &
    \includegraphics[width=0.24\linewidth]{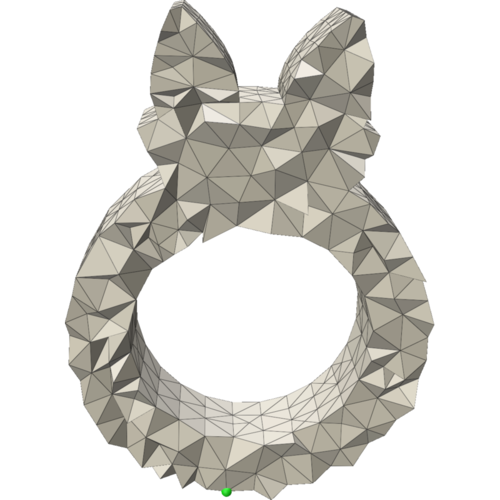} &
   \includegraphics[width=0.24\linewidth]{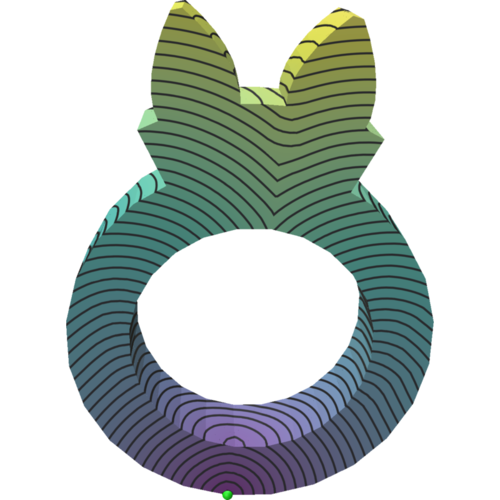} \\
  \end{tabular}

  \caption{Our method naturally handles boundaries.}
  \label{fig:boundary}
\end{figure}

\paragraph*{Boundary behaviour}
Mesh boundaries (in either 2D or 3D) are naturally handled by our algorithm without special boundary conditions. Like \DFA, we use only a gradient operator, where leaving the boundary nodal values (vertices and mid-edges) as degrees of freedom results in the correct boundary behaviour of Eq.~\eqref{eq:quadratic-convex-formulation}. Specifically, the resulting boundary condition is:
$$
\frac{\partial u}{\partial n}
    = \nabla u \cdot n \Rightarrow\ 
2d\frac{\partial d}{\partial n}
    = 2d\,\nabla d \cdot n \Rightarrow\ 
\frac{\partial d}{\partial n} = 
     \nabla d \cdot n.
$$

Note that the original \HM required an explicit fixing of boundary conditions, but that this is handled in a similar way to ours in~\cite{Feng:2024:SHM}. \figref{fig:boundary}  demonstrates our natural boundary behaviour.

\section{Evaluation}
\label{sec:experiments}

QDF is very simple to implement, we provide auxiliary code. We next provide an extensive evaluation to demonstrate its worth.  Our experiments validate three claims: 
\begin{itemize}
\item \textbf{(C1) Accuracy}: \ours is the most accurate of the finite element methods and achieves better convergence rates on both surface and volumetric meshes (\secref{subsec:benchmark-and-comparison}).
\item \textbf{(C2) Robustness}: \ours is stable under defects, noise, poor meshing, and non-manifold connectivity where exact methods are ill-posed or undefined (\secref{subsec:robustness}).
\item \textbf{(C3) Flexibility}: our formulation extends to anisotropic Riemannian metrics (\secref{subsec:anisotropic-metrics}), and allows for alternative source-configuration features (\secref{subsec:features}).
\end{itemize}

\subsection{Setup and protocol}
\label{subsec:setup}

\paragraph*{Datasets and preprocessing}
Our surface benchmark comprises $4{,}320$ triangle meshes from the Thingi10k dataset \cite{Thingi10K}, each satisfying our conditions of having a single component, no self intersections, and being a closed 2-manifold. These meshes are not remeshed and therefore vary greatly in quality.

Our volumetric benchmark contains $2{,}681$ tetrahedral meshes built from the same corpus. Each surface is first tetrahedralized by a robust constrained Delaunay method~\cite{AtteneCDT}, which reproduces the input surface exactly and applies no quality optimization.

 \paragraph*{Mesh hierarchies and query placement}
 For surfaces, we construct a refinement hierarchy using uniform 1-to-4 midpoint subdivision without deformation. The purpose of the refinement is to compute ground-truth reference solutions on fine-enough meshes, and to match degrees of freedom for the linear FEM methods (see below). We denote the original mesh by $\mathcal{M}_0$ and the mesh obtained after $\ell$ successive subdivision steps by $\mathcal{M}_\ell$.  For each input, we choose the largest available level $\mathcal{M}_K$, with $K\in\{1,2,3\}$, whose vertex count remains below approximately $1.5$M. Over the benchmark, $K=3$ for $4{,}028$ meshes, $K=2$ for $213$, and $K=1$ for $79$. The original meshes $\mathcal{M}_0$ range from $4$ to $377{,}154$ vertices. This protocol caps the computational cost of the reference solve while retaining the geometry of the original polyhedral surface.
 
For volumes, we build a ladder of refinements from the constrained Delaunay tetrahedralization using TetGen~\cite{tetgen2015}. Each rung is a single refinement run at a prescribed target volume. The base mesh $\mathcal{M}_0$ uses a target volume determined from the median edge length of the surface and is the mesh on which our method is evaluated. The mesh $\mathcal{M}_1$ is fitted so that its vertex count matches the number of nodal unknowns in our method ($|\mathcal{V}|+|\mathcal{E}|$), and is used for the piecewise-linear baselines. The reference mesh $\mathcal{M}_2$ is the finest mesh that fits within a $350$K-tetrahedron budget. To hit this target from below, we do a short bisection search over the edge-length parameters. No quality switches are used at any level. We use TetGen's refinement mode without Delaunay modification and impose only the volume constraint. Consequently,  slivers and uneven grading persist throughout the ladder, which is the regime we intend to test.  We keep 2,681 of the 4,320 meshes because the rest fail an exact self-intersection test, not conclude in TetGen within our size and time budgets, or leave no room for a valid strictly finer reference mesh. Appendix~\ref{sec:tet-corpus} details the full benchmark-production pipeline. %

For each surface mesh, we select as the source the vertex nearest to the mean of all vertex positions. For a volumetric mesh, we apply the same rule to its boundary surface. Because the constrained refinement preserves the boundary vertices, we select this source once on the common boundary and reuse the same vertex across $\mathcal{M}_0$, $\mathcal{M}_1$, and $\mathcal{M}_2$.

 \paragraph*{Reference solutions}
 When an analytic solution is available, as in the star-shaped domains in \figref{fig:closed_form}, we place the source in the kernel and compare against the closed-form Euclidean distance in both 2D and 3D. Otherwise, for surfaces, we use the output of \VTP\ on the finest available level $\mathcal{M}_K$ as the reference solution. Because subdivision does not alter the polyhedral geometry, this gives exact distances at the vertices of a denser evaluation mesh while preserving the original domain. For general volumetric examples, an exact reference method is not available. We therefore compute a high-resolution numerical reference using our ADMM solver on $\mathcal{M}_2$, with tolerance $\epsilon=10^{-4}$ and an iteration cap of $20{,}000$. Our default configuration is evaluated on $\mathcal{M}_0$ with $\epsilon=10^{-3}$ and a cap of $2{,}000$ iterations. More details are provided in Appendix~\ref{sec:tet-corpus}.
 
\paragraph*{Surface-specific methods and comparison scope} The polyhedral-exact methods are numerically accurate on well-connected 2-manifold meshes, and therefore we do not compare to them, but rather treat them as reference solutions. We compute them on the finest mesh so to have a dense representation of the correct distance conveniently on the vertices alone. We do however challenge their robustness (\figref{fig:gaussian_noise} and \figref{fig:robust_holes}) and discuss their trade-offs in a closed-form example (\figref{fig:sphere_convergence}).  The exception we make is to compare to GSP, which is designed as a fast approximate method.

\paragraph*{Baselines}
Our field-accuracy claim concerns numerical discretizations evaluated under a comparable degree-of-freedom budget. We therefore use  \HM, \DFA, and \FMM\ on surfaces and \FIM\ on volumes, for the field-error comparisons in \tabref{tab:comparison}, \figref{fig:benchmark_hist}, \figref{fig:ratio_hist}, \tabref{tab:comparison_3d}, \figref{fig:benchmark_hist_3d}, \figref{fig:ratio_hist_3d}, \figref{fig:convergence_table}, and \figref{fig:convergence_table_3d}. We additionally include \GSP\ in the surface comparisons, as discussed above. \DFA\ is an ADMM method like \ours, making it the closest baseline. In our reimplementation, we give \DFA\ the same \HM\ warm start used by our method and tune its ADMM parameters separately (\secref{subsec:admm-algorithm}), so that the comparison primarily reflects the formulation rather than the initialization. \HM\ therefore serves both as a surface baseline and as the common initialization for the two ADMM methods. \FMM\ and \FIM\ belong to the same family, but \FIM\ is designed for tetrahedral meshes and anisotropic speed functions. It extends \FMM's obtuse-angle construction to solid angles and tolerates the poorly shaped elements deliberately retained in our volume hierarchy.

\paragraph*{Degree-of-freedom matching}
For a fair comparison, we match the number of  degrees of freedom. On surfaces, our method is evaluated on $\mathcal{M}_0$, whereas competing piecewise-linear methods are evaluated on $\mathcal{M}_1$. The vertex set of $\mathcal{M}_1$ corresponds to the $|\mathcal{V}|+|\mathcal{E}|$ nodal values of our method on $\mathcal{M}_0$. In the convergence tests (\figref{fig:convergence_table}), our result on any $\mathcal{M}_N$ is compared with competing methods on the relative $\mathcal{M}_{N+1}$. On volumes, our method is evaluated on $\mathcal{M}_0$, while competing methods are evaluated on $\mathcal{M}_1$, whose vertex count is fitted to our $|\mathcal{V}|+|\mathcal{E}|$ degrees of freedom. We note that using a coarse mesh for the linear method resulted in worse errors.

\paragraph*{Error metrics}  To evaluate accuracy, we upsampled all solutions to the finest subdivision level, and measured errors \emph{only} on the finest mesh relative to the ground truth (non-squared) distance $d_{GT}$. Let $e = \frac{|d_{\text{upsample}} - d_{GT}|}{|d_{GT}|}$ denote the relative error vector. We report the \emph{relative} $L_2$ error, defined as $L_2(e)=\sqrt{e^TM_1e}$ error, where $M_1$ is the piecewise linear (PL) scalar mass matrix of the finest mesh. To ensure comparability across models of different volumes, we use the relative root-mean-square error: $RMSE(e) = \frac{L_2(e)}{L_2(\mathbf{1})}$. Furthermore, we report the $L_\infty$ error, defined as $L_\infty = \max |e|$. The reason we use relative L2 error is to weigh errors close to the source more fairly. 
	
\paragraph*{Implementation} Our experimental pipeline is implemented in Python. For geodesic-distance computations, we use the released implementation of \VTP\ for the exact surface reference, Potpourri3d (a Python binding for geometry-central~\cite{geometrycentral}) for \HM\ and \FMM, and the reference SCI-Solver\_Eikonal implementation of \citet{fim_fu_2013} for \FIM. We reimplemented \DFA. 
Performance benchmarks were conducted on a workstation with an Intel Core i9-13900K CPU and 64 GB of RAM. We cap the memory used by any single solve at $62$\,GB. Runs exceeding that cap are reported as out of memory.

\subsection{Accuracy (C1)}
\label{subsec:benchmark-and-comparison}
\begin{table}[t]
    \centering
    \small
    \begin{tabular*}{\linewidth}{@{\extracolsep{\fill}} l ccc}
        \toprule 
        Methods  & RMSE & $L_\infty$ & Time (s) \\
        \midrule
        \GSP & 9.46e-02 & 9.99e-2 & 0.30 \\
       \HM           &  1.13e-01     &  1.36e-01 & 1.67 \\
        \FMM       &    2.17e-01      &  1.49e-01  & 0.19 \\
       \DFA            &    1.98e-01      &  1.89e-01   & 4.62  \\
       Ours (\ours)                       & \textbf{ 5.18e-02 }  & \textbf{ 9.56e-02 } & 4.42   \\
        \bottomrule
    \end{tabular*}
    
    \caption{Quantitative accuracy comparison on the surface benchmark. \ours is the most accurate, averaged over the entire benchmark. Our method has superior accuracy, and while being slower, still has comparable runtime.}
    \label{tab:comparison}
\end{table}

\begin{figure}[t]
  \centering
  \begin{subfigure}[t]{0.49\columnwidth}
    \centering
    \includegraphics[width=\linewidth]{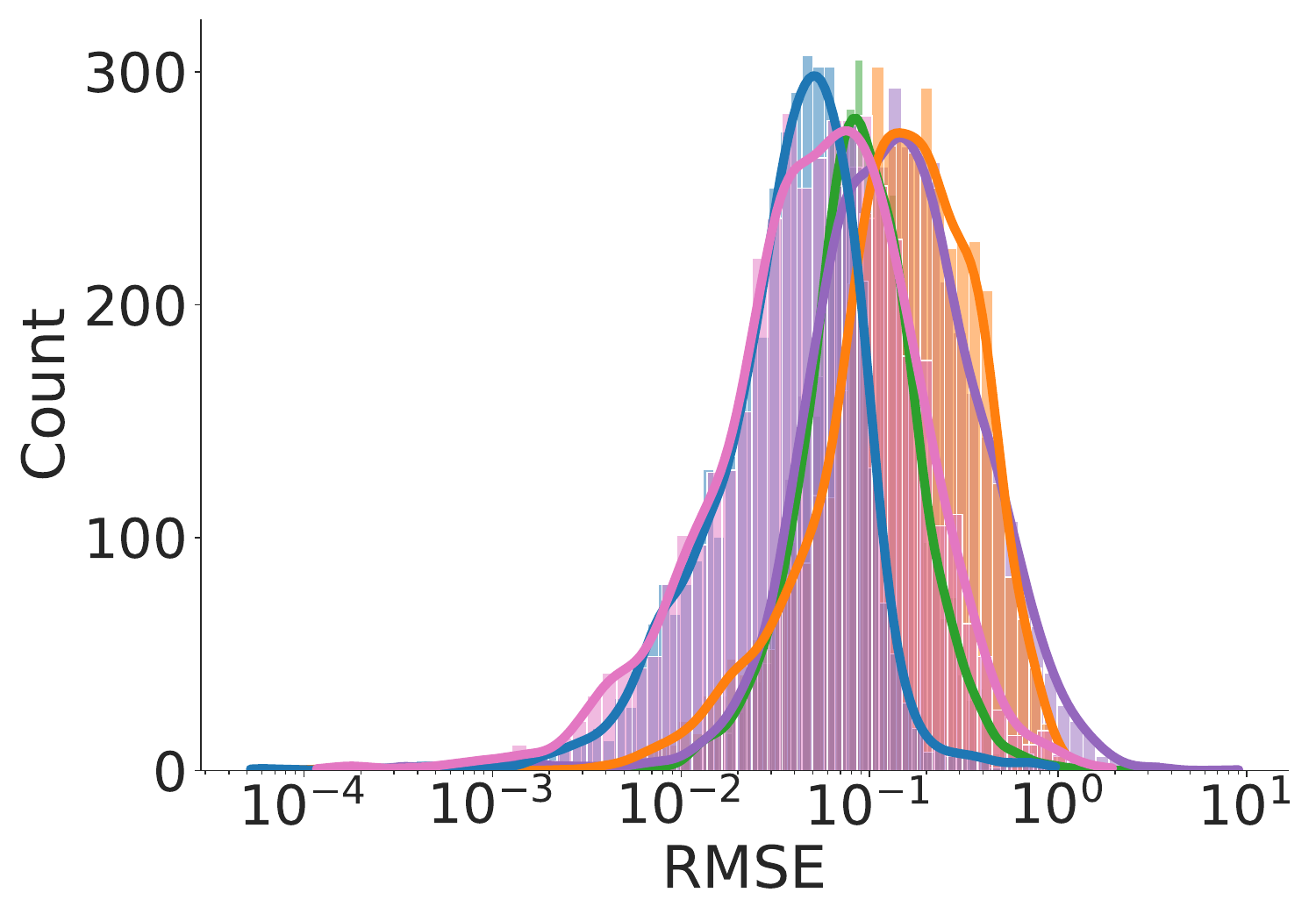}

  \end{subfigure}\hfill
  \begin{subfigure}[t]{0.49\columnwidth}
    \centering
    \includegraphics[width=\linewidth]{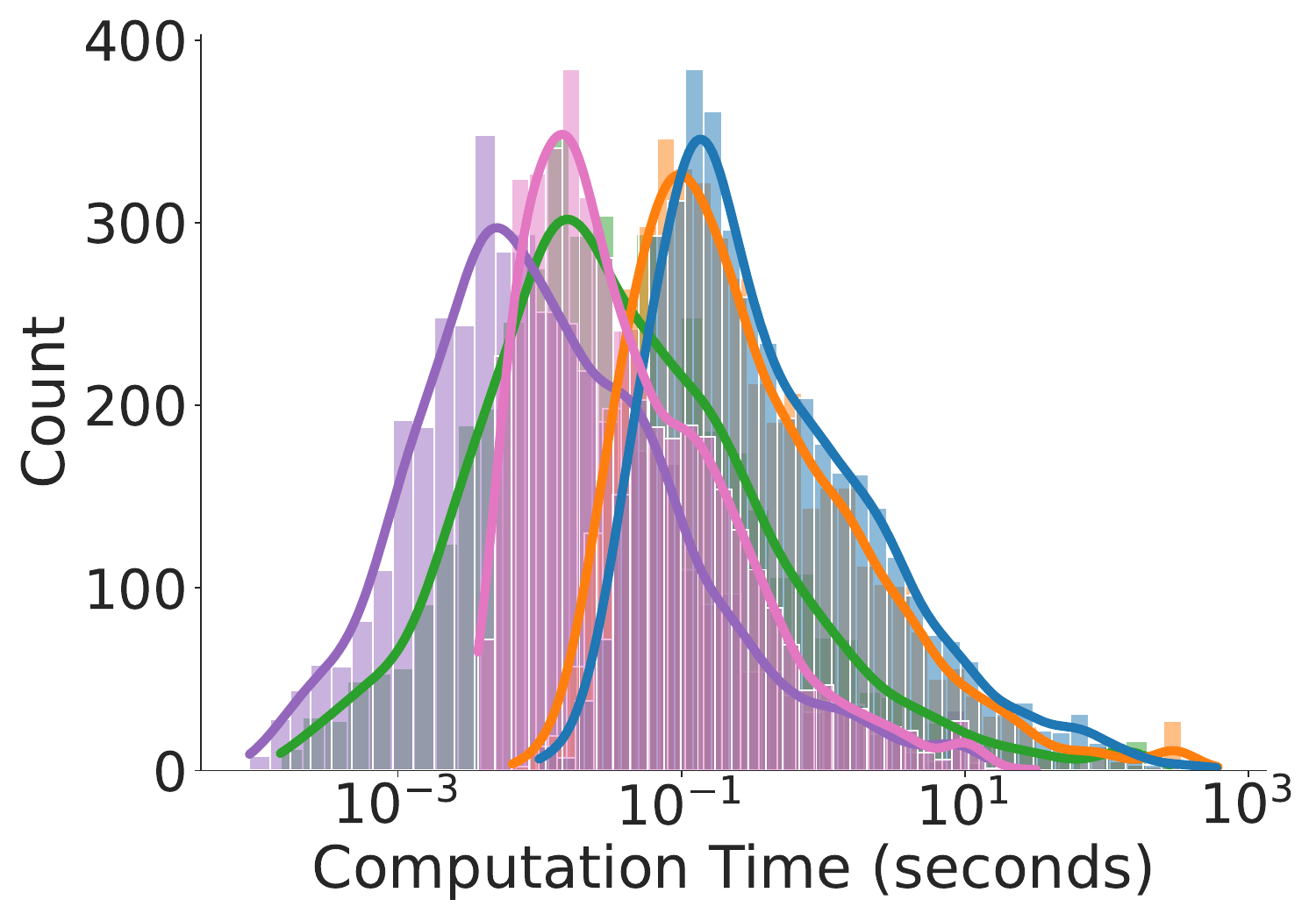}
  \end{subfigure}
    \vspace{2pt}
    \includegraphics[width=\linewidth]{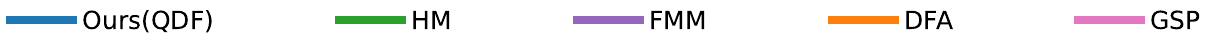}
    \vspace{-6pt}
    \setlength{\abovecaptionskip}{1pt}
  \caption{Distributions over the surface benchmark. \ours is concentrated at markedly lower error, but usually slower than the four. We explore this trade-off in \figref{fig:pareto}.}
  \label{fig:benchmark_hist}
\end{figure}

\begin{figure}[t]
  \centering
  
  \begin{subfigure}[t]{0.225\columnwidth}
    \centering
    \includegraphics[width=\linewidth]{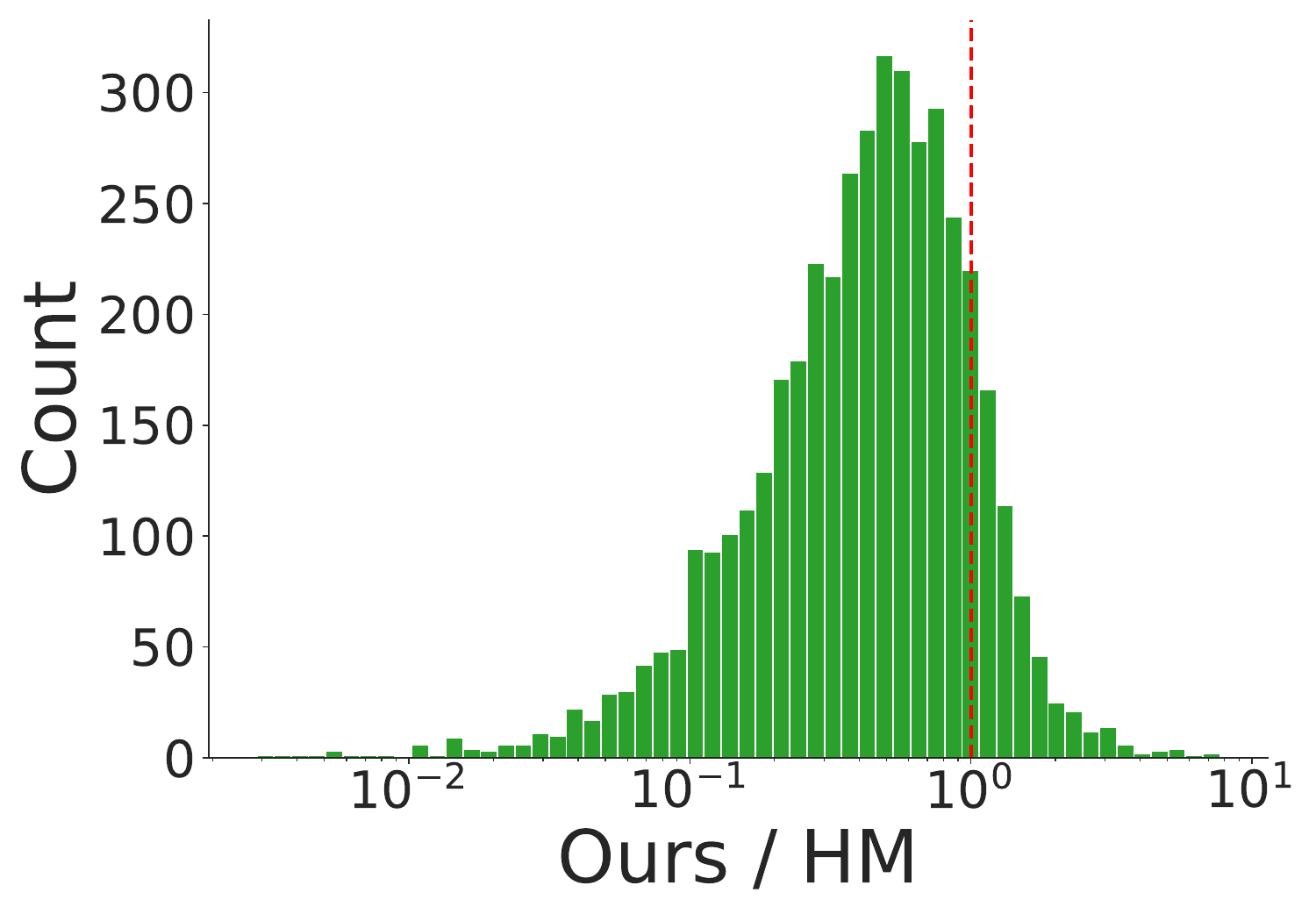}
  \end{subfigure}
  \hfill
  \begin{subfigure}[t]{0.225\columnwidth}
    \centering
    \includegraphics[width=\linewidth]{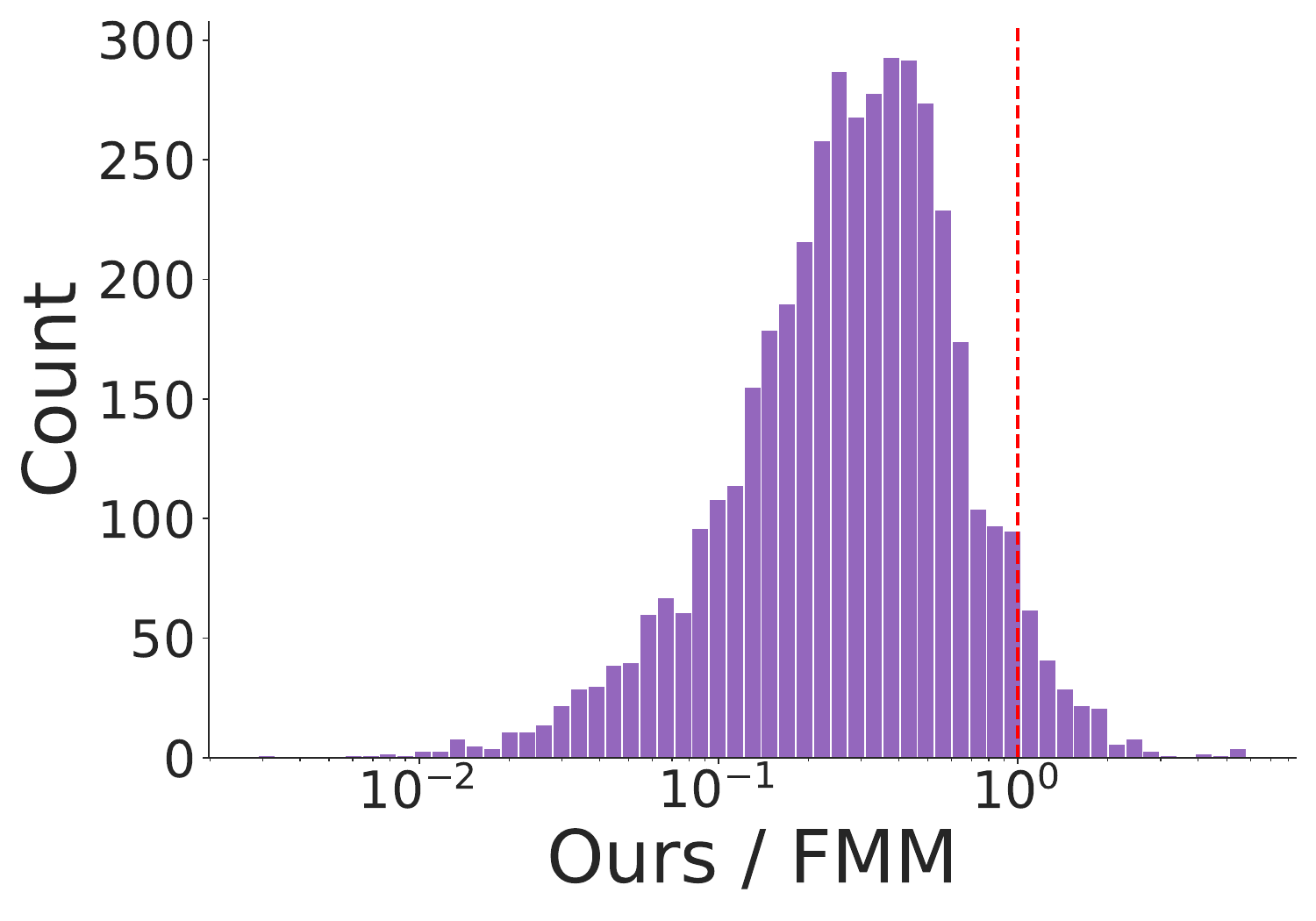}
  \end{subfigure}
  \hfill
  \begin{subfigure}[t]{0.225\columnwidth}
    \centering
    \includegraphics[width=\linewidth]{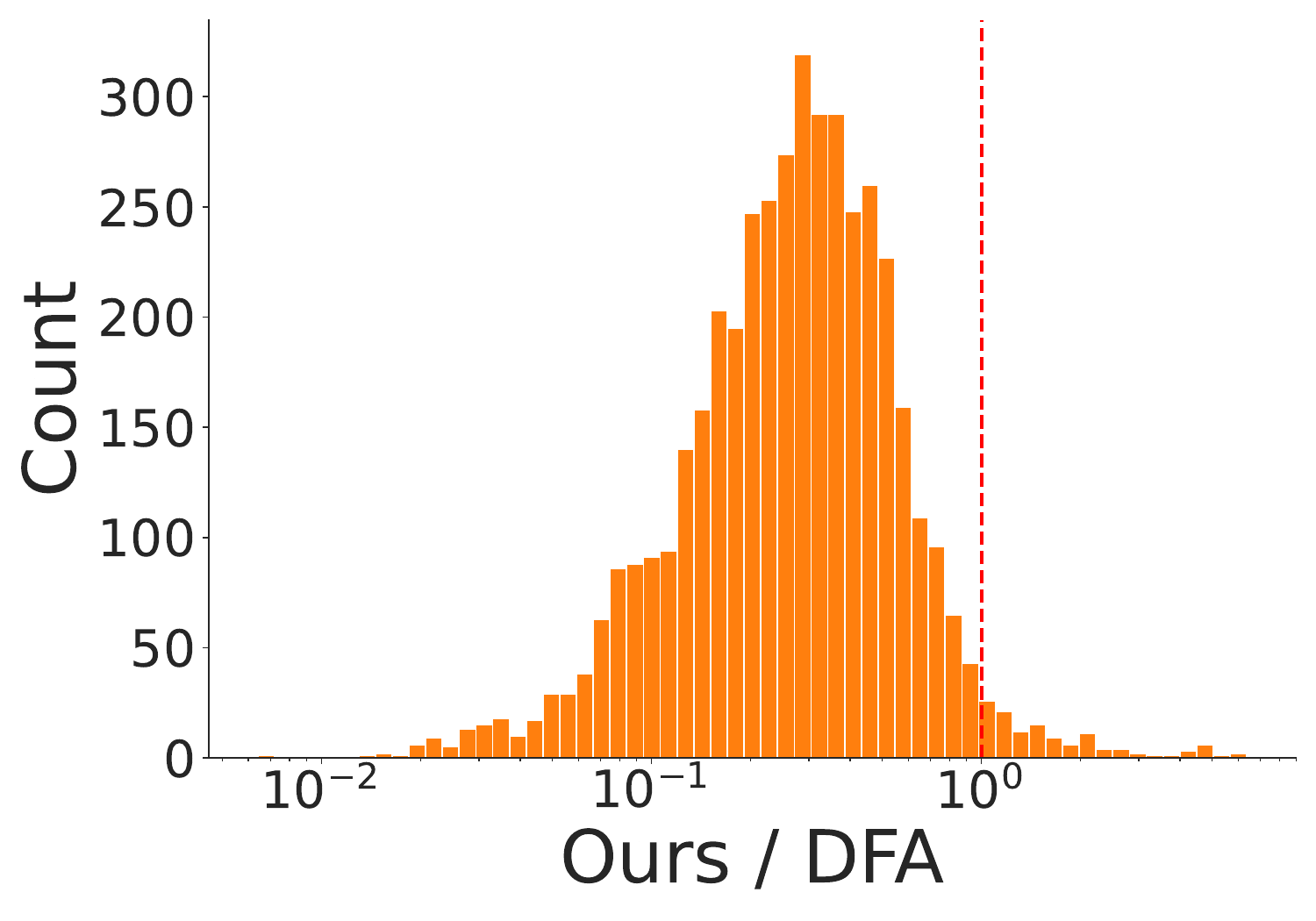}
  \end{subfigure}
 \begin{subfigure}[t]{0.225\columnwidth}
	\centering
	\includegraphics[width=\linewidth]{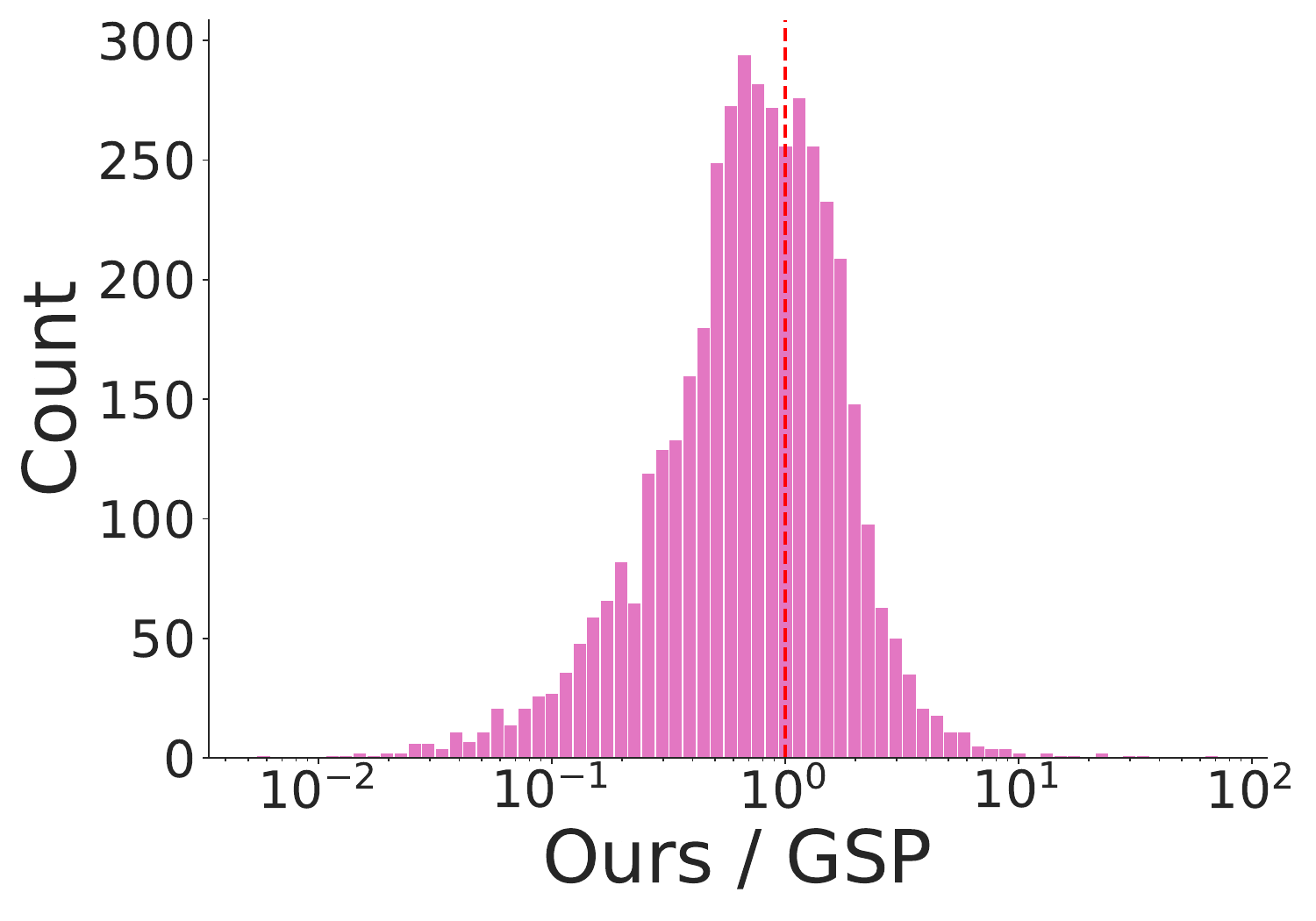}
\end{subfigure}

  \caption{Histograms of $L_2$ error ratios per mesh across finite-element methods. \QDF achieves ratio$<$1 (left of red line) on a large majority of the cases.}
  \label{fig:ratio_hist}
\end{figure}

\begin{table}[t]
    \centering
    \small
    \begin{tabular*}{\linewidth}{@{\extracolsep{\fill}} l ccc}
        \toprule 
        Methods  & RMSE & $L_\infty$ & Time (s) \\
        \midrule
        \HM     &   1.40e-01     &  5.23e-01  & 0.25\\
        \FIM   &    3.91e-02      &   6.77e-01  & 0.86\\
        \DFA  &     4.40e-02     &  5.88e-01    & 0.99 \\
        Ours (\ours) & \textbf{ 2.50e-02 }  & \textbf{2.72e-01 } & 1.14 \\
        \bottomrule
    \end{tabular*}
    
    \caption{Quantitative accuracy comparison over the volumetric benchmark. \ours is the most accurate, and only a bit slower.}
    \label{tab:comparison_3d}
\end{table}

\begin{figure}[t]
  \centering
  \begin{subfigure}[t]{0.49\columnwidth}
    \centering
    \includegraphics[width=\linewidth]{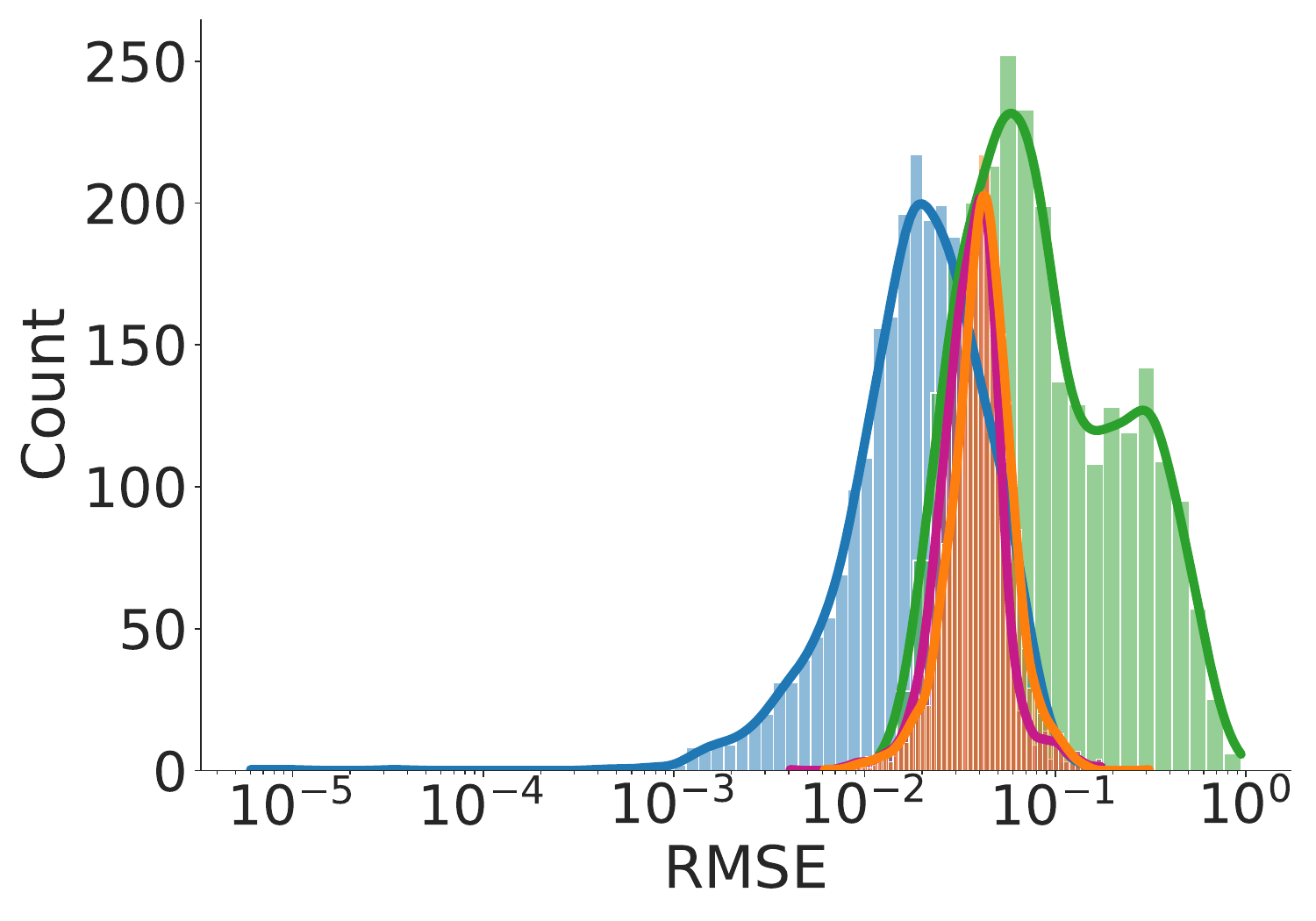}

  \end{subfigure}\hfill
  \begin{subfigure}[t]{0.49\columnwidth}
    \centering
    \includegraphics[width=\linewidth]{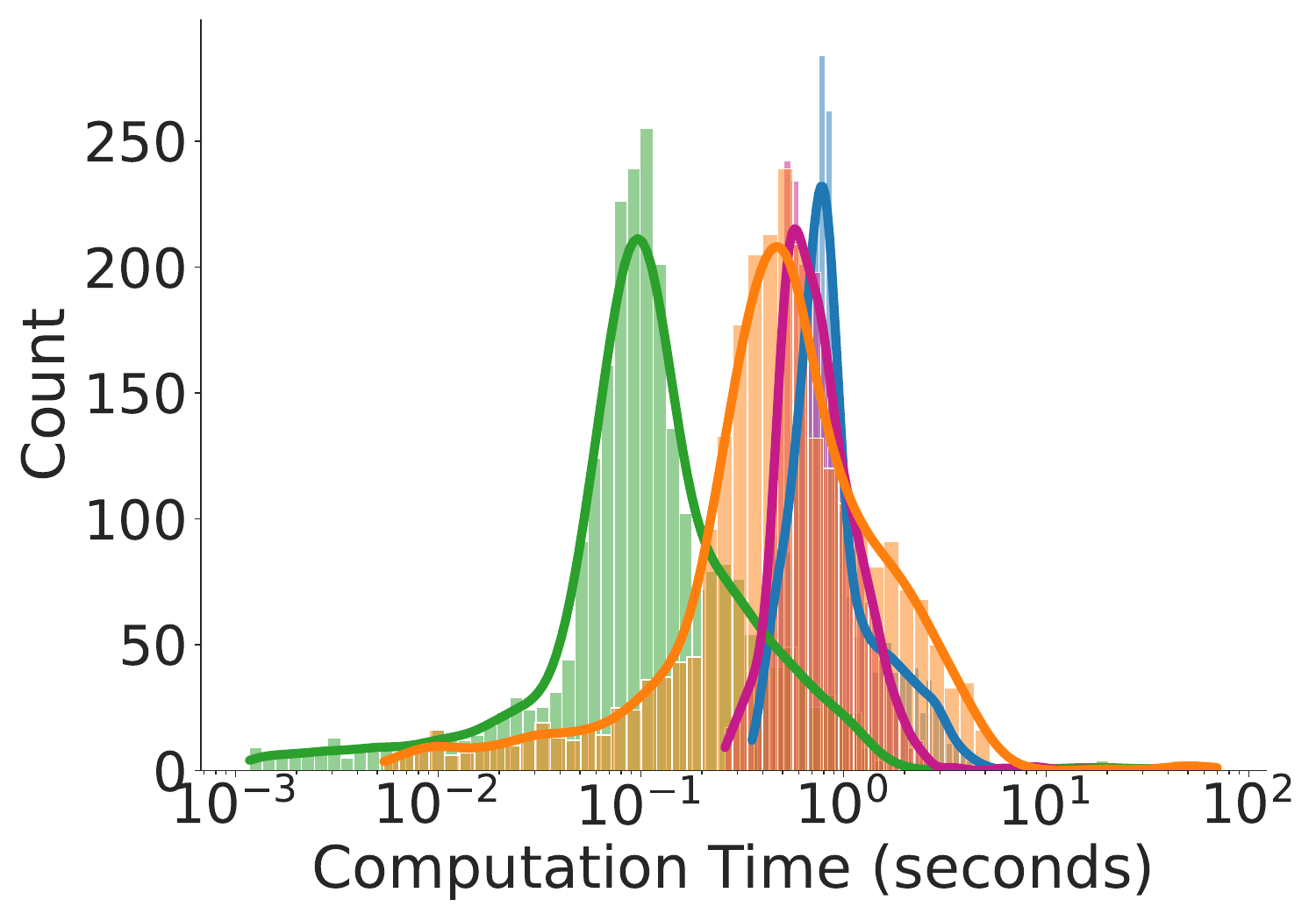}
  \end{subfigure}
    \vspace{2pt}
    \includegraphics[width=\linewidth]{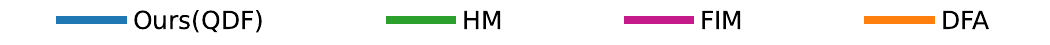}
    \vspace{-6pt}
    \setlength{\abovecaptionskip}{1pt}
  \caption{Distribution over the volumetric benchmark.  \ours consistently has better error distribution. }
  \label{fig:benchmark_hist_3d}
\end{figure}

\begin{figure}[t]
  \centering
  
  \begin{subfigure}[t]{0.325\columnwidth}
    \centering
    \includegraphics[width=\linewidth]{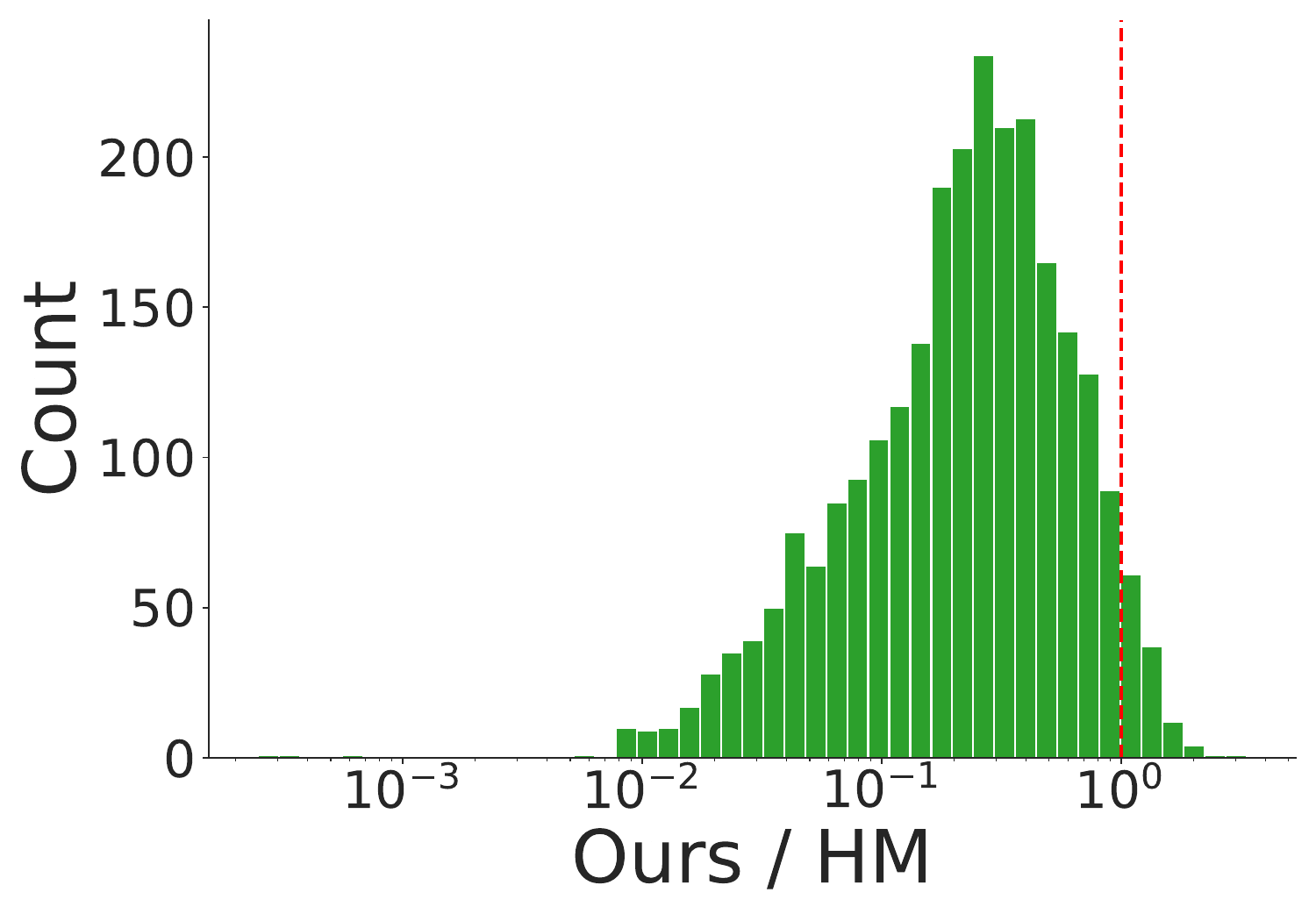}
  \end{subfigure}
  \hfill
  \begin{subfigure}[t]{0.325\columnwidth}
    \centering
    \includegraphics[width=\linewidth]{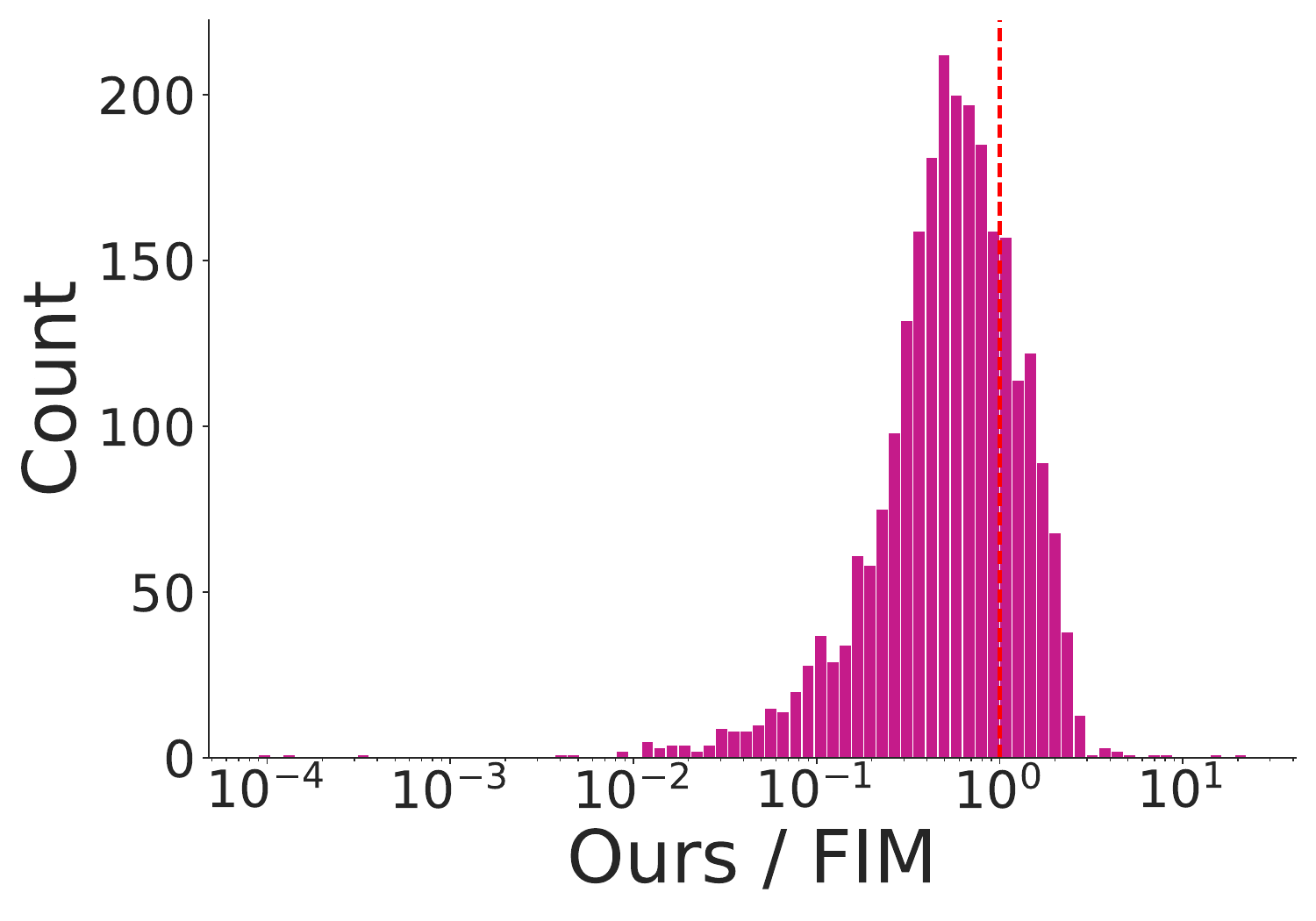}
  \end{subfigure}
  \hfill
  \begin{subfigure}[t]{0.325\columnwidth}
    \centering
    \includegraphics[width=\linewidth]{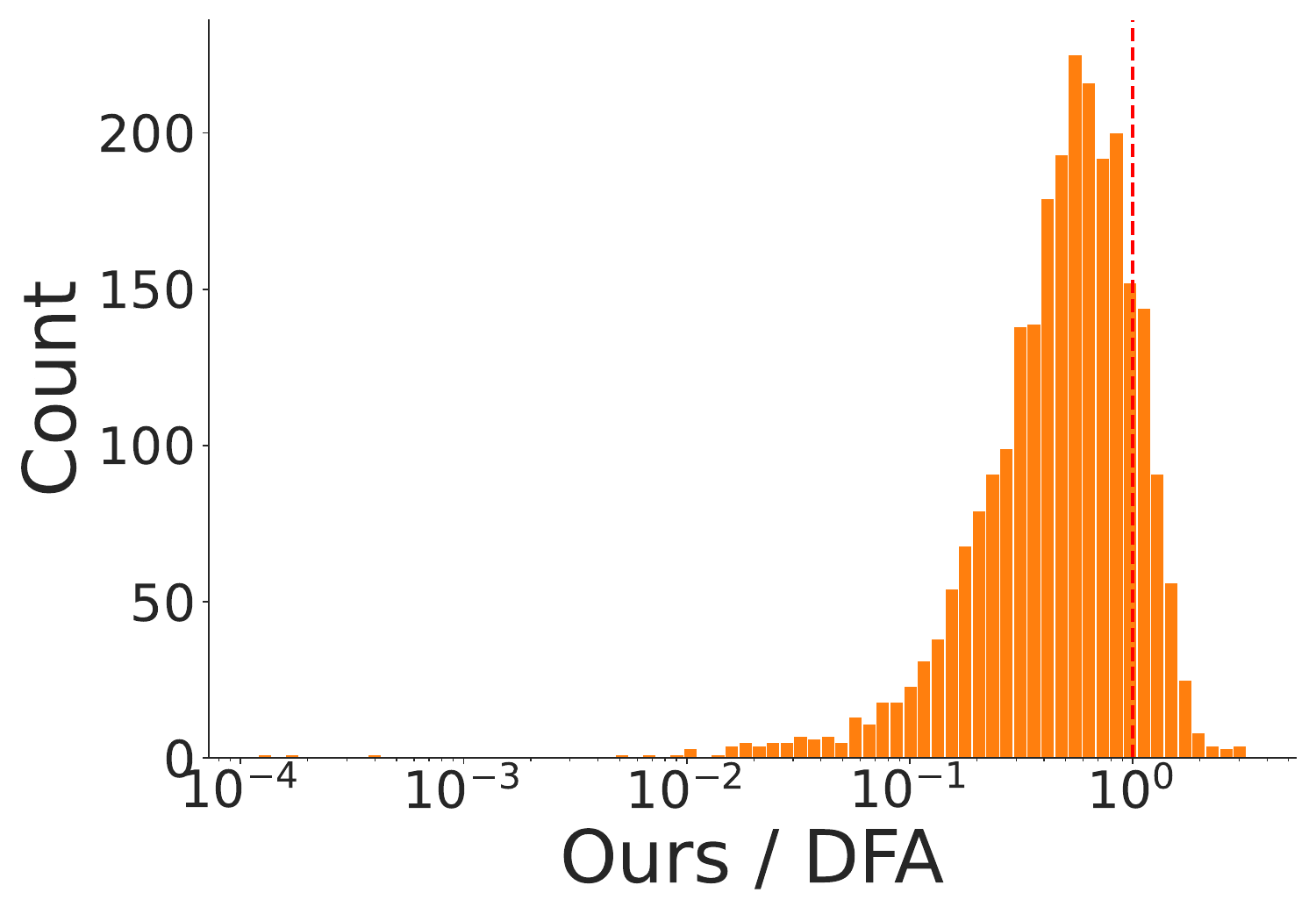}
  \end{subfigure}

  \caption{Histograms of per-mesh $L_2$ error ratios between \ours and each FE method on the volumetric benchmark. \ours result is overwhelmingly better (ratio $<$1, red line). }
  \label{fig:ratio_hist_3d}

\end{figure}

\paragraph*{Surface benchmark}
\tabref{tab:comparison} reports the benchmark-averaged errors and runtimes, while \figref{fig:benchmark_hist} and \figref{fig:ratio_hist} show their distributions over individual meshes. \ours\ achieves the lowest mean error under the degree-of-freedom-matched protocol.  Its RMSE and $L_\infty$ error are substantially lower than the competing methods across the benchmark. \figref{fig:ratio_hist} shows per-mesh error ratios against each method, demonstrating that our method outperforms the baselines in the majority of cases. As expected, our mean runtime is slower than most methods. Interesting, we are a bit fast than \DFA, but this is attributed to both \ours and \DFA running \HM as an initial solution, where \DFA has to run it on $\mathcal{M}_1$ and we run it on $\mathcal{M}_0$ (coarser). GSP is the 2nd best method, being a polyhedral method, but our error profile is still better.

\begin{figure}[t]
    \centering
    \begin{tabular}{@{}c@{\hspace{4pt}}c@{}}
        \includegraphics[width=0.48\linewidth]{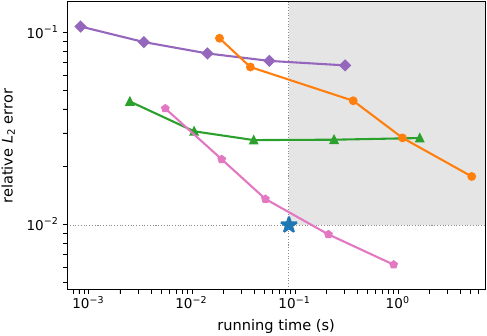} &
        \includegraphics[width=0.48\linewidth]{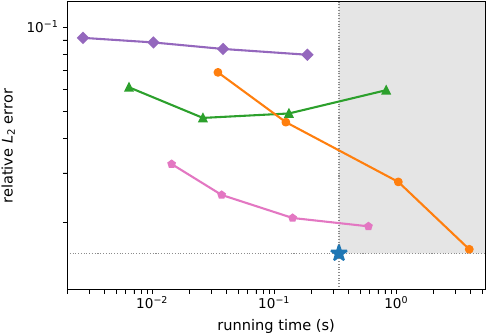} \\[2pt]
        {\small (a) Thingi10K 1417956} & {\small (b) Thingi10K 37964} \\
    \end{tabular}

    \includegraphics[width=0.7\linewidth]{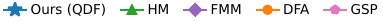}
    \caption{\textbf{Accuracy versus runtime} on two randomly sampled
    	surface meshes. The star shows \ours\ on the original mesh
    	$\mathcal{M}_0$; each curve shows a baseline over successive 1-to-4
    	midpoint refinement levels, beginning at $\mathcal{M}_1$. The shaded
    	region contains solutions that are both slower and less accurate than
    	\ours. In (a), only \GSP reaches and sightly improves upon our error with more time spent.  In (b), only \DFA reaches and slightly improves upon
    	our error, at its finest and most expensive level. }
    \label{fig:pareto}
\end{figure}

\paragraph{Accuracy versus runtime}
To get a better sense of our runtime efficiency, in \figref{fig:pareto} we ask whether the baselines can recover the accuracy gap by spending more computation on increasingly refined meshes. The star shows the result of \ours\ on the original coarse mesh $\mathcal{M}_0$, while each curve shows a competing method run over successive refinement levels beginning at $\mathcal{M}_1$ (that means the degrees of freedom are equal). We do this on two random examples from the benchmark, and measure runtime vs. accuracy reached. On the first method, only GSP reached our accuracy eventually, and on the second only \DFA (after substantial refinement).  \HM\ and \FMM\ remain less accurate throughout. This illustrate that our method is very time-efficient with regards to the accuracy gained.

\paragraph{Tetrahedral benchmark} 
The volumetric results are summarized in \tabref{tab:comparison_3d}, \figref{fig:benchmark_hist_3d}, and \figref{fig:ratio_hist_3d}. \ours\ again achieves the lowest benchmark-averaged RMSE and $L_\infty$ error. Its runtime remains comparable to those of \FIM\ and \DFA, although it is slower than \HM. As in the surface experiments, this additional cost reflects the optimization performed after initialization to obtain a more accurate solution.

\begin{figure*}[htbp]
    \centering
    \setlength{\tabcolsep}{1pt} 

    \begin{tabular}{ C{0.15\textwidth} C{0.15\textwidth} @{\hspace{2pt}}|@{\hspace{15pt}} C{0.15\textwidth} C{0.15\textwidth} @{\hspace{2pt}}|@{\hspace{9pt}} C{0.15\textwidth} C{0.15\textwidth} }
        \hline 
        \multicolumn{2}{c |}{\small \textbf{Triangulation quality: good}} & 
        \multicolumn{2}{c |}{\small \textbf{Triangulation quality: poor}} & 
        \multicolumn{2}{c}{\small \textbf{Triangulation quality: extremely poor}} \\[6pt]
        
        \imglabel{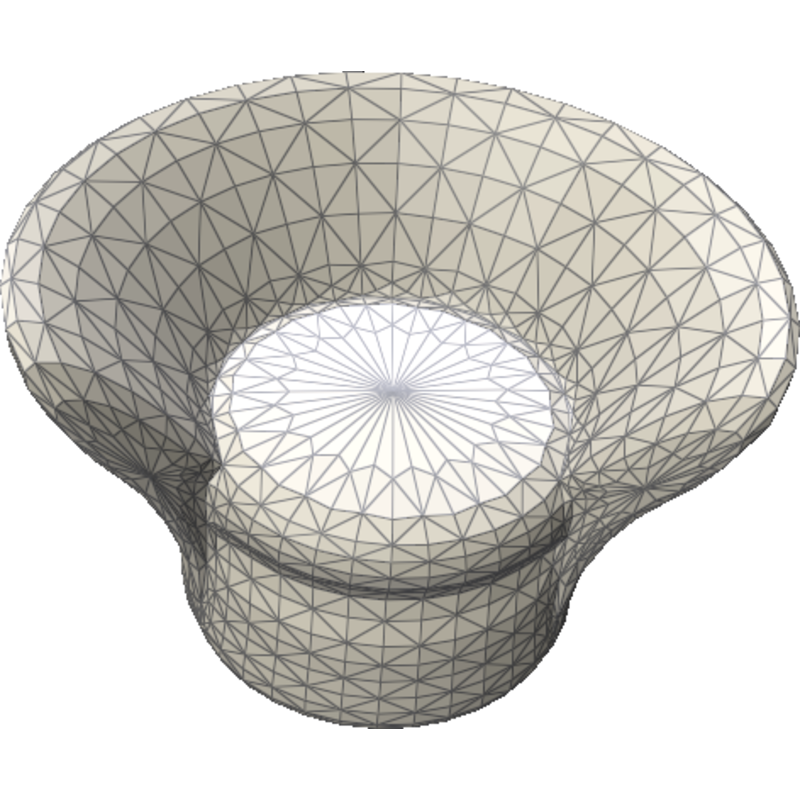}{57909} & 
        \imgwithlabeltiny{width=\linewidth}{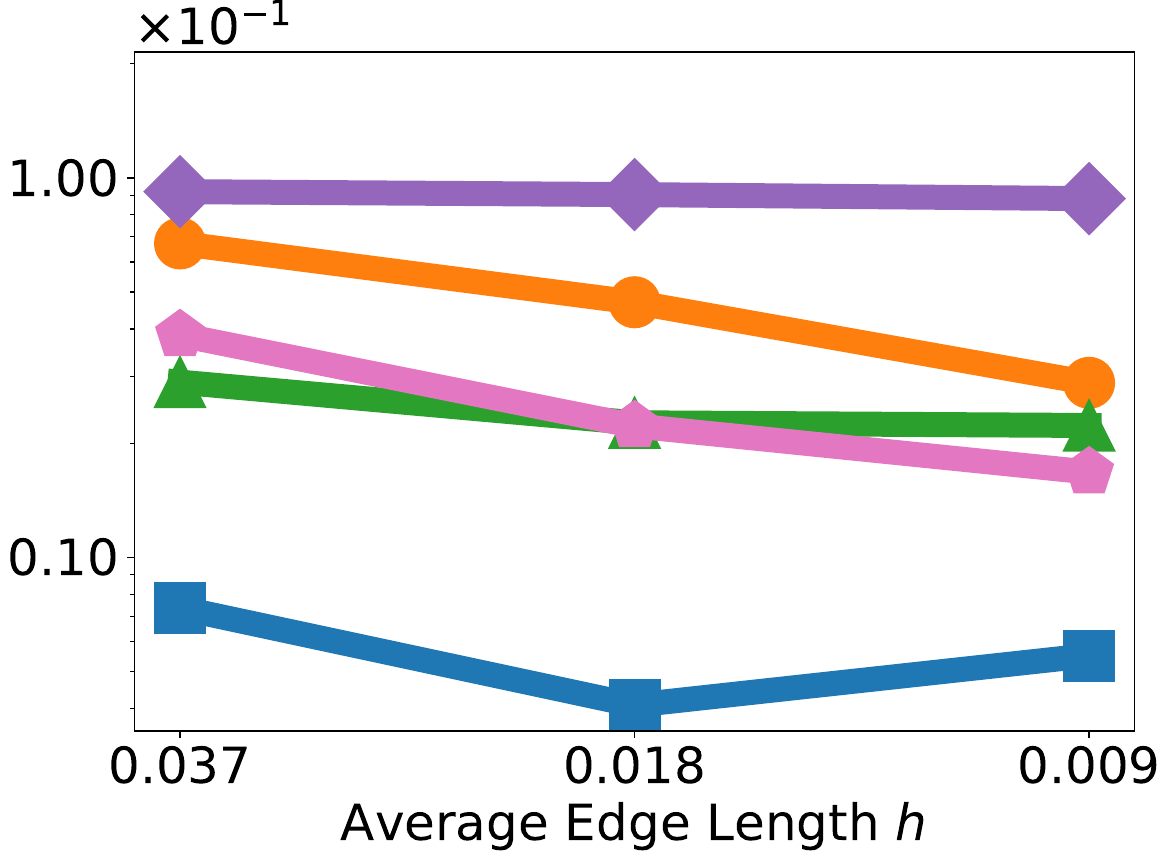}{$L_2$} & 
        \imglabel{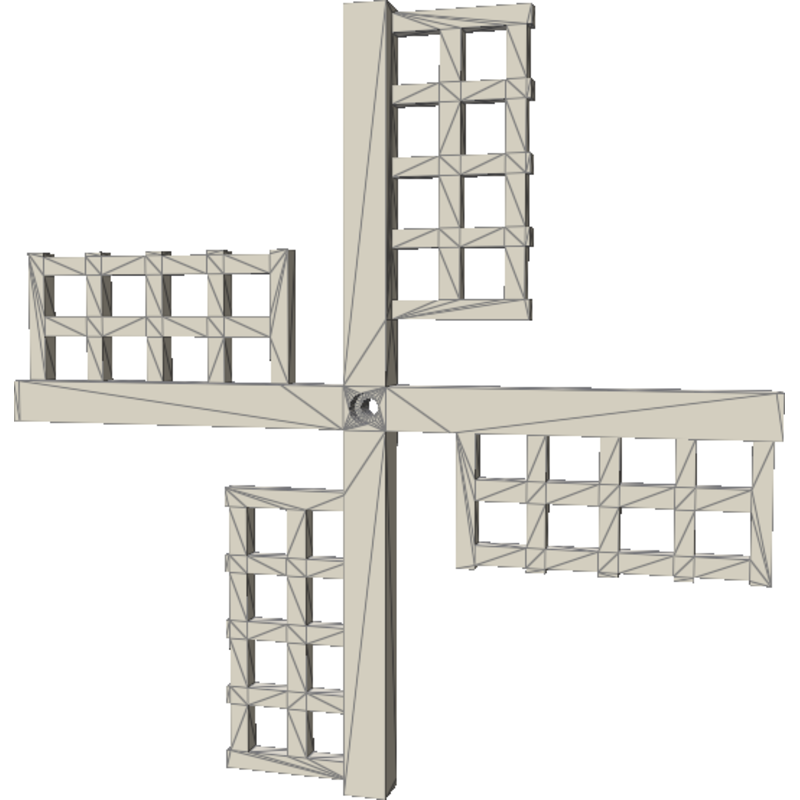}{162338} & 
        \imgwithlabeltiny{width=\linewidth}{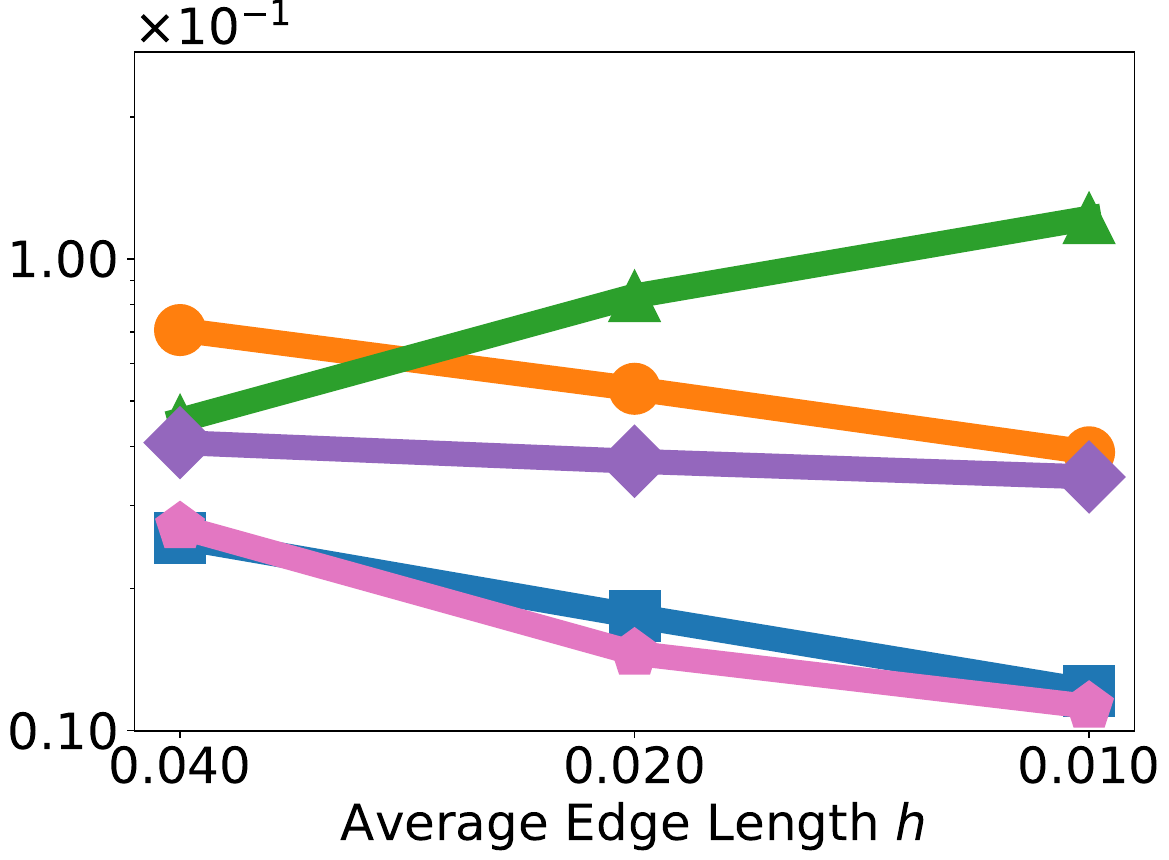}{$L_2$} & 
        \imglabel{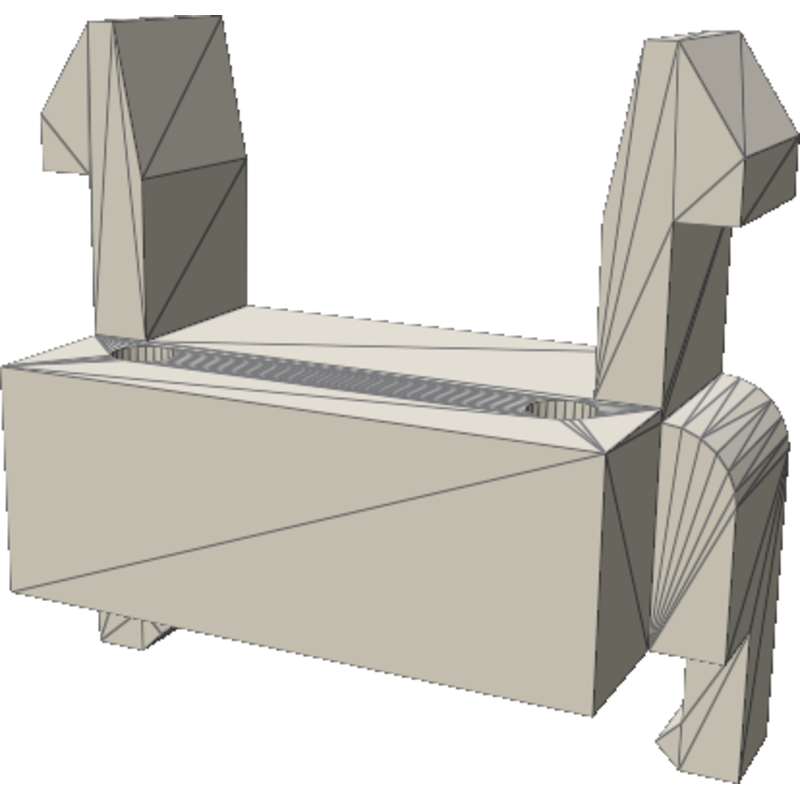}{1566089} & 
        \imgwithlabeltiny{width=\linewidth}{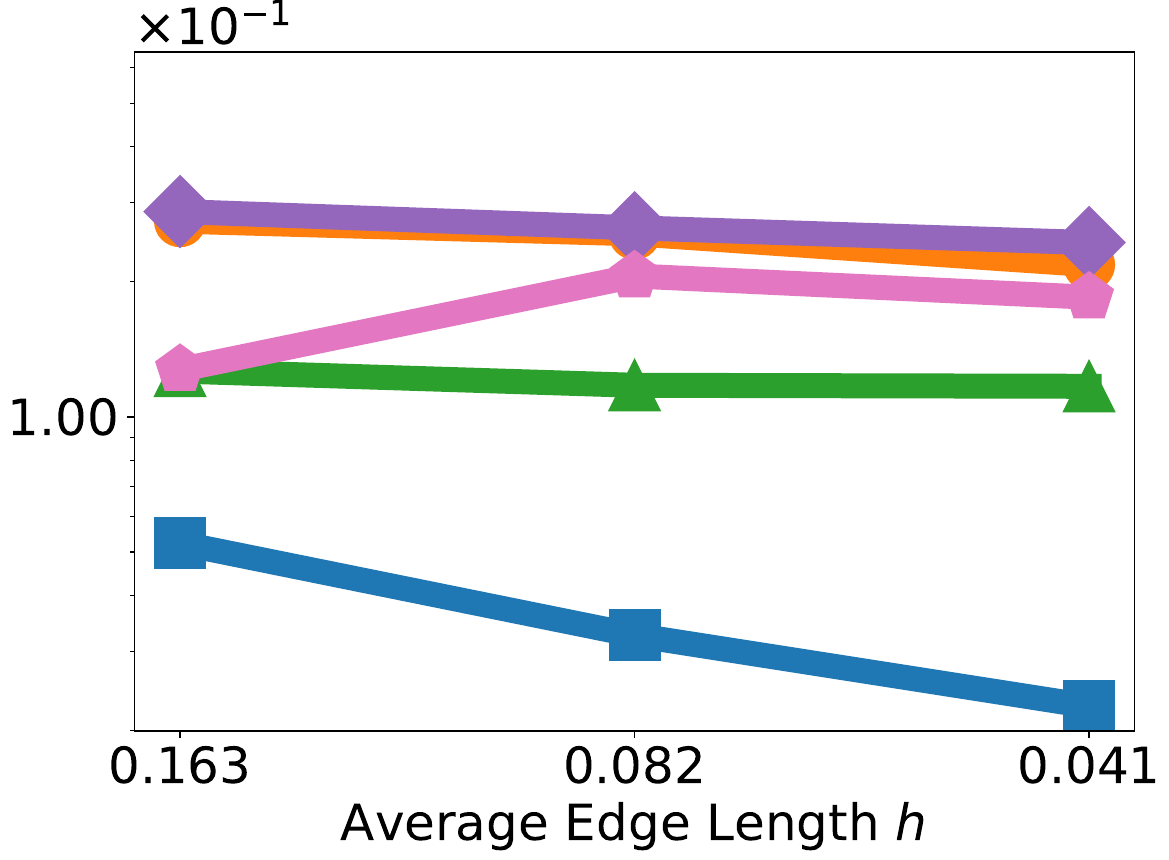}{$L_2$} \\
        
        \includegraphics[width=\linewidth]{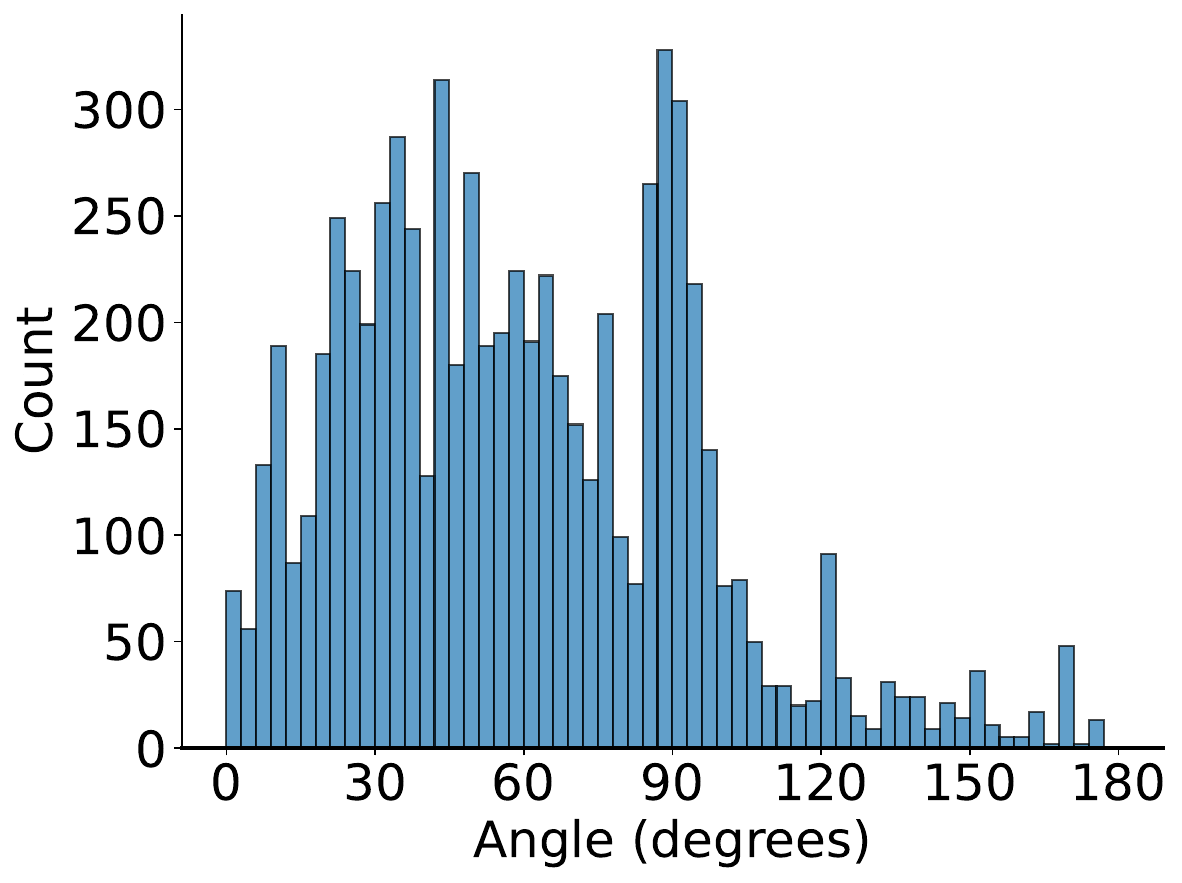} &
        \imgwithlabeltiny{width=\linewidth}{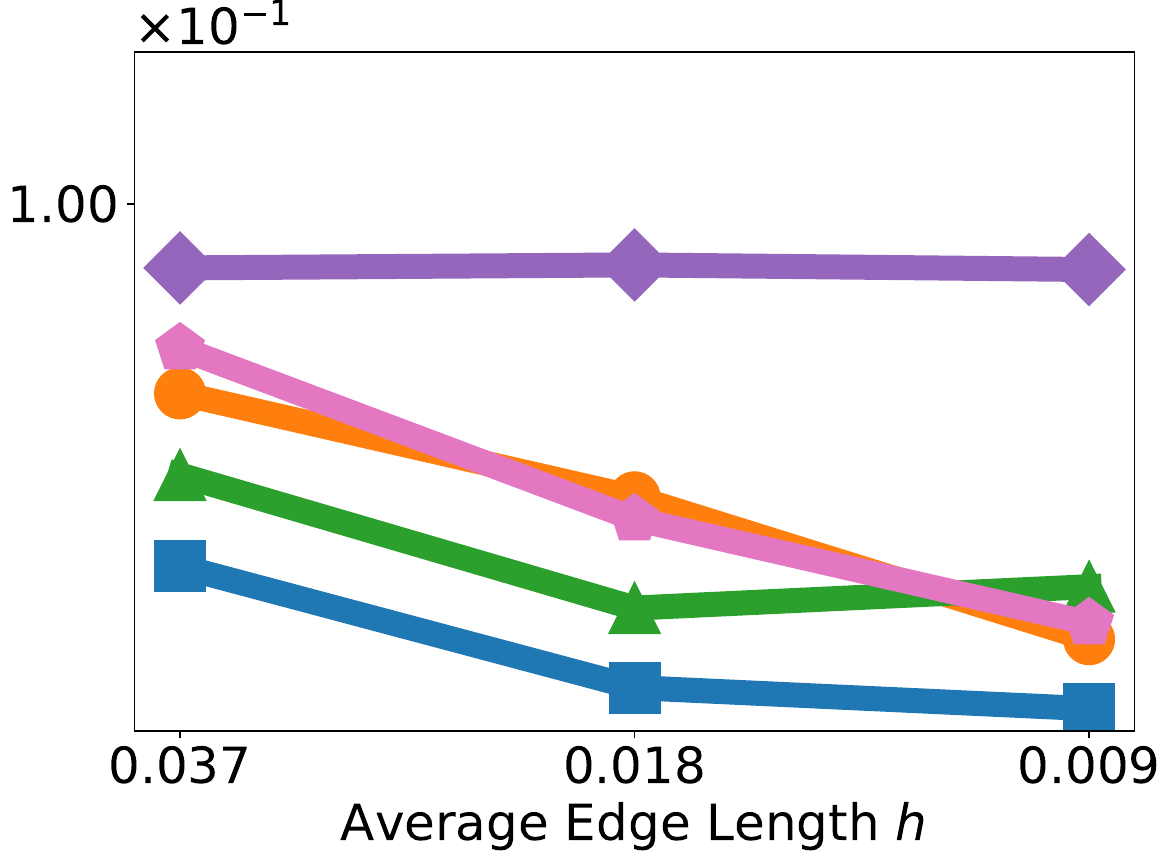}{$L_\infty$} &
        \includegraphics[width=\linewidth]{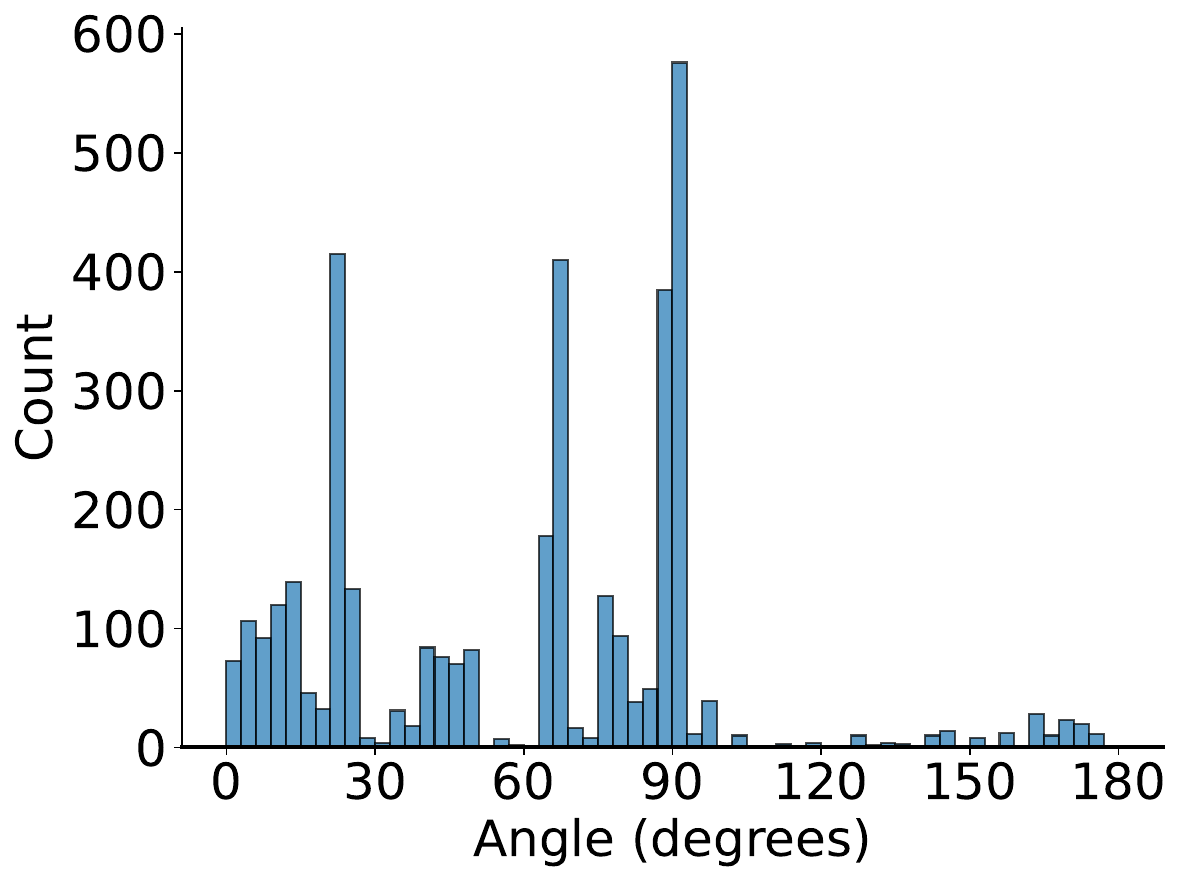} &
        \imgwithlabeltiny{width=\linewidth}{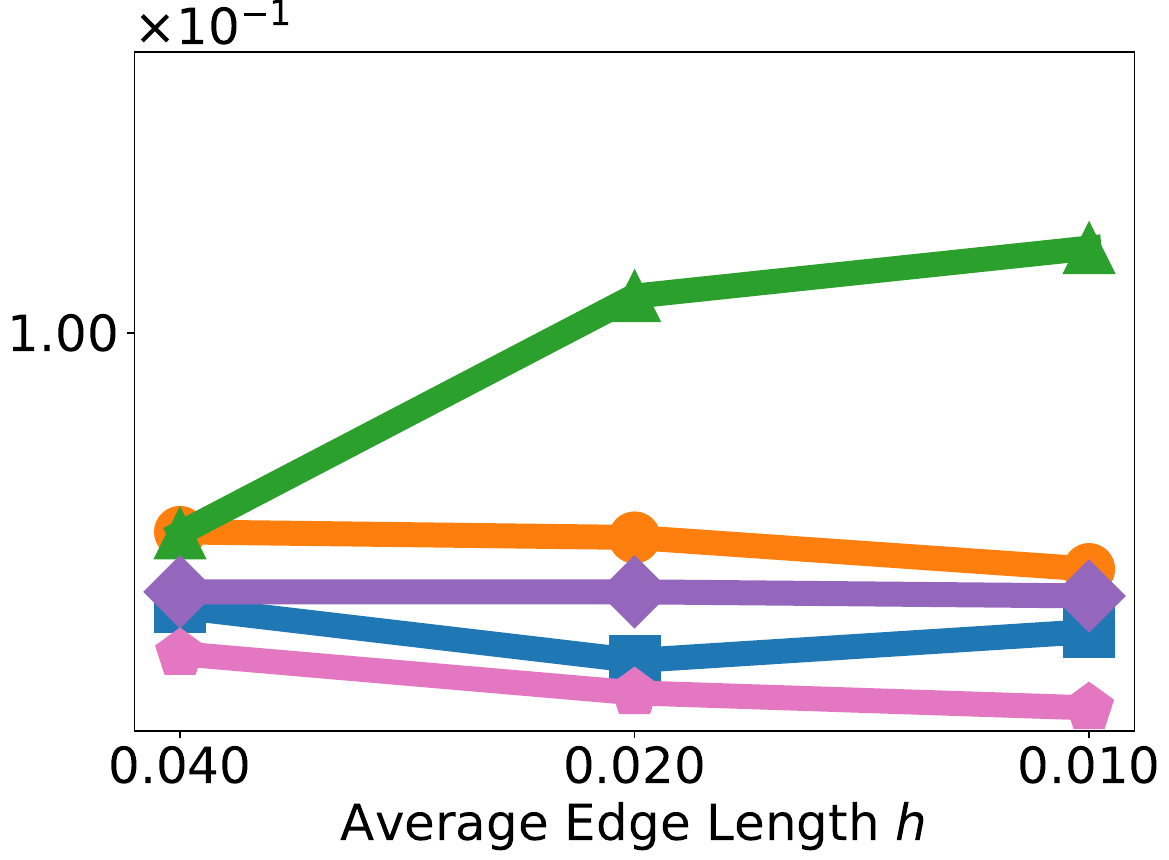}{$L_\infty$} &
        \includegraphics[width=\linewidth]{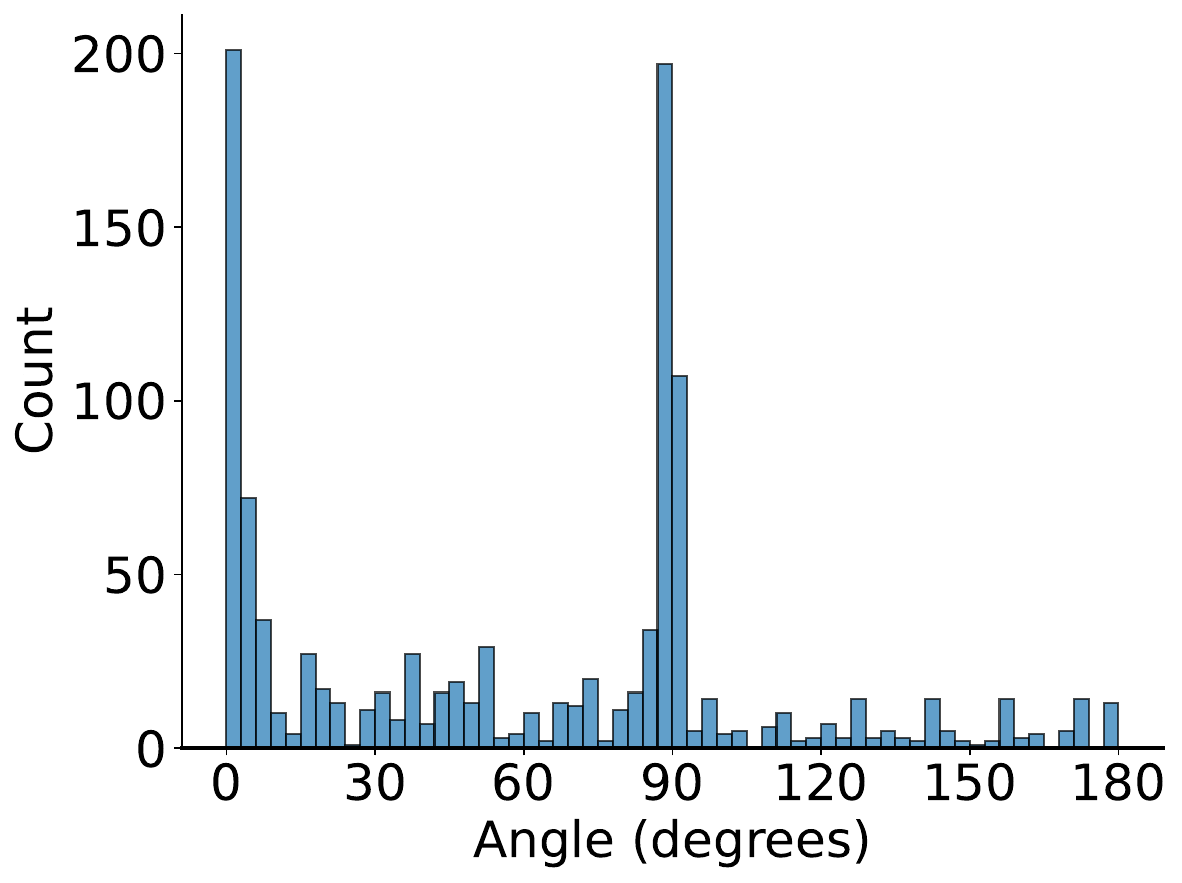} &
        \imgwithlabeltiny{width=\linewidth}{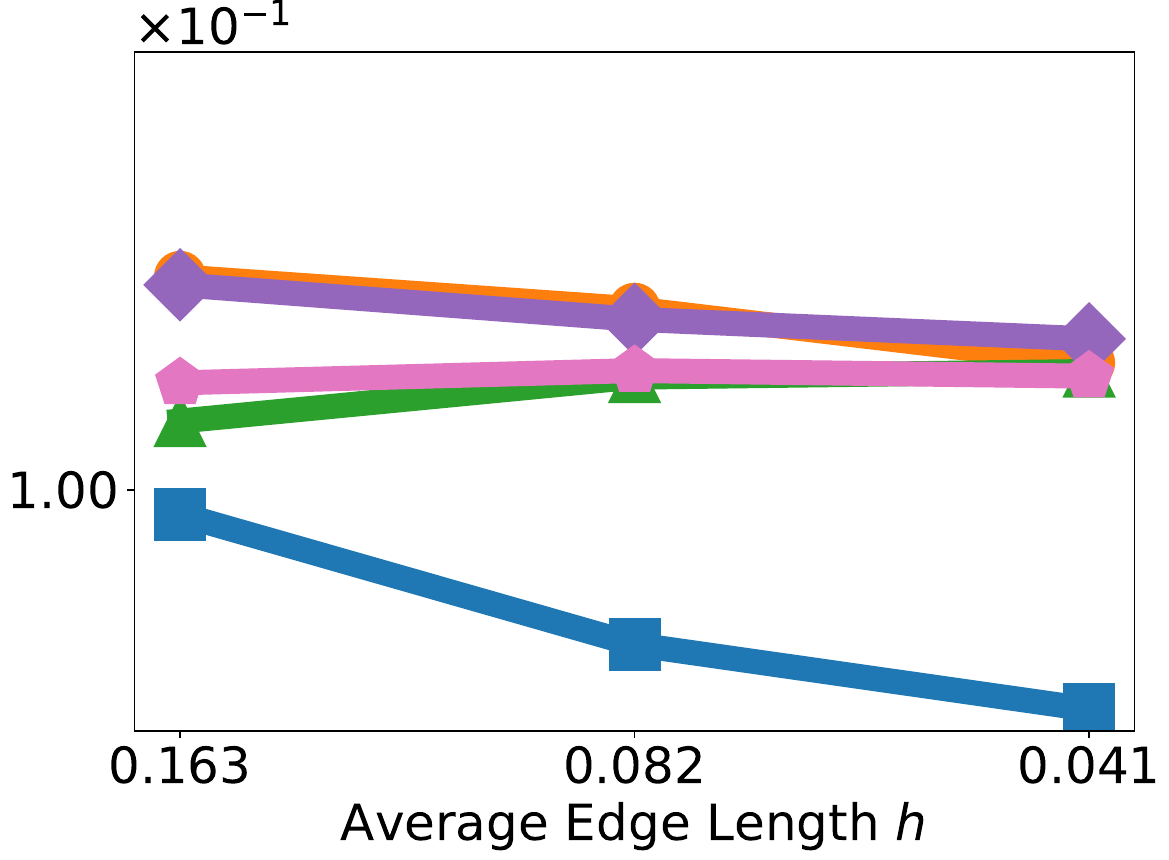}{$L_\infty$} \\

        \imglabel{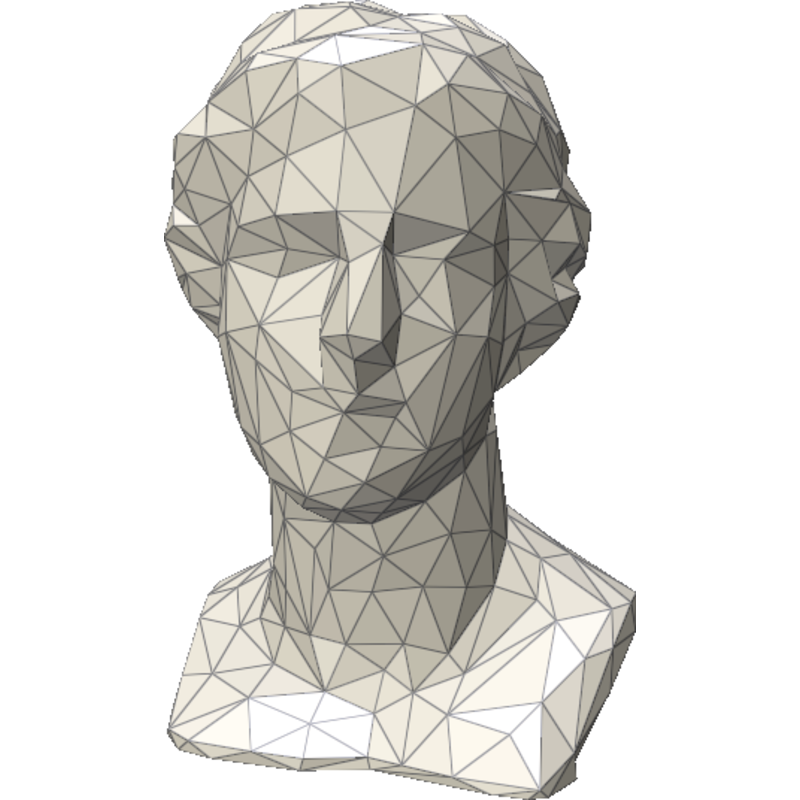}{73464} & 
        \imgwithlabeltiny{width=\linewidth}{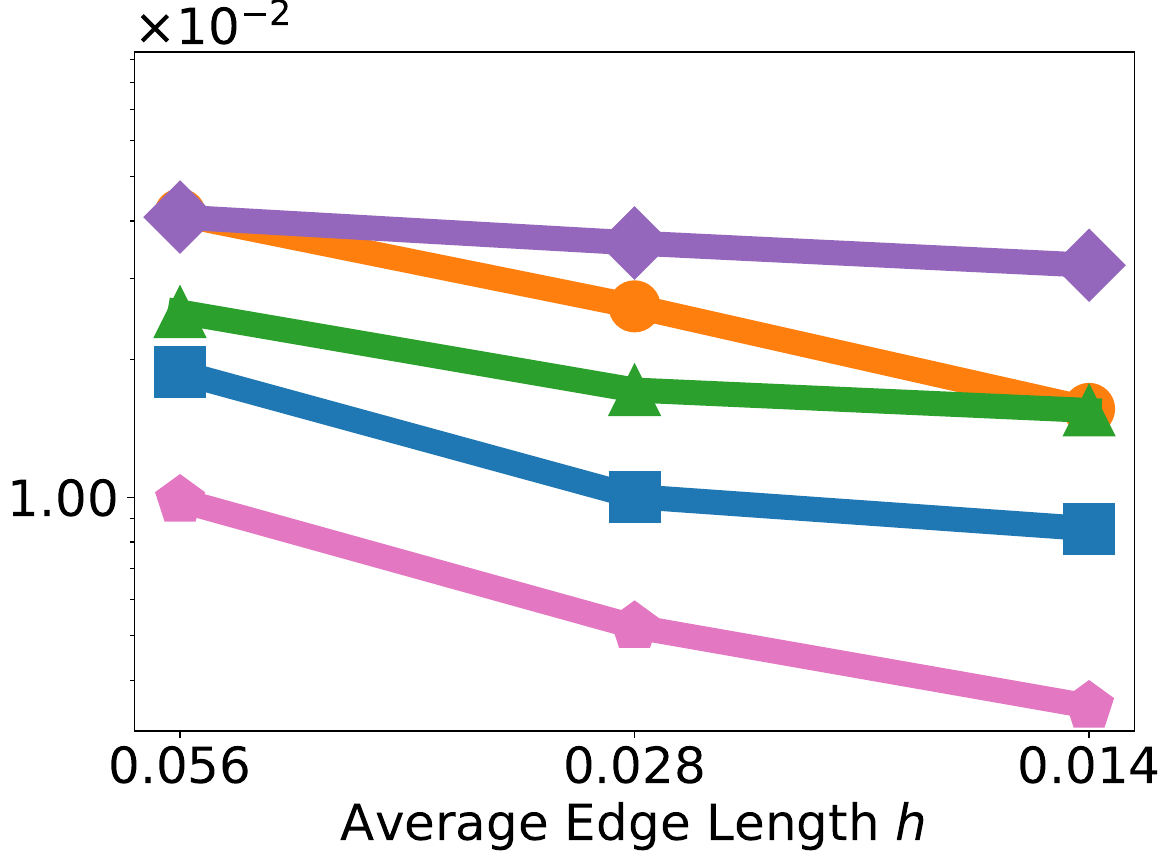}{$L_2$} & 
        \imglabel{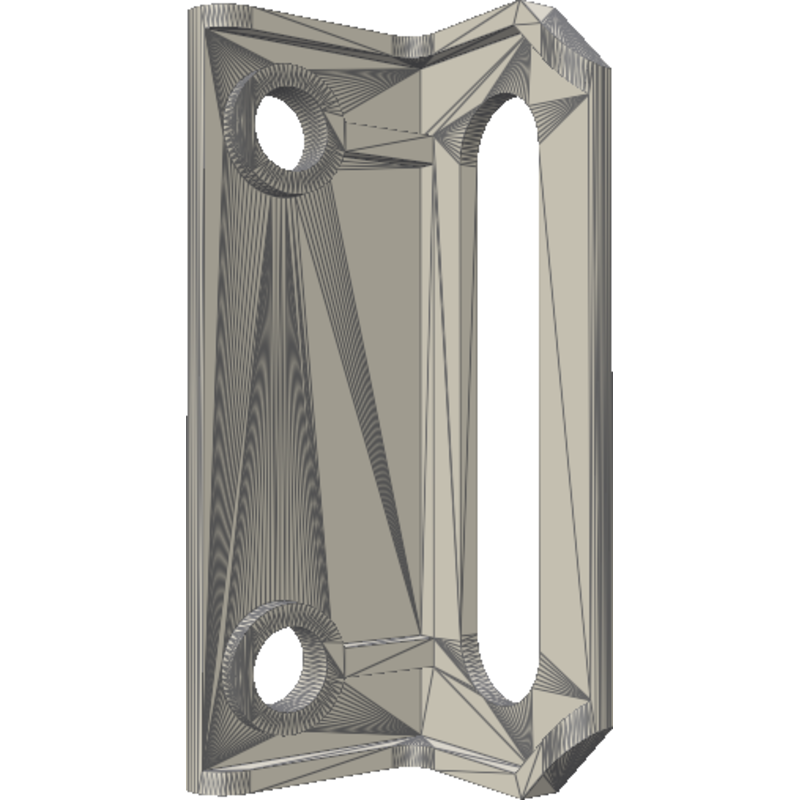}{118921} & 
        \imgwithlabeltiny{width=\linewidth}{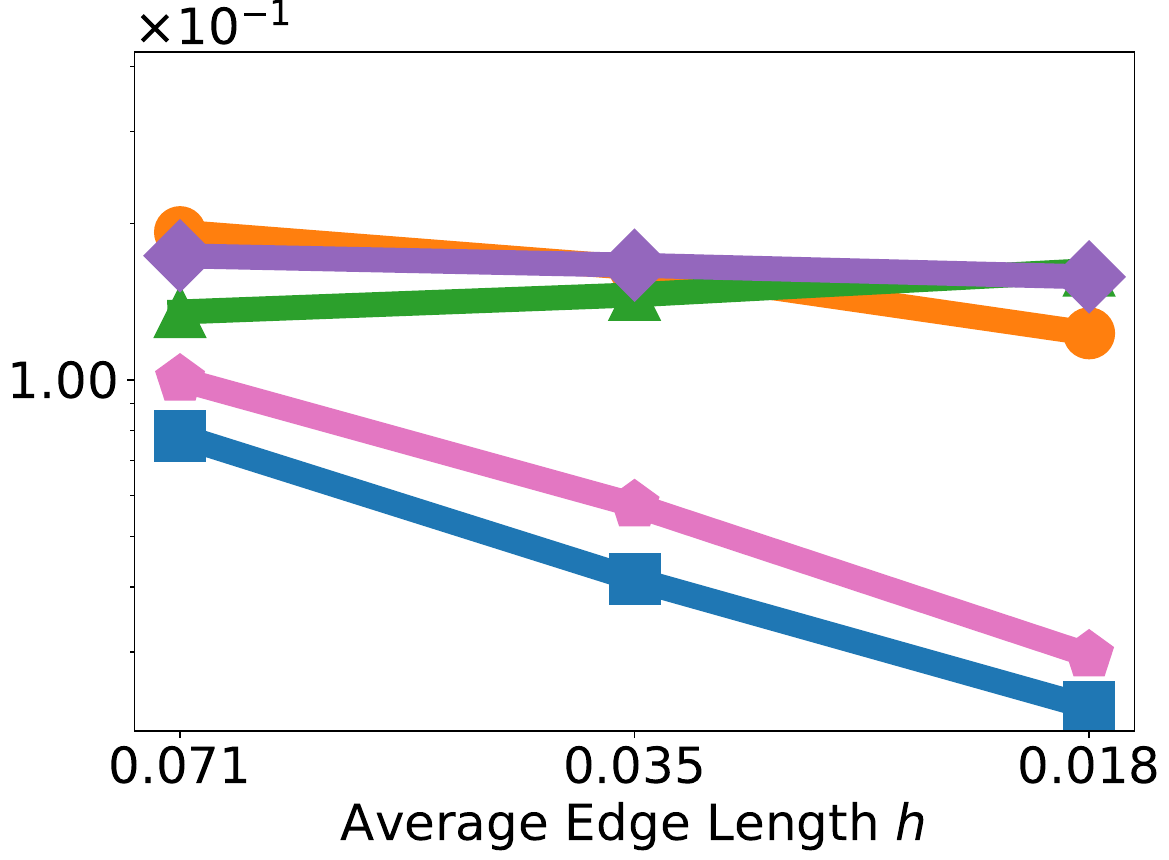}{$L_2$} & 
        \imglabel{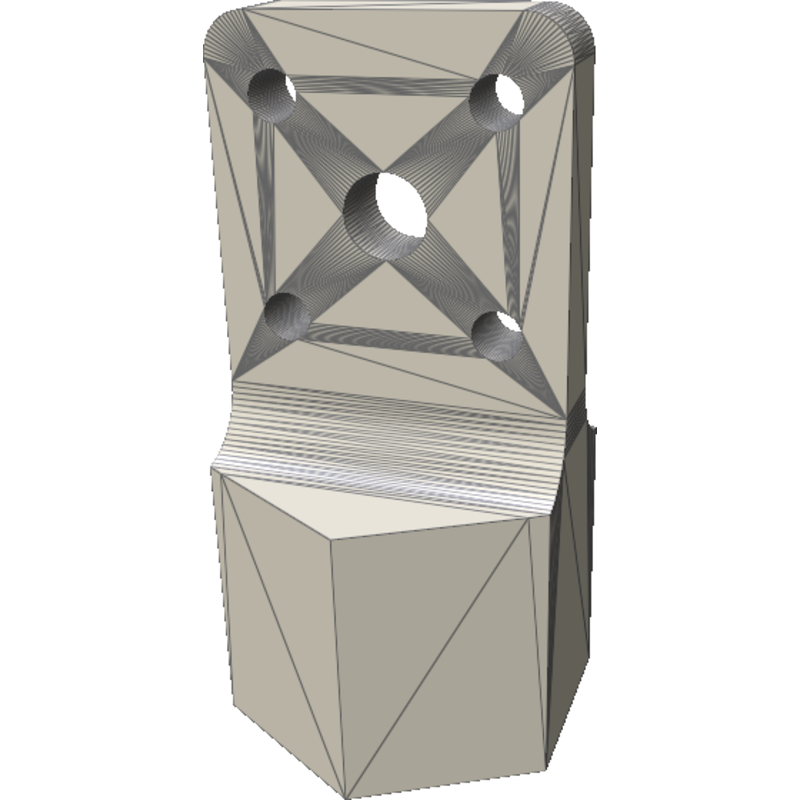}{45813} & 
        \imgwithlabeltiny{width=\linewidth}{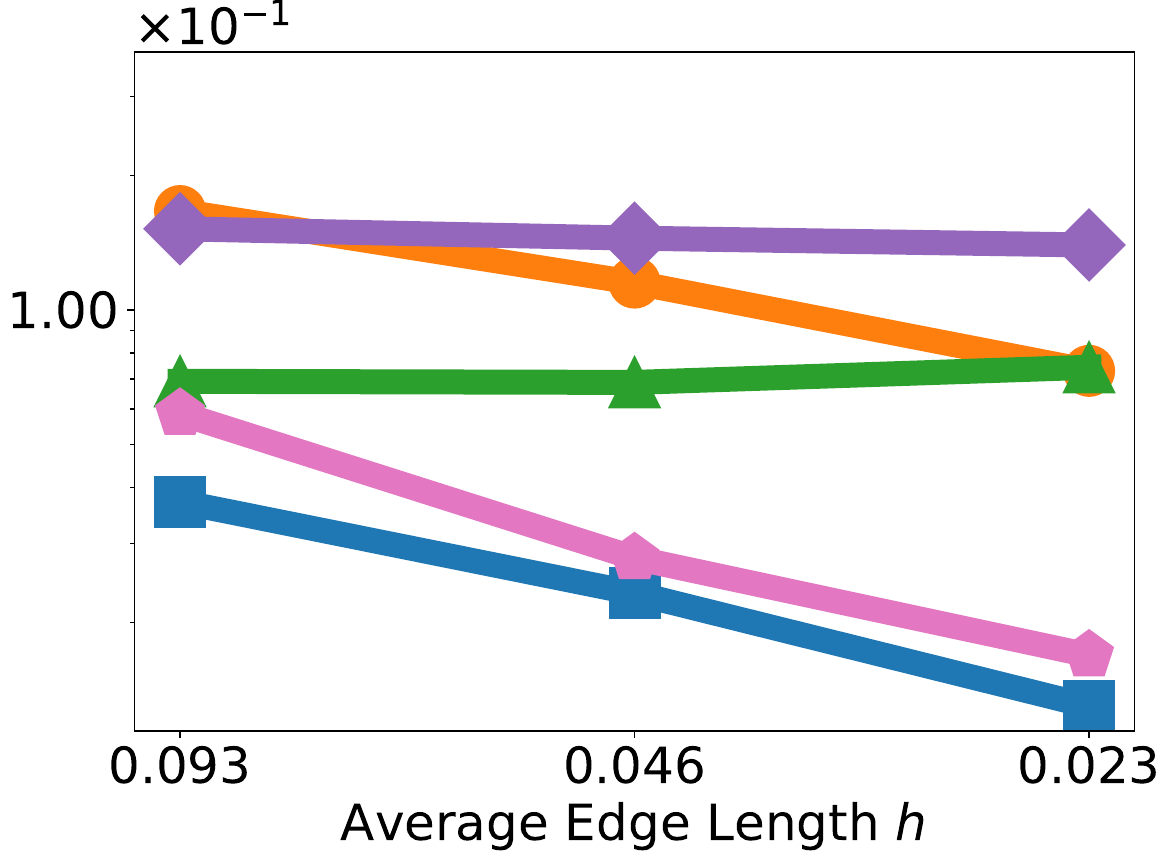}{$L_2$} \\
          
        \includegraphics[width=\linewidth]{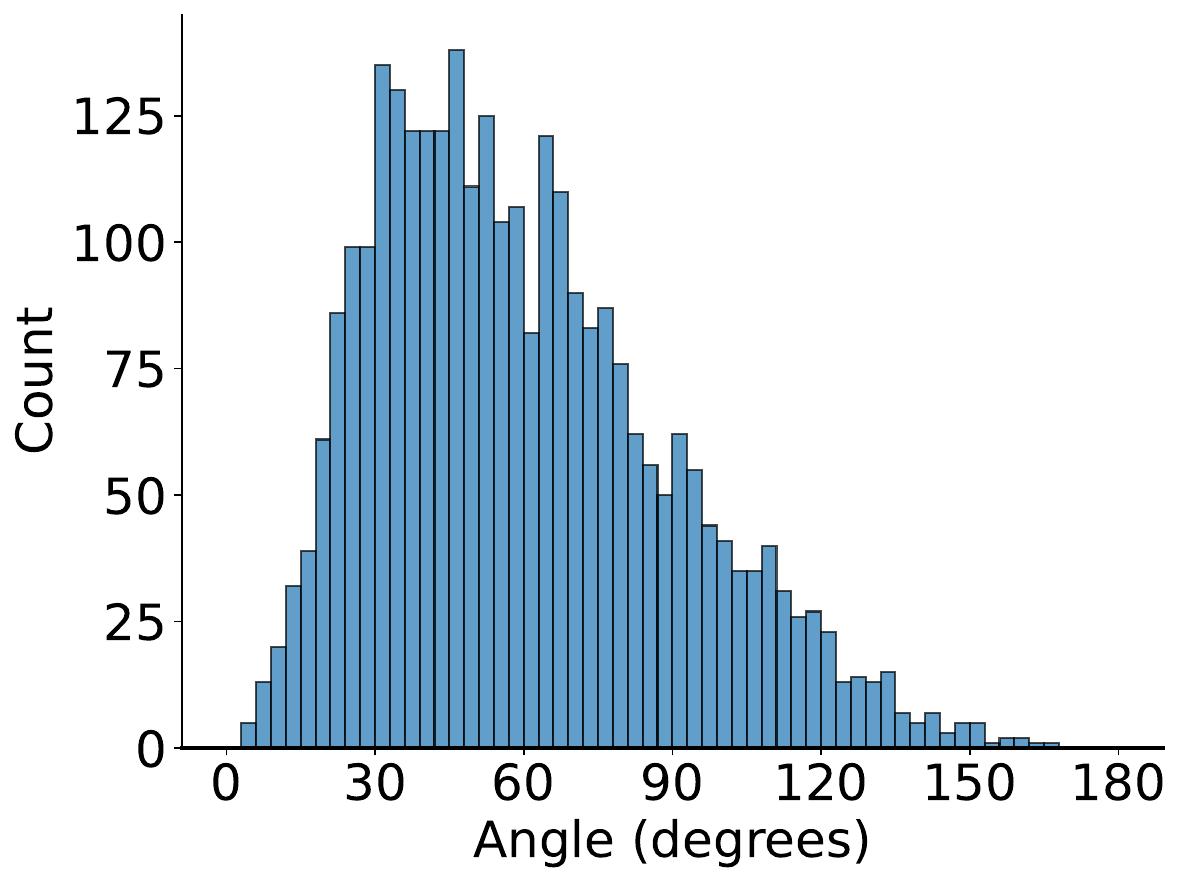} &
        \imgwithlabeltiny{width=\linewidth}{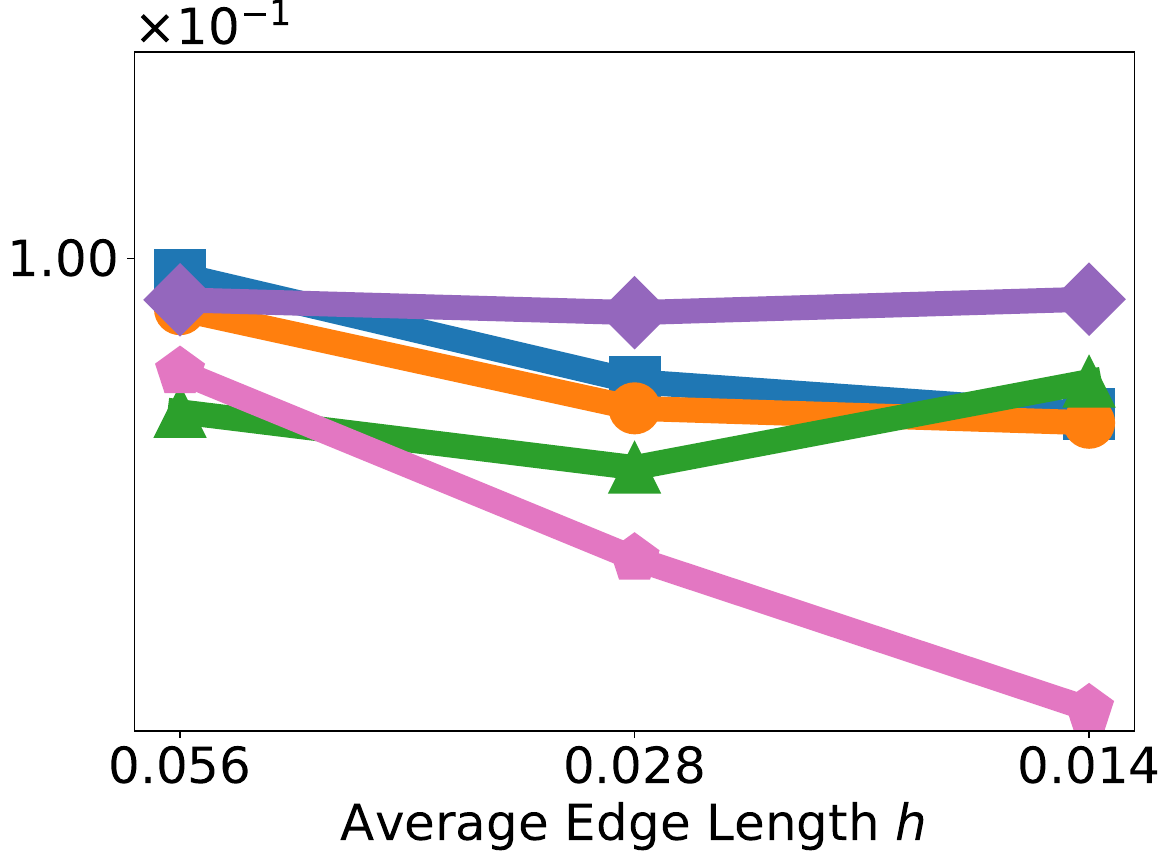}{$L_\infty$} &
        \includegraphics[width=\linewidth]{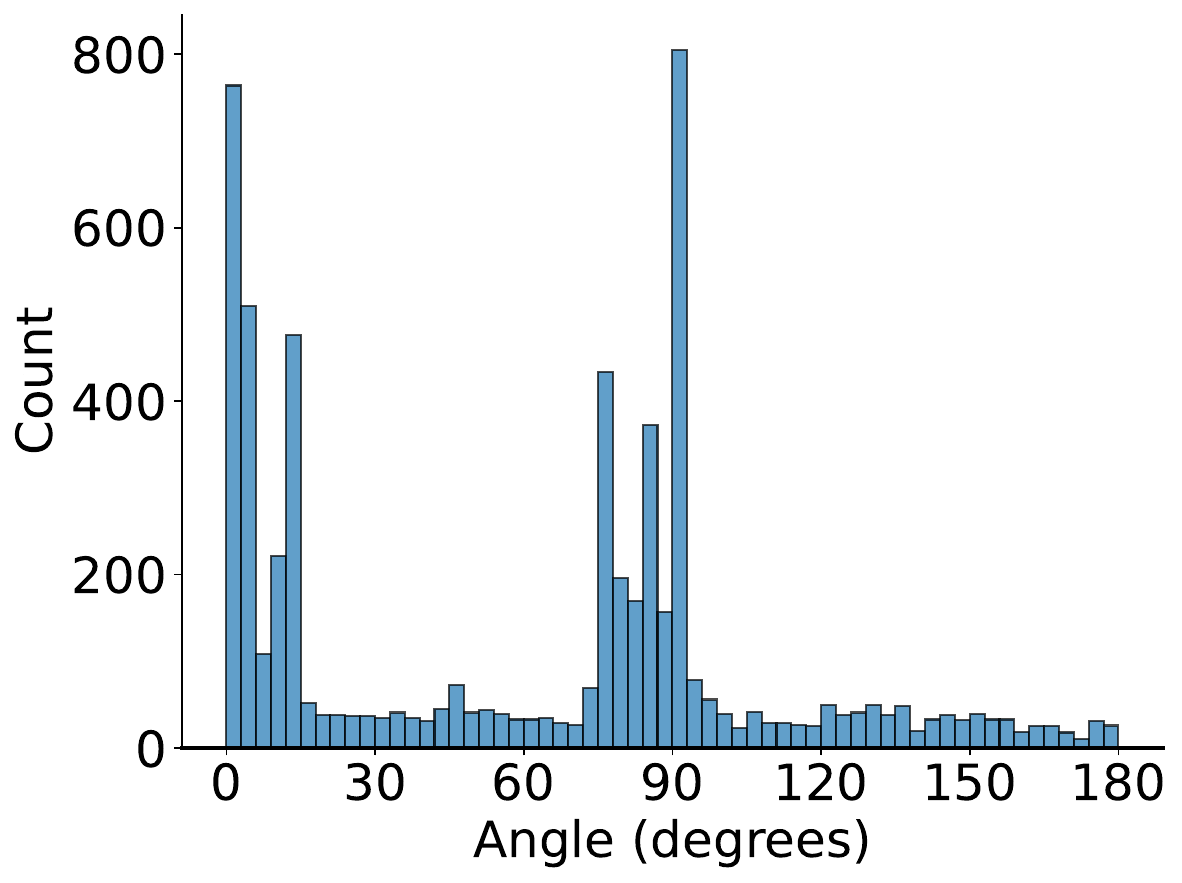} &
        \imgwithlabeltiny{width=\linewidth}{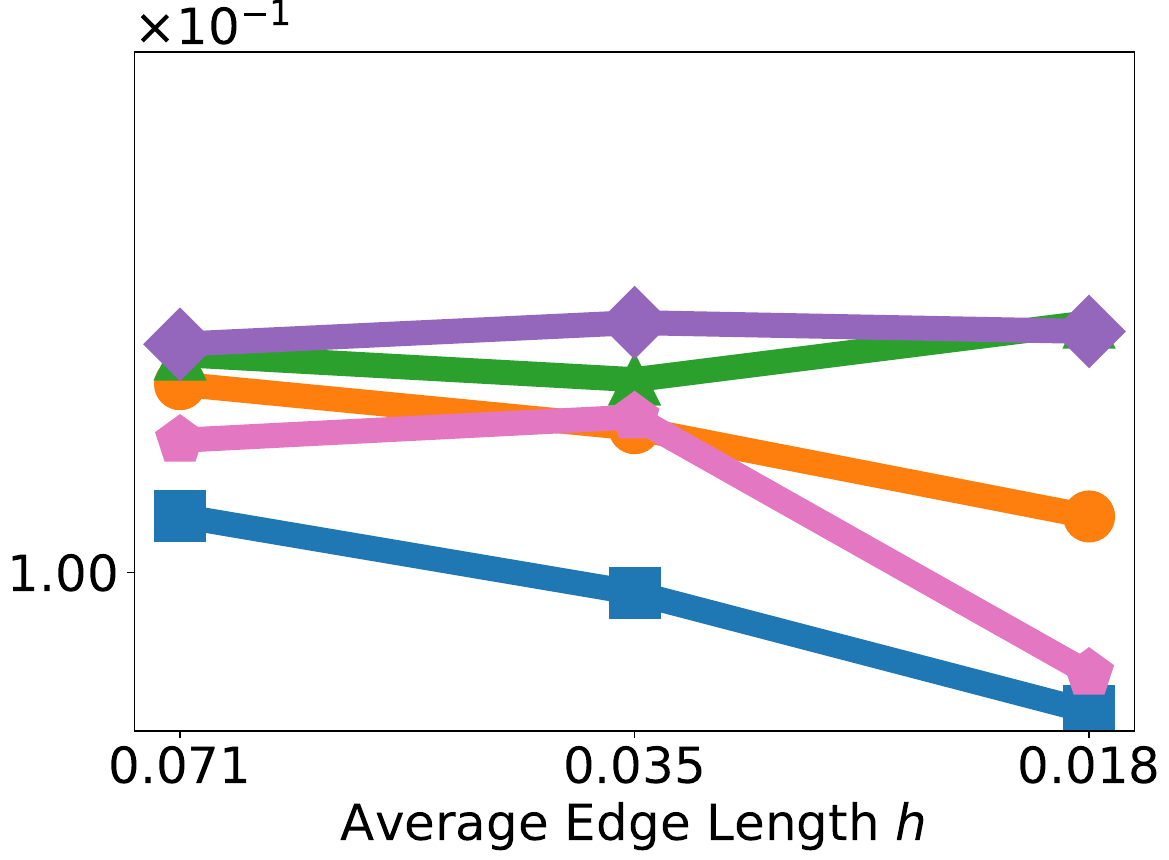}{$L_\infty$} &
        \includegraphics[width=\linewidth]{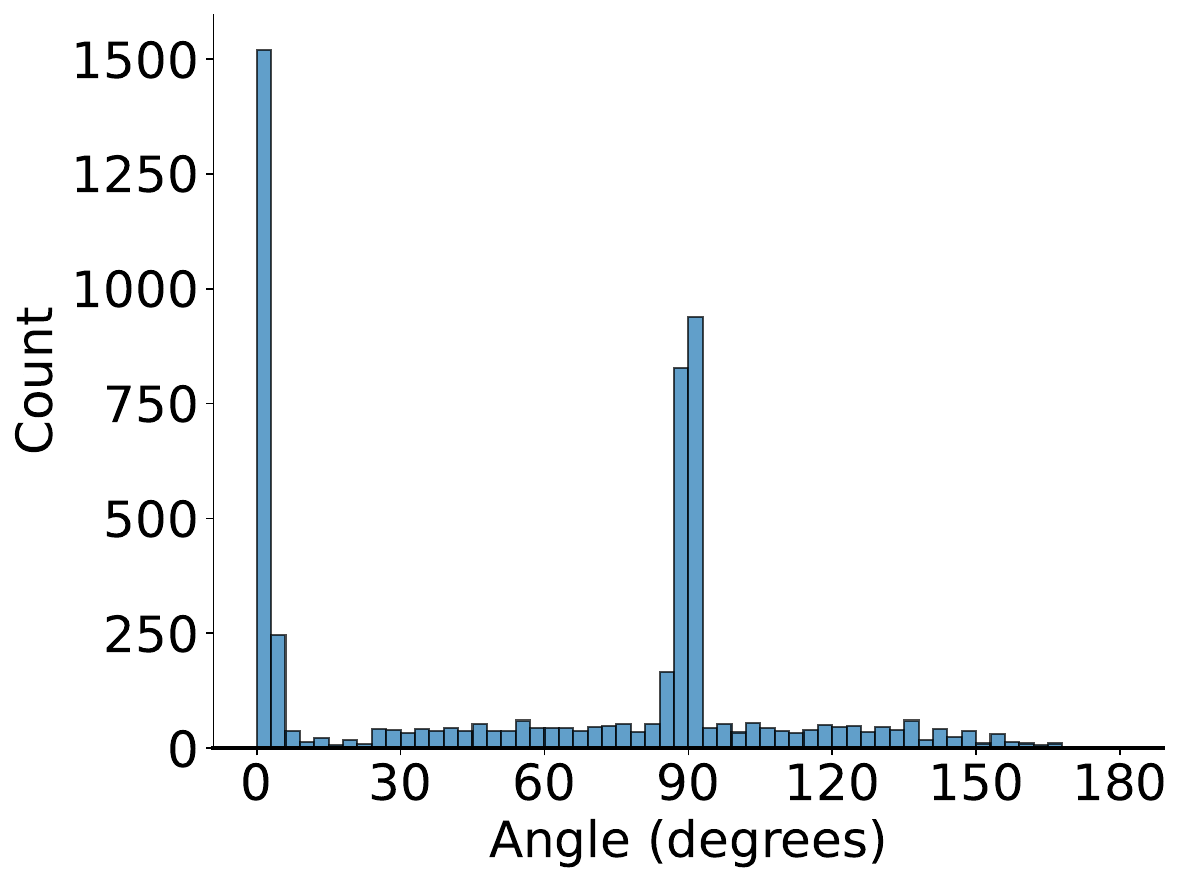} &
        \imgwithlabeltiny{width=\linewidth}{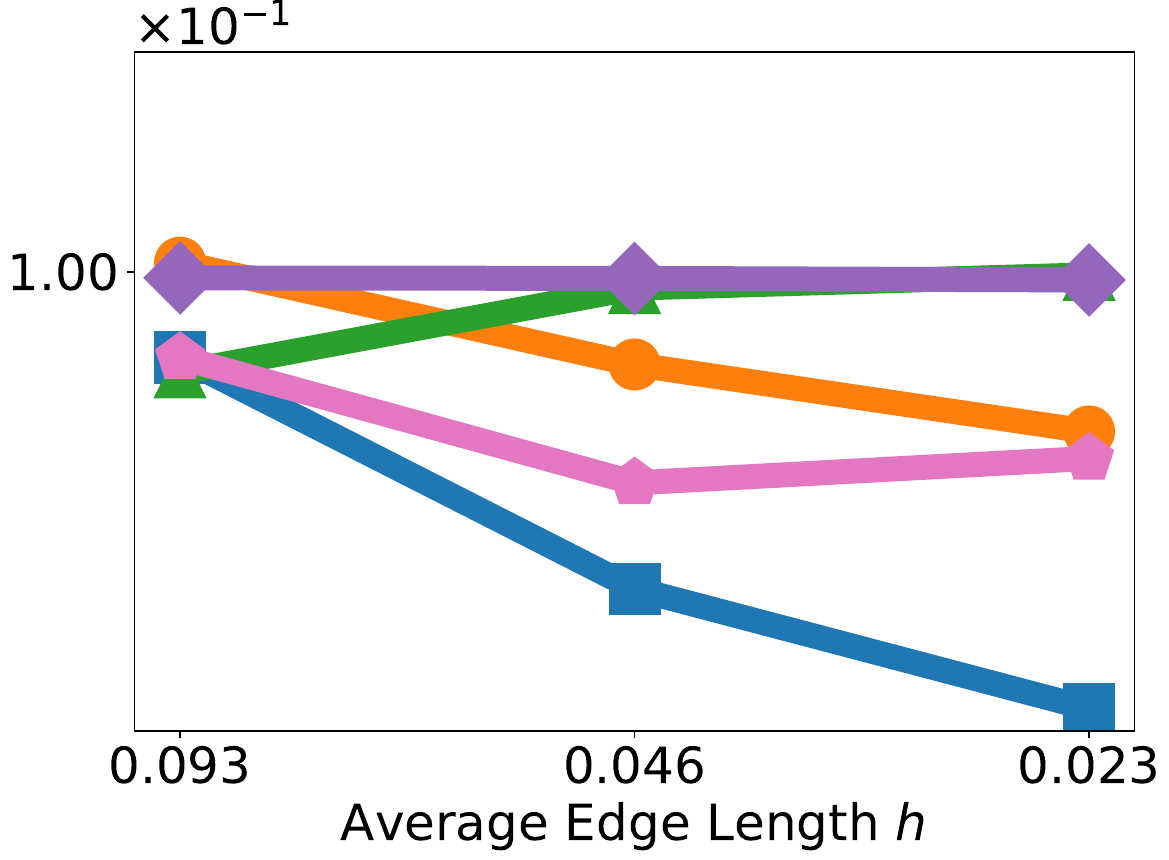}{$L_\infty$} \\
        
        \hline 
    \end{tabular}
    \vspace{2pt}
    \includegraphics[width=0.5\linewidth]{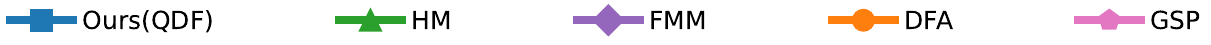}
    \vspace{-6pt}
    \setlength{\abovecaptionskip}{2pt}
    \caption{Convergence on triangular meshes with well-conditioned, poorly-conditioned, and extremely poorly-conditioned triangles. Model IDs are shown below the renderings, and the corresponding angle histograms visualize mesh quality. The x-axis in the error diagrams represents the mean edge length $h$ of the subdivided mesh, and the error is measured against 
    the reference VTP solution $d_{GT}$ on the finest level, using the degree-of-freedom matching protocol in \secref{subsec:setup}. Our error profiles and convergence slopes are consistently better for poor- and extremely poor-quality meshes and generally better for high-quality meshes as well.}
    \label{fig:convergence_table}
\end{figure*}

\begin{figure*}[htbp]
    \centering
    \setlength{\tabcolsep}{1pt} 

    \begin{tabular}{ C{0.15\textwidth} C{0.15\textwidth} @{\hspace{2pt}}|@{\hspace{15pt}} C{0.15\textwidth} C{0.15\textwidth} @{\hspace{2pt}}@{\hspace{9pt}} C{0.15\textwidth} C{0.15\textwidth} }
        \hline 
        \multicolumn{2}{c |}{\small \textbf{Tetrahedron quality: good}} & 
        \multicolumn{4}{c }{\small \textbf{Tetrahedron quality: poor}}  \\[6pt]
        
        \imglabel{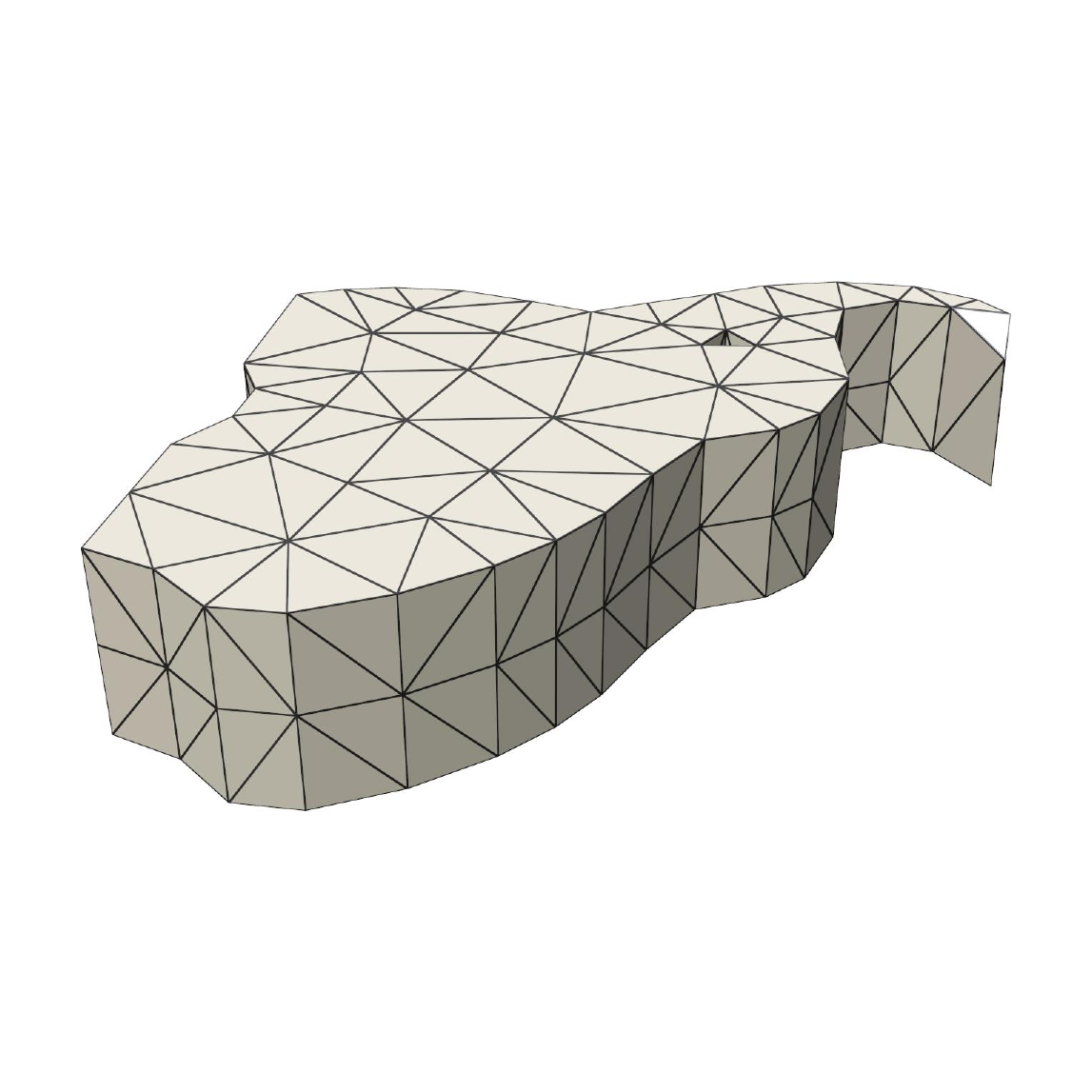}{42500} & 
        \imgwithlabeltiny{width=\linewidth}{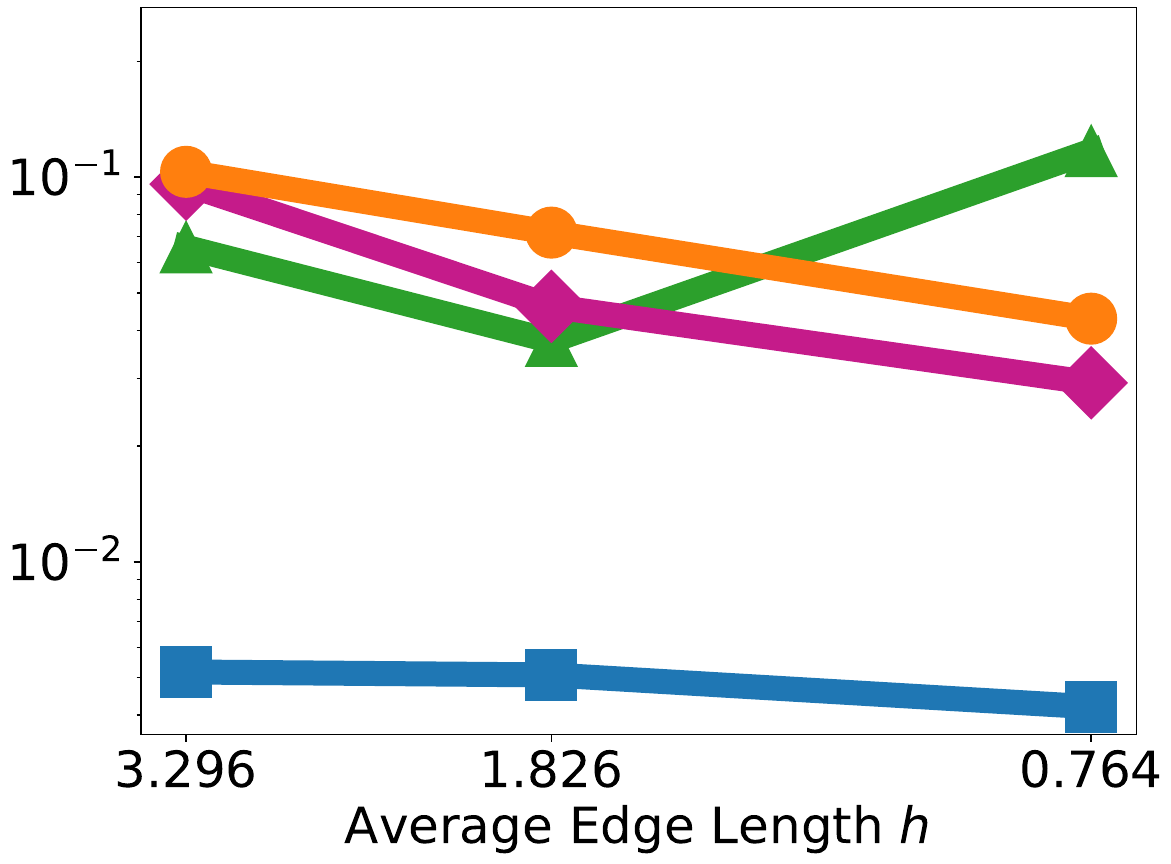}{$L_2$} & 
        \imglabel{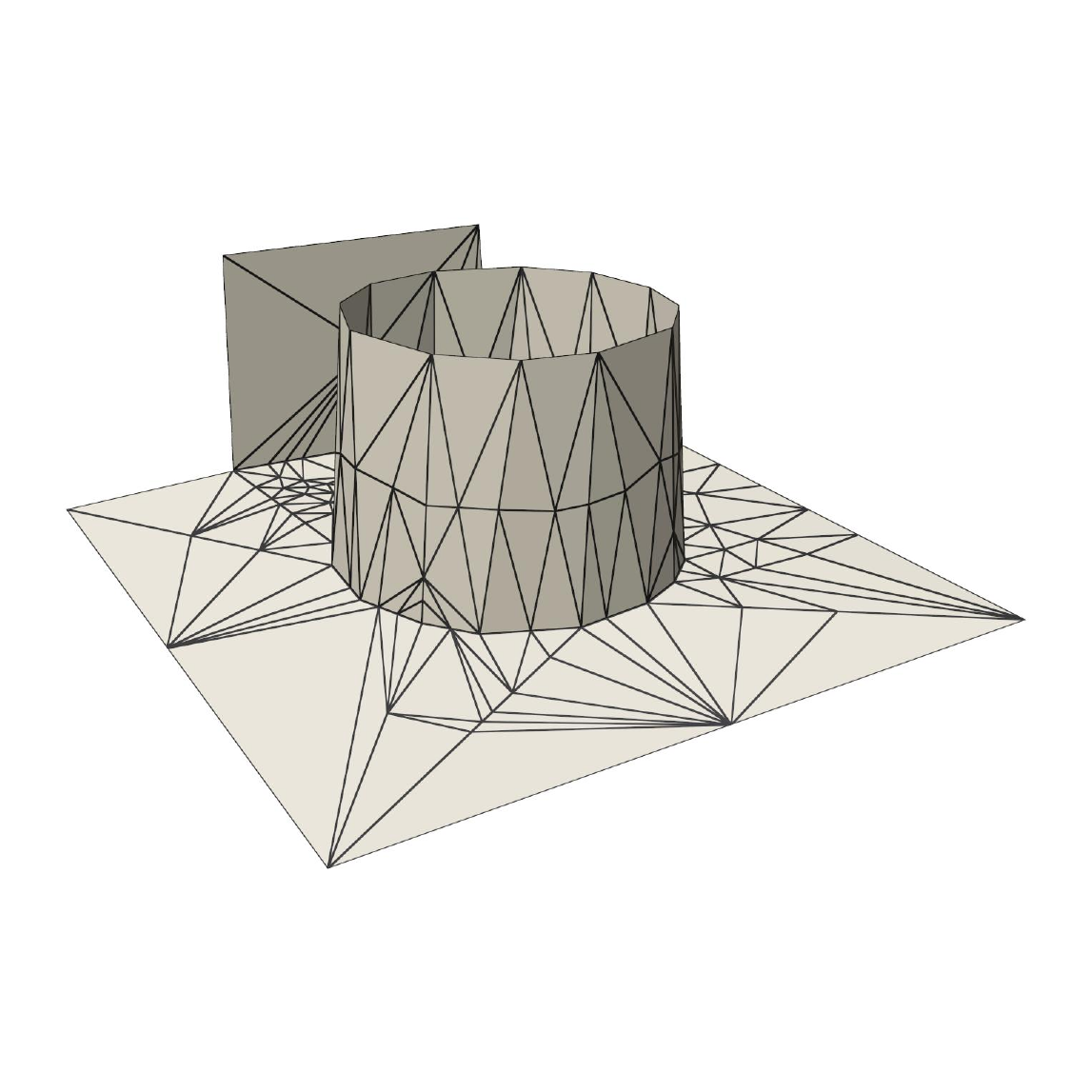}{102066} & 
        \imgwithlabeltiny{width=\linewidth}{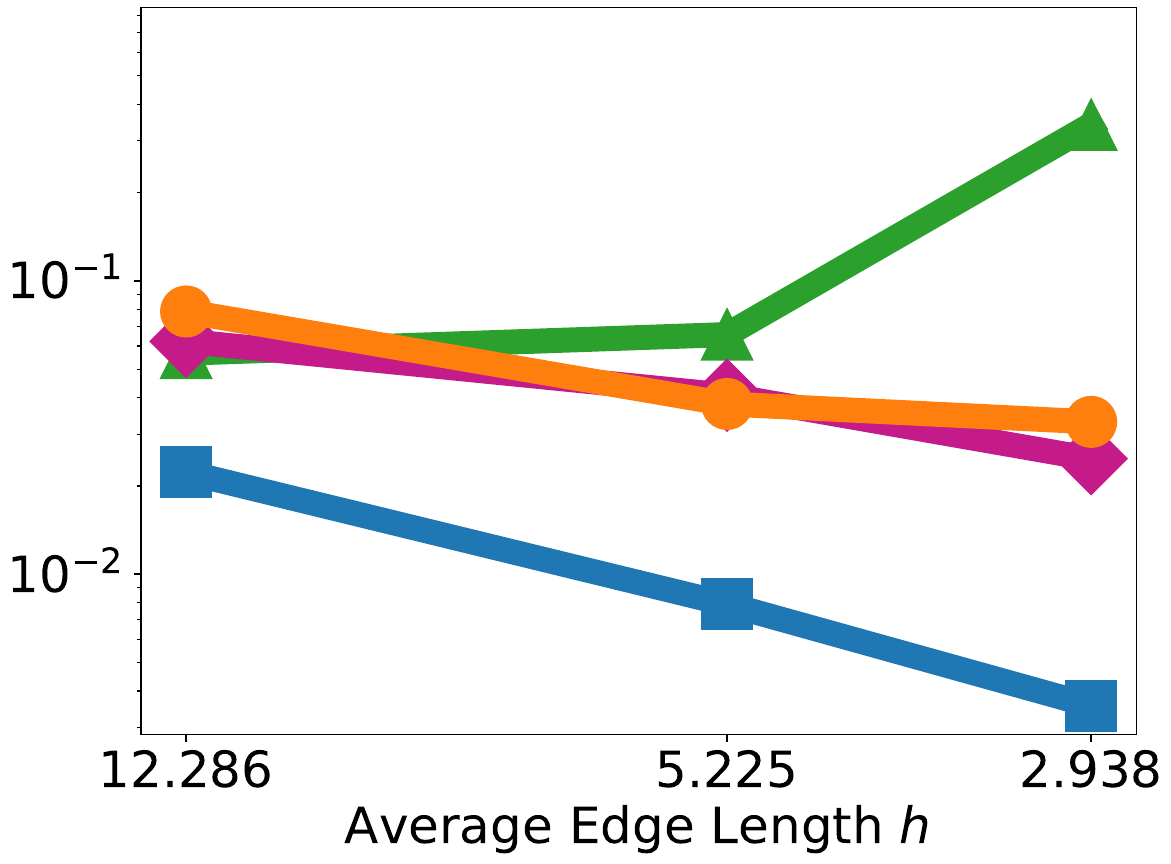}{$L_2$} & 
        \imglabel{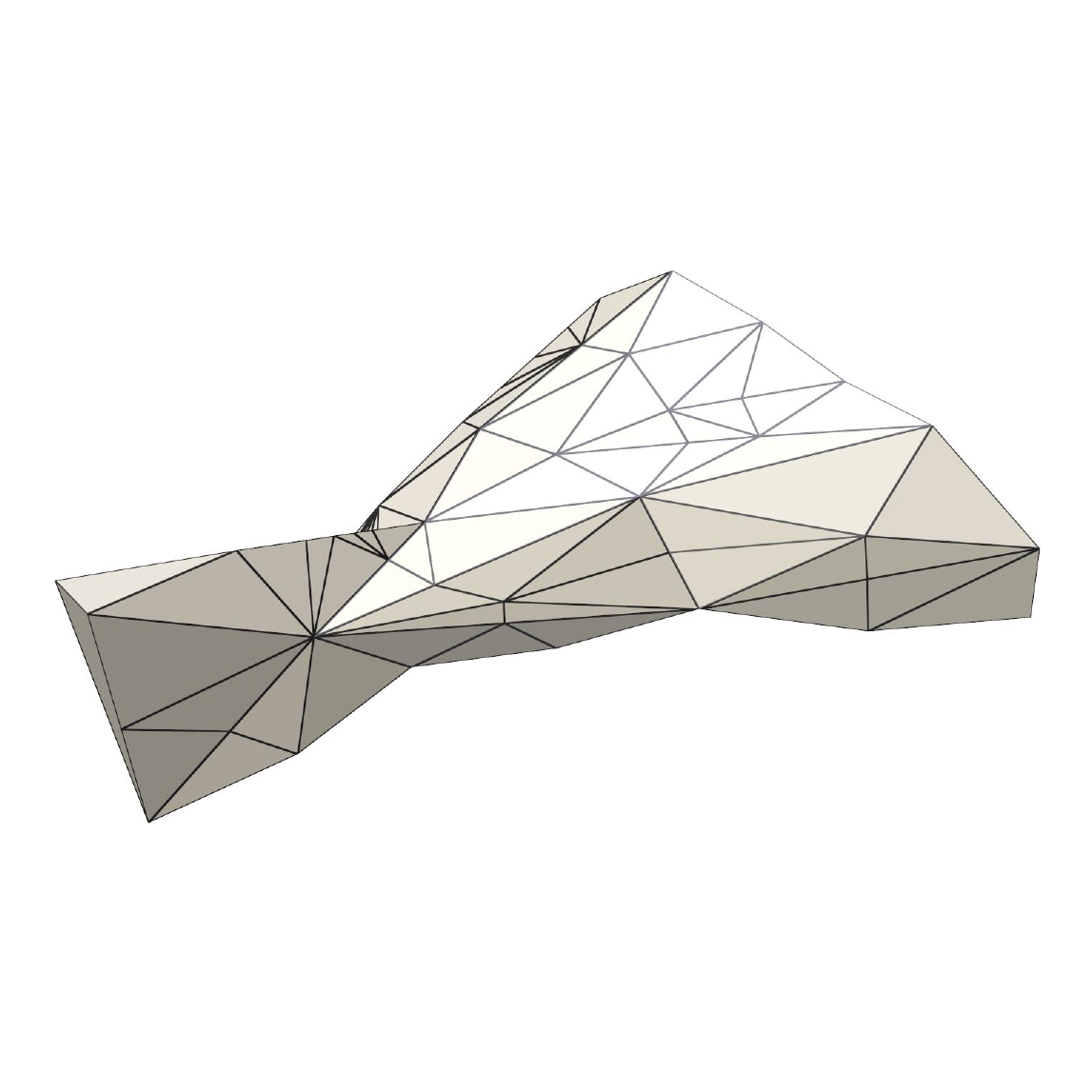}{383023} & 
        \imgwithlabeltiny{width=\linewidth}{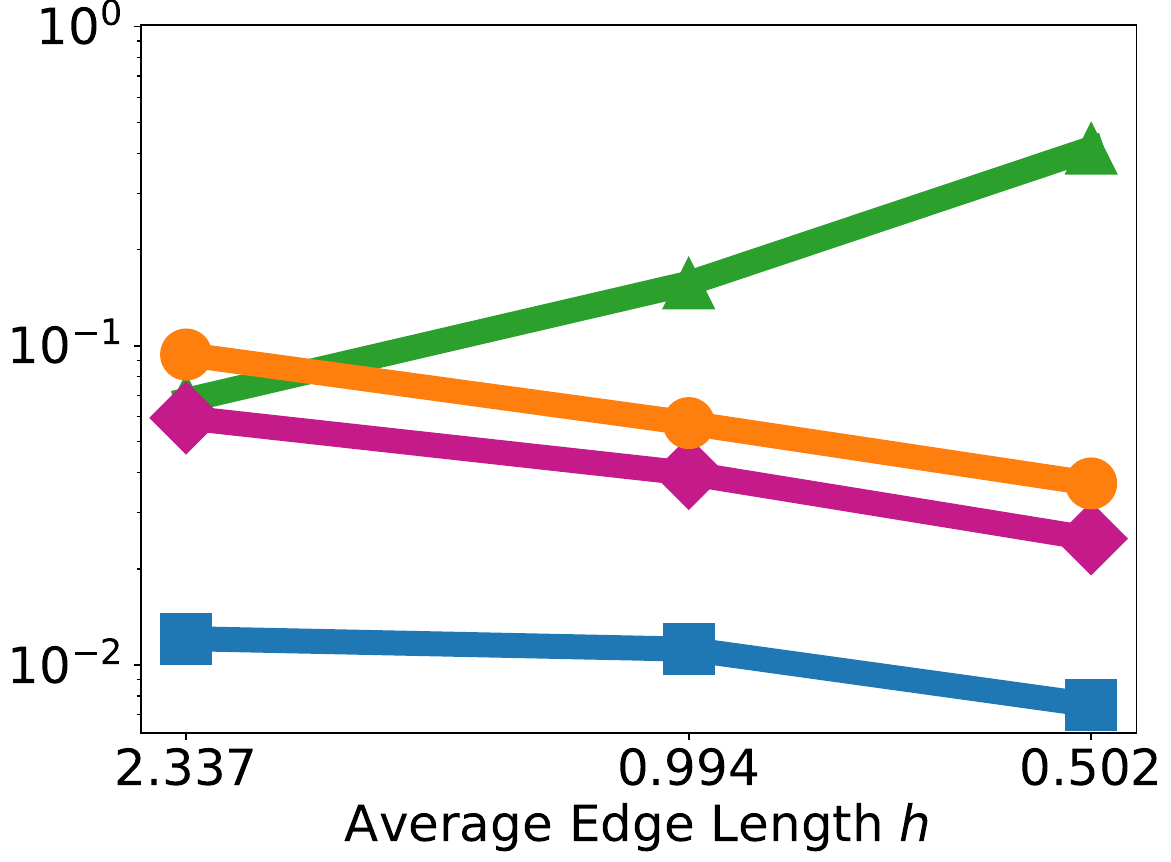}{$L_2$} \\

        \includegraphics[width=\linewidth]{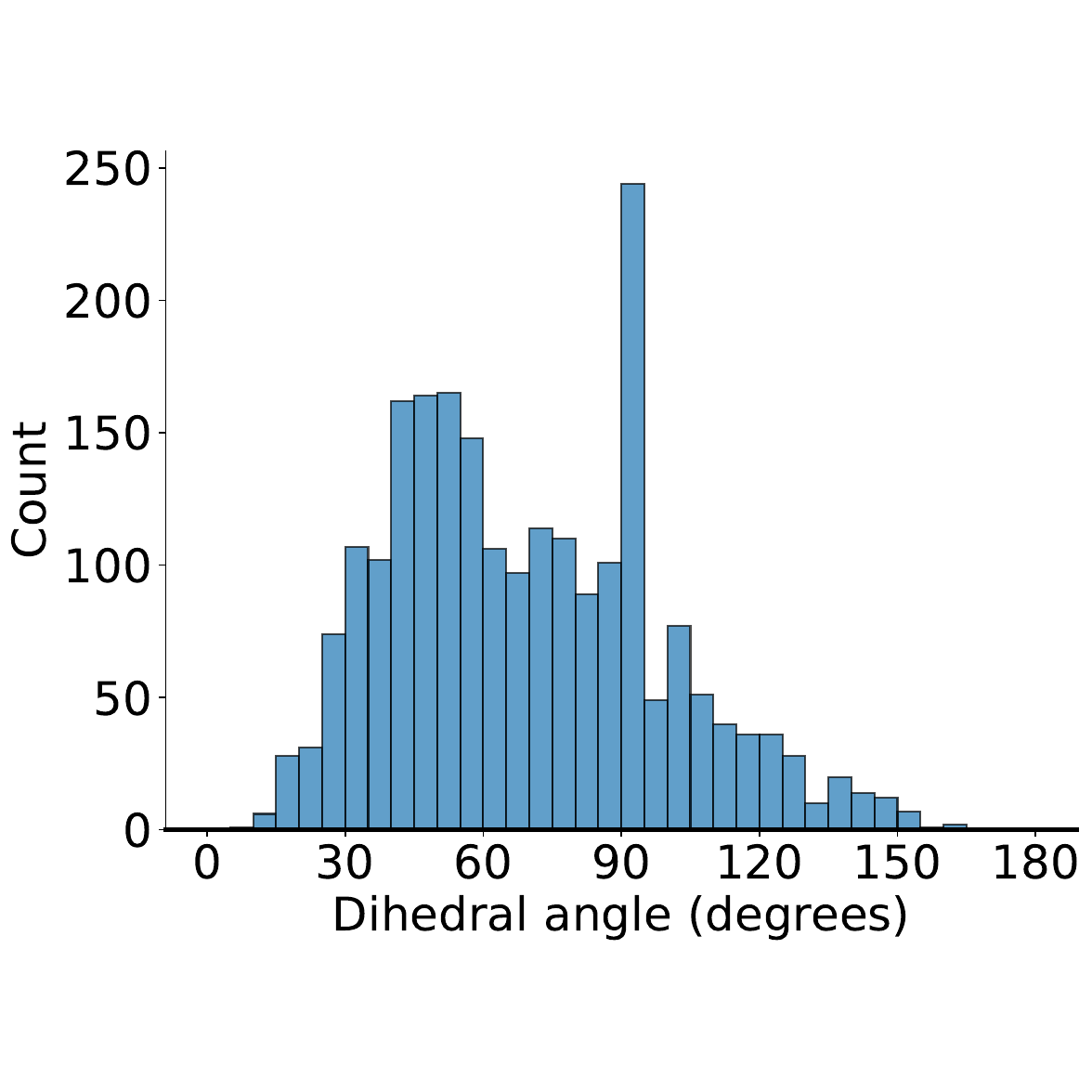} &
        \imgwithlabeltiny{width=\linewidth}{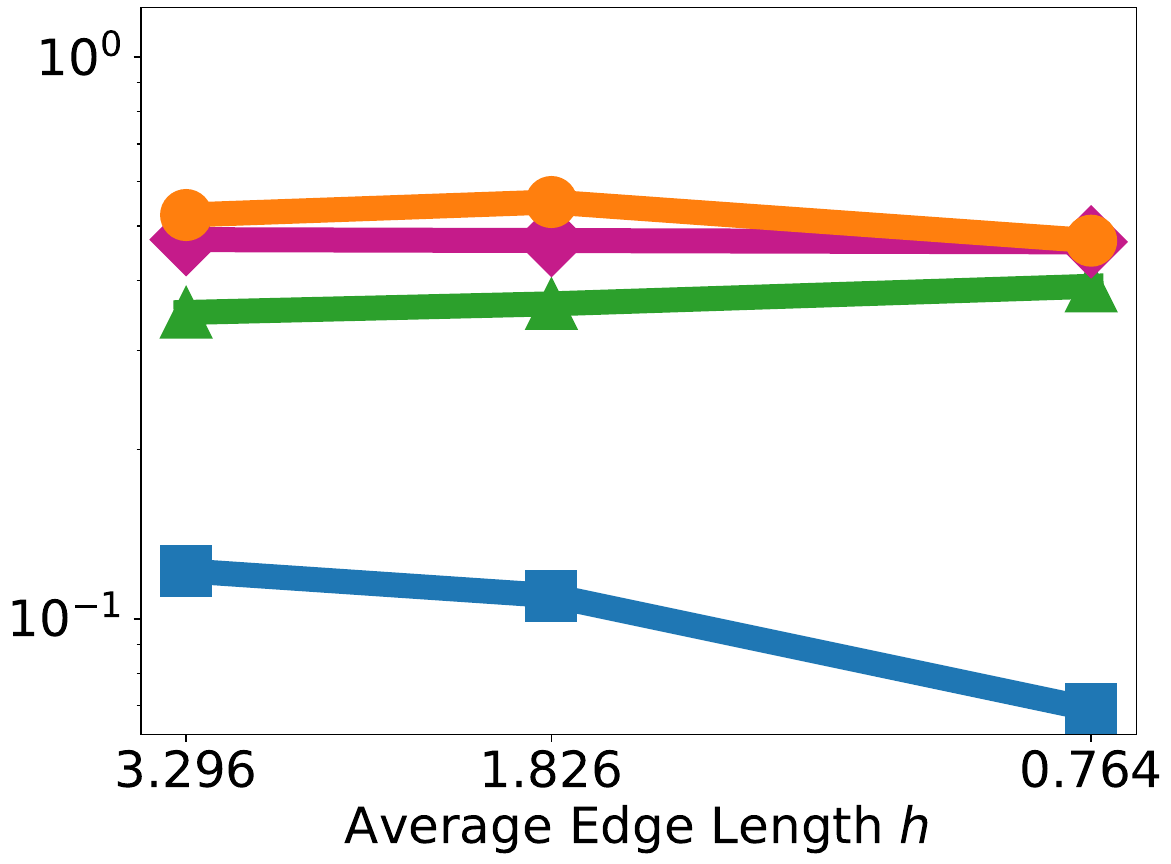}{$L_\infty$} &
        \includegraphics[width=\linewidth]{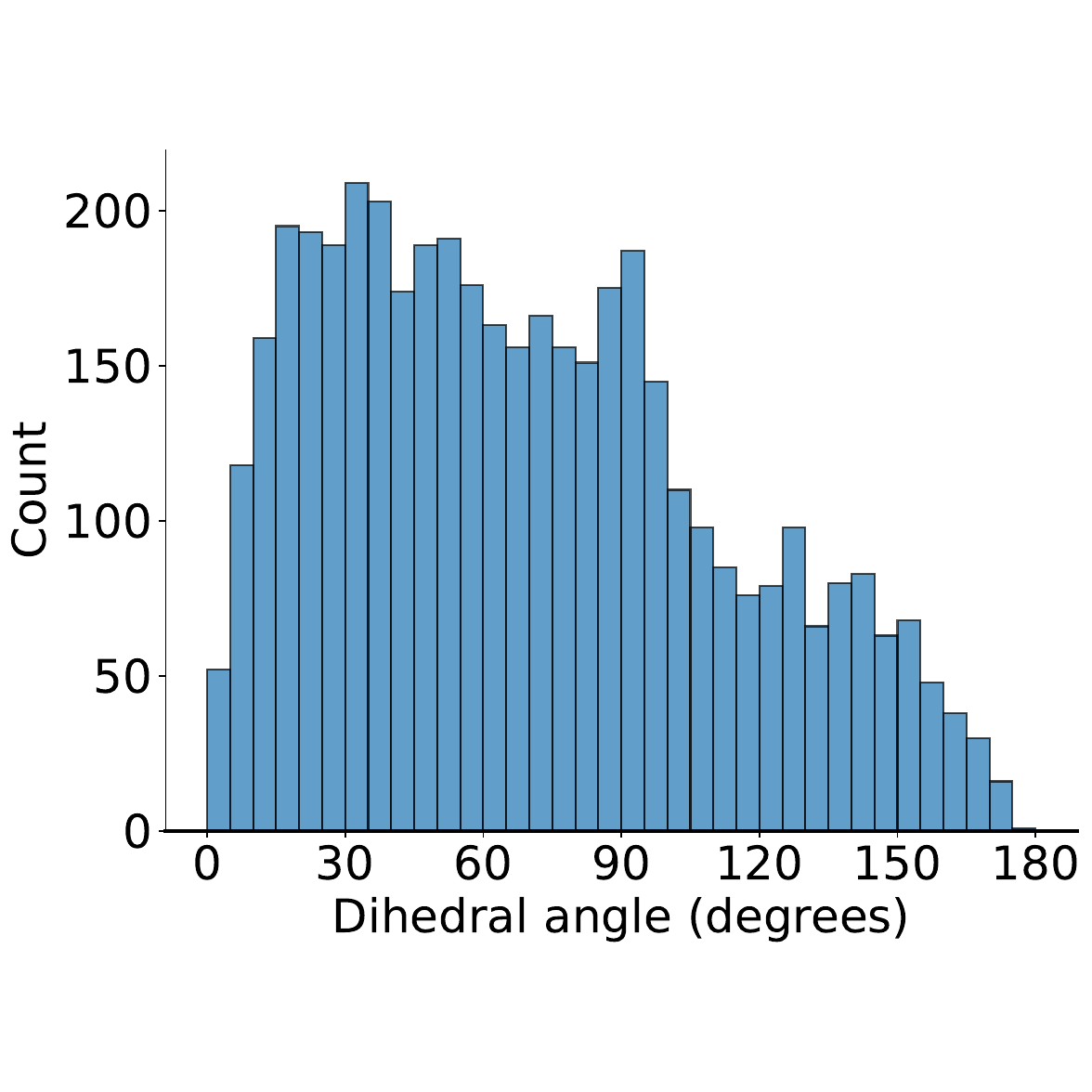} &
        \imgwithlabeltiny{width=\linewidth}{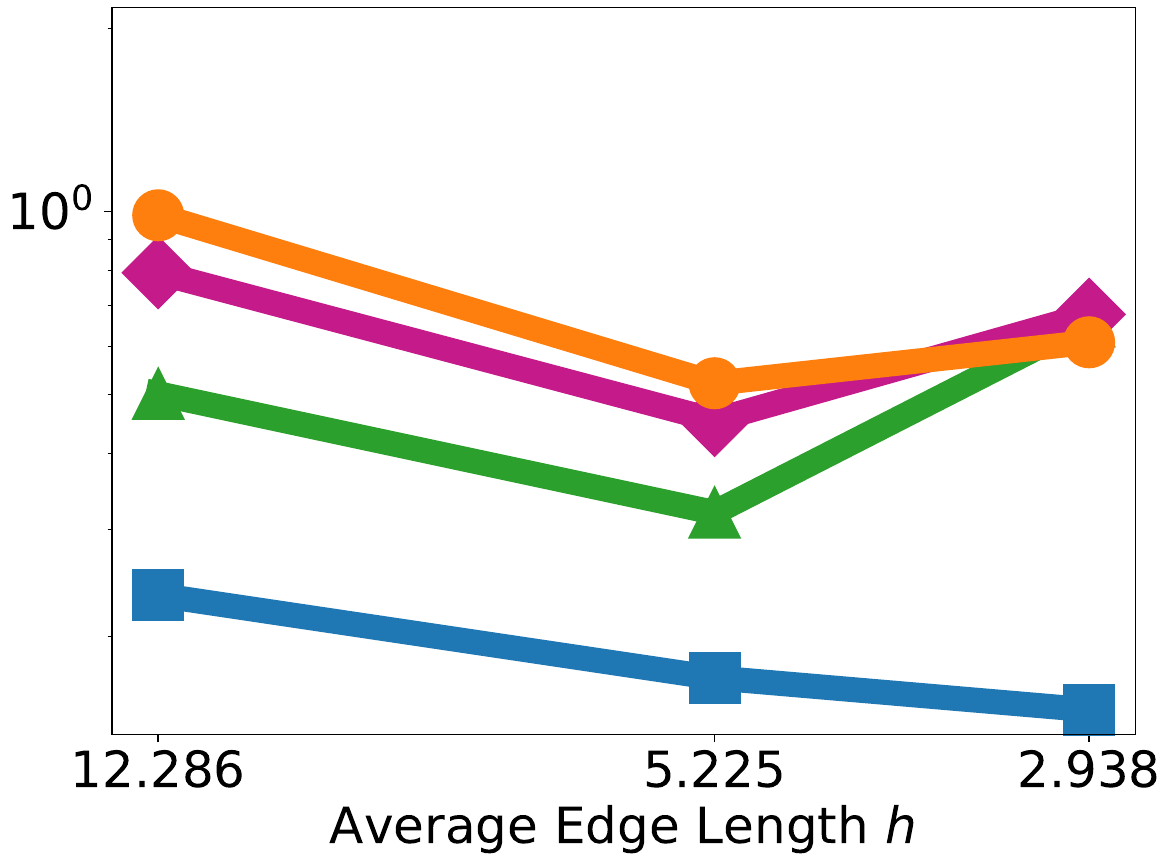}{$L_\infty$} &
        \includegraphics[width=\linewidth]{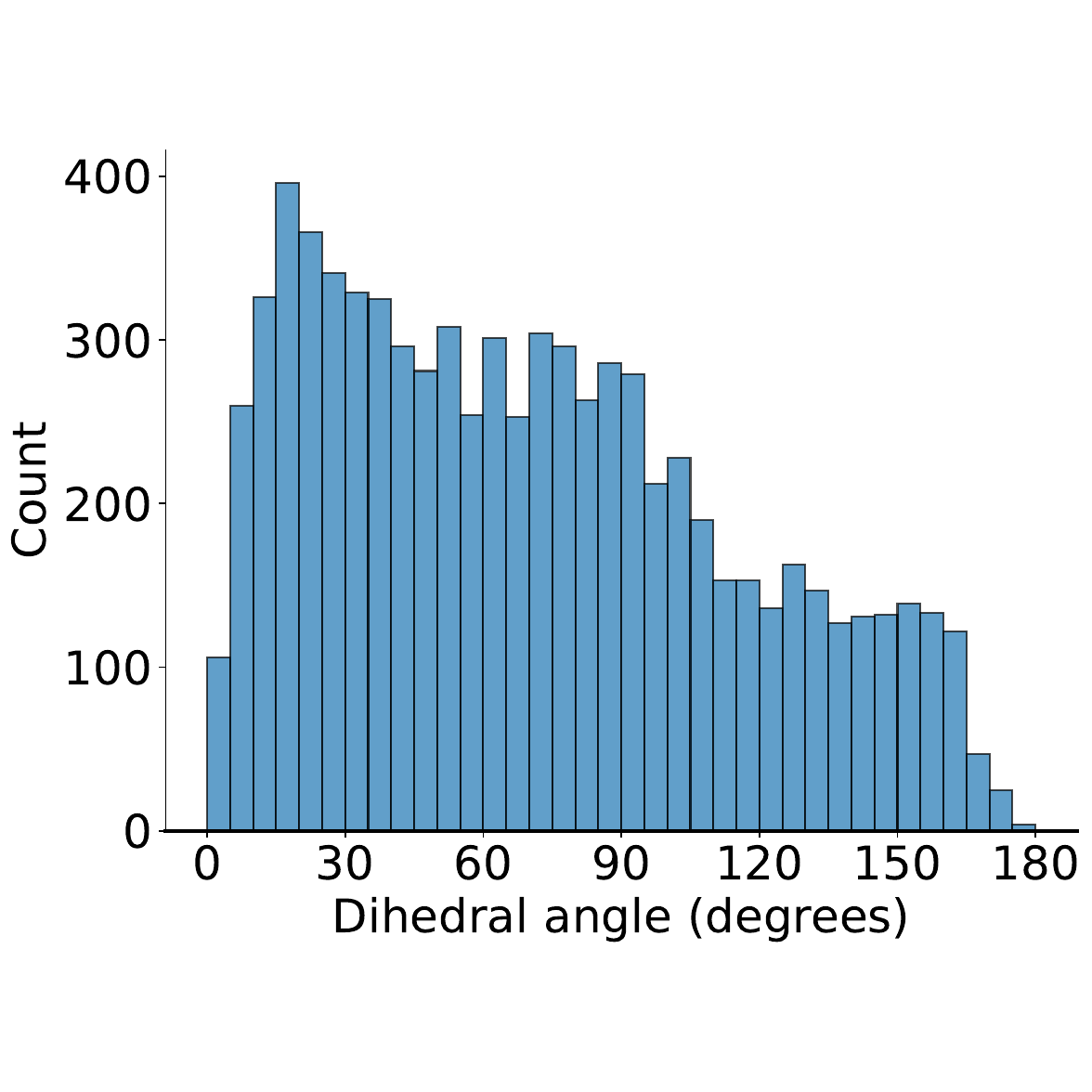} &
        \imgwithlabeltiny{width=\linewidth}{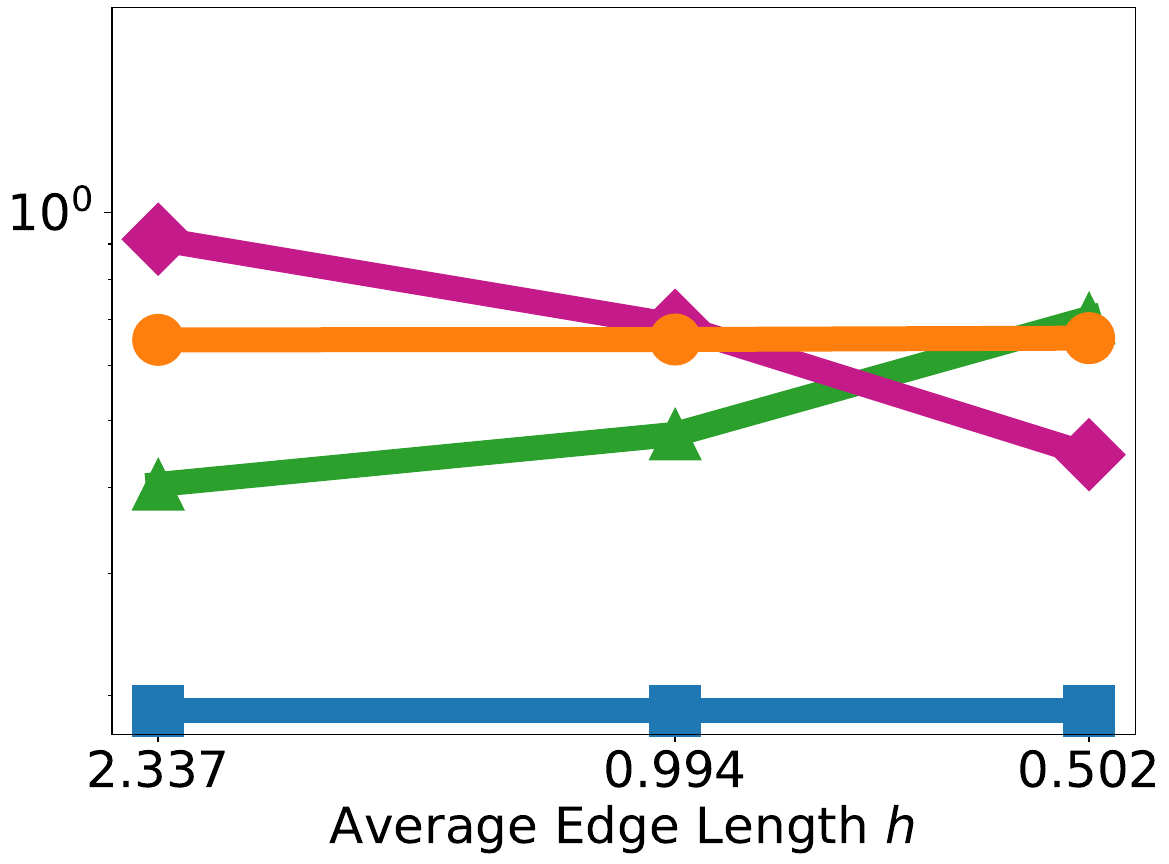}{$L_\infty$} \\
        
        \imglabel{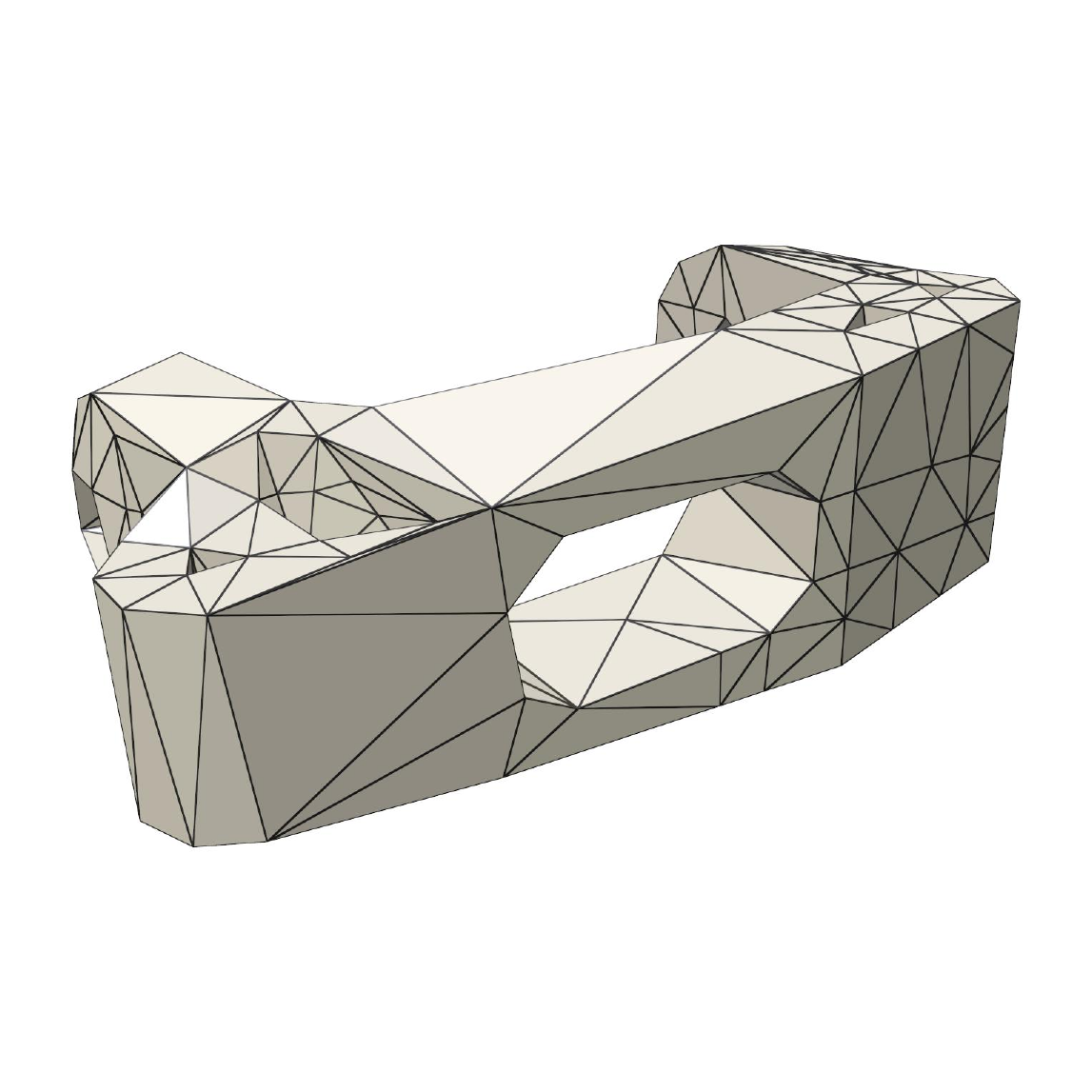}{129926} & 
        \imgwithlabeltiny{width=\linewidth}{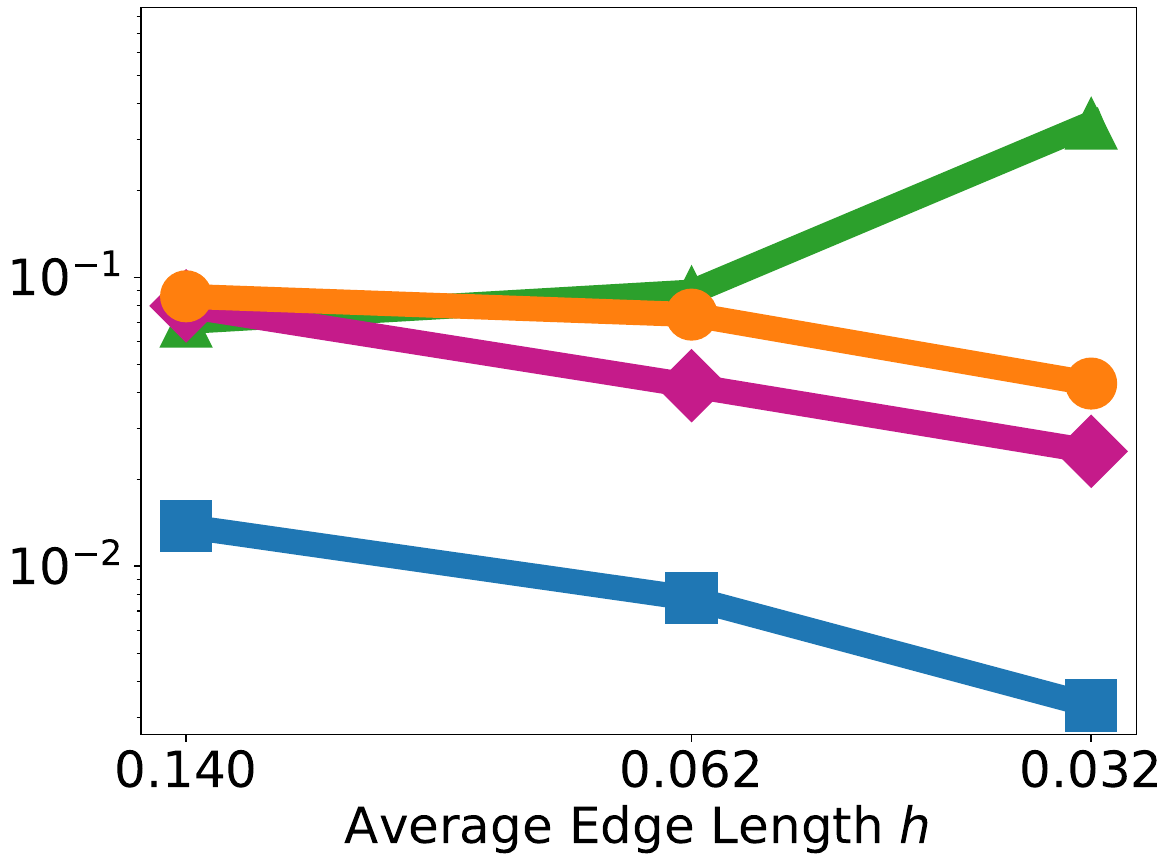}{$L_2$} & 
        \imglabel{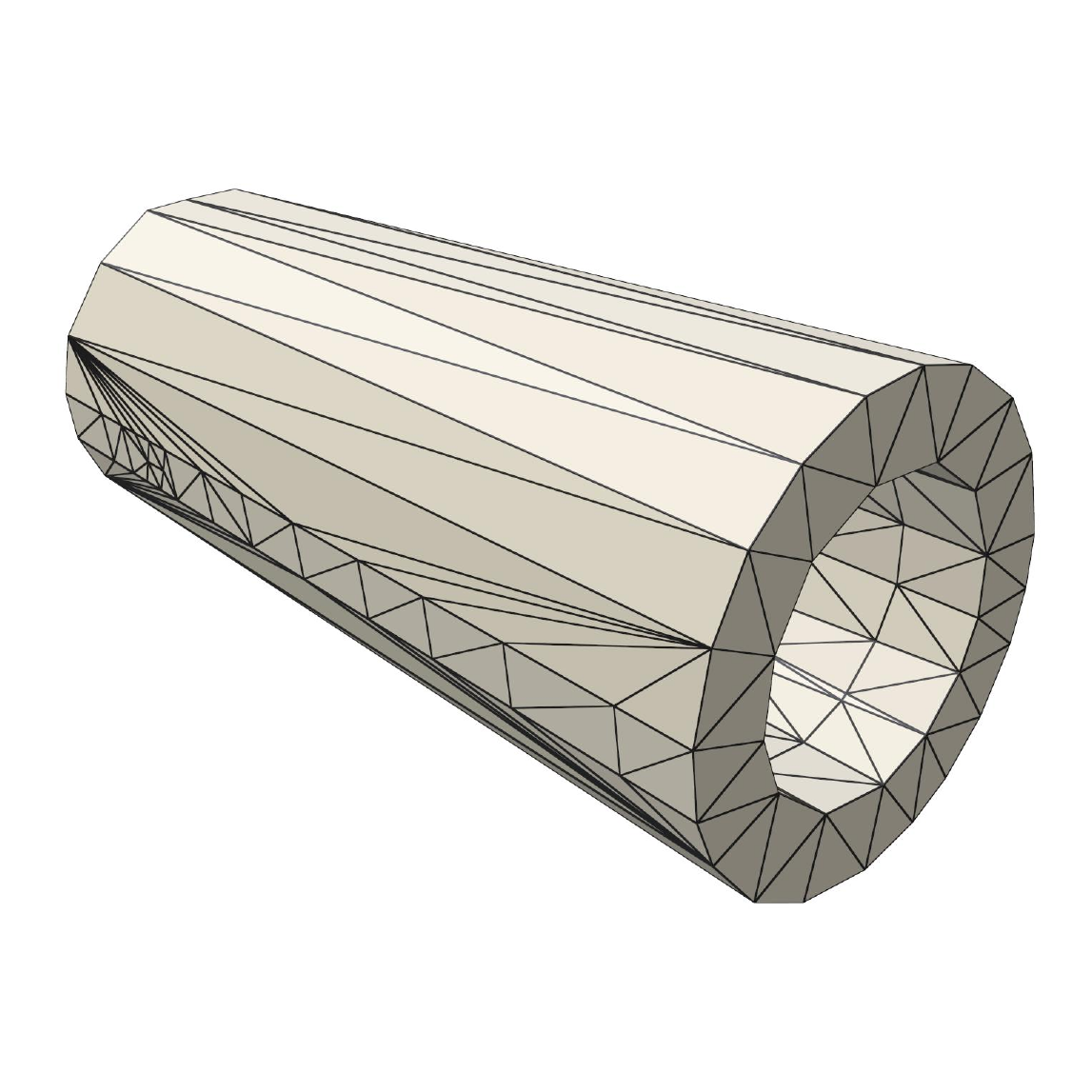}{591213} & 
        \imgwithlabeltiny{width=\linewidth}{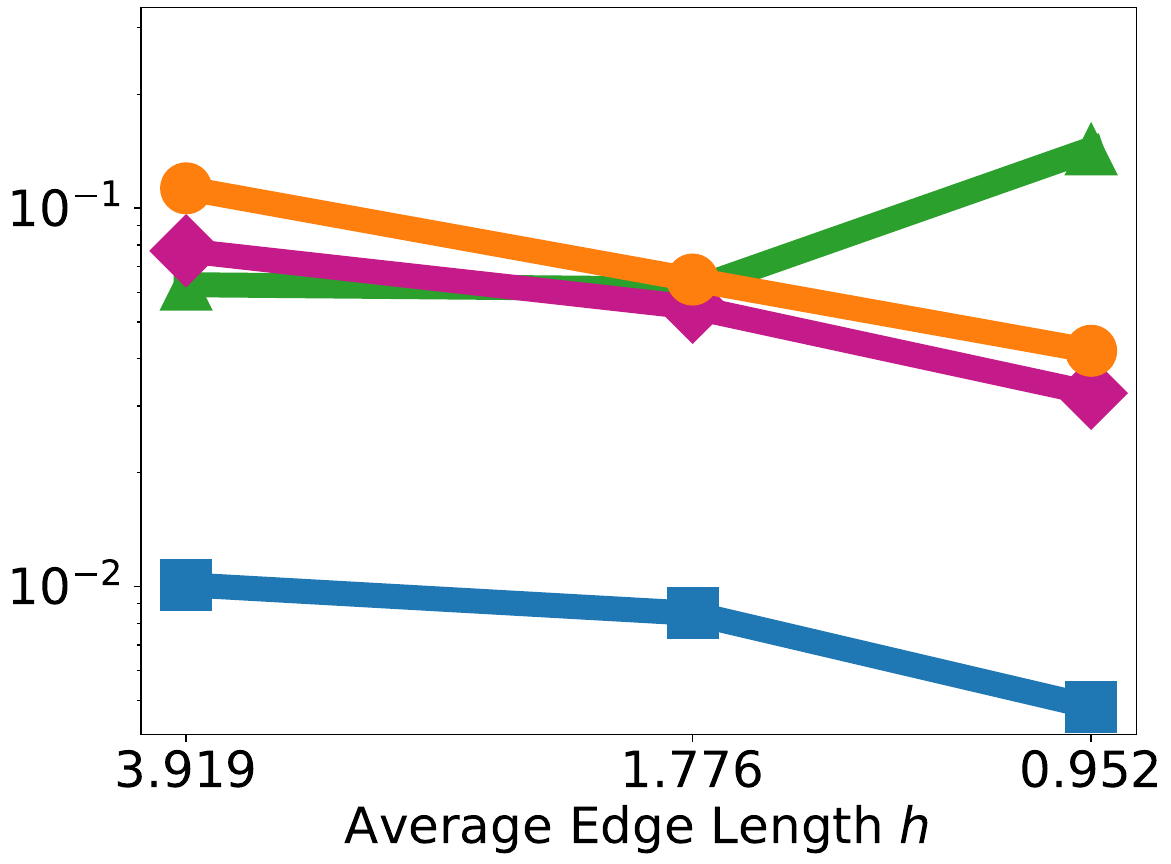}{$L_2$} & 
        \imglabel{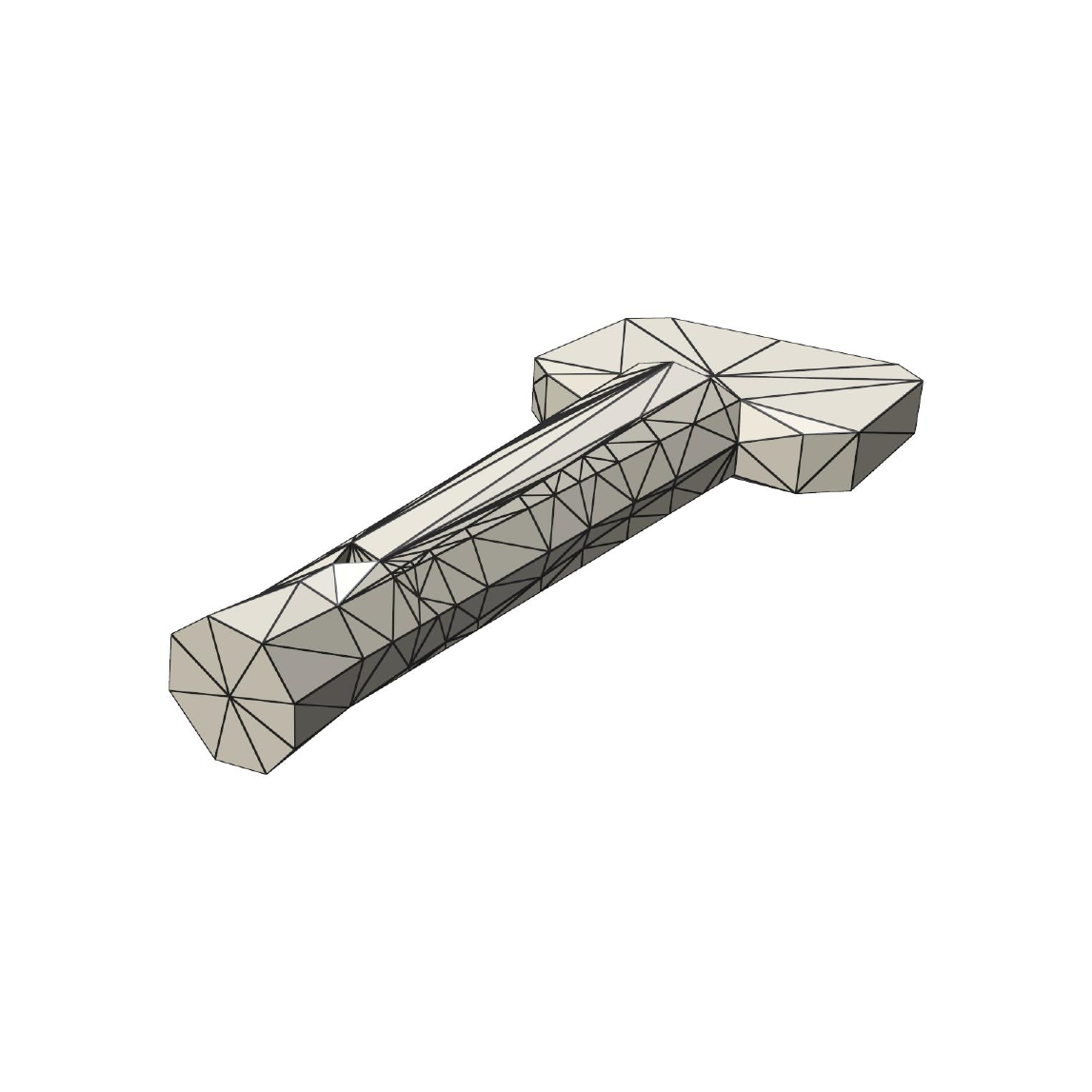}{237739} & 
        \imgwithlabeltiny{width=\linewidth}{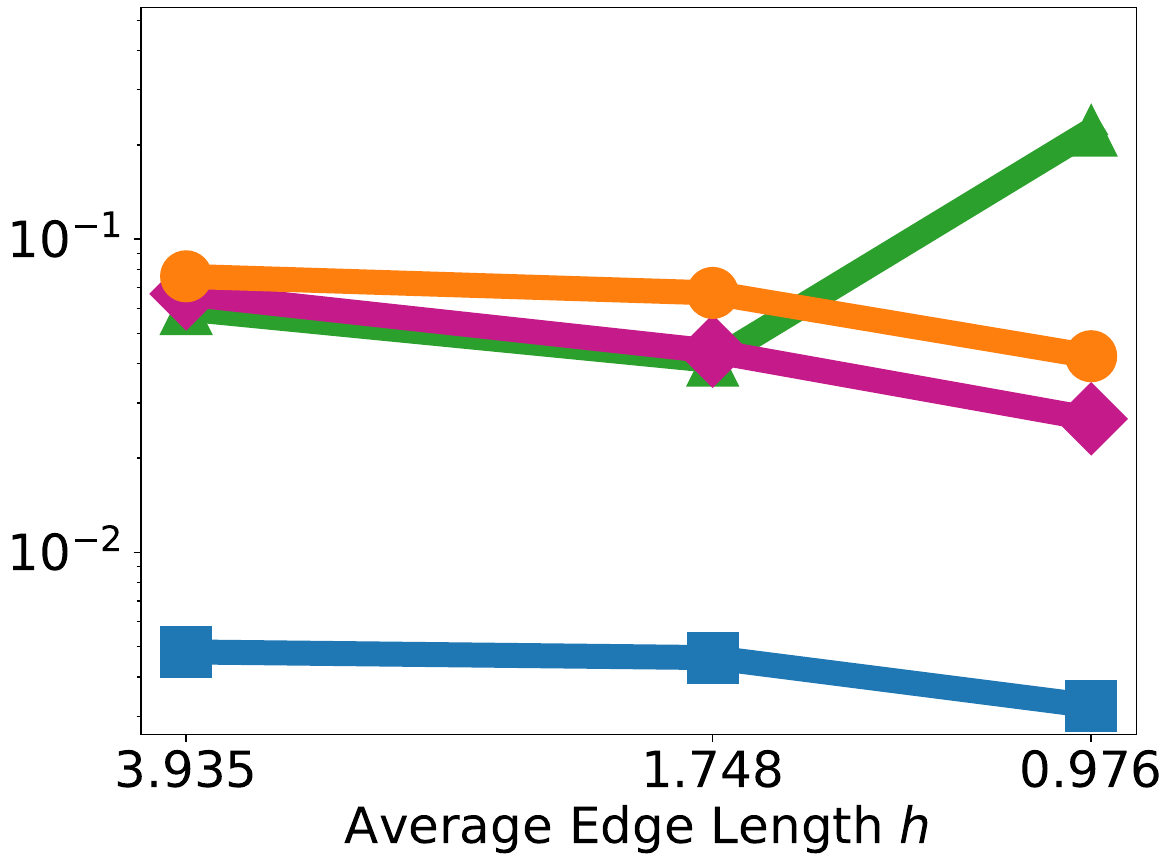}{$L_2$} \\

        \includegraphics[width=\linewidth]{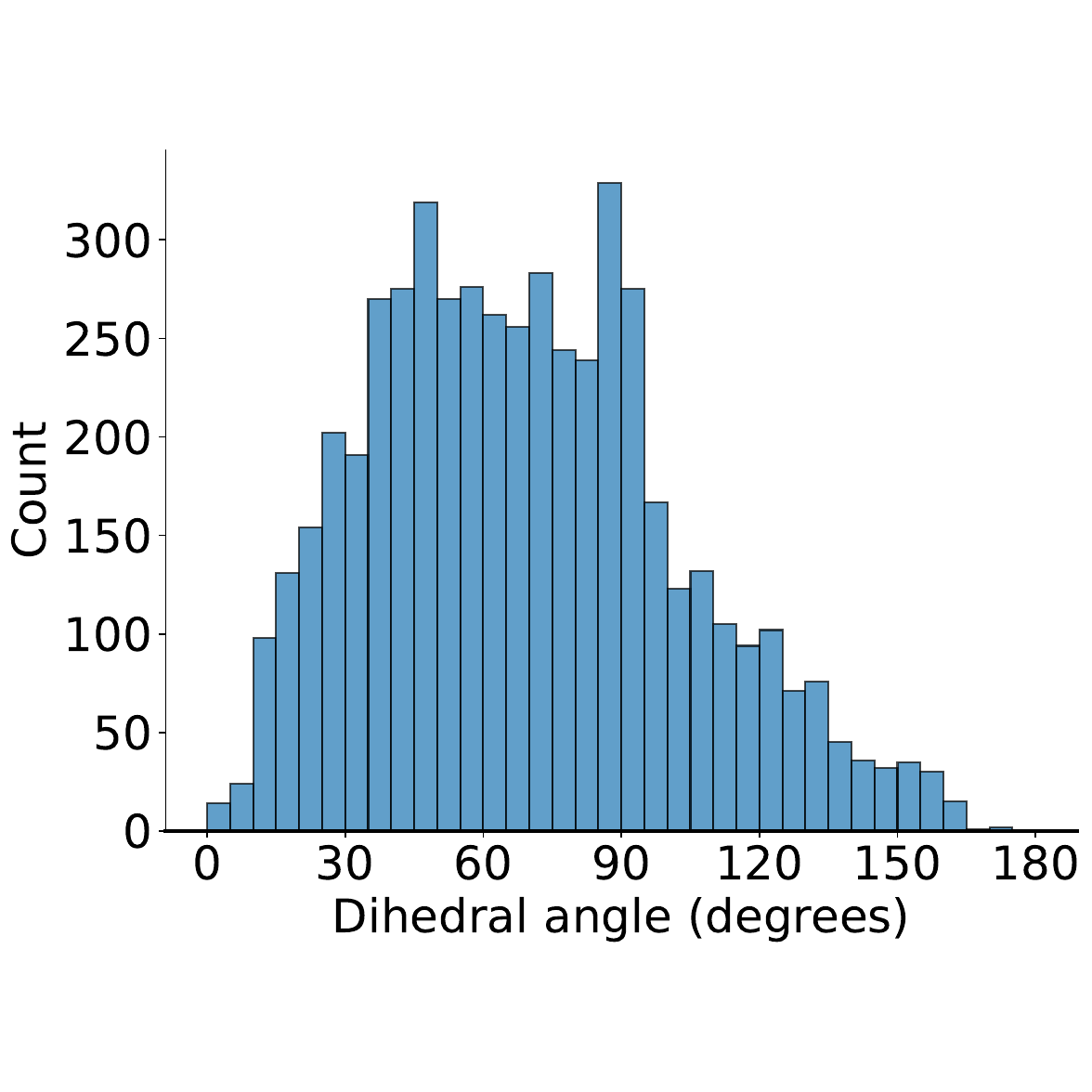} &
        \imgwithlabeltiny{width=\linewidth}{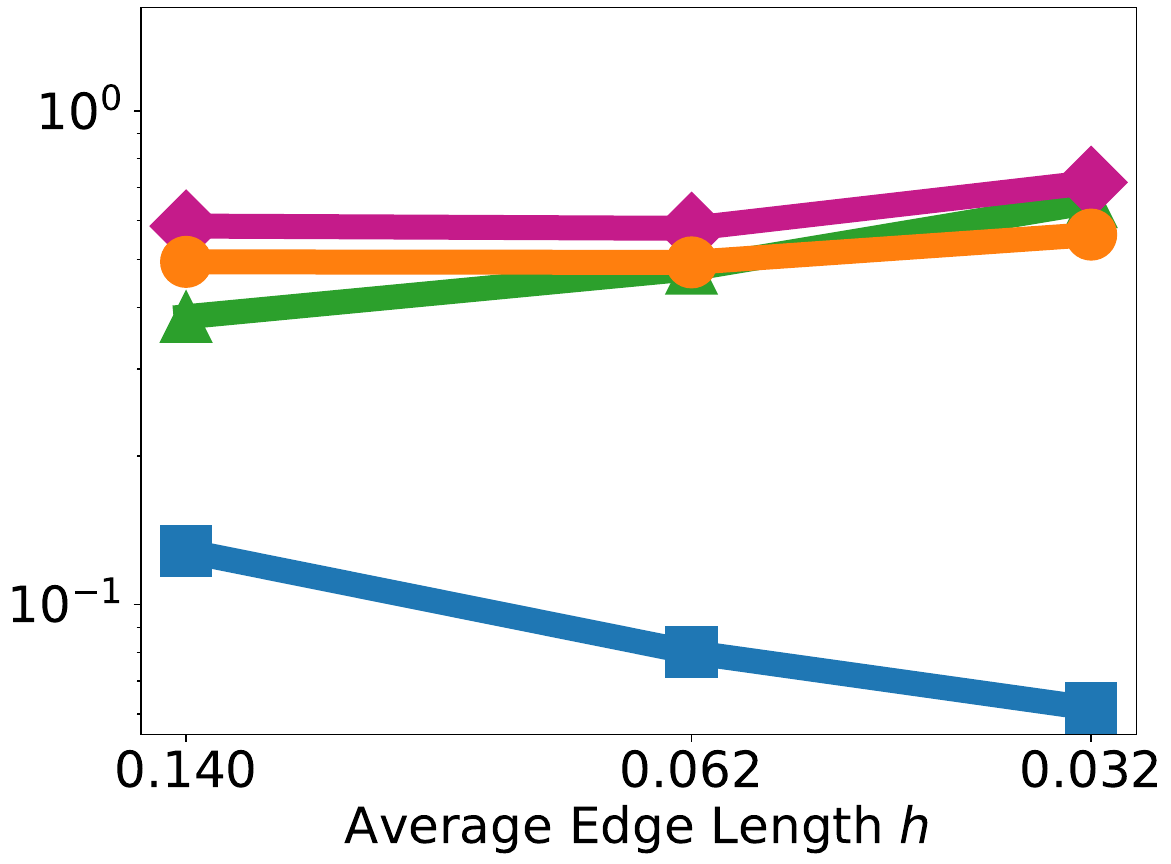}{$L_\infty$} &

         \includegraphics[width=\linewidth]{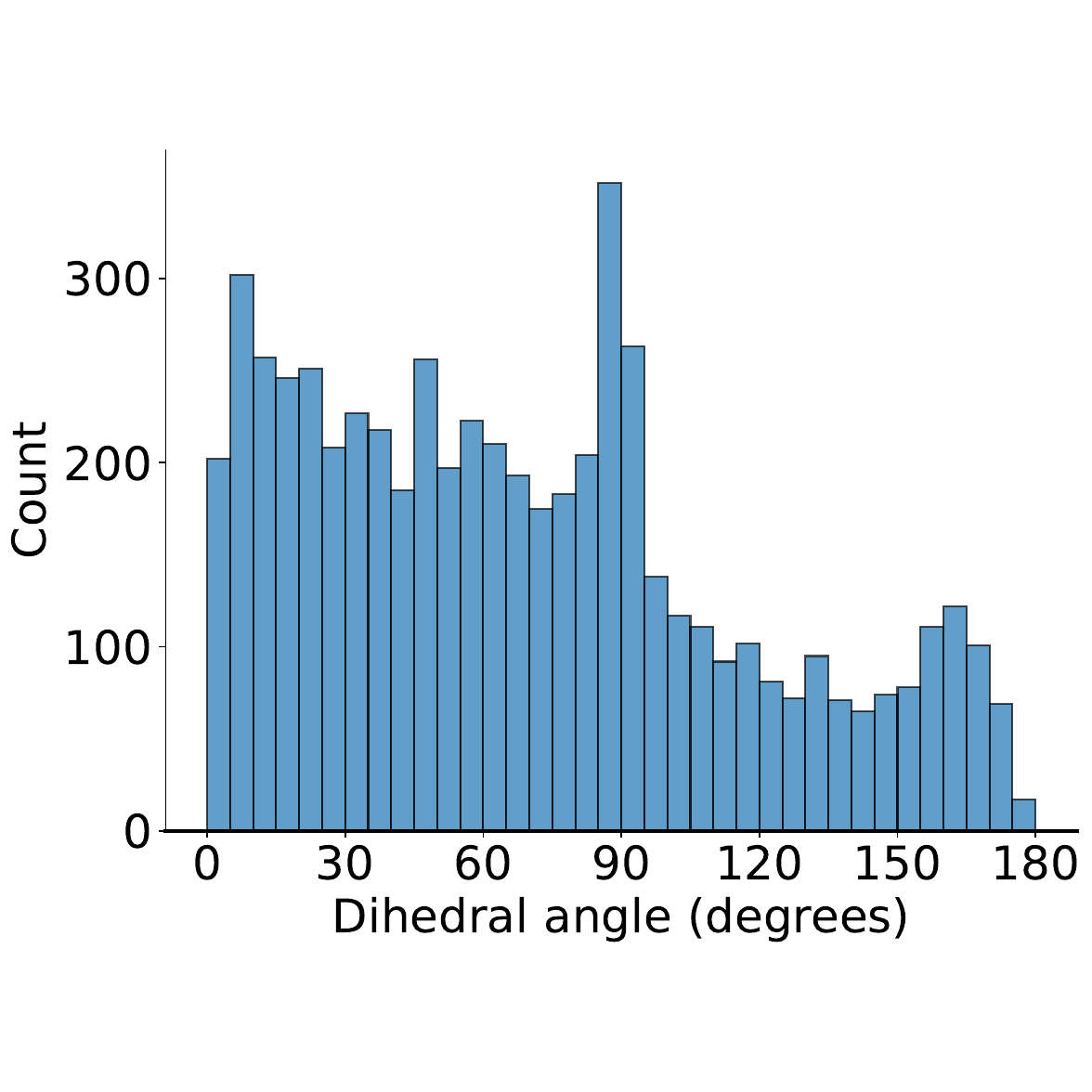} &
        \imgwithlabeltiny{width=\linewidth}{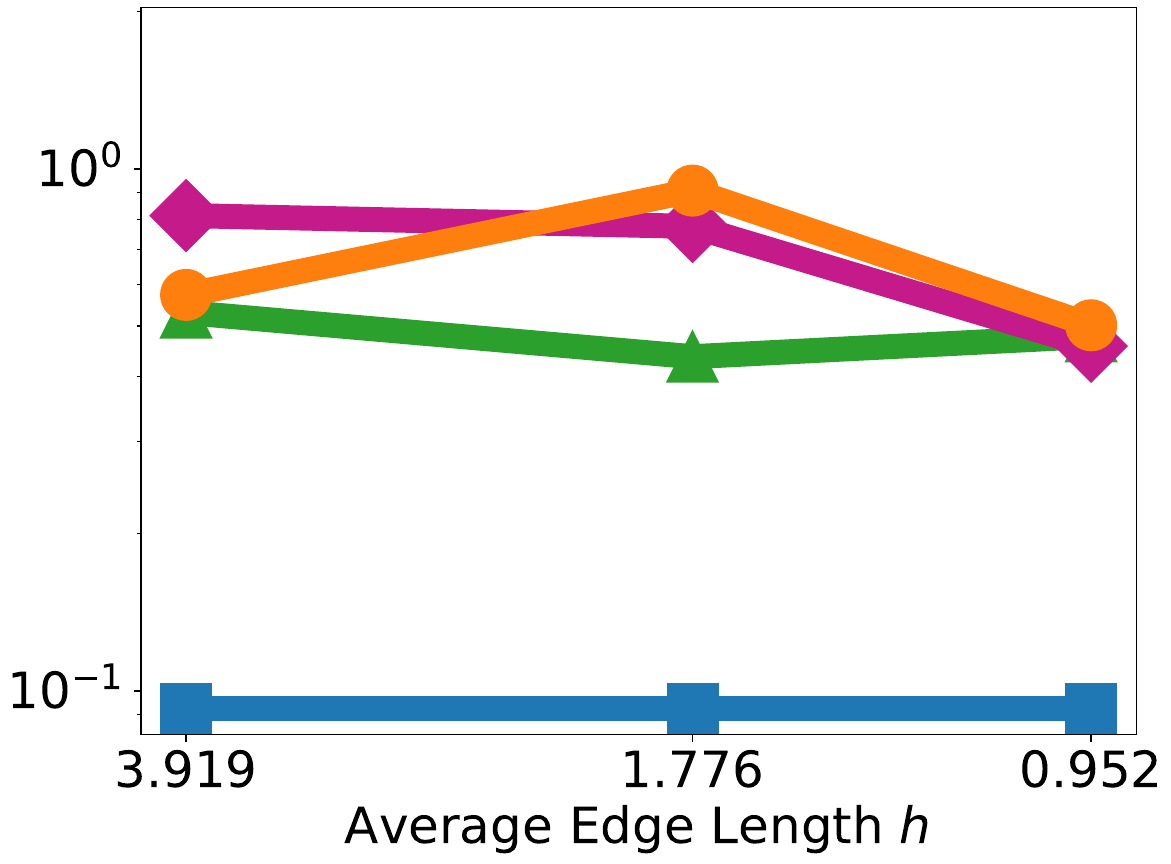}{$L_\infty$} &
        
        \includegraphics[width=\linewidth]{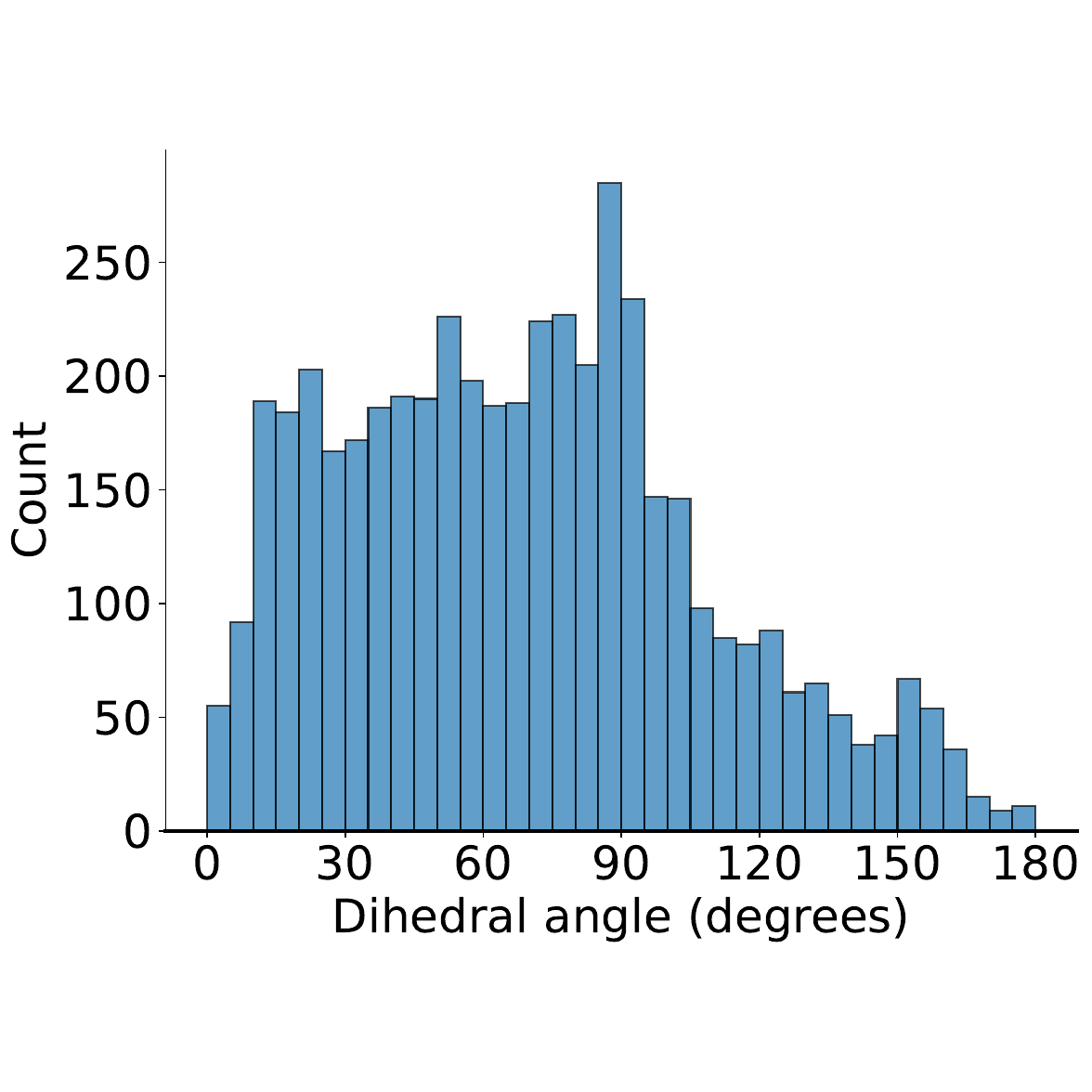} &
        \imgwithlabeltiny{width=\linewidth}{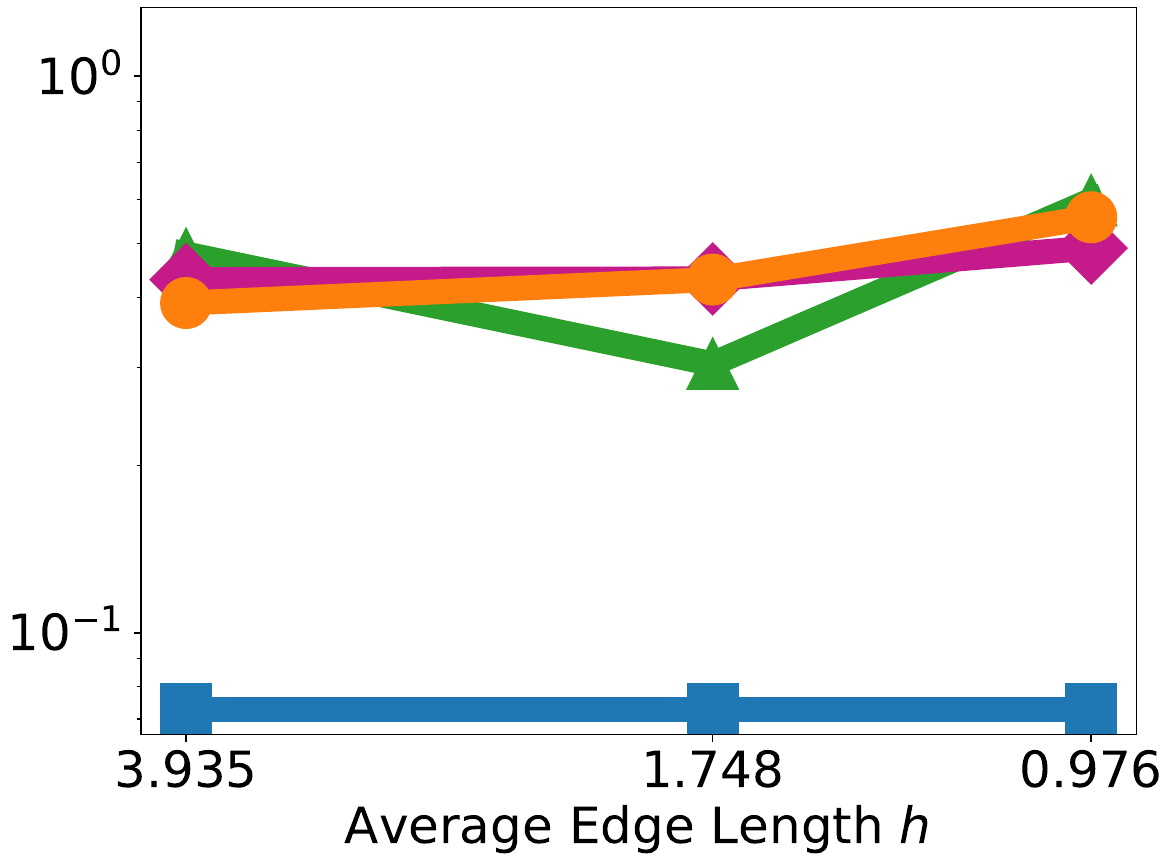}{$L_\infty$} \\
        
        \hline 
    \end{tabular}
    \vspace{2pt}
    \includegraphics[width=0.4\linewidth]{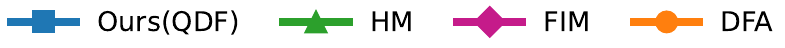}
    \vspace{-6pt}
    \setlength{\abovecaptionskip}{2pt}
    \caption{Convergence on tetrahedral meshes spanning well-shaped (near-regular) to ill-shaped (sliver) tetrahedra. The corresponding histograms show their dihedral-angle distributions. The error plots show Relative $L_2$ and $L_\infty$ error against the mean edge length $h$ of each refined mesh, using the volumetric reference and degree-of-freedom matching protocol in \secref{subsec:setup}. \ours\ maintains the lowest errors across all displayed refinement levels and quality groups. }
    \label{fig:convergence_table_3d}
\end{figure*}

\begin{figure}[t]
    \centering
    
    \begin{subfigure}[c]{0.22\columnwidth} 
        \centering
        \includegraphics[trim=12cm 1cm 12cm 1cm, clip, width=\linewidth]{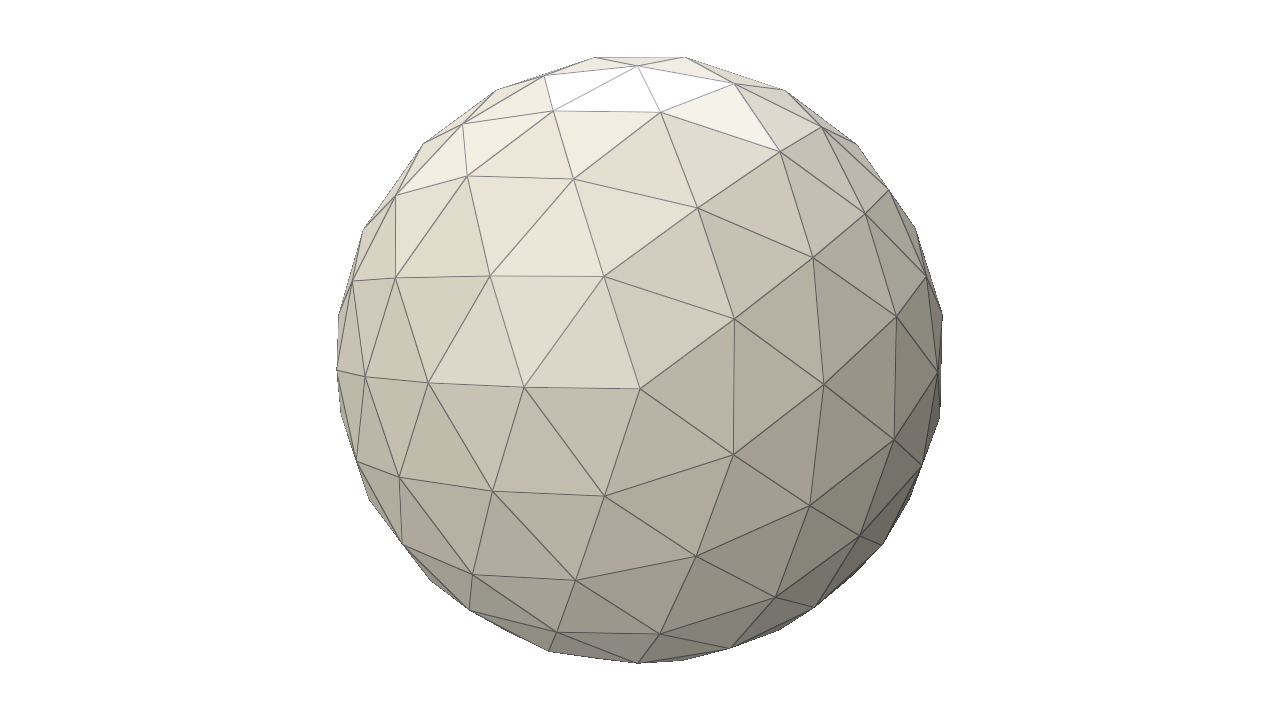}
    \end{subfigure}%
    \hfill %
    \begin{subfigure}[c]{0.35\columnwidth} 
        \centering
        \includegraphics[width=\linewidth]{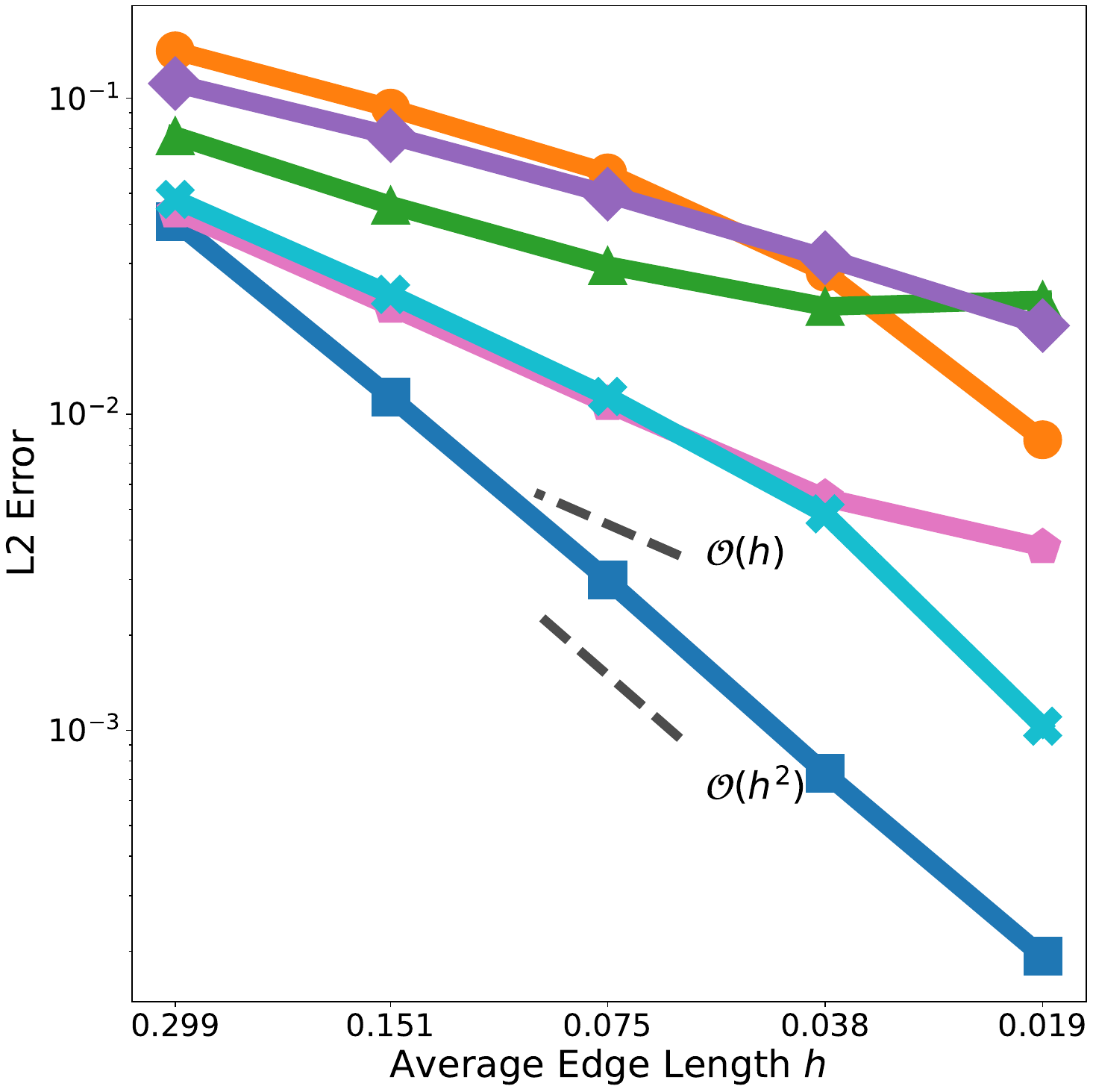}
    \end{subfigure}%
    \hfill %
    \begin{subfigure}[c]{0.35\columnwidth}
        \centering
        \includegraphics[width=\linewidth]{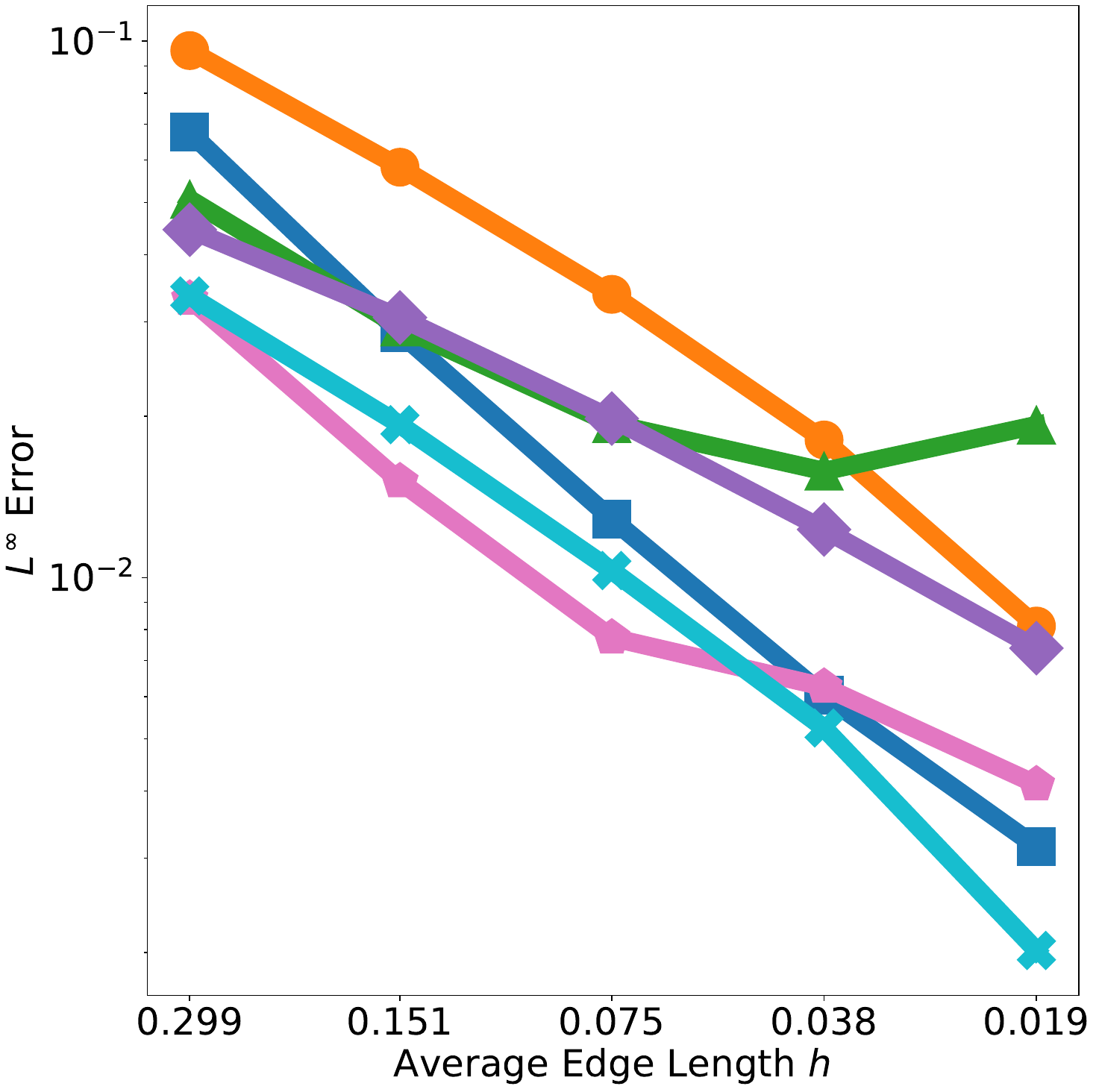}
    \end{subfigure}
    \vspace{2pt}
    \includegraphics[width=\linewidth]{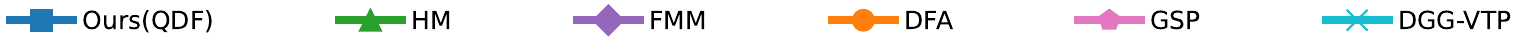}
    \vspace{-6pt}
    \setlength{\abovecaptionskip}{1pt}
    \caption{Convergence to the analytic geodesic distance on the smooth unit sphere. At each level, we apply uniform midpoint subdivision and project the new vertices radially onto the unit sphere. The plots report $L_2$ and $L_\infty$ errors against the analytic great-circle distance as functions of mean edge length $h$. \ours\ exhibits approximately second-order convergence in $L_2$ and attains the lowest $L_2$ error among the baselines. Notably \DGG-VTP is polyhedral exact, so it's ``error'' just measures the difference between the polyhedral surface to a perfect sphere.}
    \label{fig:sphere_convergence}
\end{figure}

\paragraph*{Convergence}
We evaluate convergence under surface and volumetric refinement in \figref{fig:convergence_table} and \figref{fig:convergence_table_3d}, respectively, using the reference solutions and degree-of-freedom matching protocol defined in \secref{subsec:setup}. The errors are plotted against the mean edge length of the base mesh at increasing resolutions. Our method almost always exhibits better error convergence, especially on poor quality meshes.  We note that \HM can diverge in some examples, and conjecture that this is because uniform subdivision preserves the same angles and therefore retains a possibly problematic cotangent weights.

We additionally evaluate convergence toward a known smooth-surface solution in \figref{fig:sphere_convergence}. At each level, we subdivide the sphere mesh and project the new vertices onto the unit sphere. We then compare the computed distances with the analytic great-circle distances at the projected vertices. This experiment measures convergence of the discrete surface solutions toward the geodesic distance on the smooth unit sphere.

\paragraph{Closed-form solutions}

\newcommand{\DualAutoInsetExact}[8]{%

  \def\MainImgWidth{0.35\linewidth}
  \def\InsetWidth{0.15\linewidth}
  \def\MidGap{0.2cm}
  \def\VertGap{0.2cm}

  \begin{tikzpicture}[baseline]

    \coordinate (CenterPoint) at (0,0);

    \node[anchor=south, inner sep=2pt, draw=black, line width=0.8pt] (insetL)
      at ([yshift=\VertGap/2]CenterPoint)
    {%
      \begin{tikzpicture}[x=\InsetWidth, y=\InsetWidth]
        \clip (-0.5,-0.5) rectangle (0.5,0.5);
        \pgfmathsetmacro{\z}{1/#4}
        \pgfmathsetlengthmacro{\ScaledW}{\z*\InsetWidth}
        \pgfmathsetmacro{\ax}{-\z*#2}
        \pgfmathsetmacro{\ay}{-\z*#3}
        \node[anchor=south west, inner sep=0] at (\ax,\ay)
          {\includegraphics[width=\ScaledW]{#1}};
      \end{tikzpicture}%
    };

    \node[anchor=north, inner sep=2pt, draw=black, line width=0.8pt] (insetR)
      at ([yshift=-\VertGap/2]CenterPoint)
    {%
      \begin{tikzpicture}[x=\InsetWidth, y=\InsetWidth]
        \clip (-0.5,-0.5) rectangle (0.5,0.5);
        \pgfmathsetmacro{\z}{1/#8}
        \pgfmathsetlengthmacro{\ScaledW}{\z*\InsetWidth}
        \pgfmathsetmacro{\ax}{-\z*#6}
        \pgfmathsetmacro{\ay}{-\z*#7}
        \node[anchor=south west, inner sep=0] at (\ax,\ay)
          {\includegraphics[width=\ScaledW]{#5}};
      \end{tikzpicture}%
    };

    \coordinate (mainL_pos) at ($(insetL.west |- CenterPoint) + (-\MidGap, 0)$);
    \node[anchor=east, inner sep=0] (mainL) at (mainL_pos)
      {\includegraphics[width=\MainImgWidth]{#1}};

    \coordinate (mainR_pos) at ($(insetL.east |- CenterPoint) + (\MidGap, 0)$);
    \node[anchor=west, inner sep=0] (mainR) at (mainR_pos)
      {\includegraphics[width=\MainImgWidth]{#5}};

    \begin{scope}[
        shift={(mainL.south west)},
        x={($(mainL.south east)-(mainL.south west)$)},
        y={($(mainL.north west)-(mainL.south west)$)}
    ]
      \pgfmathsetmacro{\half}{#4/2}
      \coordinate (roiL_center) at (#2,#3);
      \coordinate (roiL_sw) at ($(roiL_center)+(-\half,-\half)$);
      \coordinate (roiL_ne) at ($(roiL_center)+(\half,\half)$);
      \coordinate (roiL_se) at ($(roiL_center)+(\half,-\half)$);
      \coordinate (roiL_nw) at ($(roiL_center)+(-\half,\half)$);
      \draw[red, thick] (roiL_sw) rectangle (roiL_ne);
    \end{scope}

    \begin{scope}[
        shift={(mainR.south west)},
        x={($(mainR.south east)-(mainR.south west)$)},
        y={($(mainR.north west)-(mainR.south west)$)}
    ]
      \pgfmathsetmacro{\half}{#8/2}
      \coordinate (roiR_center) at (#6,#7);
      \coordinate (roiR_sw) at ($(roiR_center)+(-\half,-\half)$);
      \coordinate (roiR_ne) at ($(roiR_center)+(\half,\half)$);
      \coordinate (roiR_se) at ($(roiR_center)+(\half,-\half)$);
      \coordinate (roiR_nw) at ($(roiR_center)+(-\half,\half)$);
      \draw[red, thick] (roiR_sw) rectangle (roiR_ne);
    \end{scope}

    \draw[black] (roiL_ne) -- (insetL.north west);
    \draw[black] (roiL_se) -- (insetL.south west);

    \draw[black] (roiR_nw) -- (insetR.north east);
    \draw[black] (roiR_sw) -- (insetR.south east);

  \end{tikzpicture}%
}

\begin{figure}[!t]
    \centering

    \begin{subfigure}[b]{\linewidth}
        \centering
        \setlength{\tabcolsep}{2pt}
        \begin{tabular}{@{}cc@{}}
            \small PL methods & \small Ours (\ours) \\
            \includegraphics[width=0.42\linewidth]{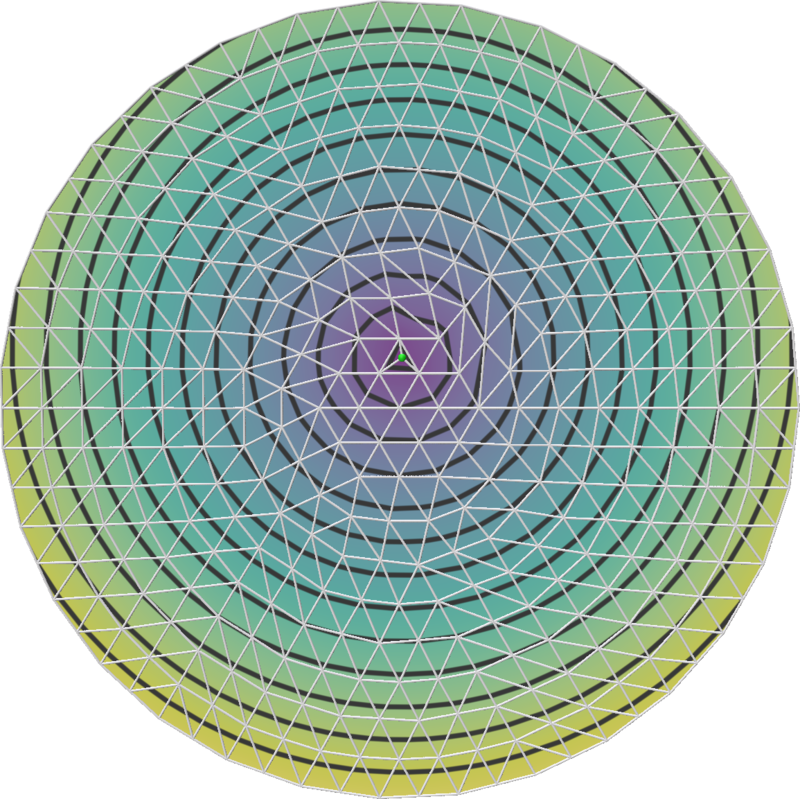} &
            \includegraphics[width=0.42\linewidth]{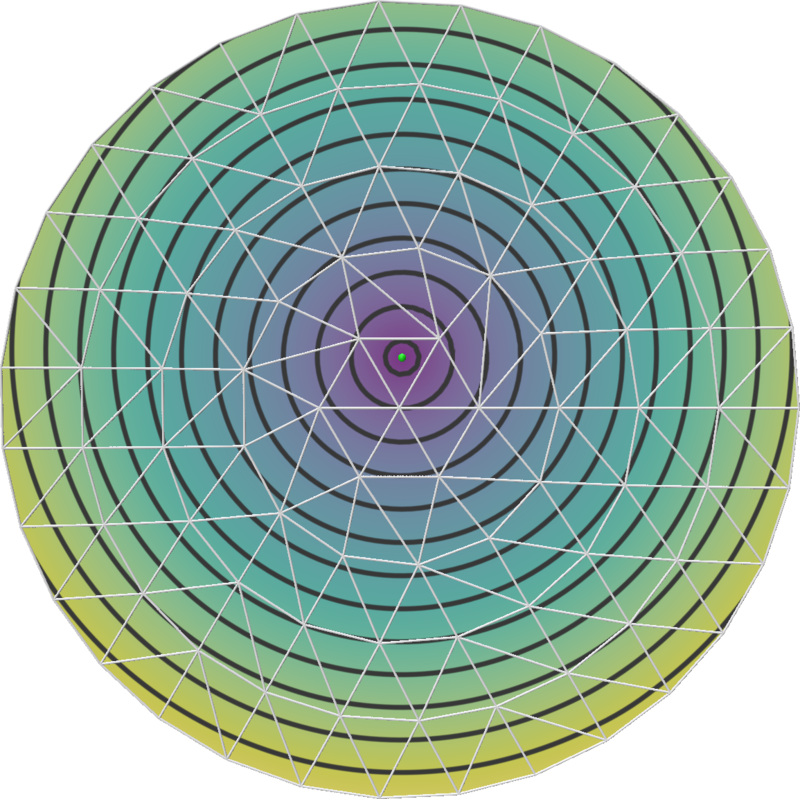} \\[1pt]
            \includegraphics[trim=0 175 0 175, clip, width=0.42\linewidth]{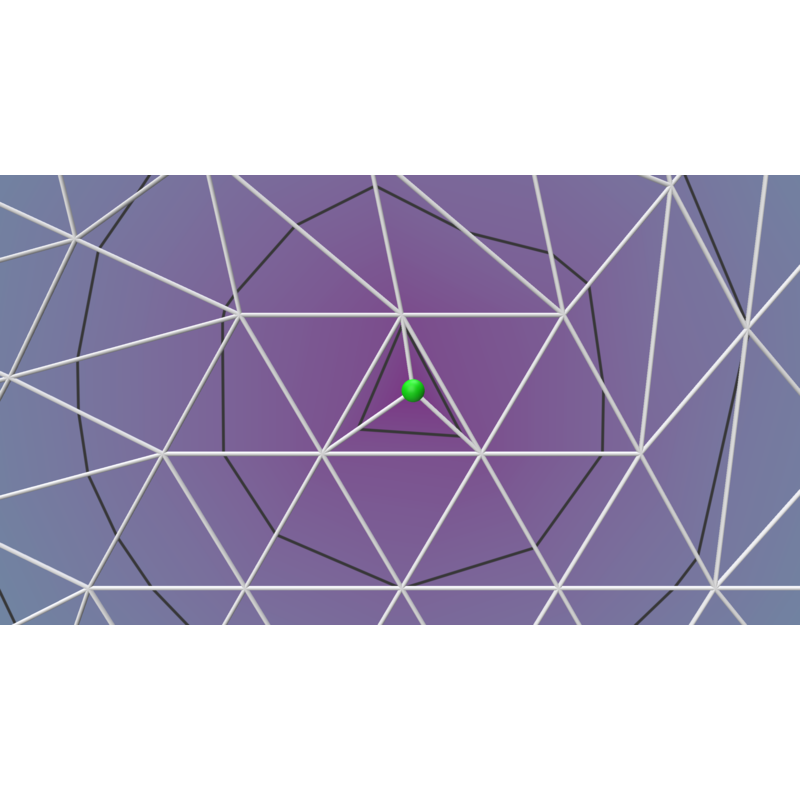} &
            \includegraphics[trim=0 175 0 175, clip, width=0.42\linewidth]{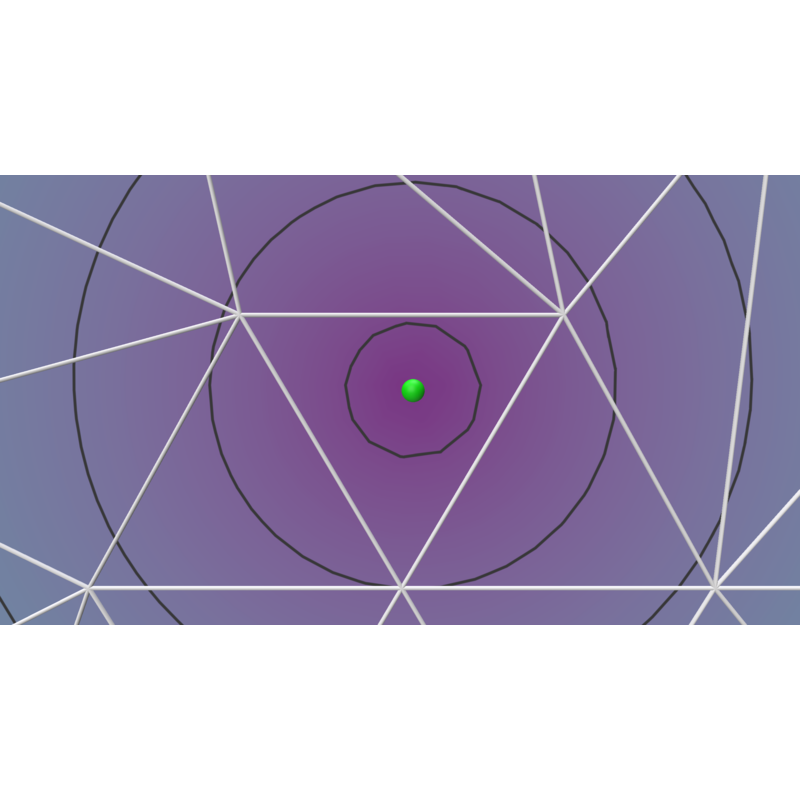} \\
        \end{tabular}
        \caption{Surface: a unit disc with a non-vertex source (zoom-in). Rel-$L_2$ errors are: \HM (\num{3.05e-02}), \FMM (\num{4.20e-02}), \DFA (\num{7.28e-02}), and \ours
        (\num{4.98e-16}). The quadratic interpolation clearly reproduces the result to machien precision, while the linear one
        introduces distortion.}
        \label{fig:disc}
    \end{subfigure}
    \\[0.8ex]
    \begin{subfigure}[b]{\linewidth}
        \centering
        \setlength{\tabcolsep}{1pt}
        \renewcommand{\arraystretch}{1}
        \begin{tabular}{@{}ccccc@{}}
        \small \textbf{Mesh} & \small \textbf{HM} & \small \textbf{FIM} & \small \textbf{DFA} & \small \textbf{Ours (\ours)} \\
          \includegraphics[width=0.19\linewidth]{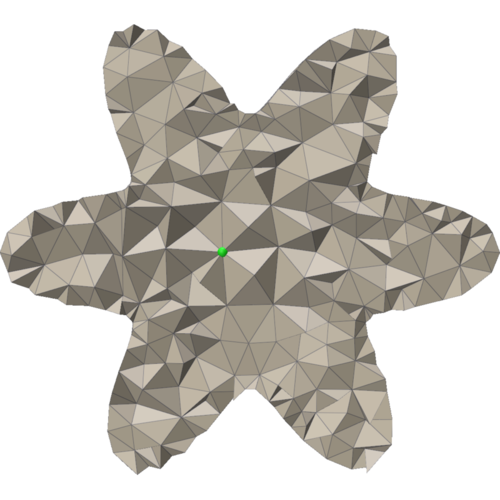} &
          \includegraphics[width=0.19\linewidth]{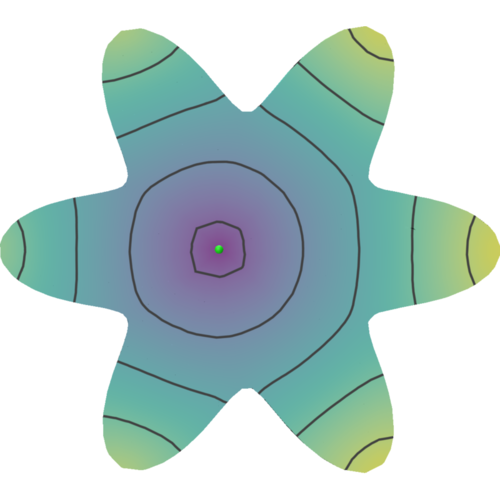} &
          \includegraphics[width=0.19\linewidth]{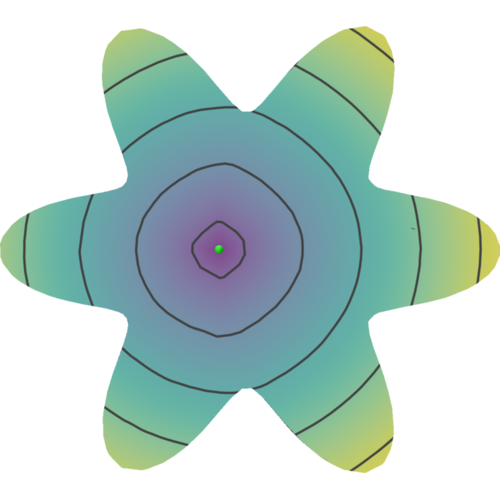} &
          \includegraphics[width=0.19\linewidth]{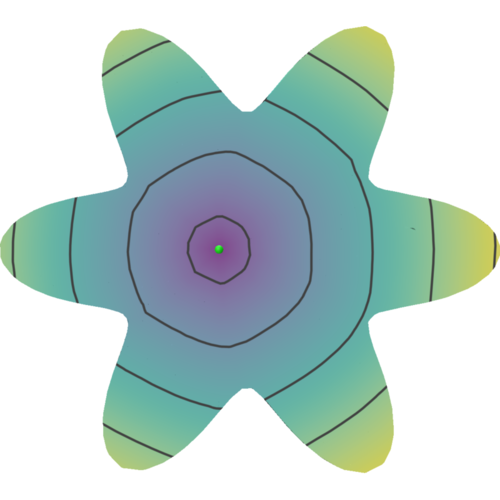} &
          \includegraphics[width=0.19\linewidth]{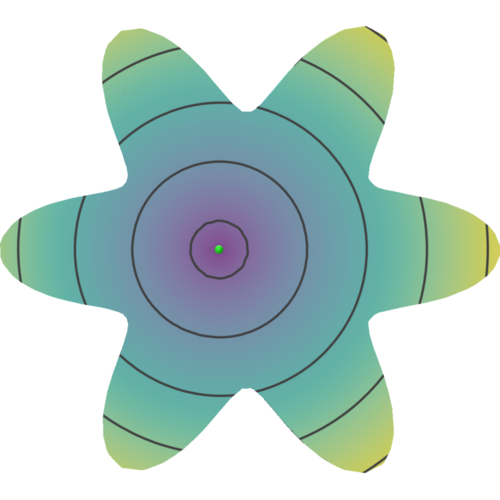} \\

          Rel. $L_2$&
          \num{7.62e-02} &
          \num{9.40e-2} &
          \num{1.32e-01} &
          \num{4.60e-09} \\

        \end{tabular}
        \caption{Volume: a star-shaped domains with the source in each kernel. }
        \label{fig:tet_number}
    \end{subfigure}

    \caption{Verification against closed-form Euclidean distances on a surface (a) and in a volume (b). In all cases, the squared distance is exactly quadratic and is therefore representable in our element space. \ours\ reproduces the disc solution to machine precision and yields substantially lower errors than the piecewise-linear baselines on the volumetric examples.}
    \label{fig:closed_form}
\end{figure}

We first consider a unit disc (\figref{fig:disc}) with a source located in the interior of a triangle rather than at an existing vertex. To make this source available to the piecewise-linear methods, we split the containing triangle by connecting the source to its three vertices. The squared Euclidean distance is exactly quadratic and therefore lies in our element space. With a stricter stopping criterion than the benchmark default, \ours\ recovers this solution to machine precision, whereas the piecewise-linear fields exhibit visible distortion near the source. We perform an analogous experiment on volumetric domains in \figref{fig:tet_number}, where the error of \ours\ is lower by several orders of magnitude.

\subsection{Robustness (C2)}
\label{subsec:robustness}
\begin{figure}[!htbp]
    \centering
    \setlength{\tabcolsep}{1pt} 
    \resizebox{\linewidth}{!}{%
    \begin{tabular}{ C{0.10\textwidth} 
    @{\hspace{3pt}}!{\vrule width 1.5pt}@{\hspace{5pt}} C{0.10\textwidth}  C{0.10\textwidth} C{0.10\textwidth} C{0.10\textwidth} }
        \toprule
        
        \scriptsize \textbf{Mesh}  & \scriptsize \textbf{HM} & \scriptsize \textbf{FMM} & \scriptsize \textbf{DFA} & \scriptsize \textbf{Ours (\ours)} \\
        \midrule

        \includegraphics[width=0.9\linewidth]{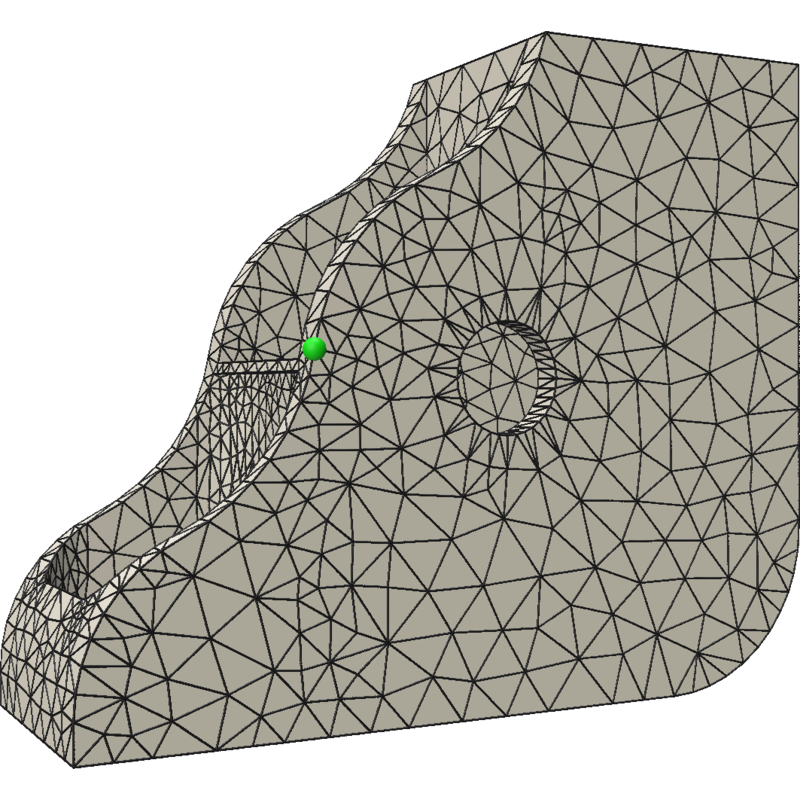} &
        \includegraphics[width=0.9\linewidth]{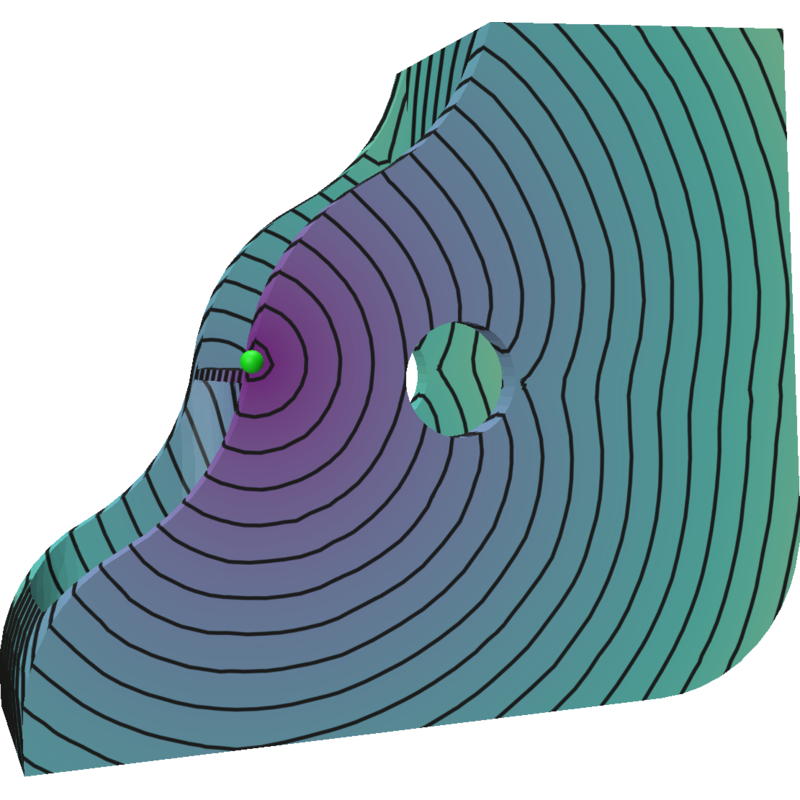} & 
        \includegraphics[width=0.9\linewidth]{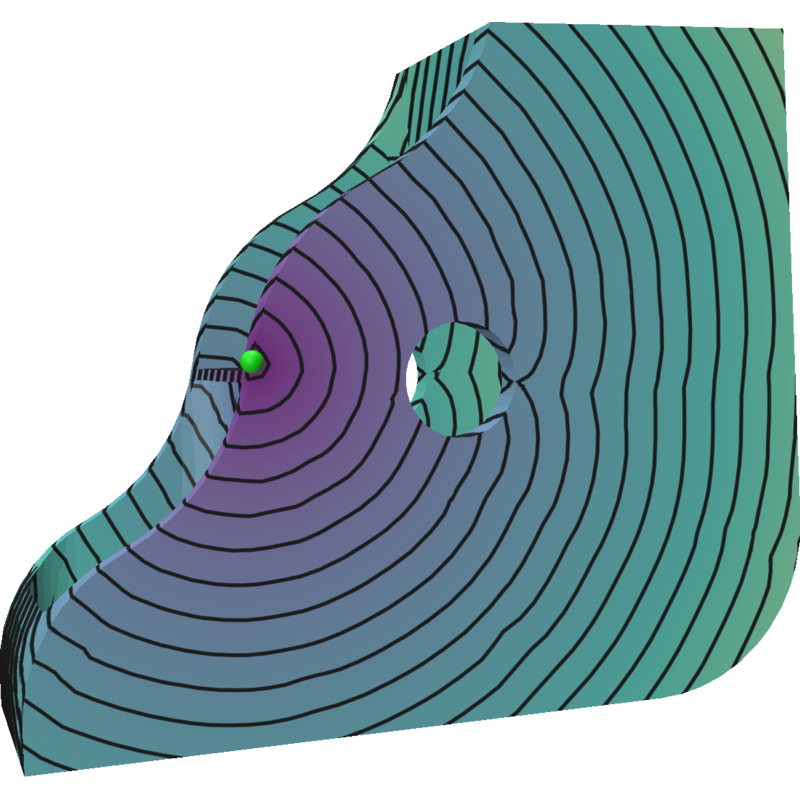} & 
        \includegraphics[width=0.9\linewidth]{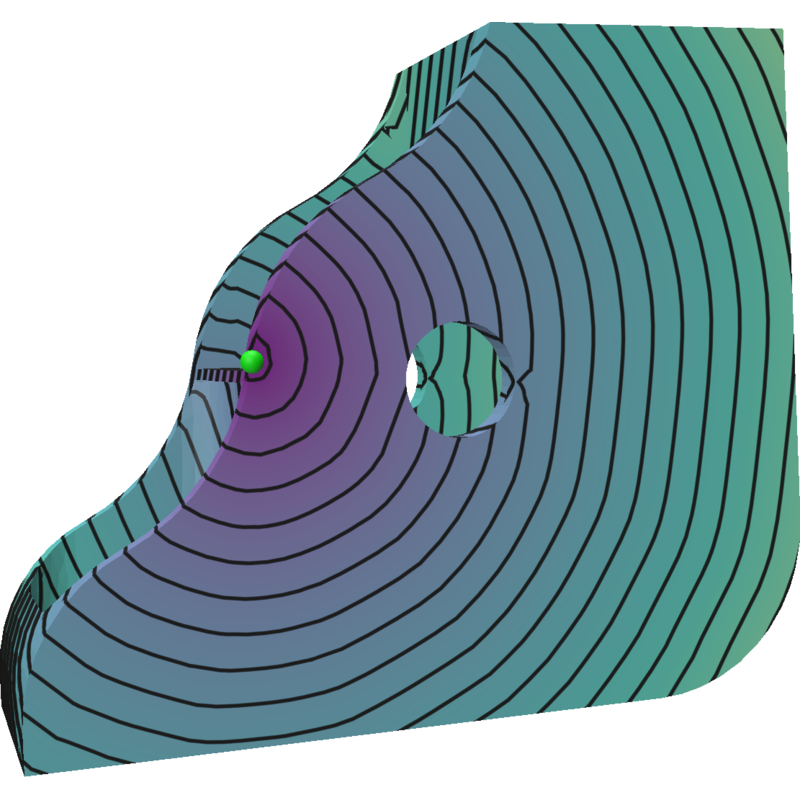} & 
        \includegraphics[width=0.9\linewidth]{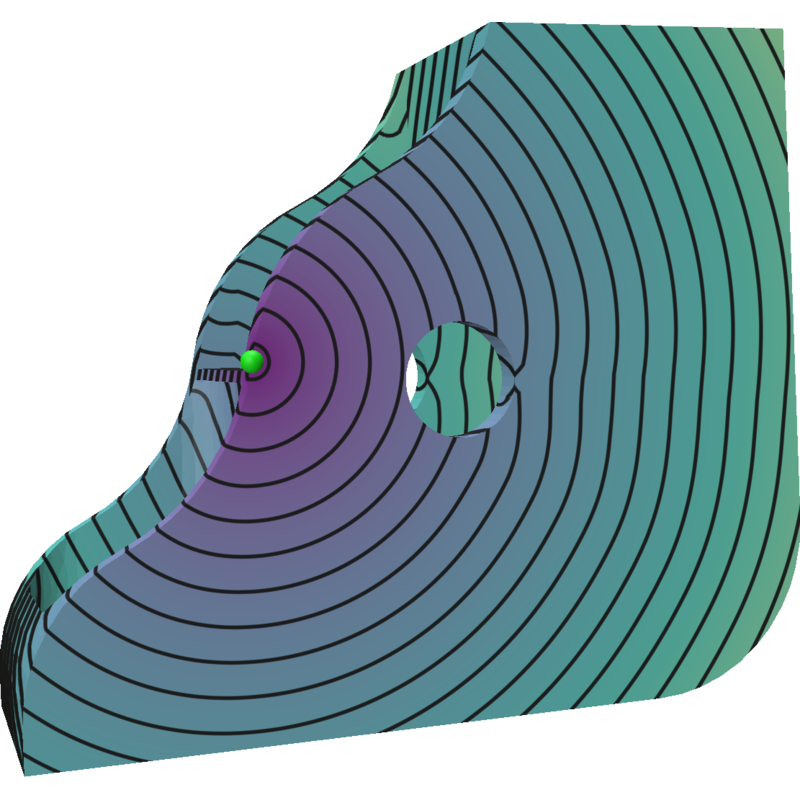} \\

        Rel-$L_2$  
        & \num{5.00e-02} 
        & \num{3.48e-02}  
        & \num{3.29e-02} 
        &  \num{1.28e-02} \\
        Time(s)  
        & 0.0335 
        & 0.0100 
        & 0.1575  
        & 0.2910  \\
        
        \AddAutoInsetExact{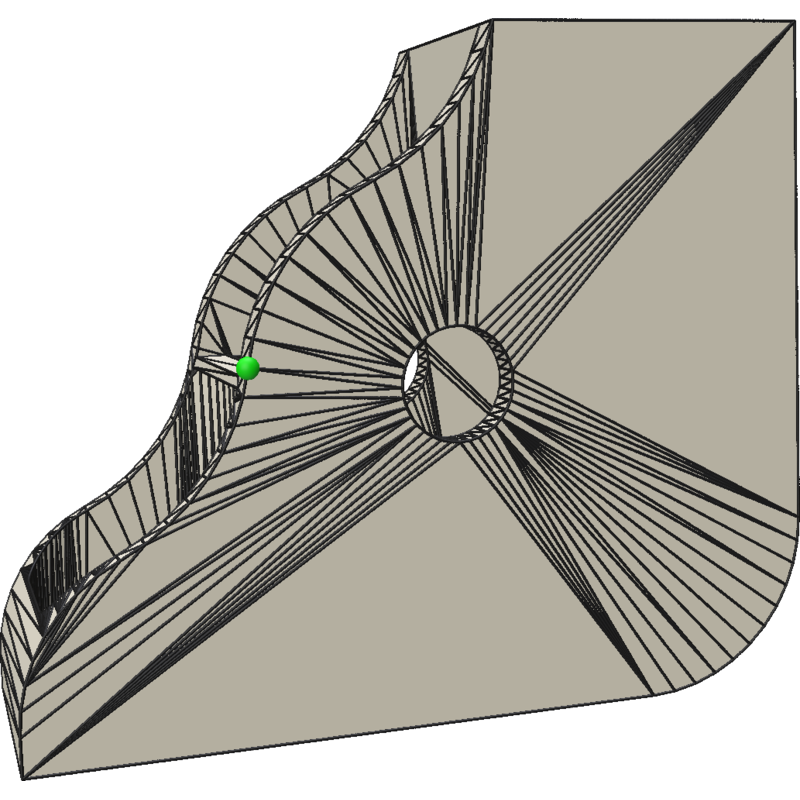}{0.45}{0.5}{0.35}{0.1cm}{0.7\linewidth} &
        \AddAutoInsetExact{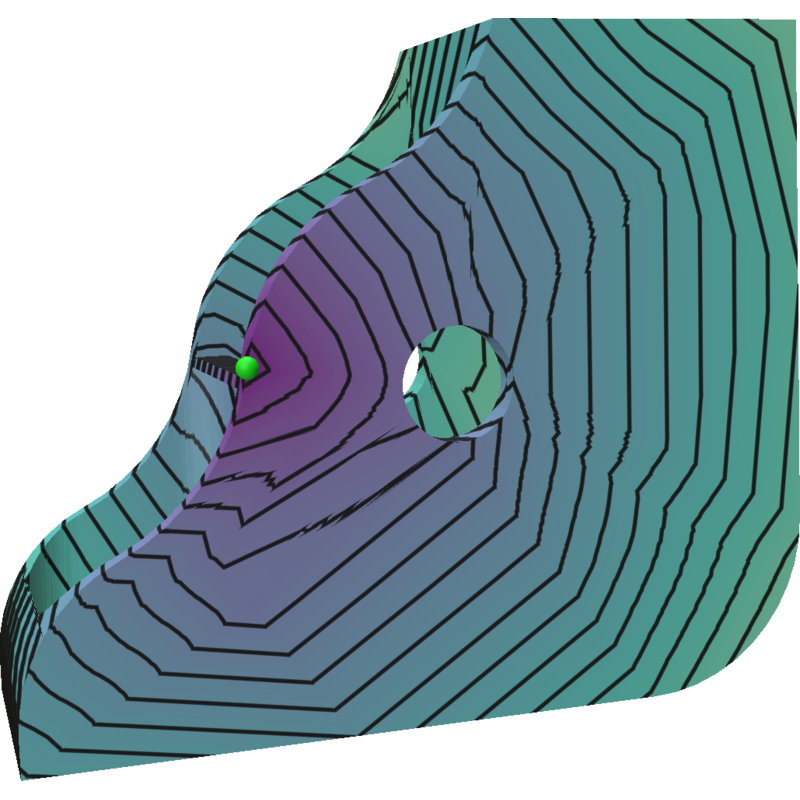}{0.45}{0.5}{0.35}{0.1cm}{0.7\linewidth} &
        \AddAutoInsetExact{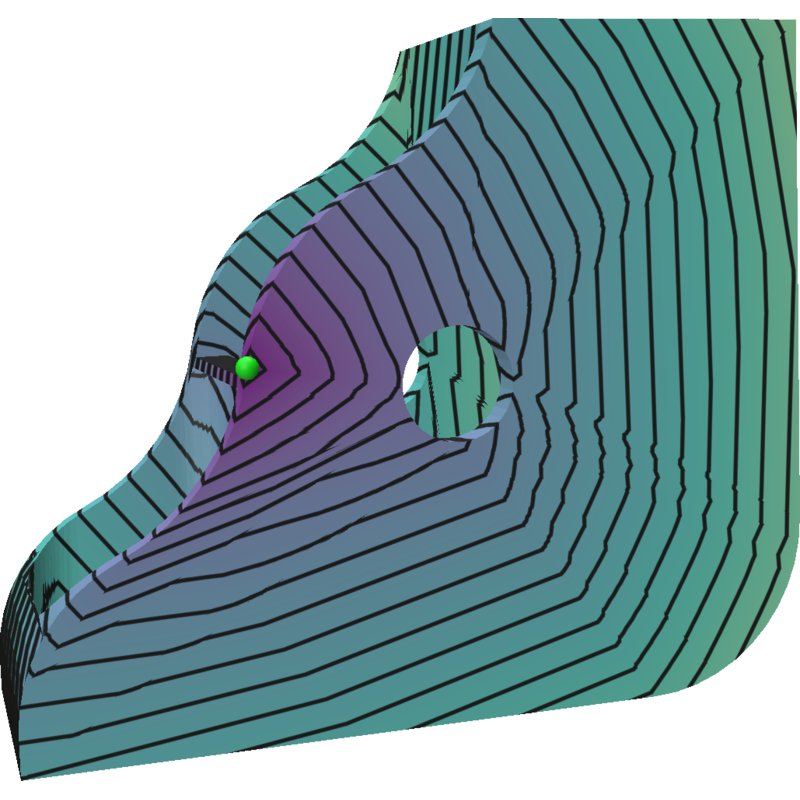}{0.45}{0.5}{0.35}{0.1cm}{0.7\linewidth} &
        \AddAutoInsetExact{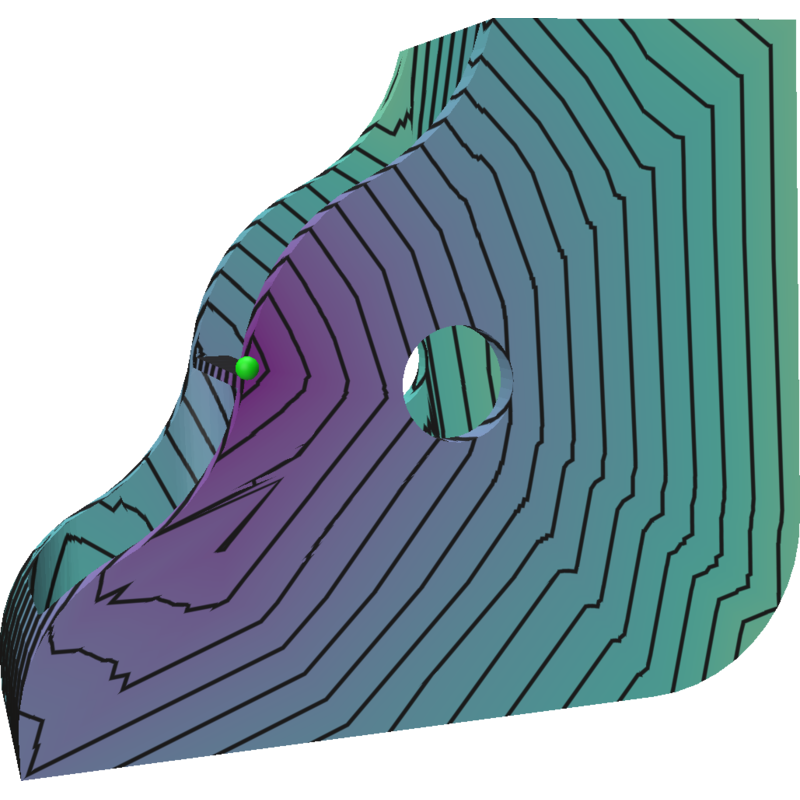}{0.45}{0.5}{0.35}{0.1cm}{0.7\linewidth} &
        \AddAutoInsetExact{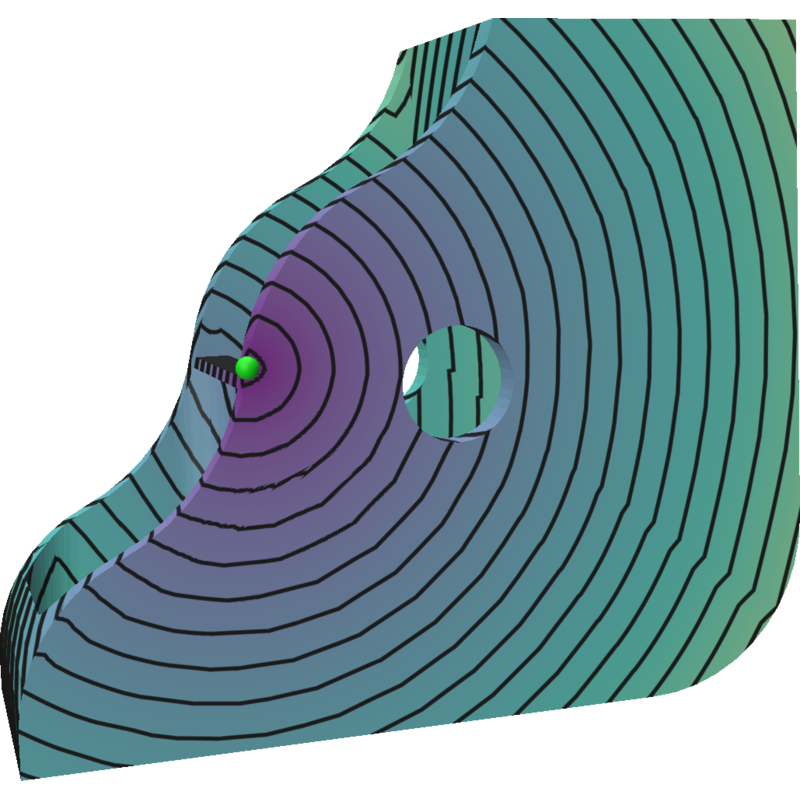}{0.45}{0.5}{0.35}{0.1cm}{0.7\linewidth} \\

        \noalign{\vspace{2pt}}
        Rel-$L_2$
        & \num{6.76e-02} 
        & \num{1.70e-01}  
        & \num{2.08e-01}  
        & \num{6.45e-02} \\
       Time(s)  
       & 0.0176
       & 0.0025 
       & 0.0468 
       & 0.0656 \\

        \midrule

        \includegraphics[trim=0 0 0 8cm, clip, width=0.9\linewidth]{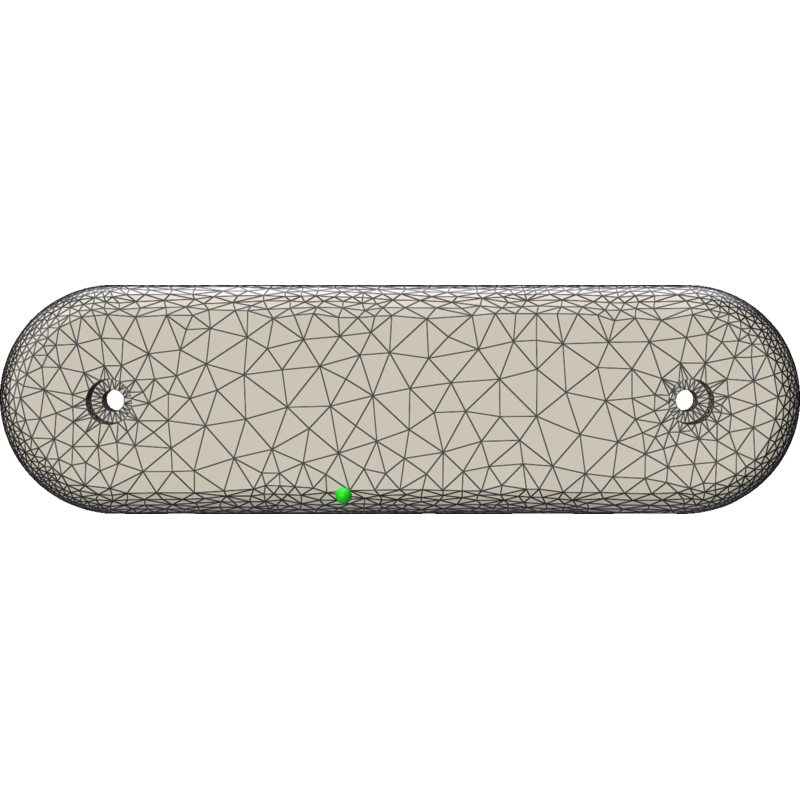} &
        \includegraphics[trim=0 0 0 8cm, clip, width=0.9\linewidth]{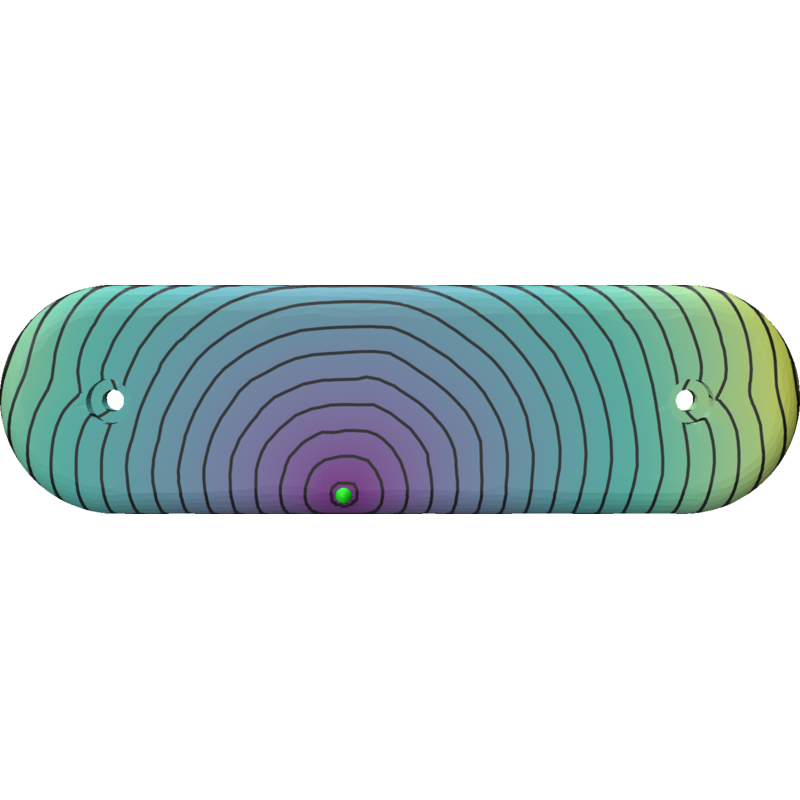} &
        \includegraphics[trim=0 0 0 8cm, clip, width=0.9\linewidth]{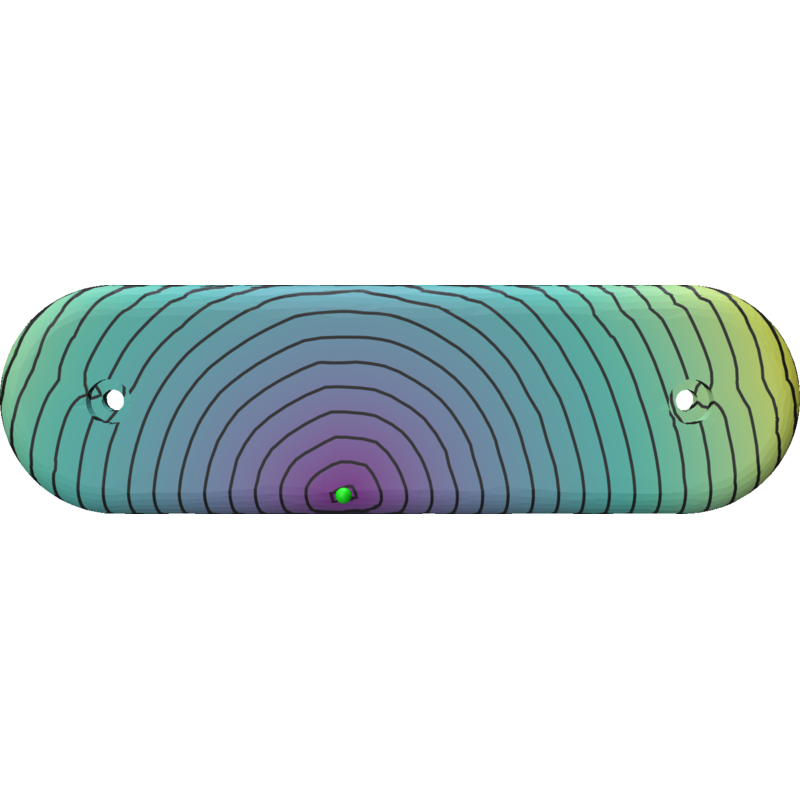} & 
        \includegraphics[trim=0 0 0 8cm, clip, width=0.9\linewidth]{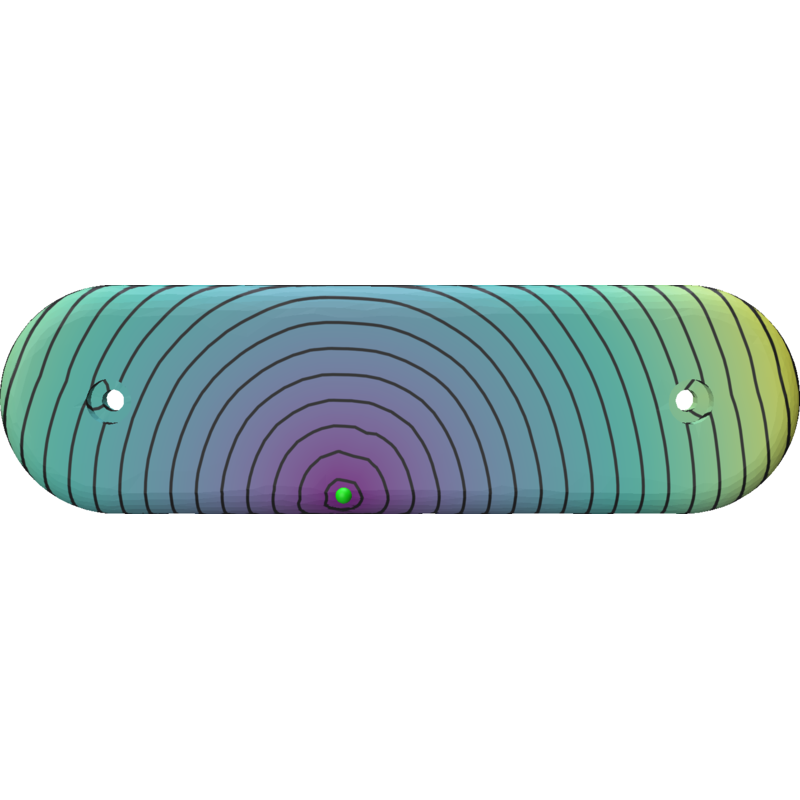} & 
        \includegraphics[trim=0 0 0 8cm, clip, width=0.9\linewidth]{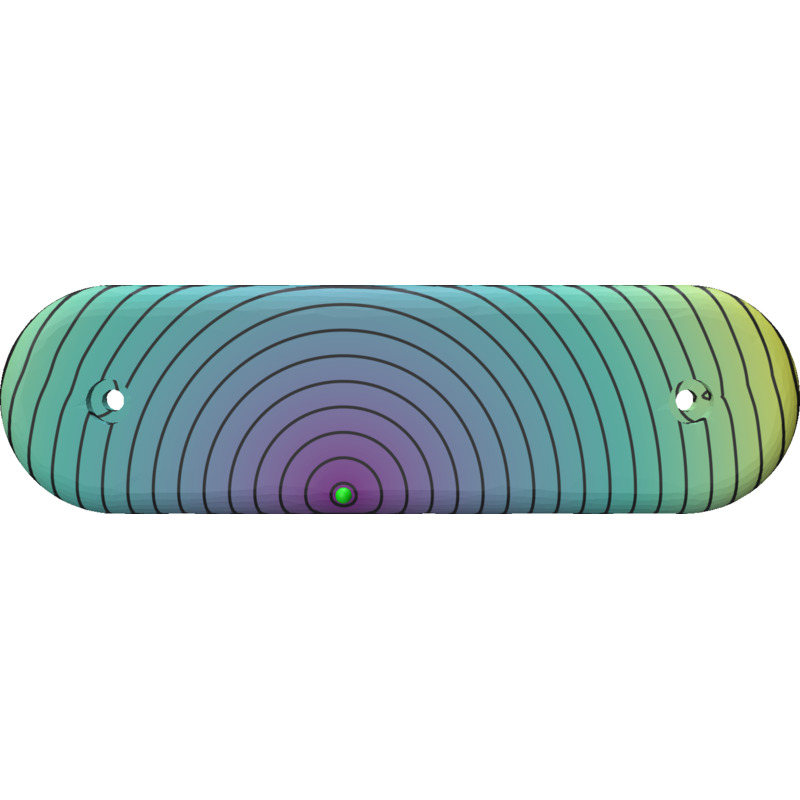} \\
        \noalign{\vspace{-18pt}}
        Rel-$L_2$  
        & \num{2.20e-02} 
        & \num{4.50e-02} 
        & \num{4.41e-02}  
        & \num{3.23e-03}  \\
        Time(s)  
        &  0.0245 
        & 0.0091 
        & 0.0973  
        & 0.1562 \\
        \AddAutoInsetExact{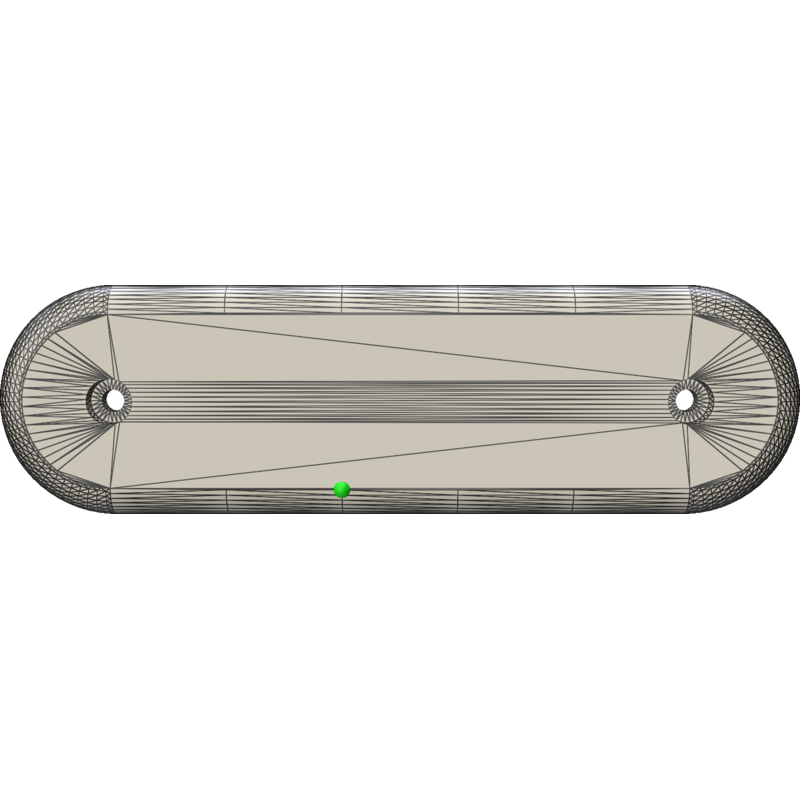}{0.45}{0.5}{0.35}{0.05cm}{0.7\linewidth} &
        \AddAutoInsetExact{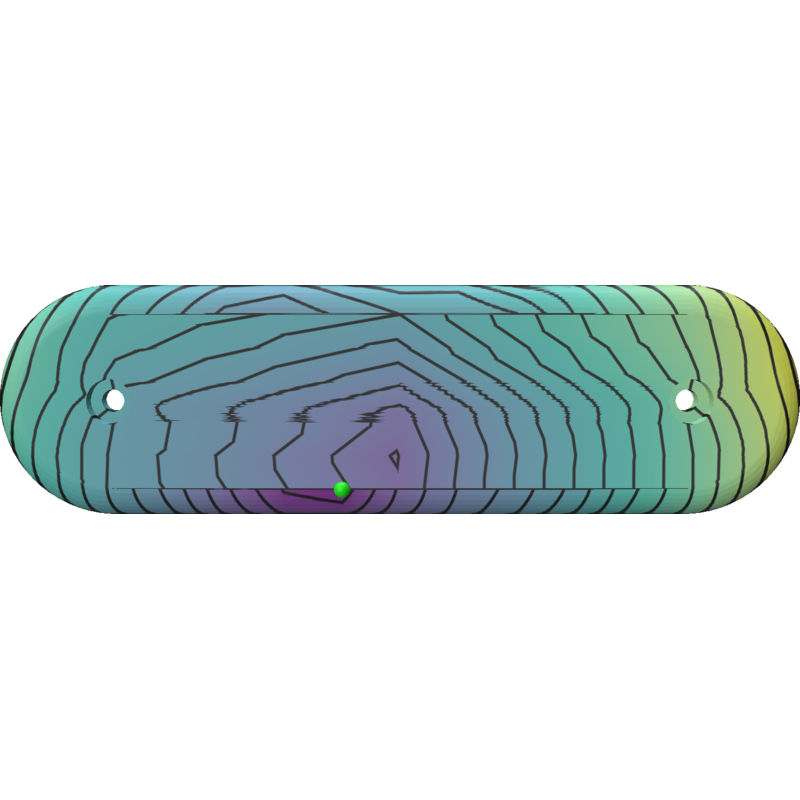}{0.45}{0.5}{0.35}{0.05cm}{0.7\linewidth} &
        \AddAutoInsetExact{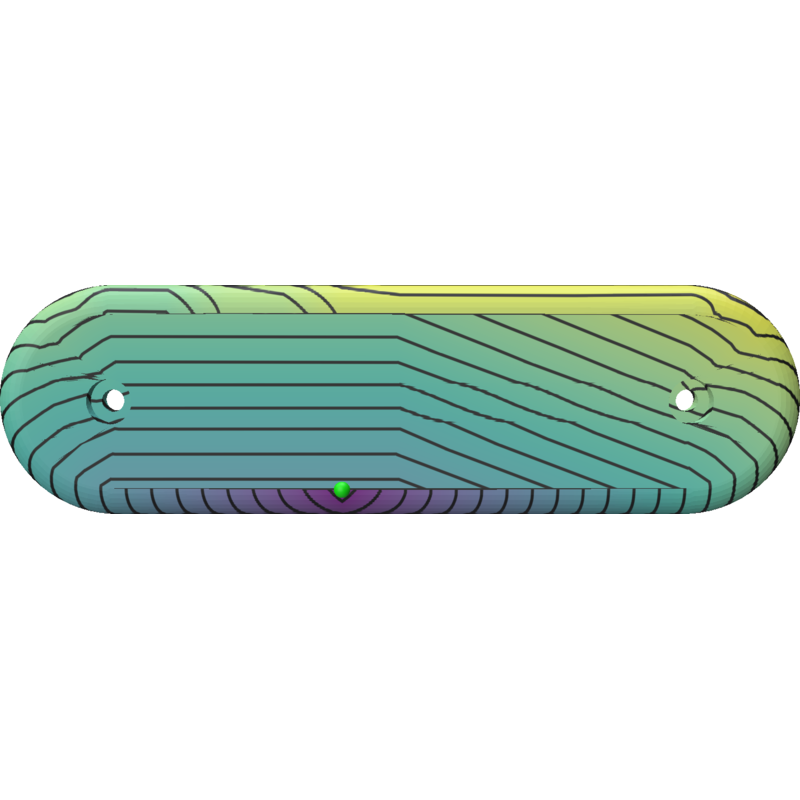}{0.45}{0.5}{0.35}{0.05cm}{0.7\linewidth} & 
        \AddAutoInsetExact{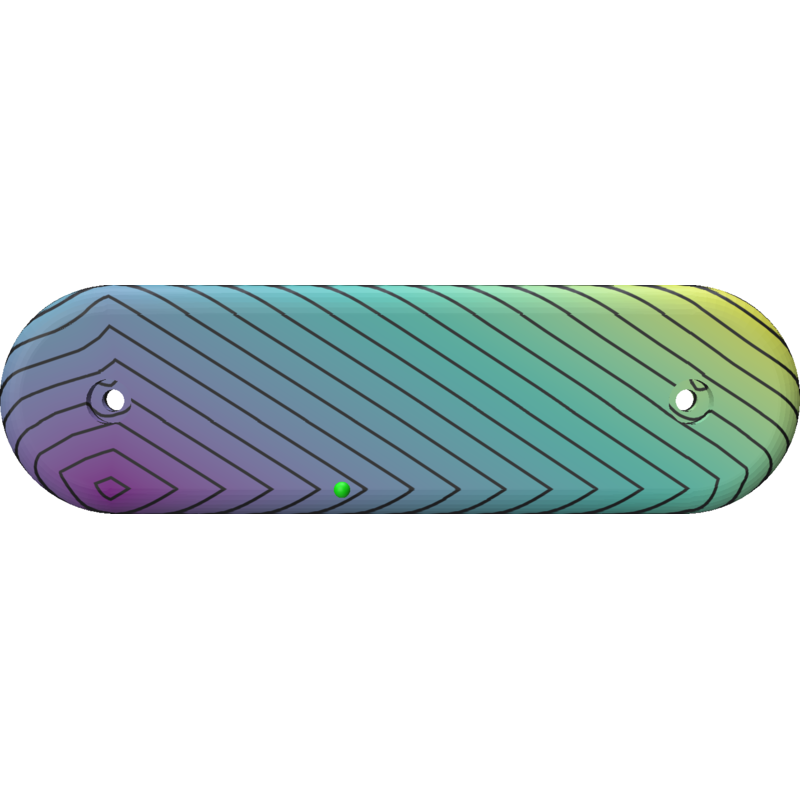}{0.45}{0.5}{0.35}{0.05cm}{0.7\linewidth} & 
        \AddAutoInsetExact{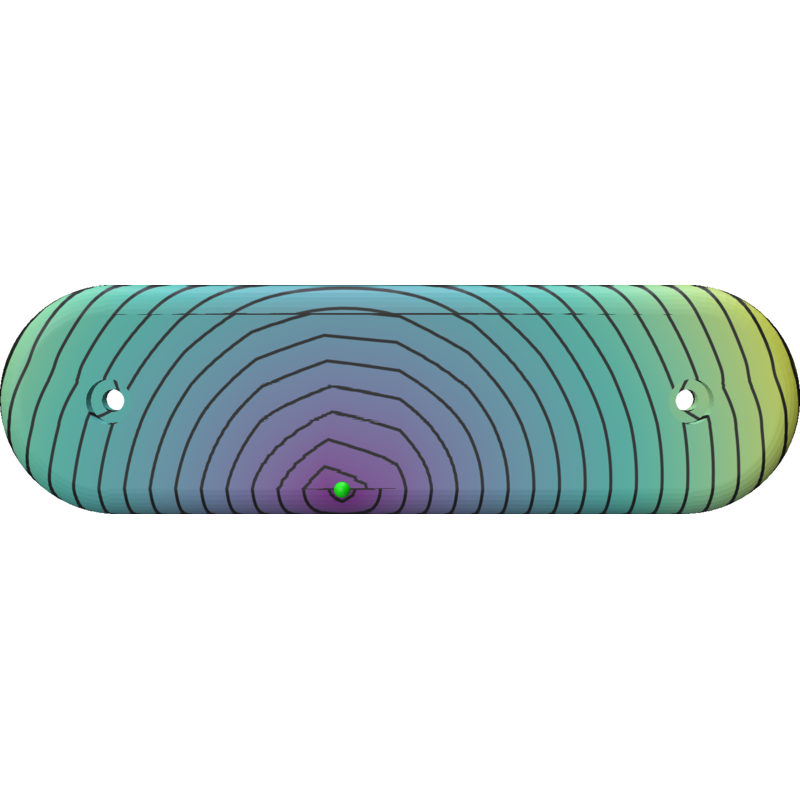}{0.45}{0.5}{0.35}{0.05cm}{0.7\linewidth} \\ 

        \noalign{\vspace{2pt}}
        Rel-$L_2$ 
        & \num{4.80e-1}
        & \num{1.14e0} 
        & \num{4.76e-1}  
        & \num{1.86e-2} \\
        Time(s)  
        &  0.0127 
        & 0.0043 
        & 0.1037  
        & 0.1899 \\
        \bottomrule
    \end{tabular}
    }
    
    \caption{Two shapes: regular triangles (relative top), and sliver triangles (relative bottom). \ours is the most accurate finite-element method and degrades least when transitioning from well-conditioned triangulation to ill-conditioned triangulation. }
    \label{fig:meshing}
\end{figure}

\begin{figure}[htbp]
\centering
\small 
\setlength{\tabcolsep}{2pt}
\begin{tabular}{cccccc}
\toprule & 
\scriptsize\textbf{HM} &\scriptsize \textbf{FMM} & \scriptsize\textbf{DFA}  &   \scriptsize \textbf{Ours (\ours)} &  \scriptsize \textbf{Ours (Isolines)} \\
\midrule
\includegraphics[width=0.14\columnwidth]{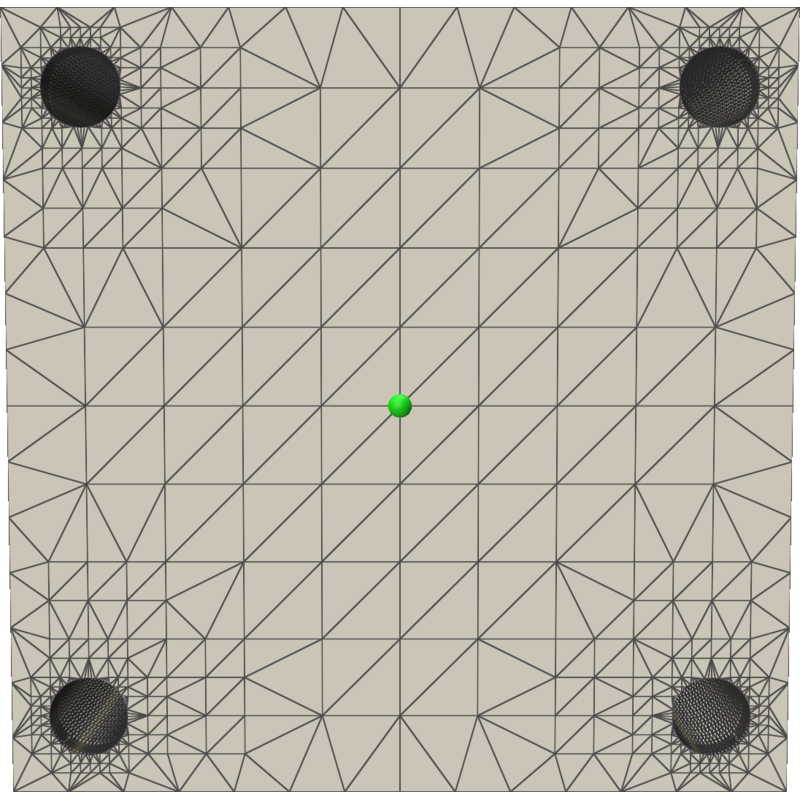} &
\includegraphics[width=0.14\columnwidth]{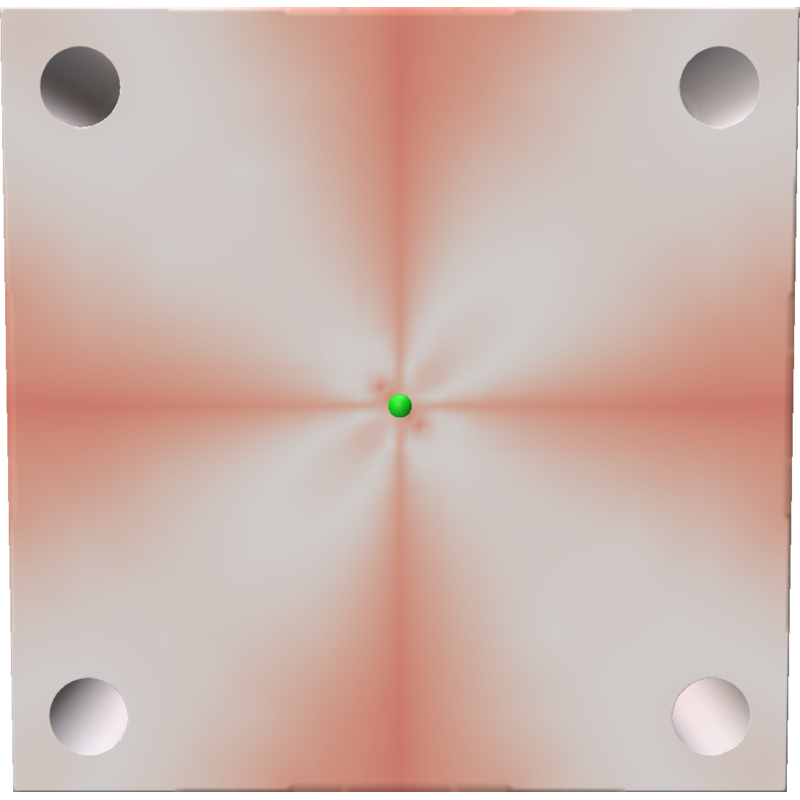} &
\includegraphics[width=0.14\columnwidth]{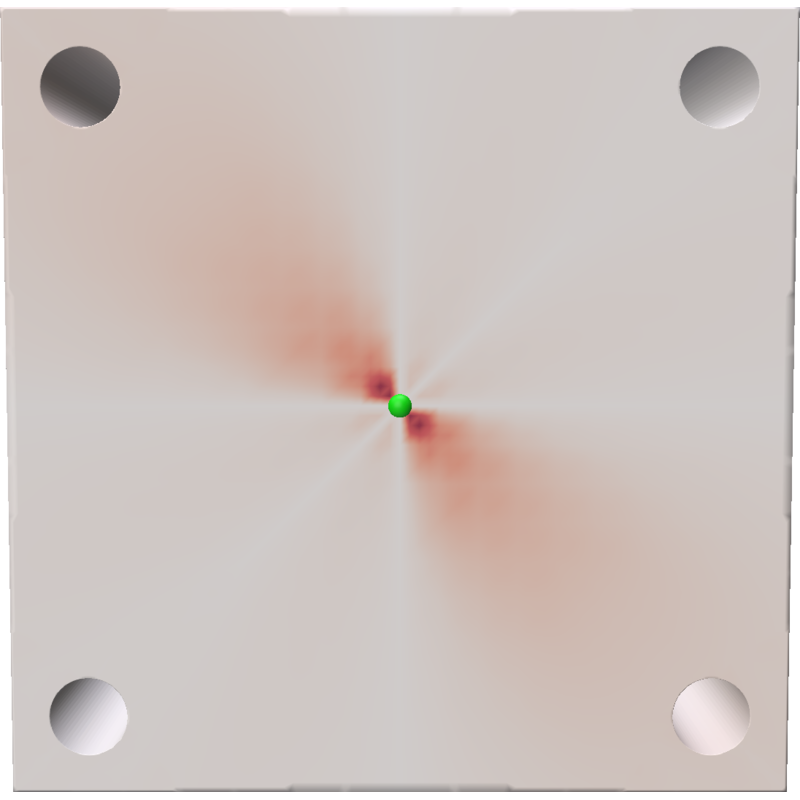} &
\includegraphics[width=0.14\columnwidth]{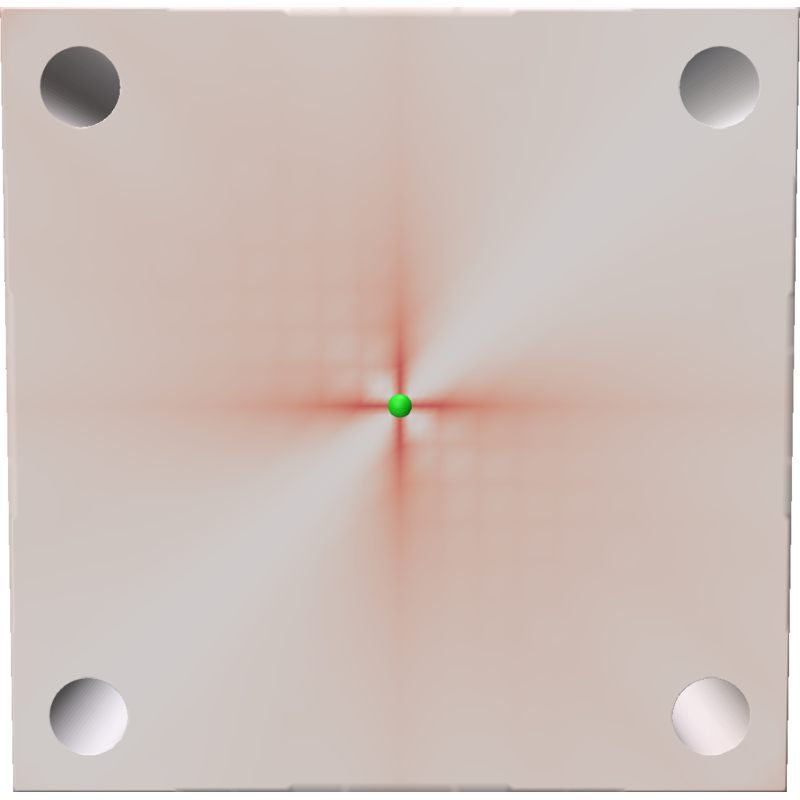} &
\includegraphics[width=0.14\columnwidth]{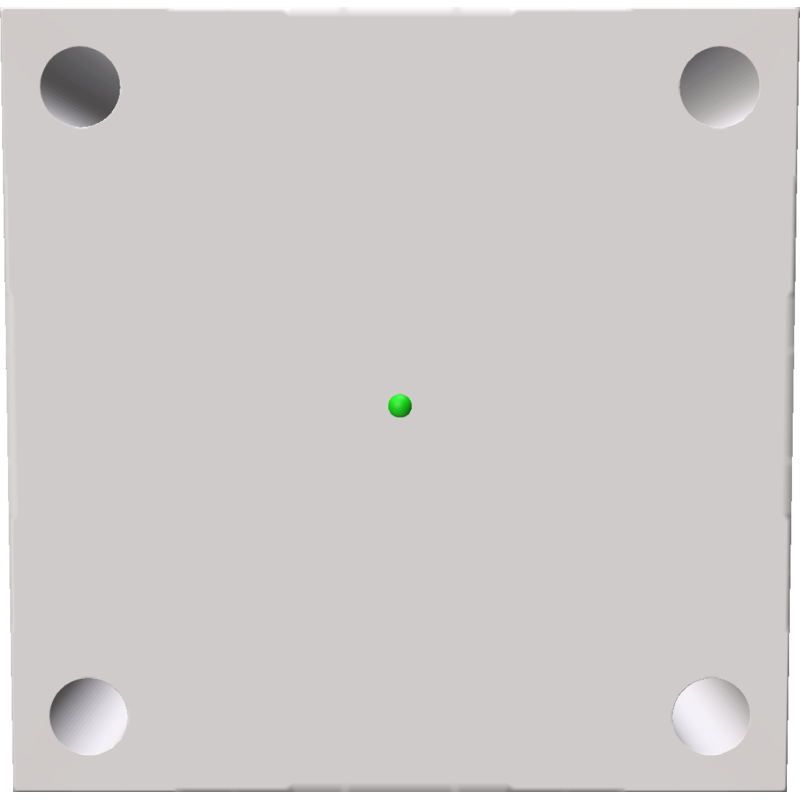} & 
\includegraphics[width=0.14\columnwidth]
{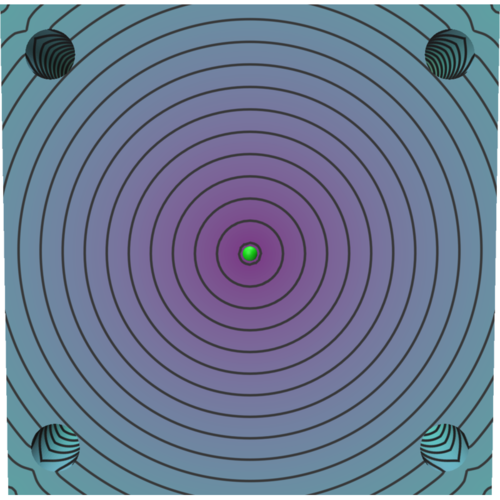} 
\\
\textbf{Rel-$L_2$} 
& $2.05 \times 10^{-1}$
& $5.70 \times 10^{-2}$
& $6.57 \times 10^{-2}$
& $\mathbf{2.44 \times 10^{-3}}$ & 
\\
\bottomrule
\end{tabular}%
\caption{An original mesh with a grid-like triangulation. The right-angle pattern of the bottom mesh causes grid artifacts in all methods except \ours. }
\label{fig:right_angle}
\end{figure}

\begin{figure}[!htb]
    \centering
    \setlength{\tabcolsep}{0.5pt} 
    \renewcommand{\arraystretch}{0.6} 
    \scriptsize

    \newcommand{\gwidth}{0.11\linewidth} 
    
    \begin{tabular}{@{} c : ccc:ccccc @{}} 
        
        \textbf{Input} & VTP & DGG-VTP & GSP  & HM & FMM & DFA & Ours (\ours) \\
        \noalign{\vspace{1pt}} 
        
        \includegraphics[width=\gwidth]{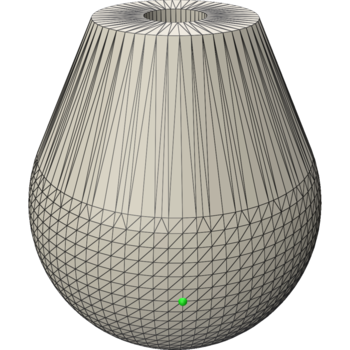} &
        \includegraphics[width=\gwidth]{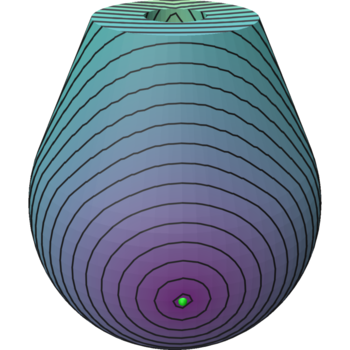} &
        \includegraphics[width=\gwidth]{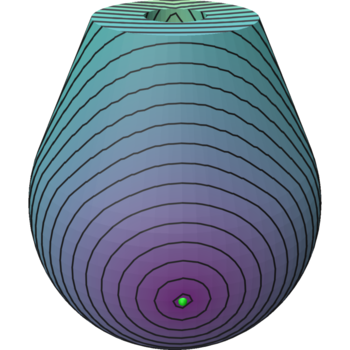} &
        \includegraphics[width=\gwidth]{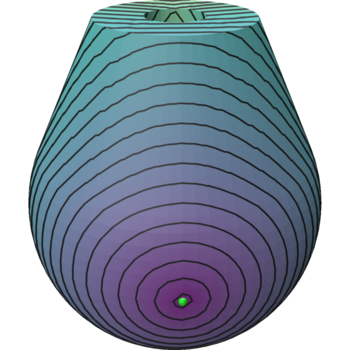} &
        \includegraphics[width=\gwidth]{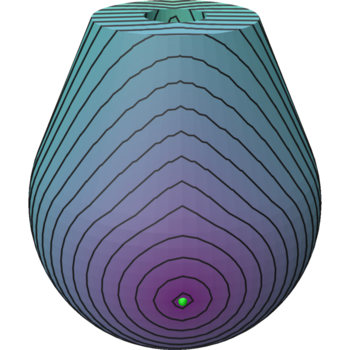} &
        \includegraphics[width=\gwidth]{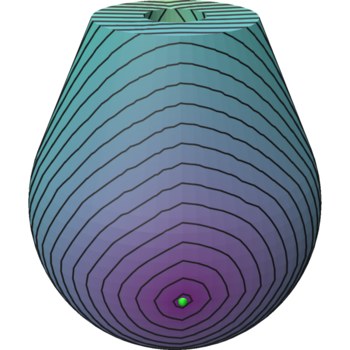} &
        \includegraphics[width=\gwidth]{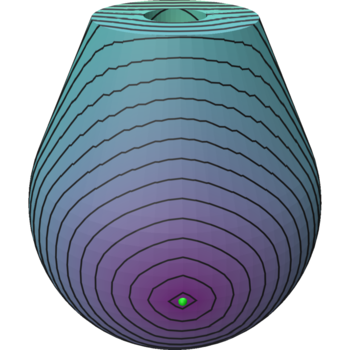} &
        \includegraphics[width=\gwidth]{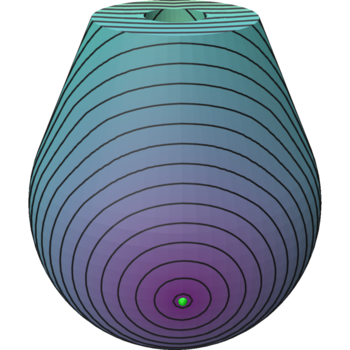} \\

        \includegraphics[width=\gwidth]{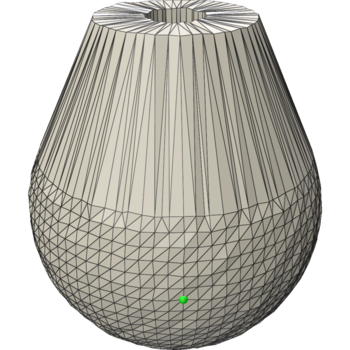} &
        \includegraphics[width=\gwidth]{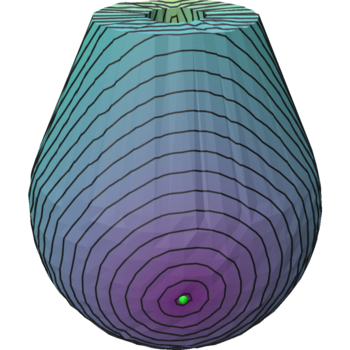} &
        \includegraphics[width=\gwidth]{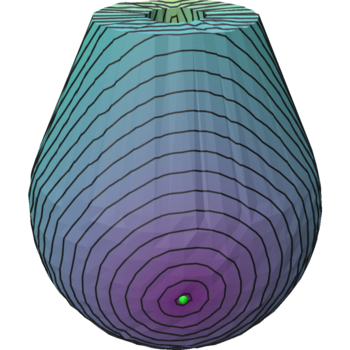} &
        \includegraphics[width=\gwidth]{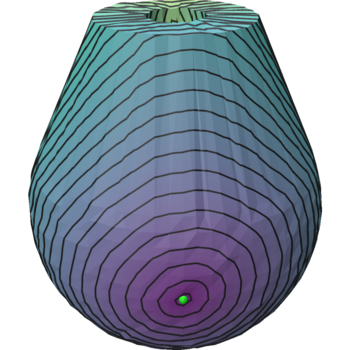} &
        \includegraphics[width=\gwidth]{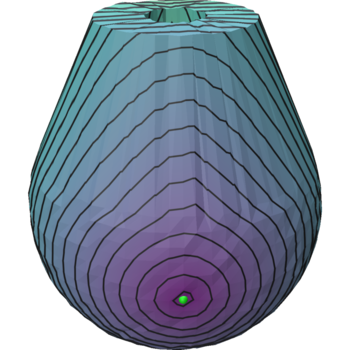} &
       \includegraphics[width=\gwidth]{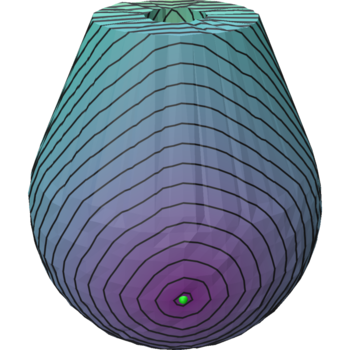} &
       \includegraphics[width=\gwidth]{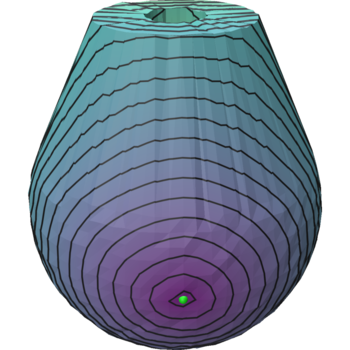} &
       \includegraphics[width=\gwidth]{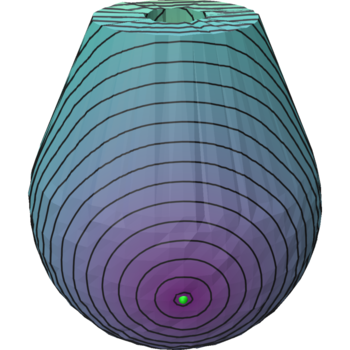}  \\

        \includegraphics[width=\gwidth]{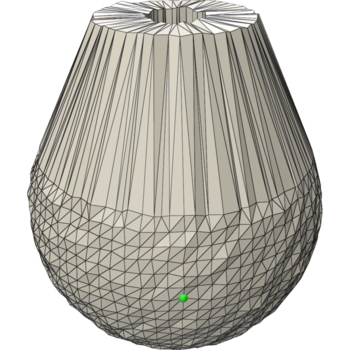} &
        \includegraphics[width=\gwidth]{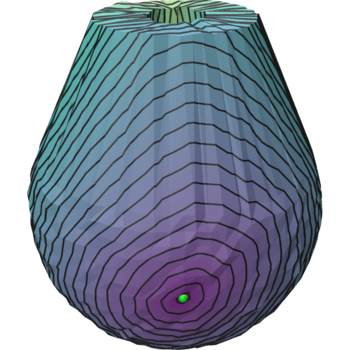} &
       \includegraphics[width=\gwidth]{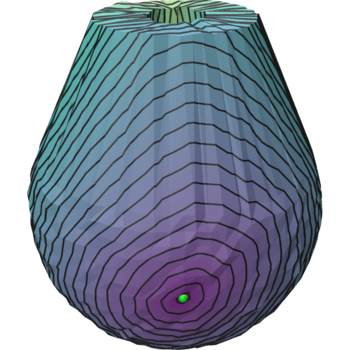} &
        \includegraphics[width=\gwidth]{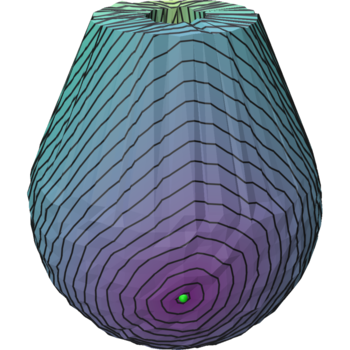} &
       \includegraphics[width=\gwidth]{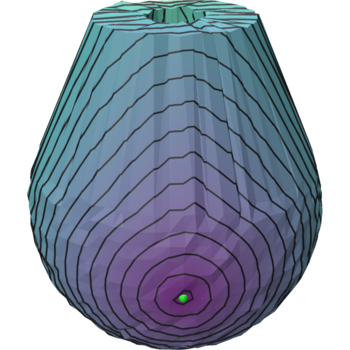} &
       \includegraphics[width=\gwidth]{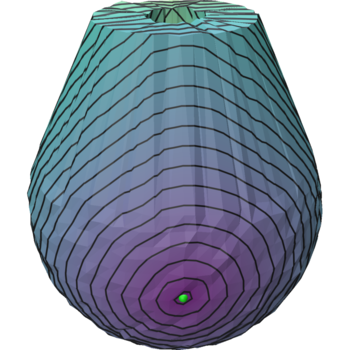} &
       \includegraphics[width=\gwidth]{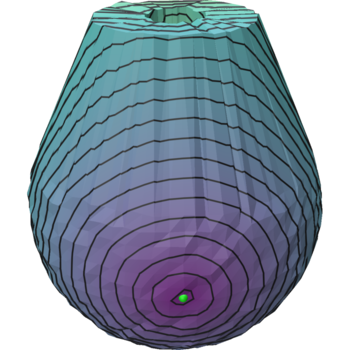} &
       \includegraphics[width=\gwidth]{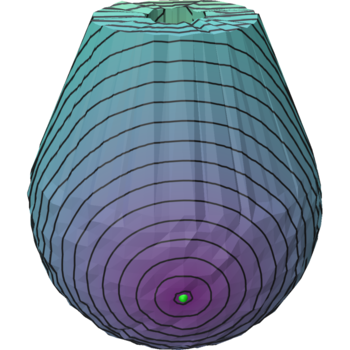} \\

    \end{tabular}

    \vspace{6pt}

    \begin{minipage}{\linewidth}
        \centering
        \vspace{2pt}

        \includegraphics[width=0.49\linewidth]{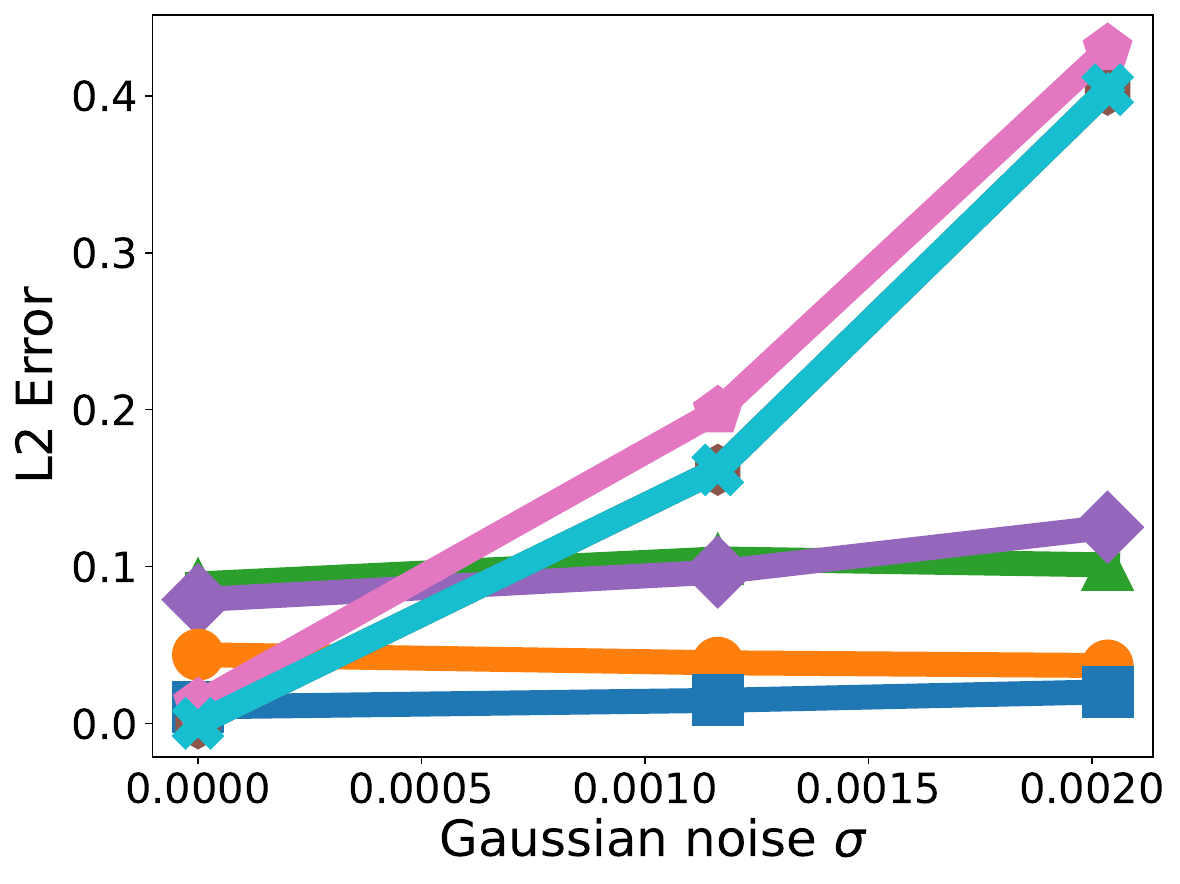}
        \hfill 
        \includegraphics[width=0.49\linewidth]{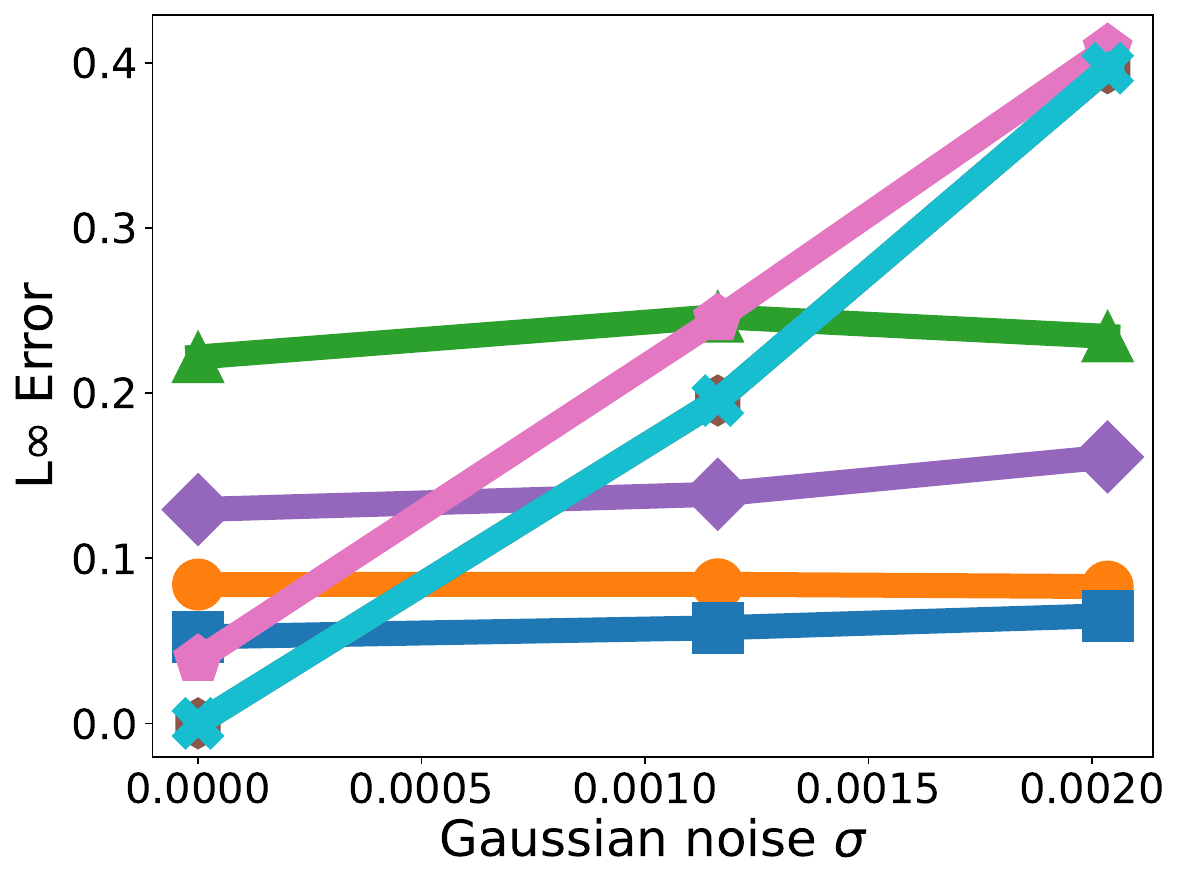}
    \end{minipage}

    \vspace{2pt}
    \includegraphics[width=\linewidth]{images/legends/legend_convergence_sphere.pdf}
    \vspace{-6pt}
    \setlength{\abovecaptionskip}{1pt}
  \caption{Robustness to Gaussian noise. Vertices are perturbed by i.i.d. Gaussian noise with $\sigma = \alpha h$, $\alpha \in \{0, 0.20, 0.30\}$, where $h$ is the mean edge length of the evaluation mesh. All errors are measured against  the exact polyhedral distance on the clean mesh. Top: distance fields at the three levels. Bottom: relative $L_2$ and $L_\infty$ error against $\sigma$. The polyhedral solvers are exact on the clean input and degrade sharply under noise. FE methods stay robust, and within, \ours is superior.} 
    \label{fig:gaussian_noise}
\end{figure}

\begin{figure*}[htbp]
\centering

\resizebox{\textwidth}{!}{%
\begin{tabular}{cccc|cccccc}
\toprule
  & \textbf{VTP} & \textbf{DGG-VTP} & \textbf{GSP} & \textbf{HM} & \textbf{FMM} & \textbf{DFA}  & \textbf{Ours (\ours)} & \textbf{Ours (\ours) isolines}\\
\midrule

\includegraphics[height=2.8cm, valign=m]{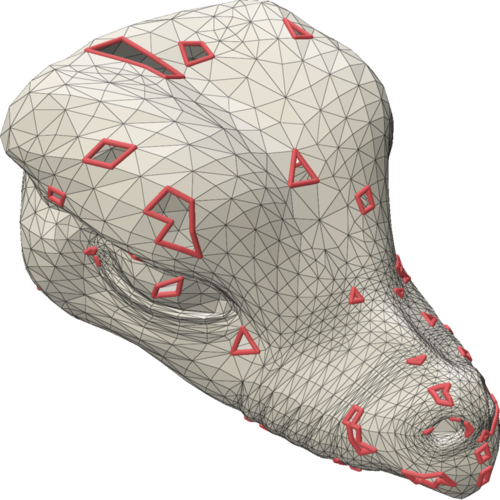} &
\includegraphics[height=2.8cm, valign=m]{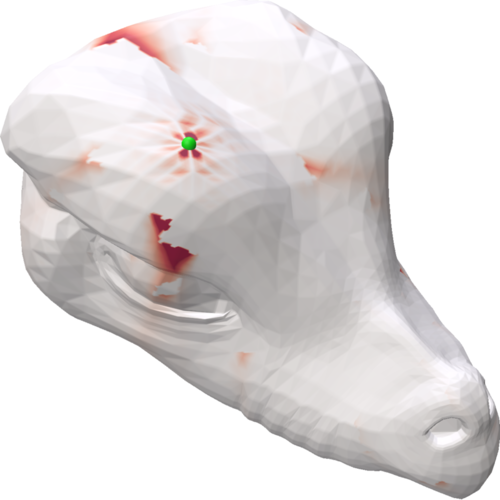} &
Failed &
\includegraphics[height=2.8cm, valign=m]{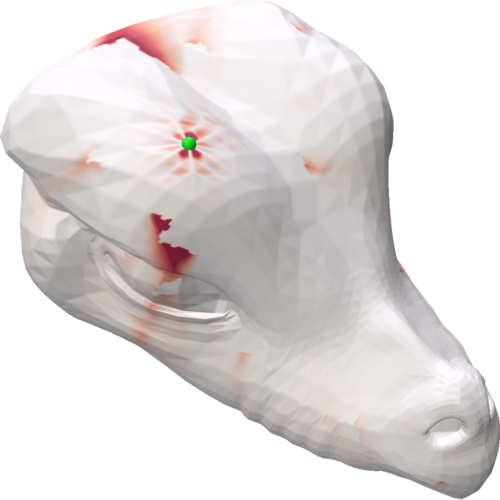} &
\includegraphics[height=2.8cm, valign=m]{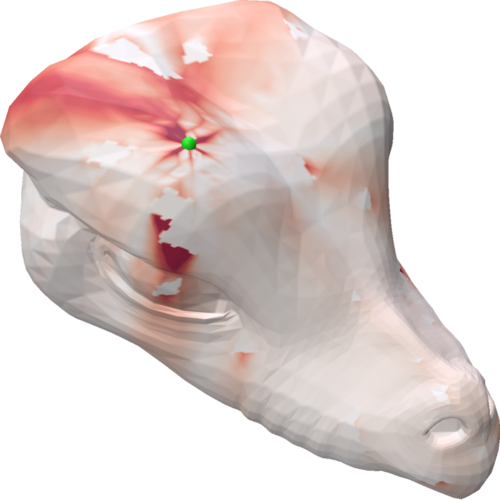} &
\includegraphics[height=2.8cm, valign=m]{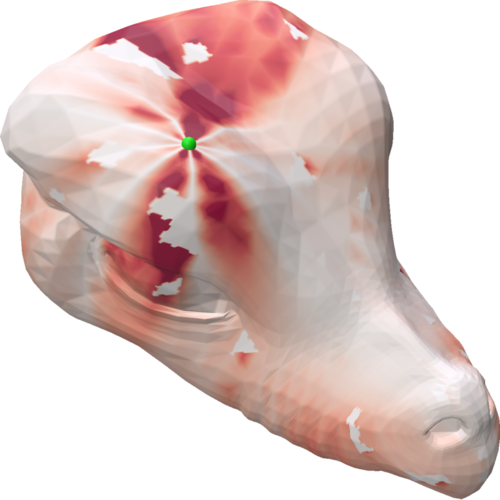} &
\includegraphics[height=2.8cm, valign=m]{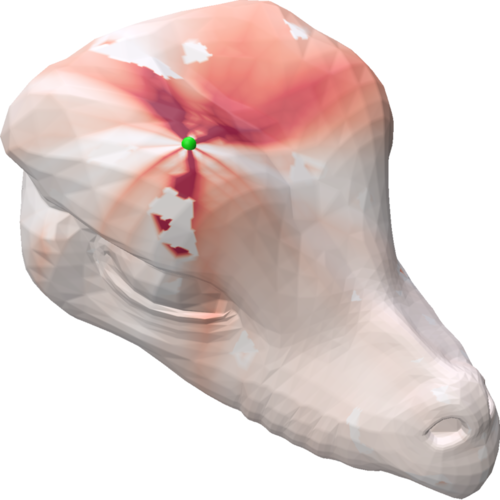} &
\includegraphics[height=2.8cm, valign=m]{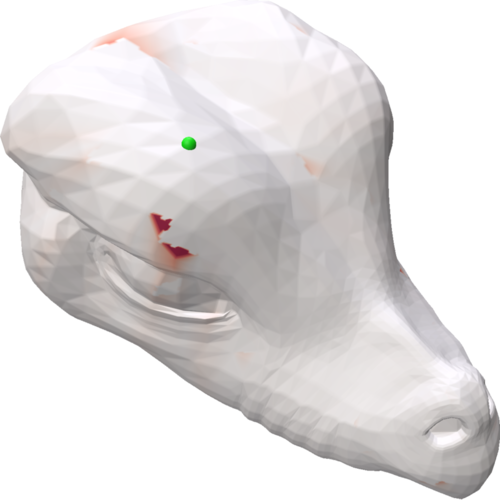} &
\includegraphics[height=2.8cm, valign=m]{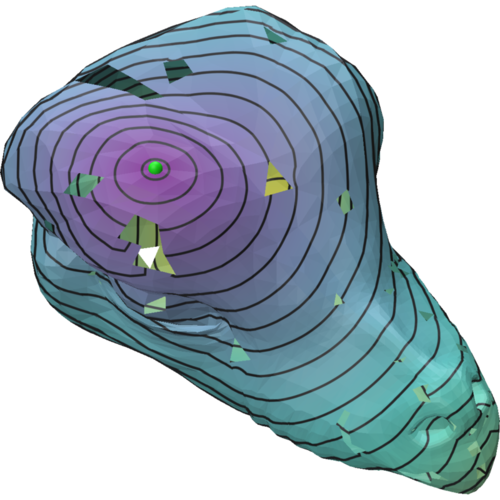} &
\\

\textbf{Rel-$L_2$} 
& $8.80 \times 10^{-3}$
& N/A
& $1.08 \times 10^{-2}$
& $2.76 \times 10^{-2}$
& $3.43 \times 10^{-2}$
& $2.30 \times 10^{-2}$
& $\mathbf{5.05 \times 10^{-3}}$ & \\

\midrule

\includegraphics[height=2.8cm, valign=m]{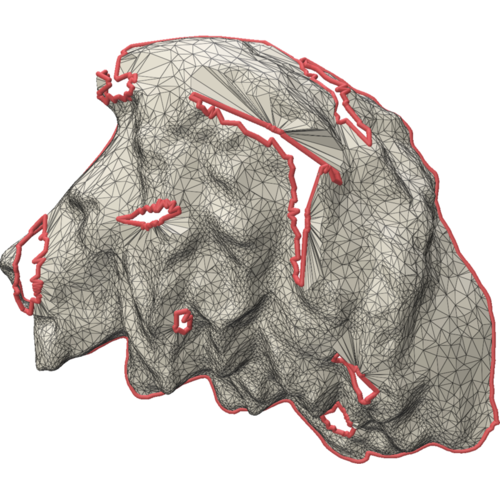} &
Failed &
Failed &
Failed &
\includegraphics[height=2.8cm, valign=m]{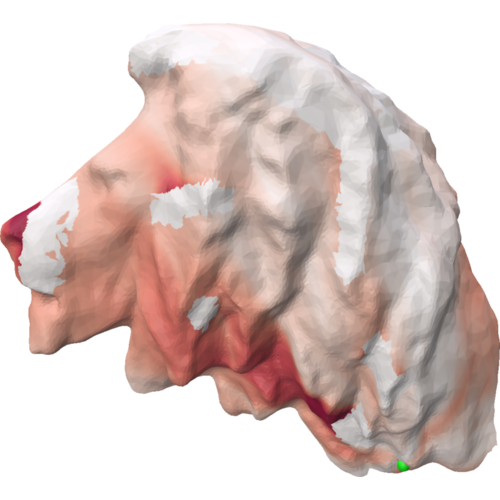} &
Failed &
\includegraphics[height=2.8cm, valign=m]{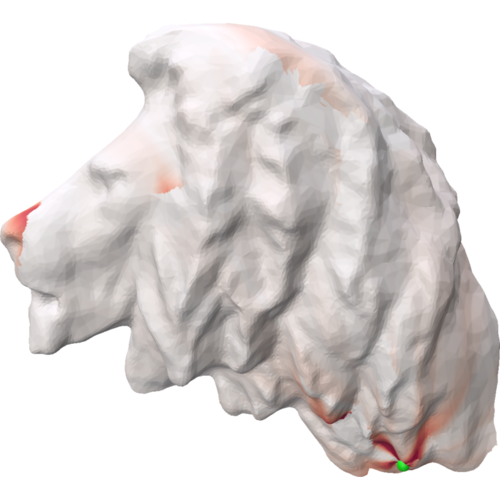} &
\includegraphics[height=2.8cm, valign=m]{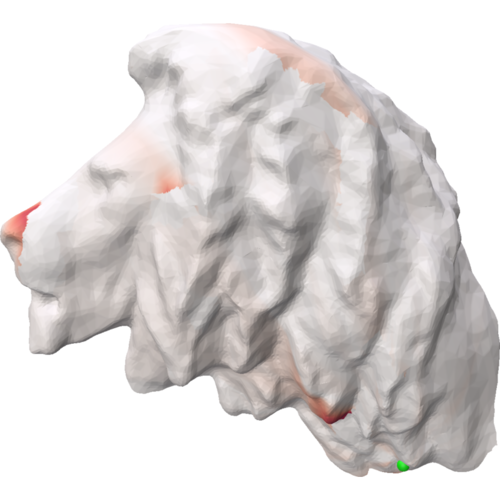} &
\includegraphics[height=2.8cm, valign=m]{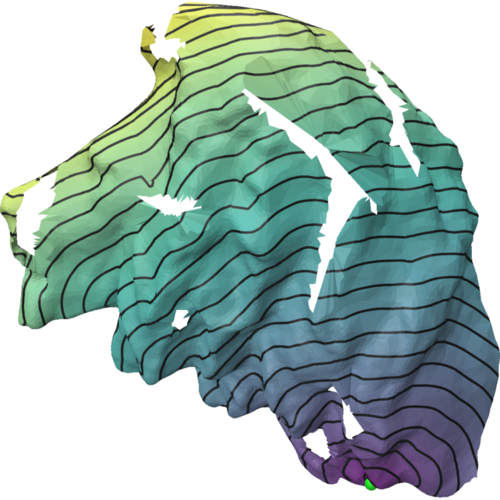} \\
\textbf{Rel-$L_2$} 
& N/A
& N/A
& N/A
& $2.61 \times 10^{-2}$
& N/A
& $1.01 \times 10^{-2}$
& $\mathbf{7.86 \times 10^{-3}}$ & 
\\

\bottomrule

\end{tabular}%
}
\caption{Robustness tests on two damaged meshes. The defects cause most vertex-exact methods to fail, and FE methods to introduce a large error where \ours remains robust.}
\label{fig:robust_holes}
\end{figure*}

We first examine failures on the unperturbed benchmark, then test
robustness to poor triangulations, geometric noise, topological
defects, and non-convex domains. We compare our method with the
 baselines and the polyhedral methods
introduced in \secref{subsec:setup}.

\paragraph*{Failure on unperturbed meshes}
Before introducing controlled perturbations, we test the released
implementations of \VTP\ and \DGGVTP\ on all $4{,}320$ benchmark
surfaces. \DGGVTP\ fails on $12.34\%$ of the meshes. Its sparse graph
construction can leave vertices unreachable from the source, producing
infinite distances, and the implementation rejects meshes containing
near-degenerate triangles with corner angles below $10^{-5}$ radians.
\VTP\ fails on $0.07\%$ of the meshes. We traced these cases to a
hardcoded degeneracy guard in the released implementation that aborts
on highly symmetric inputs, rather than to a limitation of the
underlying algorithm. These failure rates illustrate the numerical sensitivity of exact or near-exact polyhedral methods and the importance of careful implementation. 

\paragraph*{Sliver triangles} 
We examine sensitivity to mesh quality. Figure \figref{fig:meshing} compares regular and sliver-dominated triangulations on the same shapes, showing that linear FE methods degrade considerably with poor triangulation. \ours\ remains the most accurate on both.

\paragraph*{Right-angle artifacts}
\figref{fig:right_angle} shows that, even on an otherwise reasonable mesh, right angles created by the grid structure produce zero cotangent weights and considerably bias the result in flat regions. Our method avoids this artifact. We attribute this to the ability of PQ elements to represent the distance accurately in such regions.
 
 \paragraph*{Noise} In \figref{fig:gaussian_noise}, we show that \ours is robust to increasing levels of Gaussian noise. For the piecewise-linear baselines, subdivision is performed before the noise is applied. \GSP\ and \DGGVTP\ degrade sharply as the noise increases.

\paragraph*{Missing regions} In \figref{fig:robust_holes}, we remove small patches of triangles, introducing holes and  a non-manifold vertex in the second example. We then measure the degradation of each method using the relative $L_2$ error with respect to the distances on the original mesh, thereby testing robustness to topological degradation. \DGGVTP fails on both examples, and \VTP, \GSP, and \FMM\  fail on the second shape. Even when not failing, the resulting error is larger than ours.  Among the methods that return a solution, \ours\ has the lowest deviation and retains the cleanest isolines.

\begin{figure}[t]
  \centering
  \setlength{\tabcolsep}{1pt}
  \renewcommand{\arraystretch}{0.9}

  \begin{tabular}{@{}ccc@{}}
  \small \textbf{Mesh} & \small \textbf{HM} & \small \textbf{Ours (\ours)} \\
    \includegraphics[width=0.32\linewidth, trim=0 79 0 79, clip]{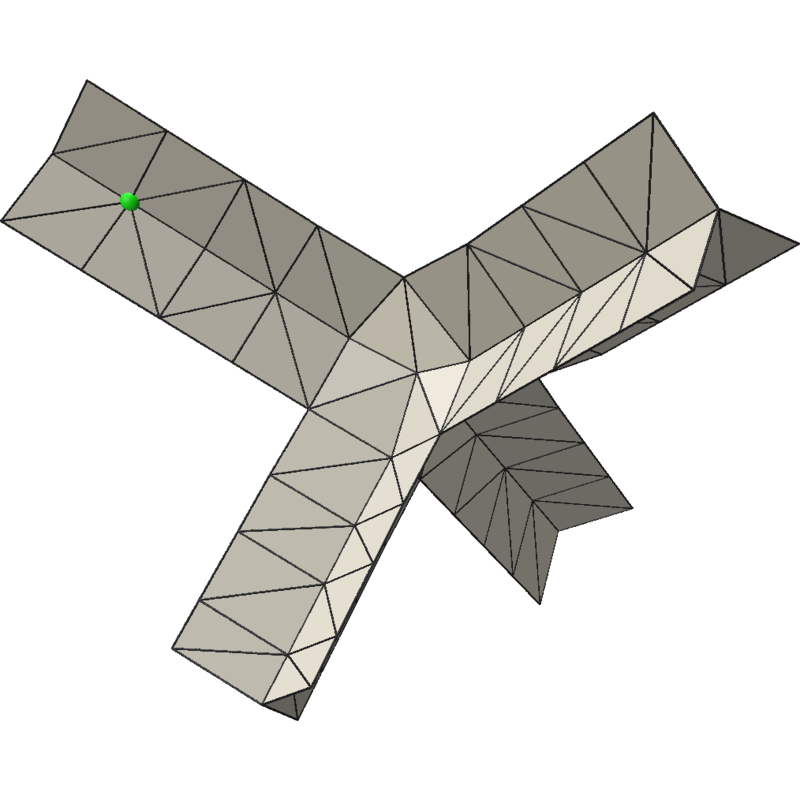} &
    \includegraphics[width=0.32\linewidth, trim=0 79 0 79, clip]{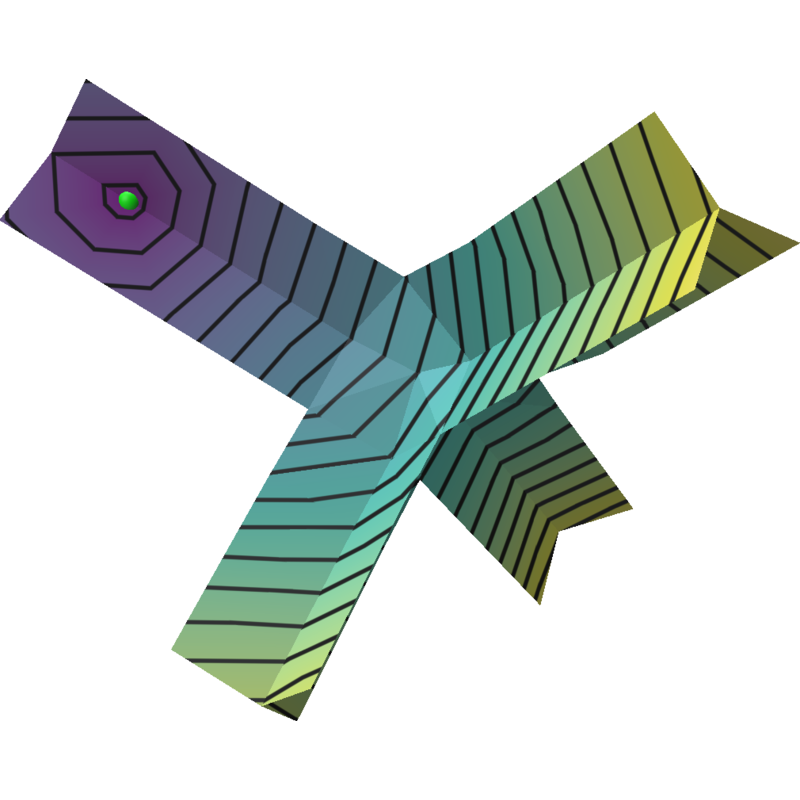} &
    \includegraphics[width=0.32\linewidth, trim=0 79 0 79, clip]{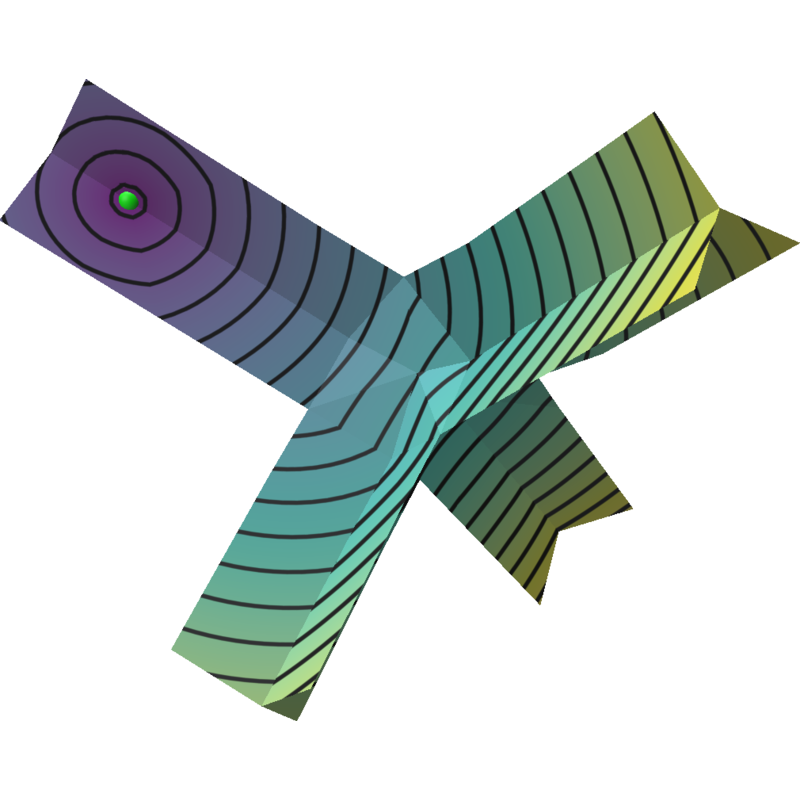} \\[1pt]
    \includegraphics[width=0.32\linewidth, trim=0 93 0 93, clip]{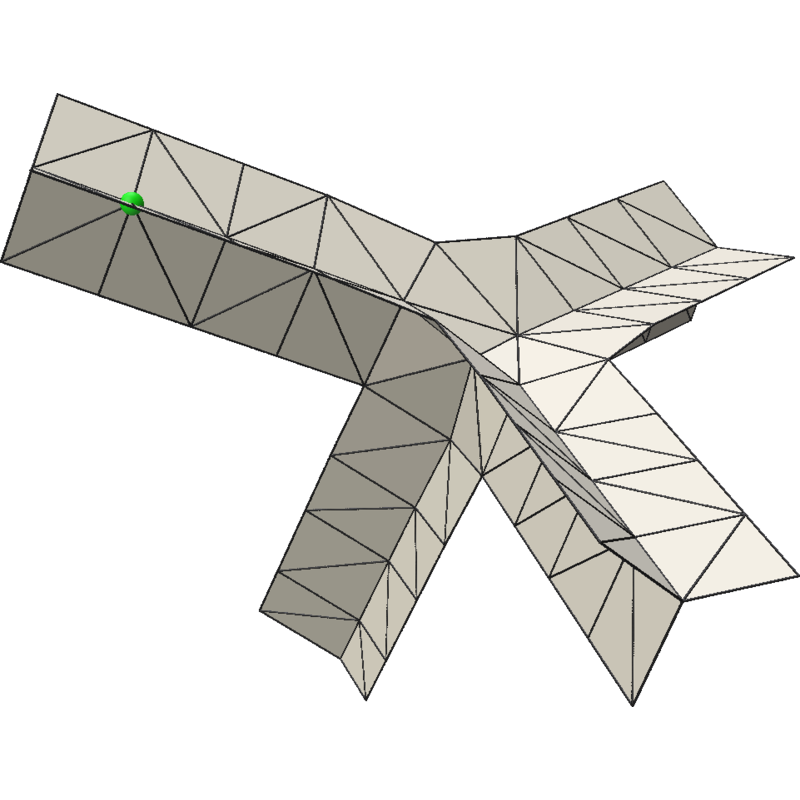} &
    \includegraphics[width=0.32\linewidth, trim=0 93 0 93, clip]{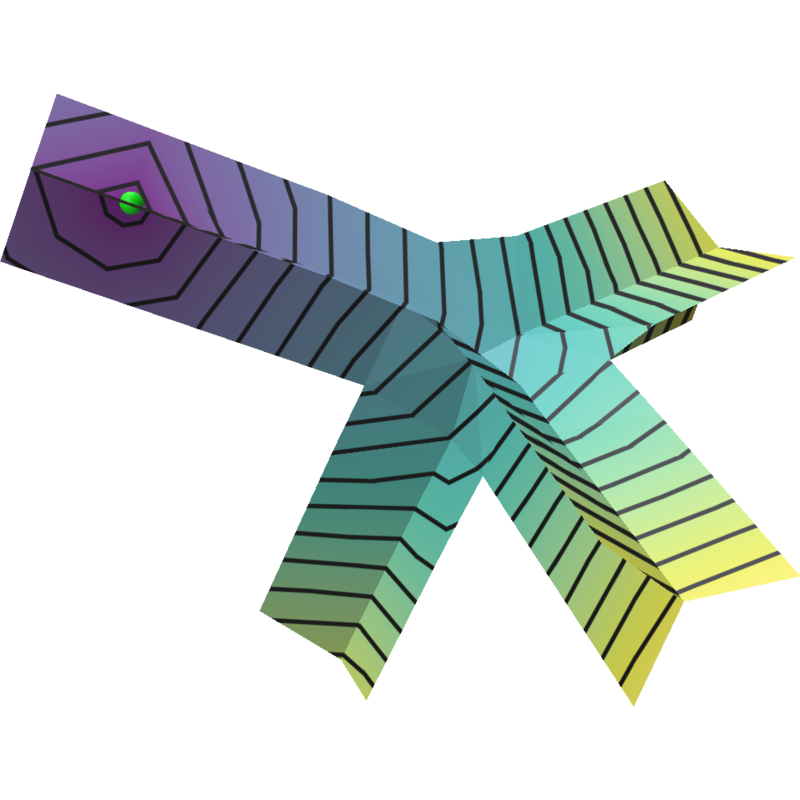} &
    \includegraphics[width=0.32\linewidth, trim=0 93 0 93, clip]{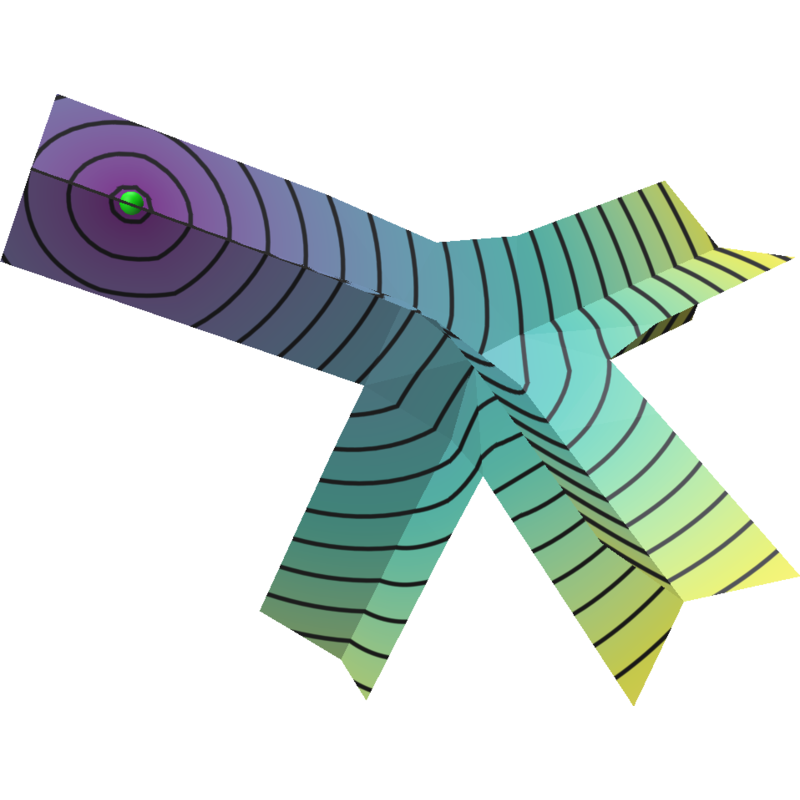} \\
  \end{tabular}

  \caption{Non-manifold robustness. The same mesh from two different views. \ours exhibits smooth and consistent  isolines across the non-manifold junctions, where \HM exhibits visible inconsistencies.}
  \label{fig:nonmanifold}
\end{figure}

\paragraph*{Non-manifold input} The polyhedral implementations used above assume manifold input and are therefore not included in this test.  We use the tufted non-manifold Laplacian~\cite{Sharp:2020:LNT} for \HM as a comparison, while \ours is run without modification. In the \textit{triple vierbein} model (\figref{fig:nonmanifold}) our isolines cross the non-manifold junctions consistently, whereas \HM with the tufted Laplacian shows visible inconsistencies.

\begin{figure}[t]
  \centering
  \setlength{\tabcolsep}{1pt}
  \renewcommand{\arraystretch}{1}

  \begin{tabular}{@{}lcccccc@{}}
     & \small Mesh & \small GT & \small HM & \small FMM / FIM & \small DFA & \small \textbf{Ours (\ours)} \\

    \small 2D &
    \includegraphics[width=0.156\linewidth]{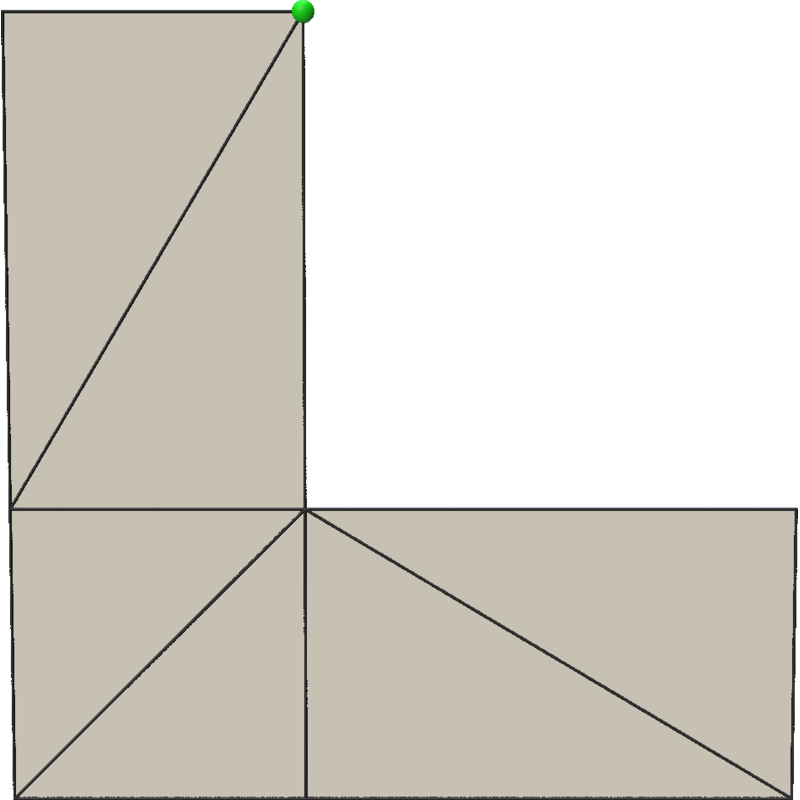} &
    \includegraphics[width=0.156\linewidth]{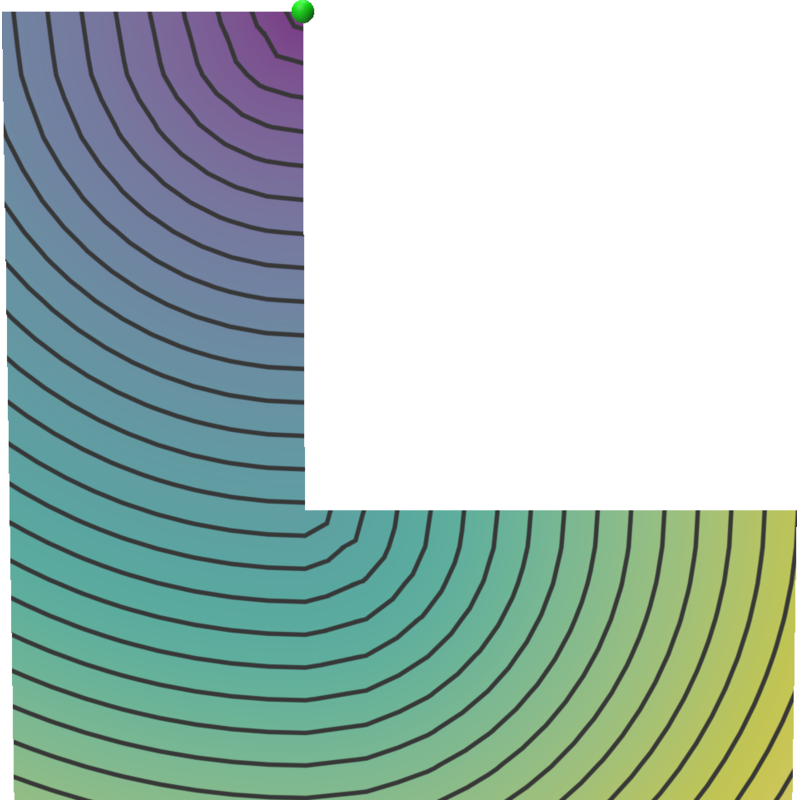} &
    \includegraphics[width=0.156\linewidth]{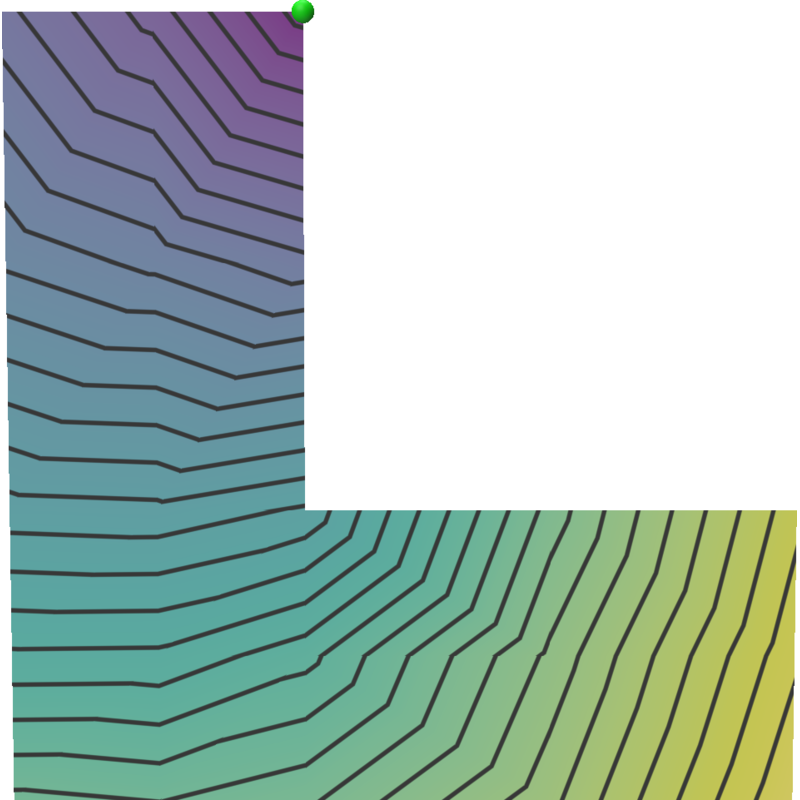} &
    \includegraphics[width=0.156\linewidth]{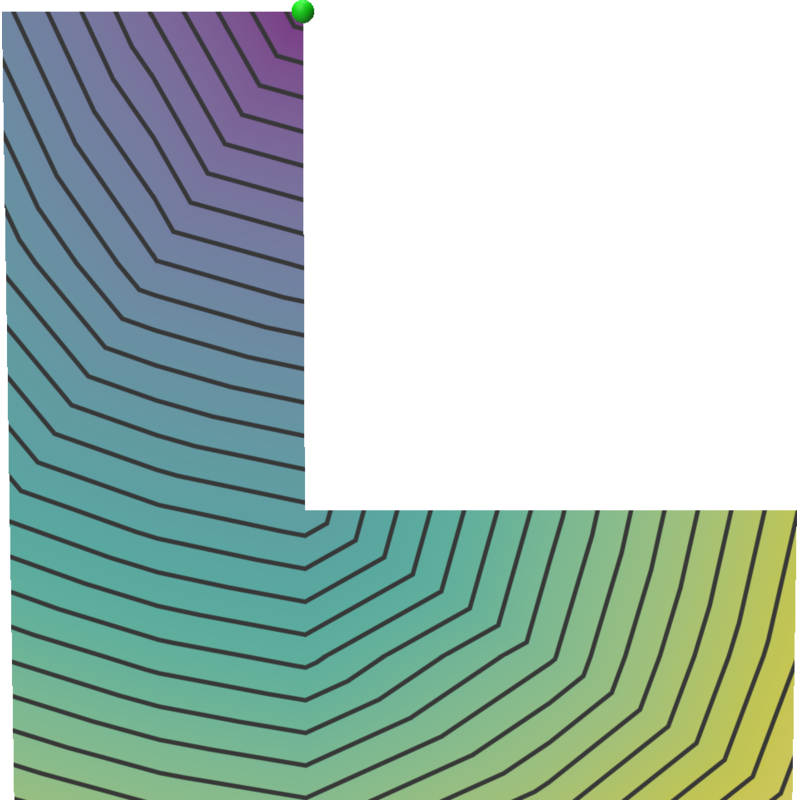} &
    \includegraphics[width=0.156\linewidth]{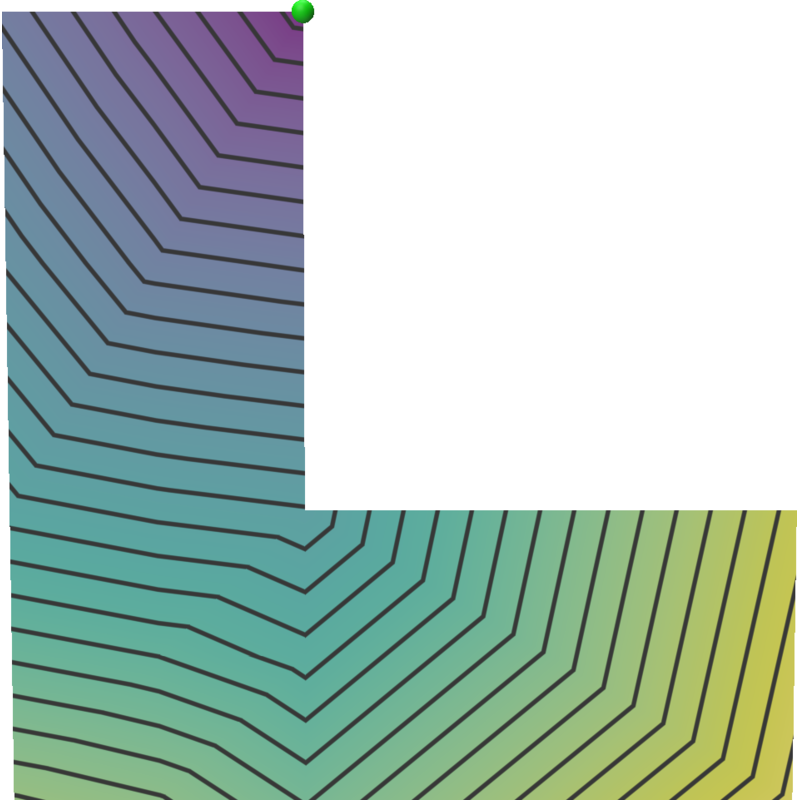} &
    \includegraphics[width=0.156\linewidth]{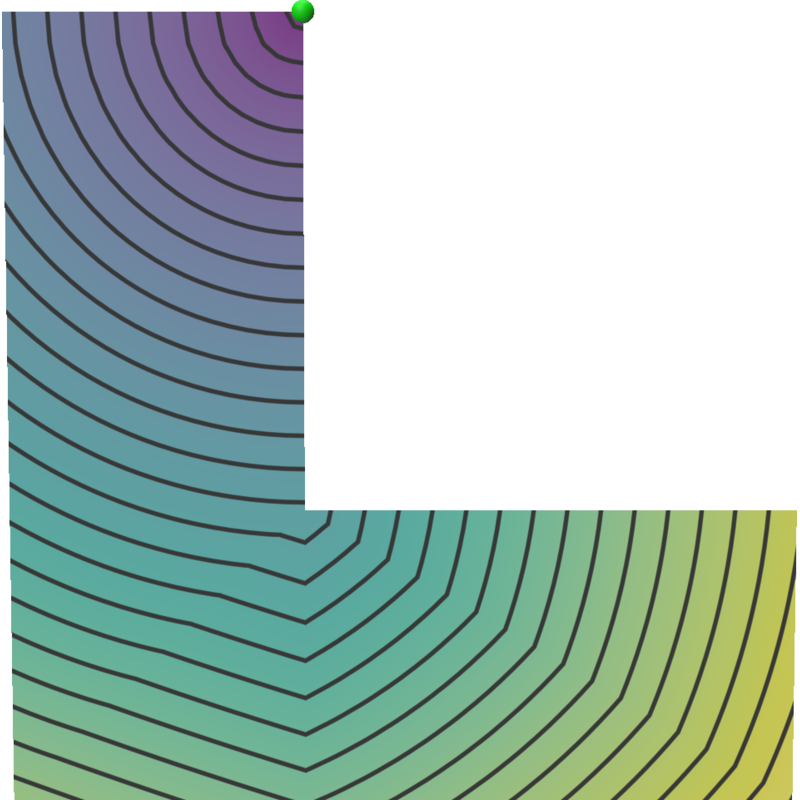} \\

     & Rel L2 &
    \small --- &
    \small \num{3.08e-2} &
    \small \num{6.25e-3} &
    \small \num{6.55e-2} &
    \small {\boldmath\num{3.59e-3}} \\[0.6ex]

    \small 3D &
    \includegraphics[width=0.156\linewidth]{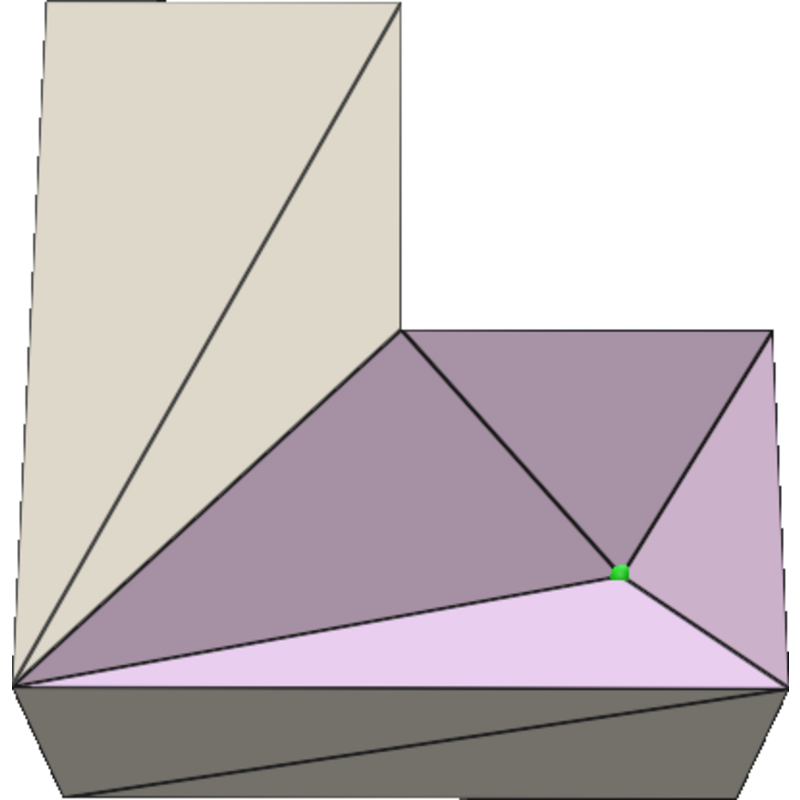} &
    \includegraphics[width=0.156\linewidth]{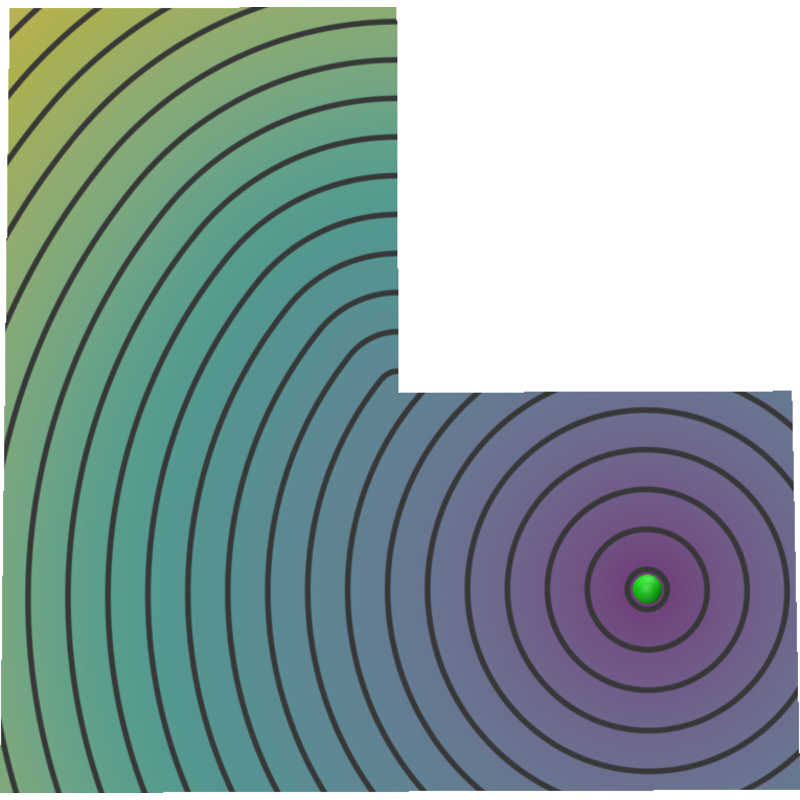} &
    \includegraphics[width=0.156\linewidth]{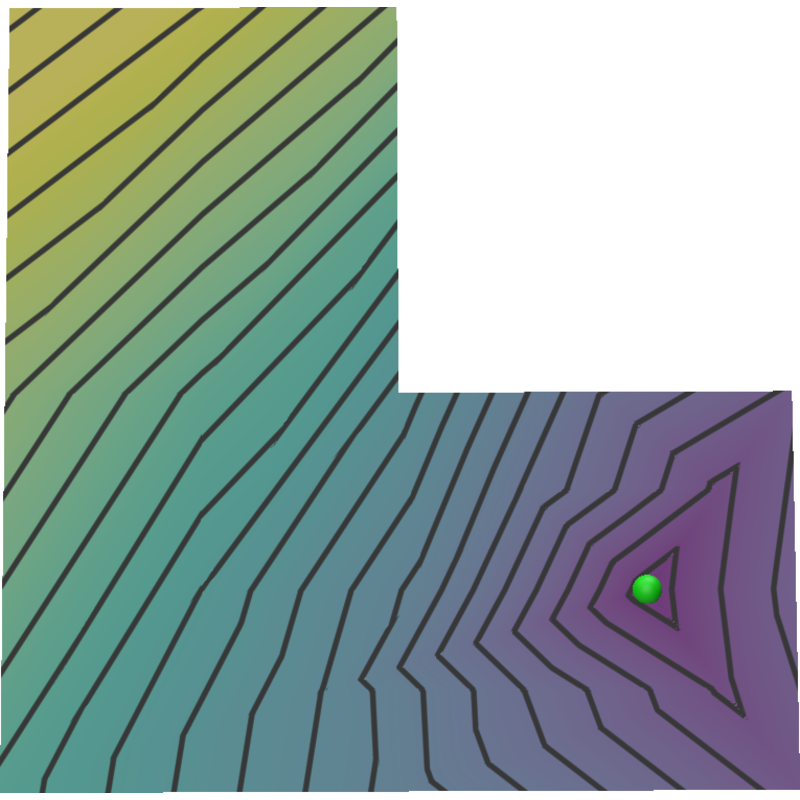} &
    \includegraphics[width=0.156\linewidth]{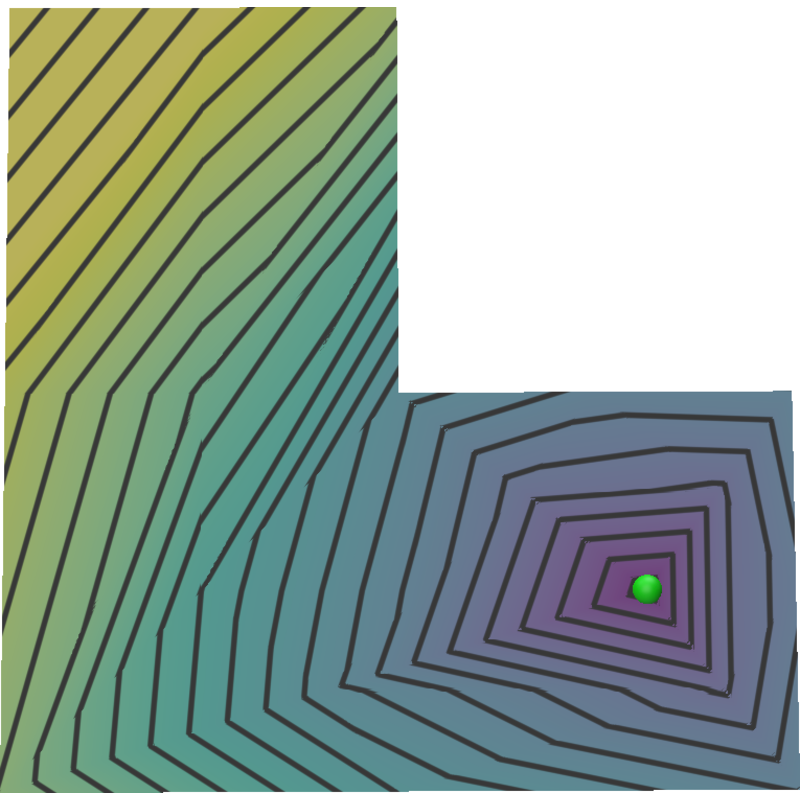} &
    \includegraphics[width=0.156\linewidth]{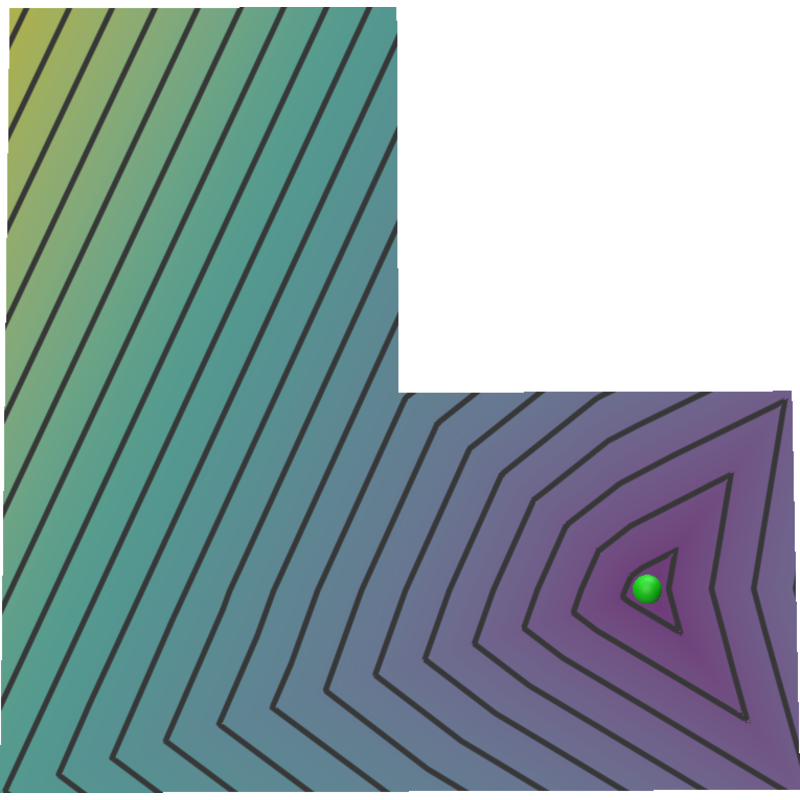} &
    \includegraphics[width=0.156\linewidth]{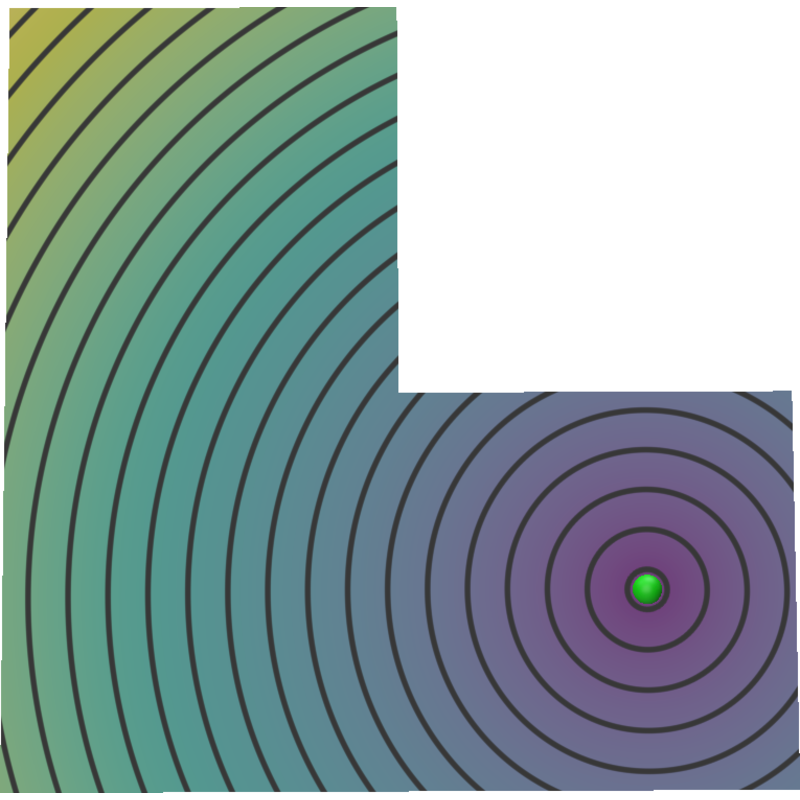} \\

     & Rel L2 &
    \small --- &
    \small \num{1.49e-1} &
    \small \num{1.80e-1} &
    \small \num{1.51e-1} &
    \small {\boldmath$1.06 \times 10^{-2}$} \\

  \end{tabular}

  \caption{Flat mesh with a non-kernel source, in 2D (top) and
  in 3D (bottom), where the ground truth is smooth but not Euclidean.
  Nevertheless, we obtain the least $L_2$ error with considerably fewer visual
  artifacts (recall that linear methods are computed on a once-subdivided
  version).}
  \label{fig:L_non_convex}
\end{figure}

 \paragraph*{Non-convex domains} Our exactness result for flat Euclidean domains assumes that the source lies in the kernel. \figref{fig:L_non_convex} tests coarse non-convex domains in 2D and 3D, where this assumption does not hold. \ours\ nevertheless has the lowest errors and substantially fewer visual artifacts.

\subsection{Flexibility (C3)}
\label{subsec:anisotropic-metrics}
\newlength{\cubepanelht}

\begin{figure}[t]
    \centering
    \settoheight{\cubepanelht}{\includegraphics[width=0.17\linewidth]{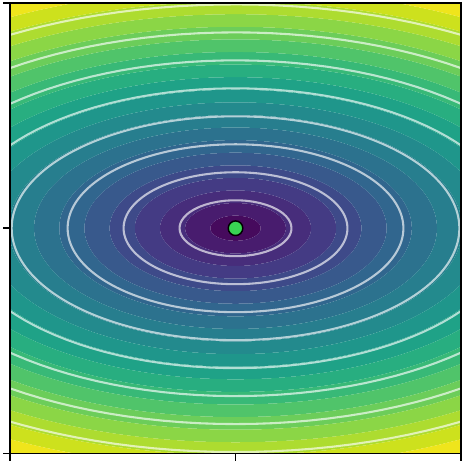}}
    \setlength{\tabcolsep}{2pt}
    \begin{tabular}{@{}ccccc@{\hspace{4pt}}c@{}}
        \includegraphics[width=0.17\linewidth]{images/cube/dist_analytic.pdf} &
        \includegraphics[width=0.17\linewidth]{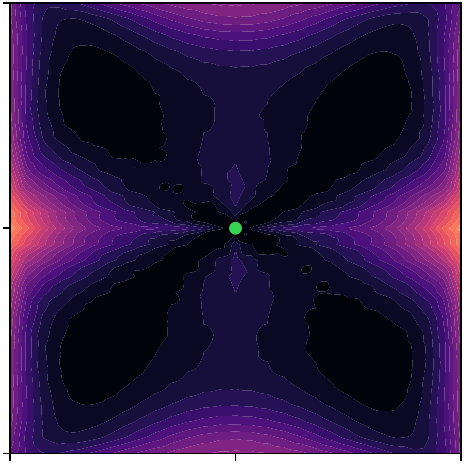} &
        \includegraphics[width=0.17\linewidth]{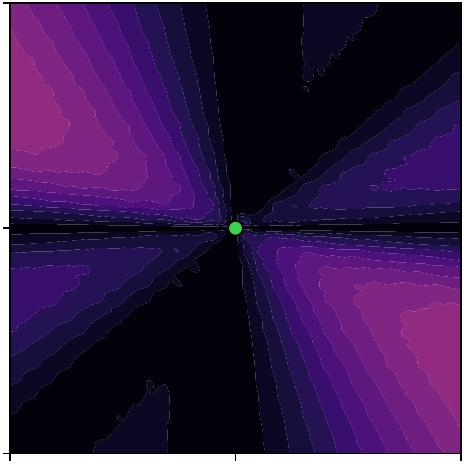} &
        \includegraphics[width=0.17\linewidth]{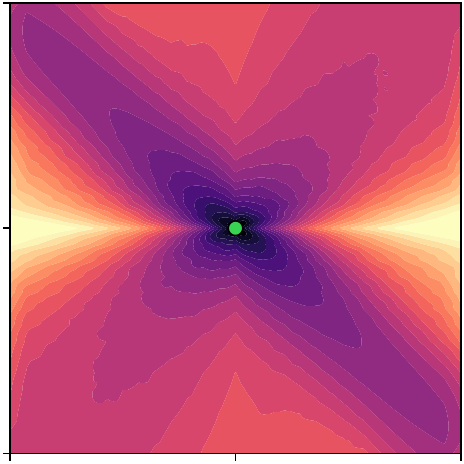} &
        \includegraphics[width=0.17\linewidth]{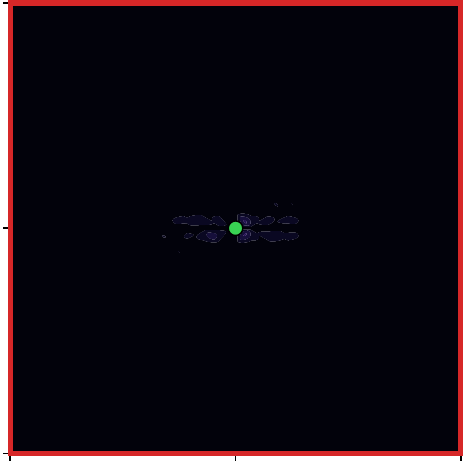} &
        \includegraphics[height=\cubepanelht]{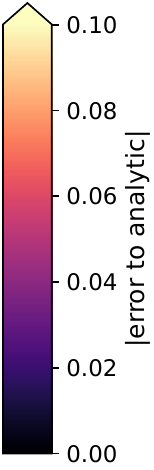} \\[2pt]

        {\small (a) Analytic} & {\small (b) HM} & {\small (c) FIM} & {\small (d) DFA} & {\small (e) Ours (\ours)} & \\
        {\small Rel $L_2$} & {\small \num{7.89e-2}} & {\small \num{1.52e-1}} & {\small \num{1.71e-1}} & {\small \num{1.35e-10}} & \\
    \end{tabular}

    \caption{Anisotropic distance in a tetrahedralized unit cube, compared
    	against the closed-form solution (Eq.~\ref{eq:aniso-gt}). We show the $z=0$ slice.  The squared distance is exactly quadratic, so our
    	error is bounded by the solver tolerance rather than by the discretization.}
    \label{fig:cube_app}
\end{figure}

\begin{figure}[t]
  \centering
  \begin{subfigure}{\linewidth}
    \centering
    \setlength{\tabcolsep}{2pt}
    \begin{tabular}{@{}rc@{}}
      \small Isotropic &
      \includegraphics[trim=288 0 290 0, clip, angle=-90, width=0.62\linewidth, valign=m]{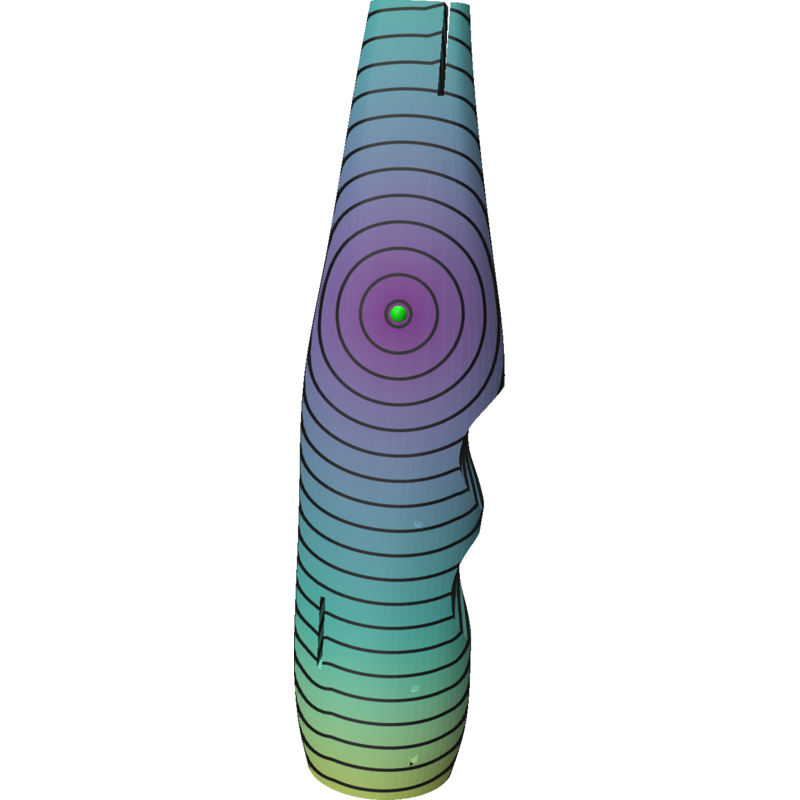} \\[2pt]
      \small Anisotropic DFA &
      \includegraphics[trim=288 0 290 0, clip, angle=-90, width=0.62\linewidth, valign=m]{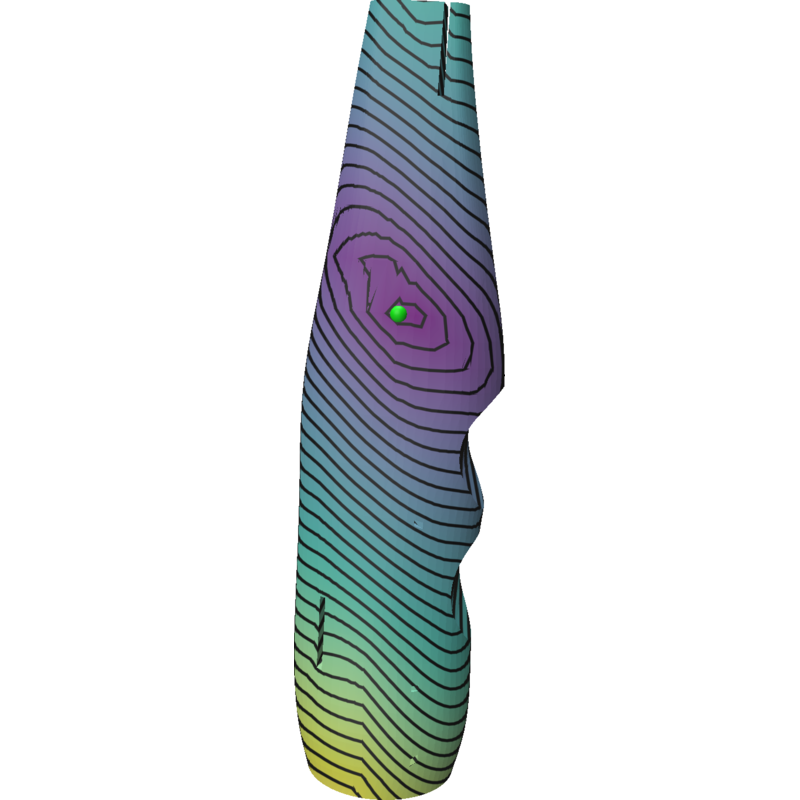} \\[2pt]
      \small Anisotropic \ours &
      \includegraphics[trim=288 0 290 0, clip, angle=-90, width=0.62\linewidth, valign=m]{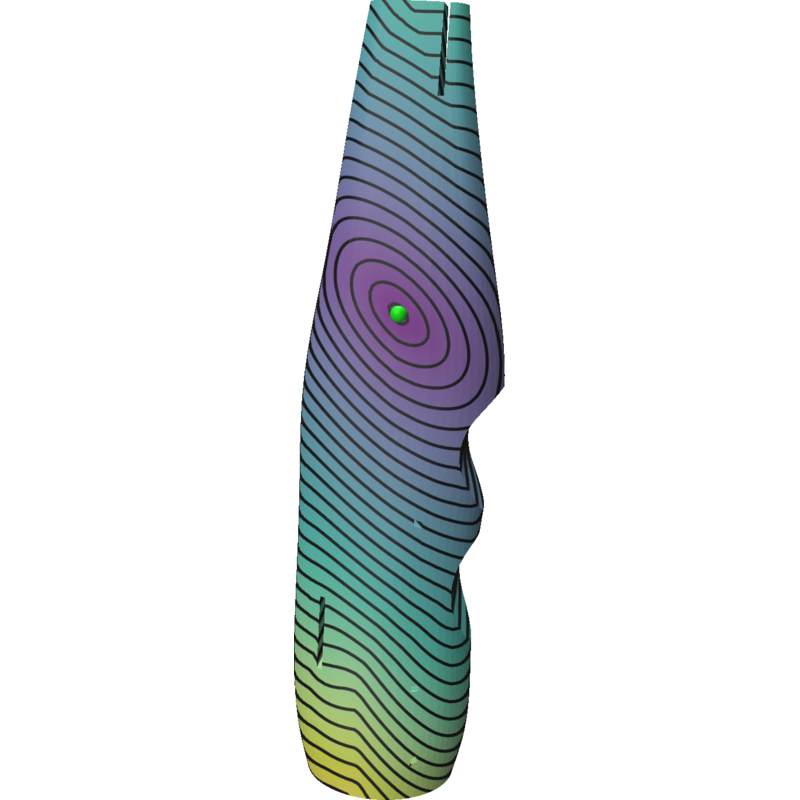} \\
    \end{tabular}
    \caption{a constant $2\!:\!1$ speed ratio at $45^\circ$}
    \label{fig:extrinsic_ani}
  \end{subfigure}\\[6pt]
  \begin{subfigure}{\linewidth}
    \centering
    \setlength{\tabcolsep}{1pt}
    \begin{tabular}{@{}ccc@{}}
      \small Isotropic & \small Anisotropic DFA & \small Anisotropic \ours \\
      \includegraphics[trim=0 180 0 180, clip, width=0.32\linewidth]{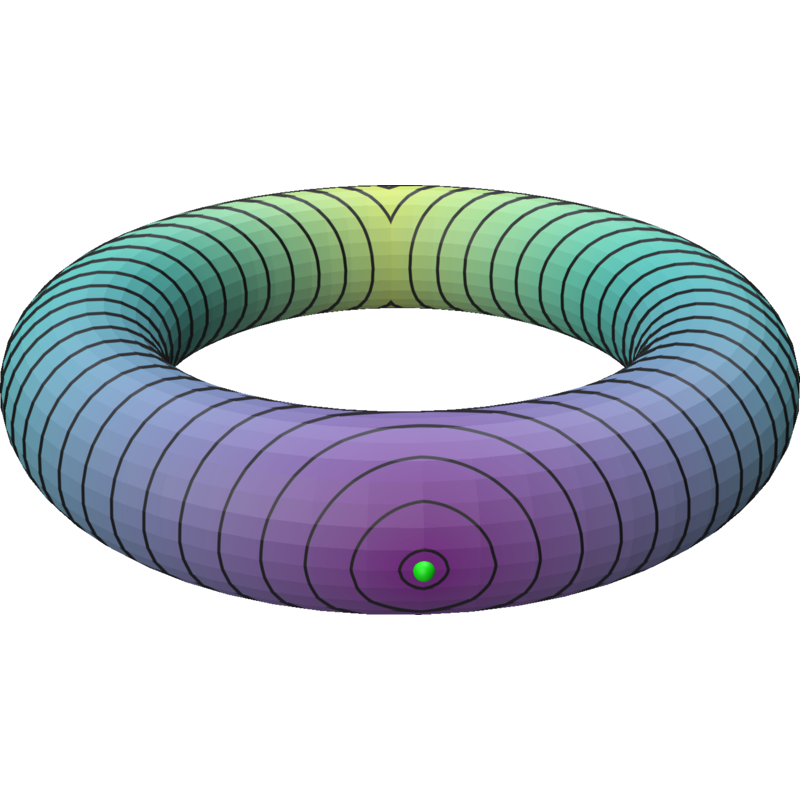} &
      \includegraphics[trim=0 180 0 180, clip, width=0.32\linewidth]{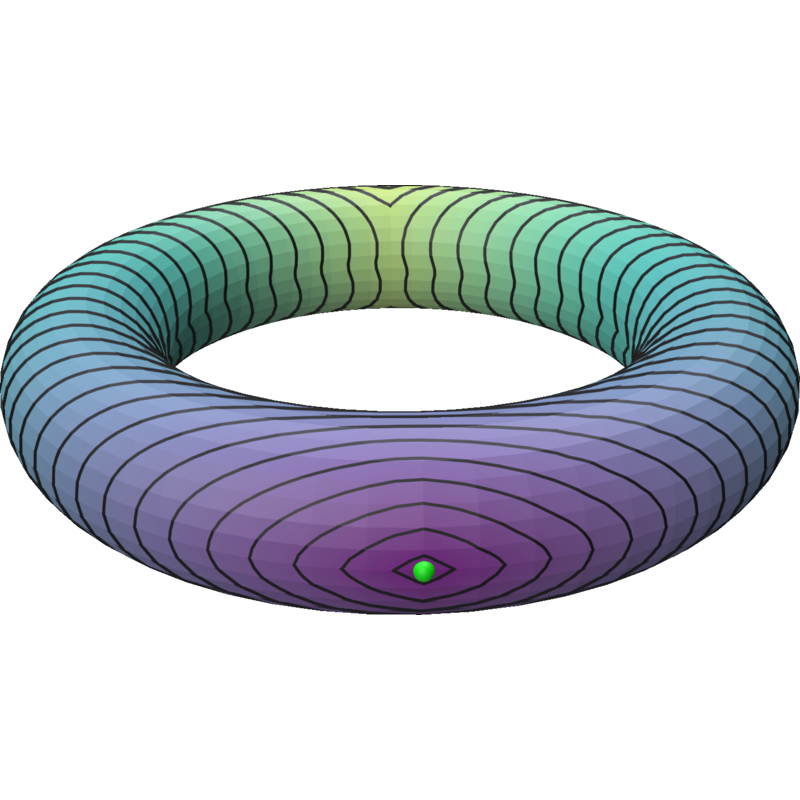} &
      \includegraphics[trim=0 180 0 180, clip, width=0.32\linewidth]{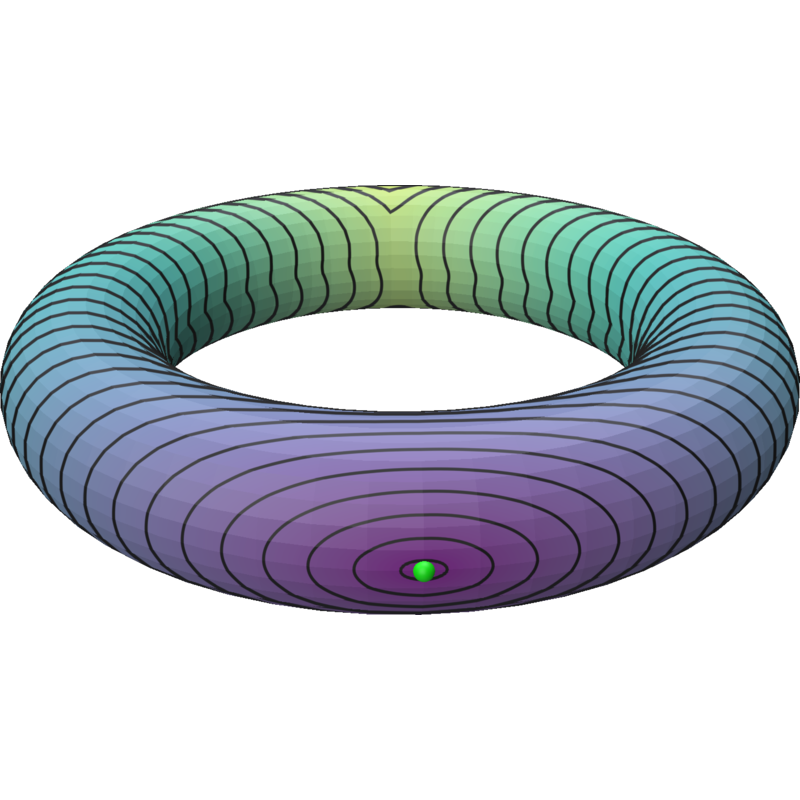} \\
    \end{tabular}
    \caption{a curvature-aligned metric on a torus}
    \label{fig:intrinsic_ani}
  \end{subfigure}

  \caption{Anisotropic metrics on surfaces. In both examples we show the isotropic field for reference, and the \emph{same} anisotropic metric solved with linear elements (DFA) and with ours. The quadratic solution represents both cleanly, whereas the linear one introduces distortion around the source.}
  \label{fig:anis_surface}
\end{figure}

\paragraph{Anisotropic metrics}

Our method easily incorporates prescribed metrics without requiring them to be integrable or embeddable. To simplify the notation, we present the discrete formulation for a single simplex. It extends to the full mesh through straightforward aggregation. We prescribe a positive definite metric tensor $\mathds{M} \in \mathbb{R}^{k \times k}$ for each simplex, where $k\in\{2,3\}$ is the dimension of the manifold. The eikonal equation becomes:

\begin{align*}
1 &= \sqrt{\nabla d^T \mathds{M} \nabla d}   = \sqrt{\nabla \sqrt{u}^T \mathds{M} \nabla \sqrt{u}} = \sqrt{\frac{1}{4u}\nabla u^T \mathds{M} \nabla u} \\\Rightarrow u &= \frac{1}{4}\nabla u ^T \mathds{M} \nabla u.
\end{align*}
As such, the convex system of Eq.~\ref{eq:discrete-convex-formulation} and the consequent adaptation of the ADMM algorithm of Eq.~\ref{eq:admm-updates} are straightforward. Given the Cholesky factorization $\mathds{M} = LL^{\!\top}$, we simply replace the gradient expression $G_\mathcal{S}u$ at each point by $L^{\!\top}G_\mathcal{S}u$ (equivalent to transforming the simplex to coordinates where the metric is flat), and the rest of \secref{subsec:admm-algorithm} proceeds as usual.

We first validate the anisotropic formulation against a closed form solution and then demonstrate it on surfaces. For a spatially constant metric $\mathds{M}=LL^{\!\top}$ on a convex domain, the anisotropic geodesic between two points is a straight line, and its length equals the Euclidean distance in the transformed coordinates.

\begin{equation}
d_\mathds{M}(x, y) = \Vert L^{-1}(x - y) \Vert.
\label{eq:aniso-gt}
\end{equation}
We prescribe the metric in the canonical orthonormal frame $v_1, v_2 (, v_3)$ with propagation speeds $\mu_1, \mu_2 (, \mu_3)$ along its directions,
\begin{equation}
	\mathds{M} = \mu_1^2\, v_1 v_1^{\!\top} + \mu_2^2\, v_2 v_2^{\!\top}
	\;(+\; \mu_3^2\, v_3 v_3^{\!\top}) ,
	\label{eq:metric-frame}
\end{equation}
and solve on a tetrahedralized cube with a single interior source and an axis-stretched metric, taking $v_1$ along the $x$ axis and $\mu_1\!:\!\mu_2\!:\!\mu_3=2:1:1$. The level sets of Eq.~\eqref{eq:aniso-gt} are then ellipsoids elongated along $v_1$
 with axis ratio $\mu_1\!:\!\mu_2$. \figref{fig:cube_app} shows the analytic field and the pointwise error of each finite-element method against Eq.~\eqref{eq:aniso-gt}. \ours is the most accurate. The squared distance  $u = (x-y)^{\!\top} \mathds{M}^{-1} (x-y)$ is exactly quadratic, hence exactly representable by our elements. 
 
On surfaces, we demonstrate this substitution in two settings (\figref{fig:anis_surface}).
First, with a metric we impose: we take $v_1$ at $45^\circ$ in the tangent
plane and $\mu_1\!:\!\mu_2 = 2\!:\!1$, which turns
the level sets elliptic (\figref{fig:extrinsic_ani}). The same metric is solved
with linear elements (\DFA) and with ours, so any difference between the two
results is due to the representation: ours generates clean ellipses, where the
linear solution distorts them and shows artifacts around the source. Second,
with a metric derived from the surface itself: we set
$v_1 = v_{f,\min}$ and $v_2 = v_{f,\max}$, the directions of minimum and
maximum curvature per face, and keep the stretches constant at
$\mu_1 :\mu_2 = 1:2$, which stretches the level sets along the curvature directions (\figref{fig:intrinsic_ani}). The metric now varies per face, and the solver of \secref{subsec:admm-algorithm} runs unchanged. Our result visually demonstrates clean anisotropic field lines, whereas artifacts are visible on the PL representation. 

\subsection{Source features}
\label{subsec:features}
\begin{figure}[h]
  \centering
  \includegraphics[width=0.24\linewidth]{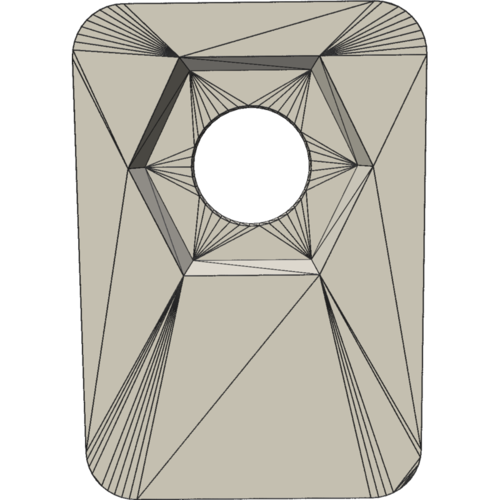} \hfill
  \includegraphics[width=0.24\linewidth]{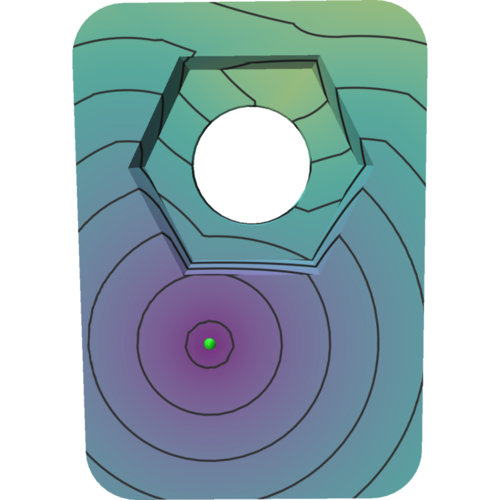} \hfill
  \includegraphics[width=0.24\linewidth]{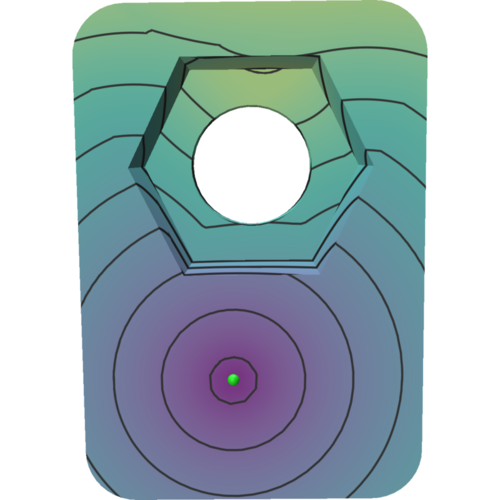} \hfill
  \includegraphics[width=0.24\linewidth]{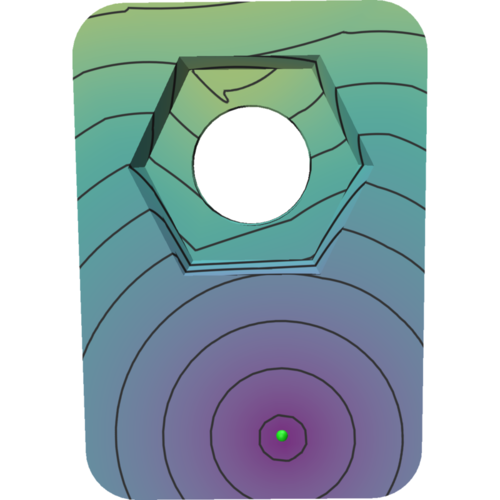}
  \caption{Our method allows placing sources anywhere on the mesh, and moving sources (even across triangle edges) creates gradually changing distance functions,.}
  \label{fig:moving_source}
\end{figure}

\begin{figure}[h] 
  \centering
    \includegraphics[width=0.24\linewidth]{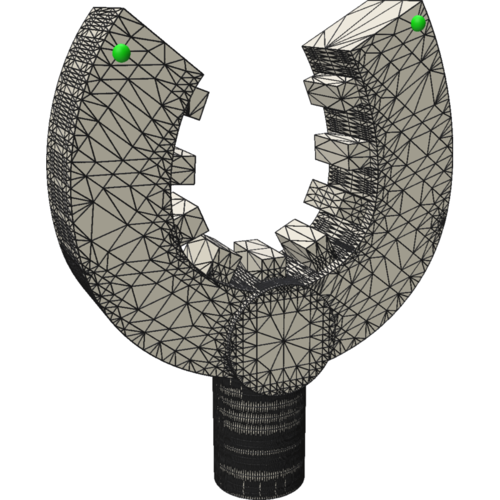}
  \hfill 
    \includegraphics[width=0.24\linewidth]{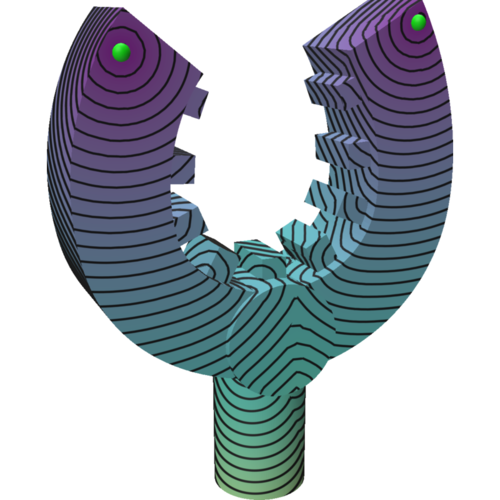}
   \hfill
   \includegraphics[width=0.24\linewidth]{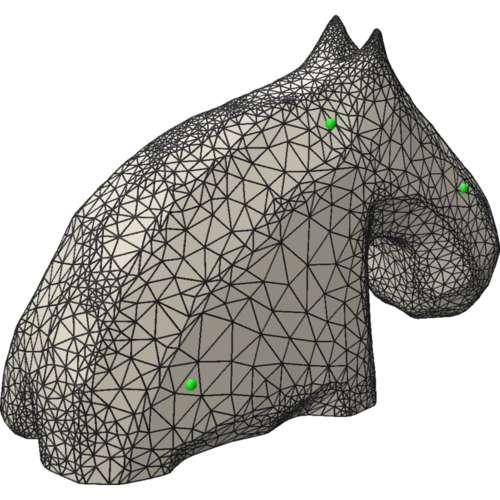}
  \hfill 
    \includegraphics[width=0.24\linewidth]{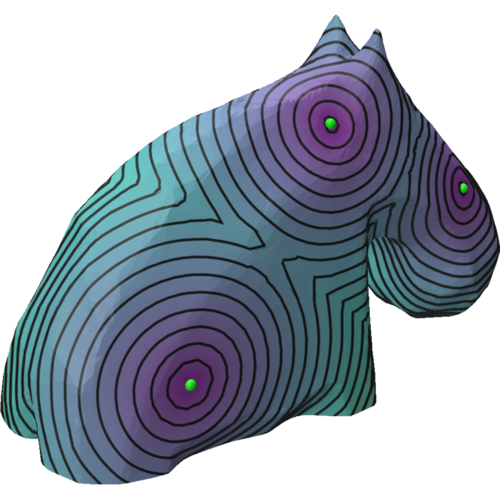}
  \caption{Our method with multiple source points: the resulting field is stable, with a clean, visually consistent medial axis between the sources.}
  \label{fig:multiple_sources}
\end{figure}

\begin{figure}[h] 
  \centering
    \includegraphics[width=0.24\linewidth]{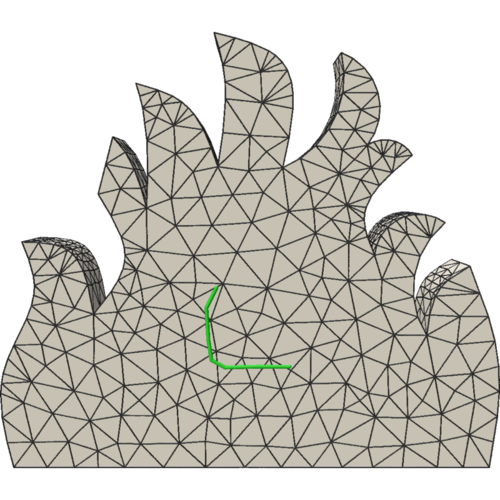}
  \hfill 
  \includegraphics[width=0.24\linewidth]{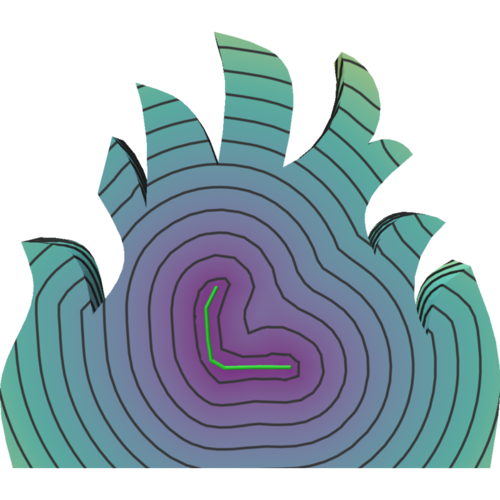}
  \hfill 
    \includegraphics[width=0.24\linewidth]{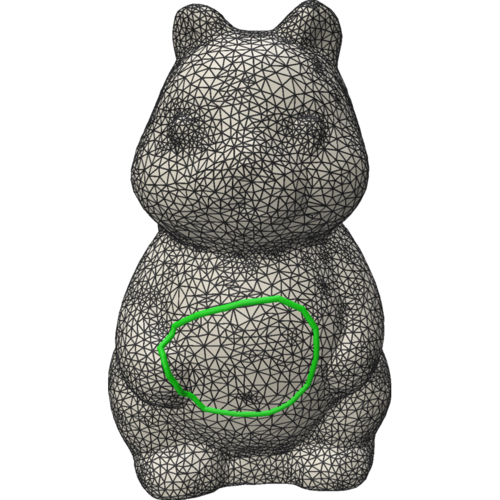}
    \hfill 
     \includegraphics[width=0.24\linewidth]{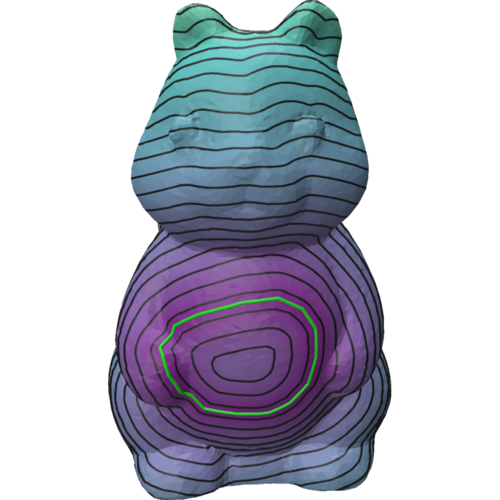}
  \caption{Our method supports continuous curve sources.}
  \label{fig:curve_source}
\end{figure}

\paragraph*{Moving source} Our method supports a source at any position on the mesh and varies continuously as the source moves. We demonstrate (\figref{fig:moving_source}) its stability and consistency by moving a point through several keyframes along a trajectory that crosses a triangle boundary.

\paragraph*{Multiple and curve sources} We evaluate our method with multiple point sources in \figref{fig:multiple_sources}. The resulting distance field is stable and has a visually consistent medial axis. Our ``anywhere source'' support also enables continuous curve sources (\figref{fig:curve_source}), which we implement by discretizing polylines into a set of point sources. Specifically, we sample curves at regular spatial intervals (by default set to half the average edge length) to ensure the sampling density is independent of the underlying triangulation.

\section{Application: Cardiac Activation}
\label{sec:cardiac}
Anisotropic accuracy was validated against the closed-form solution in \secref{subsec:anisotropic-metrics}. Here, we qualitatively demonstrate anisotropy and tetrahedral-mesh support at clinical scale on real-world data. Because the true activation sequence of a real heart is unknown, no reference solution is available, and we make no accuracy claims.

In a cardiac muscle, the electrical activation front travels several times faster
along the muscle fibres than across them. The arrival time $T$ of that front at
a point therefore \emph{is} a geodesic distance from the stimulus site, taken
in a fibre-aligned metric. Here the anisotropy comes from the tissue itself,
and the distance field under a prescribed metric is the quantity of clinical
interest.

We take the publicly available four-chamber cohort of
\citet{strocchi2020publicly}, whose tetrahedral meshes carry rule-based
ventricular fibre directions together with a node label marking the
right-ventricular endocardial pacing electrode. We use those nodes as the
source set and align the metric with the local fibre frame. The rest of the algorithm is the same as in \secref{subsec:anisotropic-metrics} with the solver of
\secref{subsec:admm-algorithm}.

\begin{figure}[t]
   \centering
   \setlength{\tabcolsep}{2pt}
   \begin{tabular}{@{}ccc@{}}
   	{\small (a) Anatomy and fibres} &
   	{\small (b) Anisotropic (\ours)} &
   	{\small (c) Isotropic (\ours)} \\[1pt]
   	
   	\includegraphics[width=0.32\linewidth]{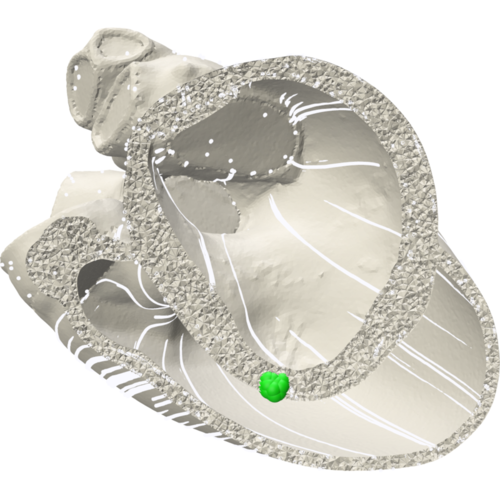} &
   	\includegraphics[width=0.32\linewidth]{images/heart/heart23_a.png} &
   	\includegraphics[width=0.32\linewidth]{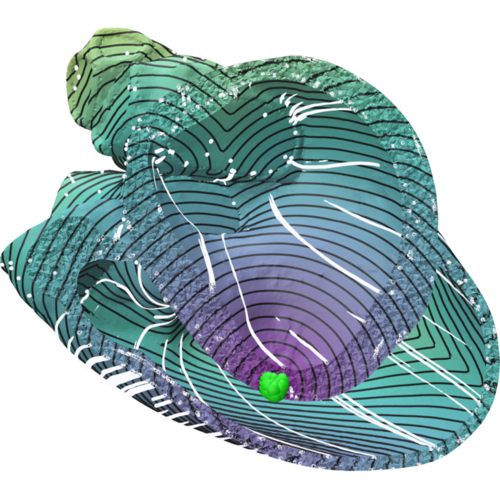} \\
   	
   	\includegraphics[width=0.32\linewidth]{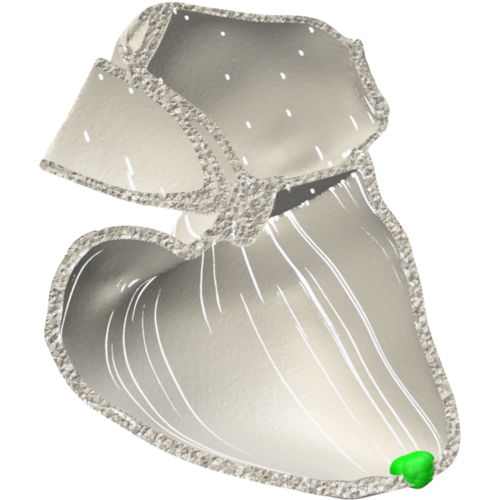} &
   	\includegraphics[width=0.32\linewidth]{images/heart/heart24_a.png} &
   	\includegraphics[width=0.32\linewidth]{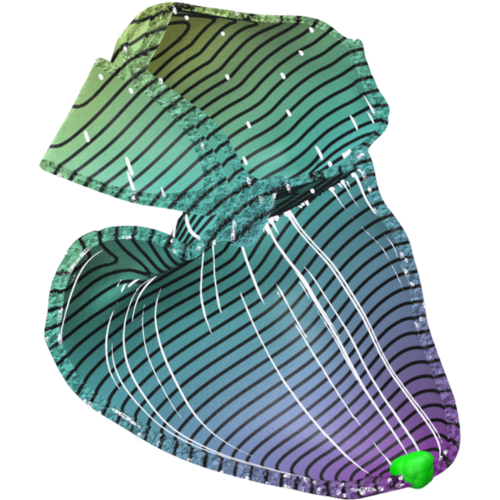} \\
   \end{tabular}

    \caption{Anisotropic activation-time field on two four-chamber cardiac
    	tetrahedral meshes of the cohort of~\citet{strocchi2020publicly}, paced at
    	the right-ventricular endocardial electrode (green). Each column is the same from a different point of view. The isotropic (ablation) solution (right) shows the regular geodesic metric, whereas the proper solution (middle) shows clear elongation of level-set distances along the white muscle fibres.}
    	
    \label{fig:heart_app}
\end{figure}

We show results for two heart meshes from the dataset in \figref{fig:heart_app}. The isochrones radiate from the electrode, remain well formed throughout the interior, and are elongated along the local fibre direction.

\section{Discussion}
\label{sec:limitations}

\paragraph*{Limited to FEM} \ours works well on uneven meshes with badly-shaped triangles, artifacts, and even non-manifolds. However, by definition, it is still limited to a finite-element concept of gradient, which requires a well-connected triangle mesh (cf. the heat method, which easily extends to e.g., point clouds~\cite{Sharp:2020:LNT}, once a Laplacian is defined). It is likely possible to extend \ours to such geometries with a similar definition of a gradient and a vector field, and we leave that for future work.

\paragraph*{Structure preservation} While the continuous $u=d^2$ problem is equivalent in the viscosity sense to the $d$ problem solved by \DFA, this equivalence does not generally hold after discretization. The analogous structure preservation issue also arises for \DFA. We empirically obtain that $u > \frac{1}{4}|\nabla u|^2$ (ideally should be equal) even away from the medial axis, and we leave an analysis of the implications to the correctness of the geodesic distance, and the challenge of creating a structure-preserving discrete geodesic condition, to future analysis.

\paragraph*{Outlook} Two interesting directions stem from our work. The first is to use our representation of the geodesic distance to compute an accurate short time heat kernel, using the distance as initialization. The second is to extend the method to distance computation on higher dimensional manifolds, including data manifolds.

\begin{figure*}[t]
\centering
\setlength{\tabcolsep}{5pt}
\begin{tabular}{ccccc}
    \includegraphics[width=0.166\linewidth]{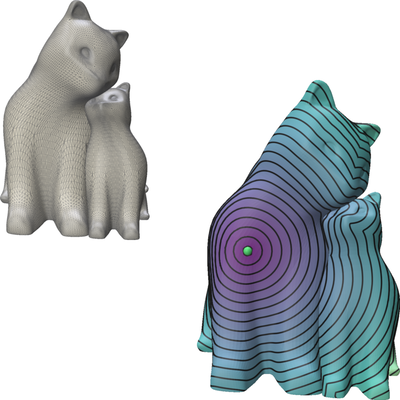} &
    \includegraphics[width=0.166\linewidth]{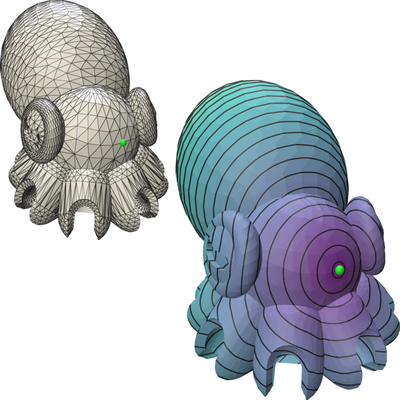} &
    \includegraphics[width=0.166\linewidth]{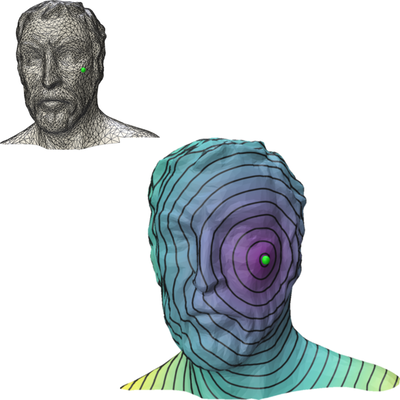} &
    \includegraphics[width=0.166\linewidth]{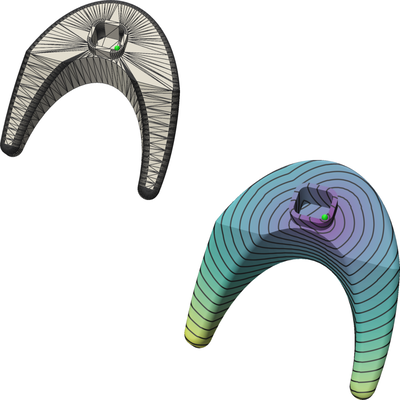} &
    \includegraphics[width=0.166\linewidth]{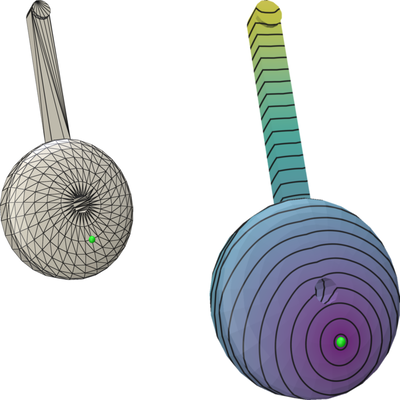} \\

    \includegraphics[width=0.166\linewidth]{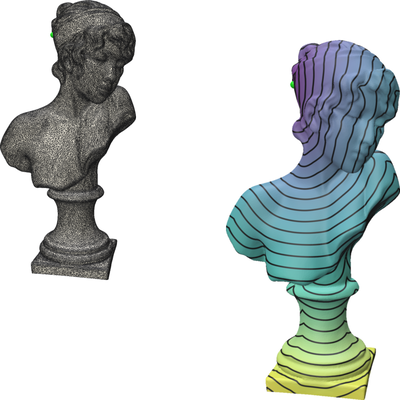} &
    \includegraphics[width=0.166\linewidth]{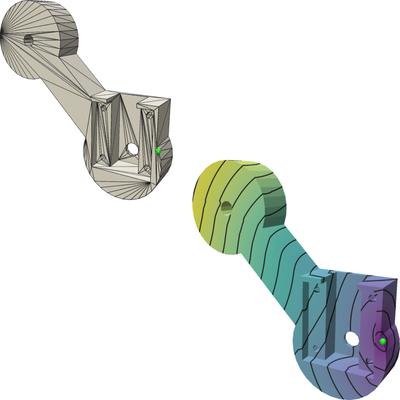} &
    \includegraphics[width=0.166\linewidth]{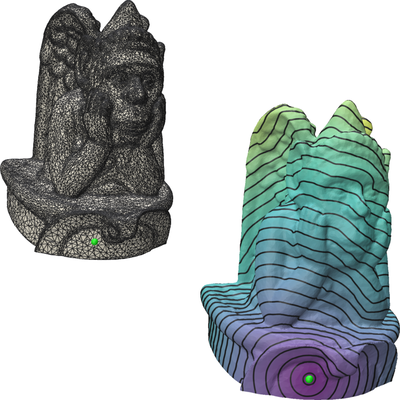} &
    \includegraphics[width=0.166\linewidth]{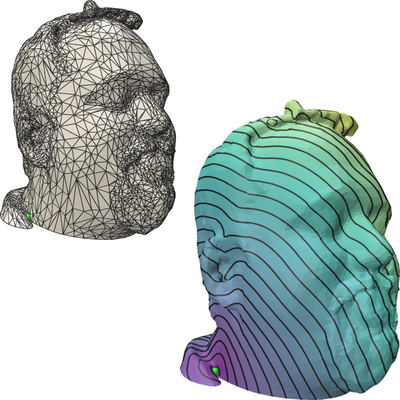} &
    \includegraphics[width=0.166\linewidth]{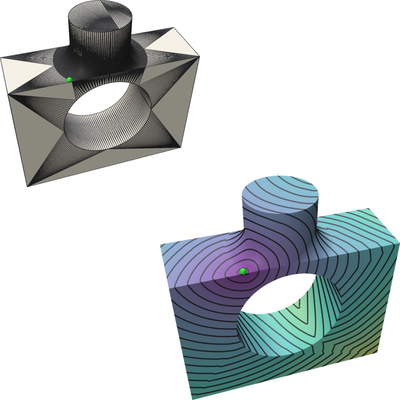} \\

    \includegraphics[width=0.166\linewidth]{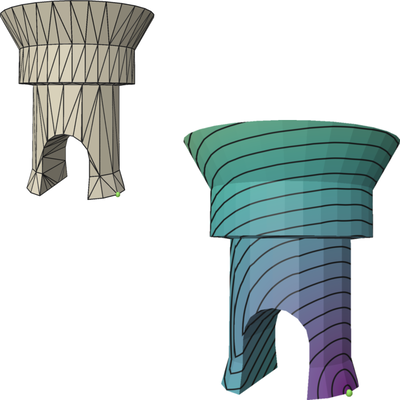} &
    \includegraphics[width=0.166\linewidth]{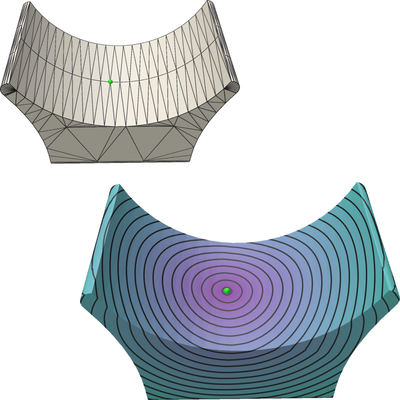} &
    \includegraphics[width=0.166\linewidth]{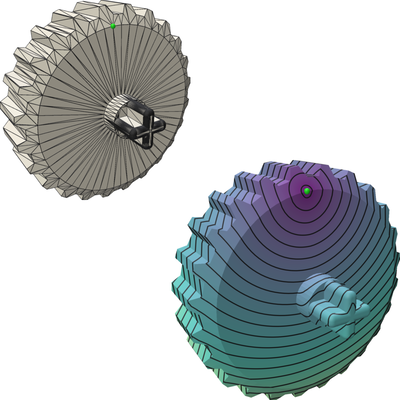} &
    \includegraphics[width=0.166\linewidth]{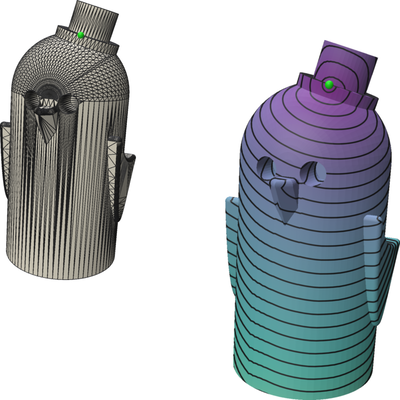} &
    \includegraphics[width=0.166\linewidth]{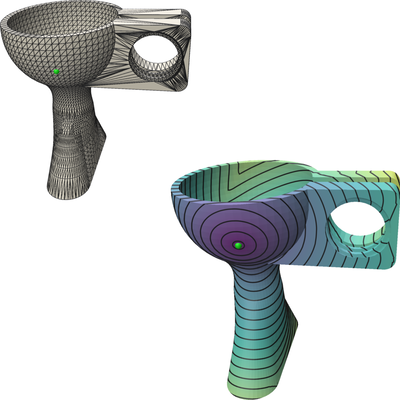} \\
\end{tabular}

\caption{Geodesic distance gallery for triangle (surface) meshes. For each example, the top inset illustrates the input mesh, while the larger image below presents the corresponding geodesic distance.}
\label{fig:gallery_2d}
\end{figure*}

\bibliographystyle{ACM-Reference-Format}
\bibliography{main}

\clearpage
\appendix
\section{Theoretical Proofs}
\subsection{Equivalence of continuous \DFA and \protect{\ours}}
\label{sec:dfa-equivalence}
Recall that the \DFA optimization problem is:
\begin{equation}
\begin{aligned}
d &= \operatorname*{arg\,max}_{d} \int_\mathcal{M} d \\
\text{s.t.}\quad |\nabla d| &\le 1,
\end{aligned}
\end{equation}
and \ours is
\begin{equation}
\begin{aligned}
u &= \operatorname*{arg\,max}_{u} \int_\mathcal{M} u \\
\text{s.t.}\quad u &\ge \tfrac14 |\nabla u|^2 .
\end{aligned}
\end{equation}
We show that the viscosity solutions to both maintain exactly $u = d^2$ everywhere. Let $S_1$ (for linear) be the set of viscosity subsolutions to \DFA, that is for any $d \in S_1$ and any point $x \in \mathcal{M}$, we have that every test function $\phi \in C^1(\mathcal{M})$ that touches $d$ from above has $|\nabla \phi| \leq 1$. Since the function  \(d\mapsto u = d^2\) is monotone on $[0,\infty)$, we have that $\phi^2$ also touches $u$ from above, and satisfies $\phi^2 \geq \frac{1}{4}|\nabla (\phi^2)|^2$, since $\nabla(\phi^2)=2\phi\nabla\phi$ gives $\frac{1}{4}|\nabla(\phi^2)|^2 = \phi^2|\nabla\phi|^2 \leq \phi^2$. As such, $u$ is in $S_2$ (for quadratic), the set of viscosity subsolutions to our optimization problem, where $u = d^2$ is a bijective map between the sets. The only thing left to show is that the maximal solutions are related by the squaring. We use two properties of Perron's method: 1) due to the Stability Property of viscosity subsolutions,  the pointwise supremum of subsolutions is a subsolution, and thus both sets $S_1$ and $S_2$ possess a unique pointwise maximal element, and 2) by the Comparison Principle, this maximal element is the unique viscosity solution that satisfies the condition with actual equality where defined. Then, the integral maximality of the objective functions induces pointwise maximality in the respective sets $S_1$ and $S_2$. Thus, a solution $d^*$ so that for all $d \in S_1$ $d^*(x)\geq d(x)$ by definition has $u^* = (d^*)^2 \geq d^2 = u \in S_2$, which makes $u^*$ the pointwise supremum of $S_2$. The $\sqrt{}$ operation is used to prove in the other direction. Note again that this equivalence only holds in the continuous setting, as neither \DFA nor our discretization is structure-preserving in the discrete setting. Finally, we note that this proof is generalizable to any monotone transformation of Hamilton-Jacobi type equations.

\subsection{Deriving the squared geodesic distance from Varadhan's Formula}
\label{sec:squared_geodesic_proof}

Squaring both sides of Varadhan's formula \eqref{eq:VaradhanFormula} yields
\begin{align}
\label{eq:SquaredVaradhan}
    u(x) = \lim_{t\to 0}-4t\log k_t(x)
\end{align}
where \(u(x) = d^2(x,y)\) is the squared distance between \(x\) and \(y\), and \(k_t(x) = k_t(x,y)\) is the heat kernel with a source at \(y\), satisfying the heat equation
\begin{align}
\label{eq:HeatEquation}
    {\frac{\partial}{\partial t}}k_t(x) = \Delta k_t(x)
\end{align}
for \(t>0\). Throughout, we suppress the dependence on \(y\) in the notation.  Define \(v_t(x)\coloneqq -4t\log  k_t(x)\), which satisfies \(\lim_{t\to 0}v_t(x) = u(x)\) by \eqref{eq:SquaredVaradhan}.  Moreover, by parabolic regularity, this convergence extends to spatial derivatives: for \(x\neq y\) and away from conjugate points, \(\nabla v_t(x)\to \nabla u(x)\) and \(\Delta v_t(x)\to \Delta u(x)\) as \(t\to 0\).  In fact, \(v_t\) admits the short-time asymptotic \(v_t(x)\sim u(x) + O(t\log(t))\), which implies in particular that \(\lim_{t\to 0}t{\frac{\partial v}{\partial t}}=0\).
The heat kernel on a surface admits the short-time expansion \(k_t(x) = {1\over 4\pi t}e^{-{u(x)\over 4t}}\sum_{j=0}^\infty \phi_j(x)t^j\) where \(\phi_0 = 1\) \cite{kannai1977off}.  Consequently, \(v_t(x) = -4t\log k_t(x) = u(x) +4t\log(t) + 4t\log(4\pi) - 4t\log(1+\sum_{j=1}^\infty\phi_j(x) t^j)\). 

We now rewrite the heat equation \eqref{eq:HeatEquation} in terms of \(v_t(x)\).  This is achieved by first expressing \(k = e^{-{v\over 4t}}\) followed by taking its derivatives, \({\partial\over\partial t}k = ({v\over 4t^2} - {1\over 4t}{\partial v\over\partial t})e^{-{v\over 4t}}\) and \(\Delta k = (-{\Delta v\over 4t} + {|\nabla v|^2\over 16 t^2})e^{-{v\over 4t}}\), and substituting these derivative expressions into \eqref{eq:HeatEquation}.  After rearrangement and canceling the common exponential factor, we obtain
\begin{align}
    t\left(\frac{\partial v}{\partial t}-\Delta v\right) + {1\over 4}|\nabla v|^2 - v = 0.
\end{align}
Taking the limit \(t\to 0\), using \(v_t\to u\), \(\nabla v_t\to \nabla u\), \(\Delta v_t\to \Delta u\), and  \(t{\partial v_t\over \partial t}\to 0\), we get
\begin{align*}
    {1\over 4}|\nabla u|^2 - u = 0.
\end{align*}

\section{PQ Elements}
\label{sec:pq_matrices}

\subsection{2D Triangle Meshes}

\begin{figure}[t]
	\centering
	\begin{subfigure}[b]{0.36\linewidth}
		\centering
		\includegraphics[width=\linewidth]{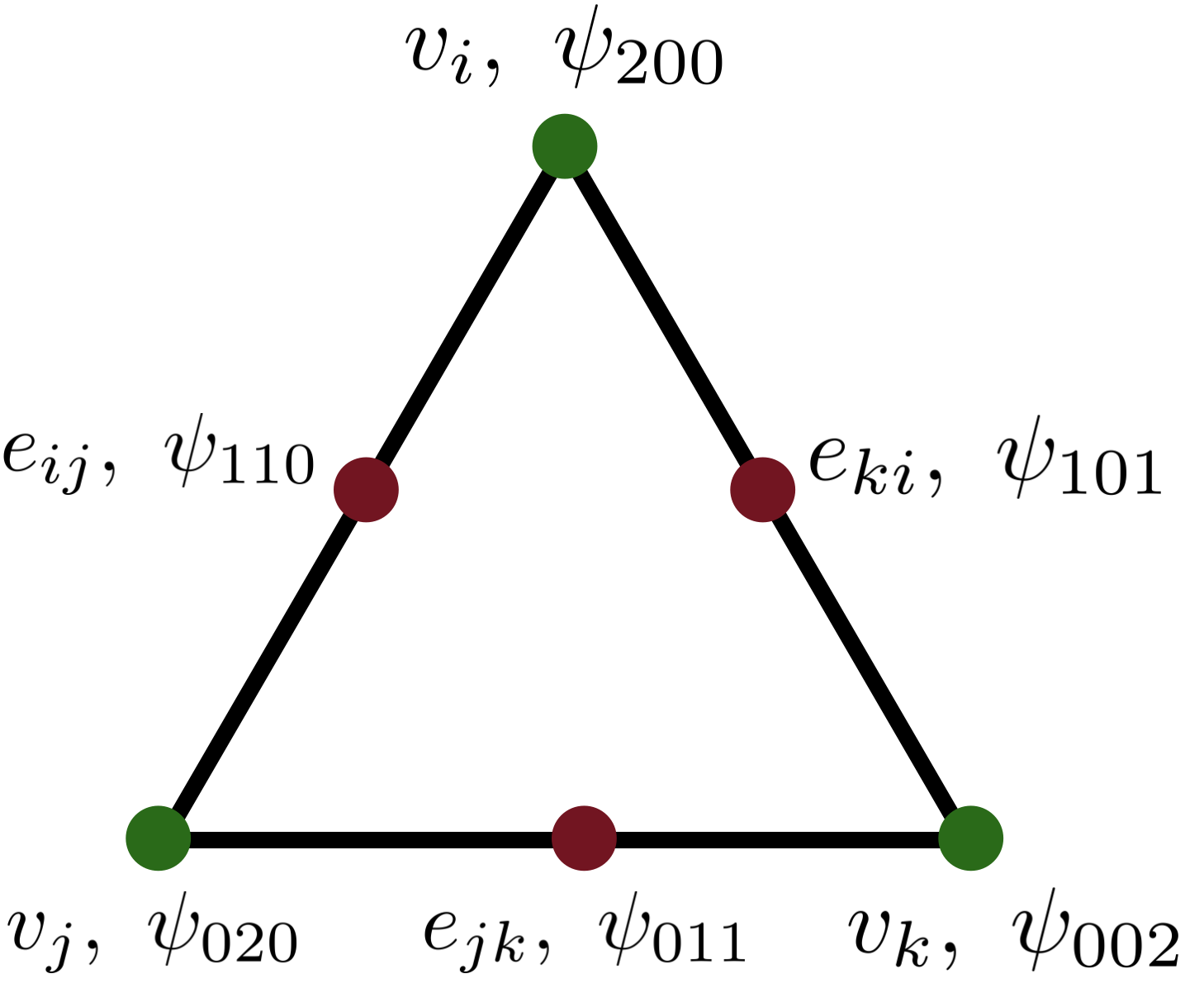}
		\caption{Triangle}
		\label{fig:p2-illustration-tri}
	\end{subfigure}
	\hfill
	\begin{subfigure}[b]{0.48\linewidth}
		\centering
		\includegraphics[width=\linewidth]{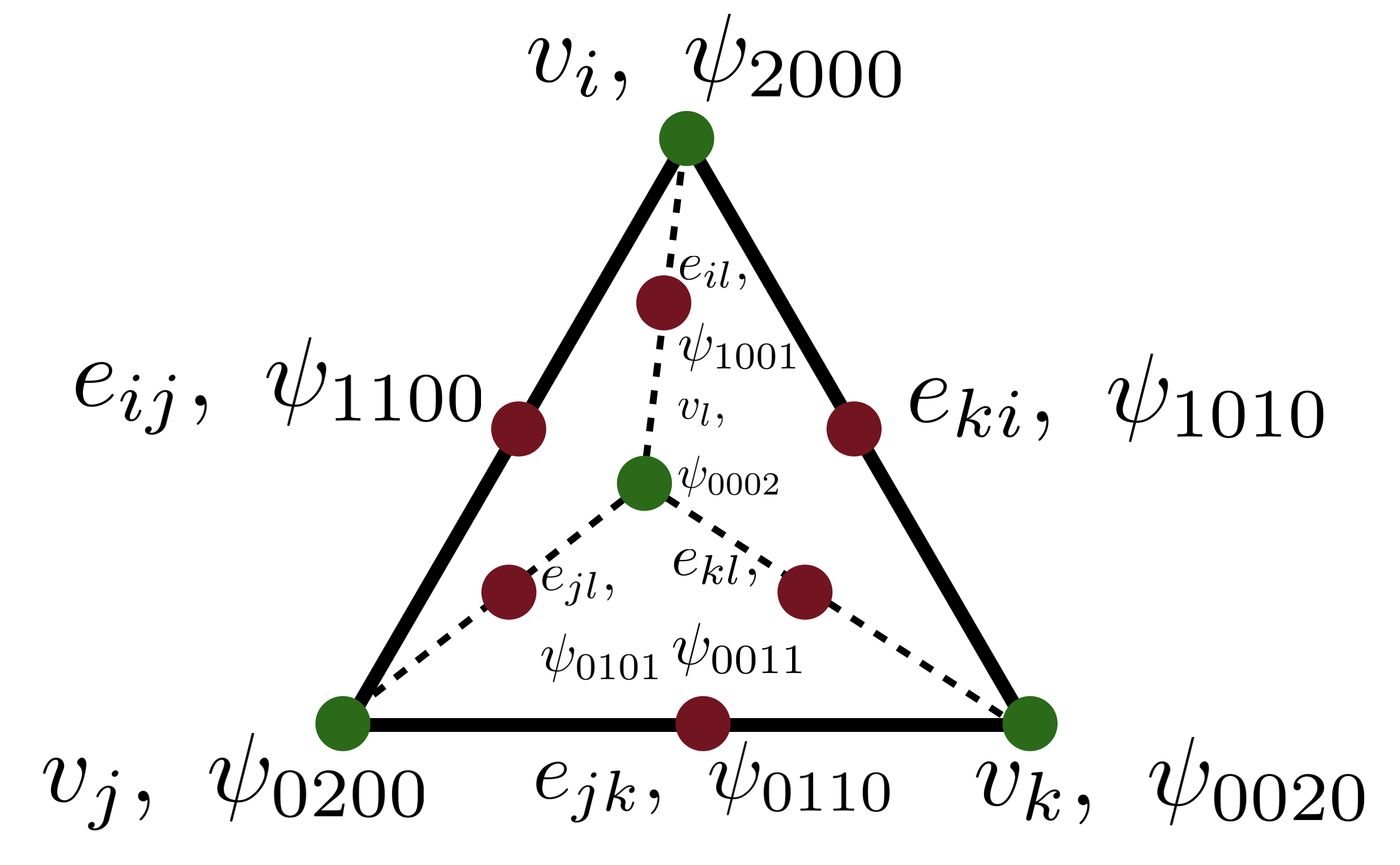}
		\caption{Tetrahedron}
		\label{fig:p2-illustration-tet}
	\end{subfigure}
	\caption{P2 element node layout on a triangle (left) and a tetrahedron (right).}
	\label{fig:p2-illustration}
\end{figure}
For full reproducibility, we include details of the construction of all PQ matrices. Assume the nodes as in the~\figref{fig:p2-illustration-tri} for a single triangle $t=ijk$, marking the vertex and midedge nodes. Consider the associated basis functions $\psi^2$, and the area $A_{ijk}$ of each triangle. Finally,  consider the barycentric coordinate functions $\lambda_i(p),\lambda_j(p),\lambda_k(p),\ \forall p \in t$, and their gradients:
$$
\nabla \lambda_{i|j|k} = \frac{e^{\perp}_{jk|ki|ij}}{2A_{ijk}},
$$
where $e^{\perp}$ is the rotated vector of the edge by $\frac{\pi}{2}$ around the normal of $t$. We have that:
\begin{align*}
\psi^2_{200} &= 2\lambda_i^2-\lambda_i,\\
\psi^2_{110} &= 4\lambda_i\lambda_j.
\end{align*}
The rest are obtained respectively. The expressions for the gradients are then (complete by symmetry):
\begin{align*}
\nabla \psi^2_{200} &=  4\lambda_i\nabla \lambda_i - \nabla \lambda_i\\
\nabla \psi^2_{110} &=4(\lambda_i\nabla \lambda_j + \lambda_j \nabla \lambda_i)
\end{align*}

For PL elements, we simply have $\psi^1_{i|j|k}=\lambda_{i|j|k}$. The gradients are used to create the gradient matrix $G$.

\subsection{Mass matrices}
\label{sec:mass_matrices}

We describe the mass matrices $M_2$ and $M_\chi$ per triangle, where the global mass matrices are composed by conjugation with $Q_2$ and $Q_\chi$, respectively. In general, an element of the intra-triangle scalar mass matrix $M_r$ for $r=1,2$ is computed as:
$$
M(a,b) = \int_t{\psi^r_a(p)\psi^r_b(p)dp}.
$$
We get that $M_1:3\times 3$ is:
\begin{align*}
    M_1(t)[a,b] = A_{ijk} \cdot \begin{cases} \frac{1}{6} & a=b\\
    \frac{1}{12} & a \neq b
    \end{cases} 
\end{align*}
and (assuming order $\left(\psi_{200},\psi_{020},\psi_{002}, \psi_{110},\psi_{011},\psi_{101}\right)$) we have $M_2:6 \times 6$ where:
$$
M_2(t) = \frac{A_{ijk}}{180}\begin{pmatrix}
    6 & -1 &   -1 & 0  & -4 & 0\\
   -1 & 6 & -1 & 0 & 0  &  -4\\
   -1 & -1 & 6 & -4 & 0 & 0\\
   0 & 0 & -4 & 32 & 16 & 16\\
   -4 & 0 & 0 & 16 & 32 & 16\\
   0 & -4 & 0 & 16 & 16 & 32
\end{pmatrix}
$$

To obtain the vector mass matrix $M_\chi: 12 \times 12$, we just duplicate the matrix to treat the $(x,y)$ coordinates of an intrinsic vector field separately (a Kronecker product with the identity $I_{2 \times 2}$). To obtain the global mass matrices, one concatenates all per-triangle matrices as block diagonal matrices (and then conjugates with the respective $Q$ matrix).

\subsection{3D Tetrahedral Meshes}

Assume the nodes as in the~\figref{fig:p2-illustration-tet}, with $V_{ijkl}$ as the volume of the tet. The gradients of the barycentric functions (for vertex $i$ and without loss of generality):%
$$
\nabla \lambda_{i}= \frac{A_{jkl}\,n_{jkl}}{3V_{ijkl}},
$$
where $n_{jkl}$ is the unit normal of face $jkl$ facing inside the tet and $A_{jkl}$ its area (mirroring the 2D expression, in which $e^{\perp}$ carries the edge length). The basis functions $\psi^1$ and $\psi^2$ are identical to the 2D case, and thus also their gradients $\nabla\psi^2$ and $\nabla\psi^1$. The linear mass matrix basic block $M_1:4 \times 4$ is

$$
M_1[a,b] = V_{ijkl} \cdot 
\begin{cases}
    \frac{1}{10} & a=b\\
    \frac{1}{20} & a \neq b.
\end{cases}
$$

The P2 mass matrix $ M_2:10\times 10$ is (can also be found in~\cite{hanukah2014consistent})
$$
M_2[a,b] = \frac{6V_{ijkl}}{2520}\begin{cases}
    6 & a=b,\ \text{vertex nodes}\\
    1 & a \neq b,\ \text{vertex nodes}\\
    -4 & a \text{ vertex node}, b \text{ adjacent edge node}\\
    -6 & a \text{ vertex node}, b \text{ non-adjacent edge node}\\
    32 & a=b\ \text{edge nodes}\\
    16 & a \neq b\ \text{adjacent edges}\\
    8 & a \neq b\  \text{non-adjacent edges}
\end{cases}
$$

For tetrahedra, the nodal rows of the sampling operators are assembled from the same per-element P2 reference blocks used for the mass and stiffness matrices; the barycentric row evaluates the PQ shape functions at $\lambda=(\tfrac14,\tfrac14,\tfrac14,\tfrac14)$, giving weights $-\tfrac18$ on each of the four vertex DOFs and $+\tfrac14$ on each of the six edge DOFs, with the gradient there equal to the mean of the four vertex-sample gradients.

Finally, we note that a general formula for the mass matrix of a simplex (triangle or tetrahedron) is derived from the following formulas for barycentric product integration (e.g.,~\cite{Vermeulen:2018}):
\begin{align*}
\int_{ijk}{\lambda_i^a\lambda_j^b\lambda_k^c}dA &= \frac{a!b!c!}{(2+a+b+c)!}(2A_{ijk})\ \ \text{in 2D}\\
\int_{ijkl}{\lambda_i^a\lambda_j^b\lambda_k^c\lambda_l^d}dV &= \frac{a!b!c!d!}{(3+a+b+c+d)!}(6V_{ijkl})\ \ \text{in 3D}\\
\end{align*}
The mass matrices above follow directly from these identities.

\section{Solver Details}
\label{sec:solver-details}
A single set of parameters serves the surface benchmark and volumetric benchmark, collected in \tabref{tab:admm-settings}.

\begin{table}[t]
	\centering
	\caption{Settings for our ADMM solver. The algorithm is identical; only these differ. The first column is the setting for quantative accuracy comparison against PDE-based methods on surface benchmark. The second column is on tetrahedral benchmark. The third is the configuration used only to produce the volumetric reference fields. }
	\label{tab:admm-settings}
	\begin{tabular}{lccc}
		\toprule
		& \multirow{2}{*}{triangle} & \multicolumn{2}{c}{tetrahedron} \\
		\cmidrule(lr){3-4}
		&                           & method & reference \\
		\midrule
		samples / element                      & $7$ & $11$ & $11$ \\
		penalty $\rho$                     & $3$      & $1$       & $1$       \\
		over-relaxation $\alpha$               & $1.5$& $1.0$ & $1.6$  \\
		warm start                             & HM / FMM & HM   & FIM \\
		tolerance $\varepsilon_{\mathrm{field}}$ & $10^{-4}$ & $10^{-3}$   &  $10^{-4}$  \\
		iteration cap                          & $500$      & $2{,}000$ & $20{,}000$   \\
		\bottomrule
	\end{tabular}
\end{table}

\subsection{The tetrahedral benchmark and its references}
\label{sec:tet-corpus}
No closed-form distance is available on a raw tetrahedral domain, so the field
every method is scored against is produced by our own solver, on the finest rung
of the mesh ladder of \secref{subsec:setup} and in the tighter configuration of
\tabref{tab:admm-settings}.

The corpus starts from the same $4{,}320$ Thingi10K meshes as the surface
benchmark, and a model is kept only if it clears all four of the following: 
\begin{enumerate}
	\item \textbf{Exact self-intersection test} --- $121$ ($2.8\%$) fail.
	\item \textbf{Tetrahedralization.} The surface is filled by the constrained
	Delaunay method of \citet{AtteneCDT} and refined into the L0--L3 ladder
	with TetGen. $632$ ($14.6\%$) fail outright, to an internal error, an
	out-of-memory or a timeout, and a further $555$ ($12.8\%$) are never
	attempted because their projected size exceeds our $350$K-tetrahedron
	budget. 
	\item \textbf{Room for a reference mesh.} The reference must live on a
	strictly finer mesh than the one the methods solve on. For $242$
	($5.6\%$) that finer rung would itself exceed the size budget, leaving
	nothing to measure against.
	 \item \textbf{Non-degenerate meshes.} TetGen emits a numerically                                                            zero-volume tetrahedron in $89$ ($2.1\%$) of the remaining ladders.            
\end{enumerate}
$2{,}681$ models remain, each with a complete four-method comparison
against a certified reference. 

\subsubsection{Adpative Penalty Scaling}
\label{sec:adaptive_penalty_scaling}
 In the volumetric benchmarks we replace the uniform ADMM penalty by a spatially adaptive one: 
 $$s_t = \min\big((\bar u/\bar u_t)^{\beta}, s_{\max}\big)$$ 
 with $\beta=1$, where $\bar u_t$ is the mean of  the warm-start field over the vertices of tetrahedron $t$ and $\bar u$ its median over the mesh. For one iteration, the u-update, g-update and w-update, the only change is the diagonal  $S=diag(s_t)$ multiplied into $M_\mathcal{S}$.

This adaptation is essential because $u = d^2$ varies from the source to the bulk. A uniform penalty under-enforces the cone constraint exactly where $u$ is smallest. On finely meshed thin plates, the objective term then drives $u$ negative near the source, causing the cone radius $2\sqrt{u^+}$ to vanish and trapping the algorithm in a degenerate state with $g = 0$ and $u \le 0$. 

Weighting the penalty by $1/\bar{u}_t$ makes the consensus error scale relative to the local field magnitude, keeping near-source iterates feasible from the first sweep and eliminating the collapse. 
On the surface benchmark, adaptive weighting produces only marginal differences, so we retain the uniform penalty there.

\subsubsection{Hybrid CPU/GPU Implementation}
\label{sec:gpu-implementation}

The implementation below is used only on tetrahedral meshes, where  the number of
iterations dominates the running time. An iteration contains exactly one sequential
step, the two triangular solves against the Cholesky factor of
$\rho\,G_\mathcal{F}^\top M_\chi G_\mathcal{F}$; the sparse matrix--vector
products, the per-sample cone projections and the vector updates are all data
parallel. We therefore keep the factorization and the triangular solves on the
CPU and move everything else to the GPU, where operators and state stay
resident, so that only the right-hand side and the solution cross the bus,
$2N_\mathcal{F}$ doubles per iteration. 
The split buys $3$--$4\times$ per iteration beyond $300$k tetrahedra.

\subsection{Solver scalability}
\label{sec:scalability-curves}

\tabref{tab:scalability} compare our ADMM
solver against an interior-point solver (\textsc{Mosek}) on the same mesh. The interior-point cost grows super-linearly: at $575\,434$ degrees of
freedom, the largest instance it completes, it already needs $44.5$\,GB, where
we reach comparable accuracy in $3.0\times$ less running time and $6.1\times$
less memory. Past that size it runs out of machine rather than out of time,
while we continue --- $997\,950$ degrees of freedom in $204$\,s and $12.6$\,GB.
That ceiling is where our sampled meshes end, not where our solver does: the
largest problem anywhere in this paper is mesh $65942$ at base resolution, with
$1\,509\,186$ piecewise-quadratic degrees of freedom. The point is not to beat
exact polyhedral methods or the heat method in raw speed, but to make the
higher-order convex formulation available at all at these resolutions.

\begin{table}[htbp]
    \centering
    \small
    \setlength{\tabcolsep}{5pt}
    \begin{tabular}{@{}lrrrrrr@{}}
        \toprule
        & & & \multicolumn{2}{c}{Time (s)} & \multicolumn{2}{c}{Peak memory (GB)}
           \\
        \cmidrule(lr){4-5} \cmidrule(lr){6-7} 
        Mesh & $|\mathcal{V}|$ & DOF & \textsc{Mosek} & Ours (\ours) & \textsc{Mosek} & Ours (\ours)
             \\
        \midrule
        472088  &      995 &     3\,992 &   1.6 &   0.1 &  0.6 &  0.1 \\
        822698  &   3\,978 &    15\,936 &   6.8 &   0.9 &  1.4 &  0.3 \\
        113855  &  14\,884 &    59\,644 &  34.7 &   3.6 &  4.4 &  0.8 \\
        1706463 &  39\,588 &   158\,352 &  71.0 &  11.9 & 12.0 &  2.1 \\
        351479  &  79\,941 &   319\,806 & 240.4 &  35.8 & 22.0 &  4.1 \\
        99269   & 143\,860 &   575\,434 & 318.3 & 104.4 & 44.5 &  7.3 \\
        \midrule
        372056  & 249\,489 &   997\,950 & \textsc{oom} & 204.5 & \textsc{oom} & 12.6 \\
        \bottomrule
    \end{tabular}
    \caption{Solver scalability, on Thingi10K manifold meshes spanning three
    orders of magnitude in problem size. The interior-point solver
    (\textsc{Mosek}) and our ADMM solver run on the \emph{identical} convex
    program, including the same pinned source neighbourhood, both on the CPU.
    Above the rule both solvers complete, and below it the interior point exceeds
    the $62$\,GB memory cap and returns nothing, whereas ours runs every instance. }
    \label{tab:scalability}
\end{table}

\section{Demonstrated Scope}
\label{sec:capability}
\newcommand{\omark}{$\circ$}                 %
\newcommand{\na}{---}                        %
\newcommand{\adapt}{\textsuperscript{a}}     %
\newcommand{\indirect}{\textsuperscript{i}}  %

\begin{table*}[t]
	\centering
	\small
	\setlength{\tabcolsep}{4pt}
	\renewcommand{\arraystretch}{1.15}
	\begin{tabular*}{\textwidth}{@{\extracolsep{\fill}}l ccc@{}}
		\toprule
		Method
		& \shortstack{Non-\\manifold}
		& \shortstack{Tet.\\volumes}
		& \shortstack{Anisotropic\\metric} \\
		\midrule
		MMP / \FEG / \VTP~\cite{mitchell1987discrete,Surazhsky_2005,Qin2016}
		& \xmark          & \xmark        & \xmark        \\
		\DGG~\cite{Wang:2017}               & \xmark          & \xmark        & \xmark        \\
		FastDGG~\cite{Adikusuma:2020}       & \xmark          & \xmark        & \cmark\adapt  \\
		\DGGVTP~\cite{Adikusuma:2022}       & \xmark          & \xmark        & \xmark        \\
		\GSP~\cite{trettner2021geodesic}    & \omark          & \omark        & \cmark\adapt  \\
		Geodesic tracks~\cite{meng2021geodesic}
		& \cmark          & \xmark        & \cmark        \\
		Convex QP~\cite{Chen:2024}          & \cmark          & \xmark        & \xmark        \\
		Defect-tolerant~\cite{campen2011}   & \cmark\indirect & \na           & \na           \\
		STVD~\cite{campen2013}              & \na             & \cmark\adapt  & \cmark        \\
		\FMM~\cite{kimmel1998computing}     & \na             & \na           & \cmark\adapt  \\
		\FIM~\cite{fim_fu_2013}             & \na             & \cmark        & \cmark        \\
		\HM~\cite{geodesics_in_heat}        & \cmark\adapt    & \cmark        & \cmark\adapt  \\
		Generalized SDF~\cite{Feng:2024:SHM}
		& \cmark          & \cmark           & \na           \\
		\DFA~\cite{Belyaev-Fayolle}         & \na             & \cmark        & \na           \\
		\midrule
		\textbf{\ours}                      & \cmark          & \cmark        & \cmark        \\
		\bottomrule
	\end{tabular*}
	
	\vspace{2pt}
	\begin{minipage}{\linewidth}\footnotesize\raggedright
		\cmark~demonstrated \quad
		\omark~claimed or formulated, no result shown \quad
		\xmark~stated or shown not to work \quad
		\na~not addressed by the paper \\[2pt]
		\textsuperscript{a}~Reached only by adapting the method for that setting. The
		adaptations for \FMM and \HM are due to~\citet{campen2013} 
		and~\citet{Sharp:2020:LNT}, the others are in the paper cited on that row. \\
	\textsuperscript{i}~Indirectly: distance is computed on a simplicial mesh or
	regular grid that approximates the original domain~\cite{geodesics_in_heat}.
	\end{minipage}
	
		\caption{What each method has been \emph{demonstrated} to handle. The marker
		\textsuperscript{a} records where a method had to be modified for that
		setting. Verdicts are taken from the cited papers only, and record what was
		shown rather than what is possible.}
	\label{tab:capability}
\end{table*}

\clearpage

\end{document}